\documentclass[11pt, twoside, english, listtotoc, tablecaptionabove, bibtotoc, parskip=full, headings=small, numbers=noenddot]{report}

\usepackage[T1]{fontenc}
\usepackage[latin9]{inputenc}
\usepackage{import}
\usepackage[a4paper]{geometry}
\usepackage{fancyhdr}
\usepackage{float}
\usepackage{dsfont}
\usepackage{amsmath}
\usepackage{amssymb}
\usepackage{graphicx}
\usepackage{axodraw4j}
\usepackage{pstricks}
\usepackage{cite}
\usepackage[linktocpage]{hyperref}
\hypersetup{
  pdfpagelayout=TwoColumnRight, 
  pdfpagemode=UseNone, 
  bookmarksopen, 
  bookmarksopenlevel=2, 
  pdftitle={Perturbation theory of non-perturbative Yang-Mills theory: a massive expansion from first principles},
  pdfauthor={Giorgio Comitini}
  }

\makeatletter
\usepackage{fancyhdr}
\renewcommand{\footnoterule}{\kern+5\p@ \hrule \@width 5.9in \kern2.6\p@}

\@ifundefined{showcaptionsetup}{}{%
 \PassOptionsToPackage{caption=false}{subfig}}
\usepackage{subcaption}
\makeatother

\usepackage{babel}

\usepackage{braket}
\usepackage{mathrsfs}

\newcommand{\cbar}{\overline{c}}
\newcommand{\psibar}{\overline{\psi}}
\newcommand{\avg}[1]{\left\langle#1\right\rangle}
\newcommand{\intr}{\text{int}}
\newcommand{\BE}{\begin{equation}}
\newcommand{\EE}{\end{equation}}
\newcommand{\mc}[1]{\mathcal{#1}}
\newcommand{\mf}[1]{\mathfrak{#1}}
\newcommand{\tx}[1]{\text{#1}}
\newcounter{count}
\numberwithin{equation}{count}

\begin{document}
\begin{titlepage}
    \begin{center}
        \vspace*{1cm}
 
        \Huge
        \textbf{Perturbation Theory of Non-Perturbative Yang-Mills Theory}
 
        \vspace{0.5cm}
        \LARGE
        A massive expansion from first principles
 
        \vspace{1.5cm}
 
        \textbf{Giorgio Comitini}
        
        \vfill
        
        \includegraphics[width=0.32\textwidth]{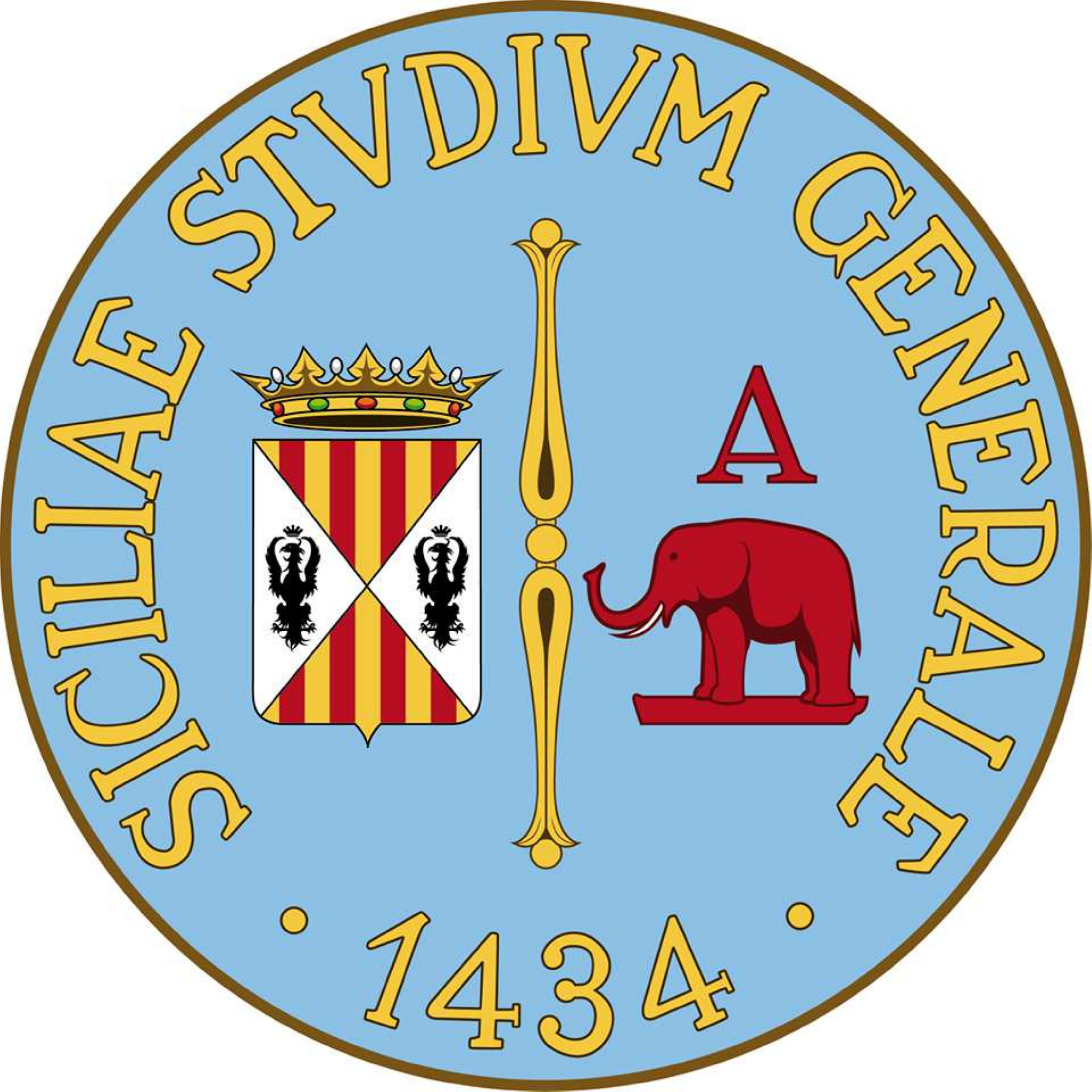}
 
        \vfill
 
        A thesis presented for the Master's Degree in Physics
 
        \vspace{2cm}
 
        \Large
        Dipartimento di Fisica e Astronomia ``E. Majorana''\\
        Universit\`{a} degli Studi di Catania\\
        Italy\\
        October 22, 2019
 
    \end{center}
\end{titlepage}
\newpage
\thispagestyle{empty}
\
\newpage

\thispagestyle{empty}
\
\\
\\
\\
\\
\\
\\
\\
\begin{flushright}\textit{
Pauli asked, "What is the mass of this field B?" I said we did not know.\\
Then I resumed my presentation but soon Pauli asked the same question again.\\
I said something to the effect that it was a very complicated problem, we had worked\\
on it and had come to no definite conclusions. I still remember his repartee:\\
"That is not sufficient excuse".}\\
\
\\
C.N. Yang, Princeton (1954)
\end{flushright}

\clearpage
\thispagestyle{empty}
\
\clearpage
\pagenumbering{Roman}

\chapter*{Preface\index{Preface}}

The main objective of this thesis is to present a new analytical framework for low-energy QCD that goes under the name of massive perturbative expansion. The massive perturbative expansion is motivated by the phenomenon of dynamical mass generation, by which the gluons acquire a mass of the order of the QCD scale $\Lambda_{\tx{QCD}}$ in the limit of vanishing momentum. It is a simple extension of ordinary perturbation theory that consists in a shift of the expansion point of the Yang-Mills perturbative series with the aim of treating the transverse gluons as massive already at tree-level, while leaving the total action of the theory unchanged. The new framework will be formulated in the context of pure Yang-Mills theory, where the lattice data is readily available for comparison and the perturbative results have been perfected by enforcing the gauge invariance of the analytical structure of the gluon propagator.\\
\\
This thesis is organized as follows. In the Introduction we review the definition and main computational approaches to QCD and pure Yang-Mills theory, namely, ordinary perturbation theory and the discretization on the lattice. In Chapter 1 we address the issue of dynamical mass generation from a variational perspective by employing a tool known as the Gaussian Effective Potential (GEP). Through a GEP analysis of Yang-Mills theory we will show that the massless perturbative vacuum of the gluons is unstable towards a massive vacuum, implying that a non-standard perturbative expansion that treats the gluons as massive already at tree-level could be more suitable for making calculations in Yang-Mills theory and QCD than ordinary, massless perturbation theory. In Chapter 2 we set up the massive perturbative framework and use it to compute the gluon and ghost dressed propagators in an arbitrary covariant gauge to one loop. The propagators will be shown to be in excellent agreement with the lattice data in the Landau gauge, despite being explicitly dependent on a spurious free parameter which needs to be fixed in order to preserve the predictive power of the method. In Chapter 3 we fix the value of the spurious parameter by enforcing the gauge invariance of the analytic structure of the gluon propagator, as required by the Nielsen identities. The optimization procedure presented in this chapter will leave us with gauge-dependent propagators which are in good agreement with the available lattice data both in the Landau gauge and outside of the Landau gauge.\\
The contents of Chapter 1-3 are original and were presented for the first time in \cite{com1,com2,sir2,com3,sir3,sir4}.
\clearpage
\
\thispagestyle{empty}


\tableofcontents{}

\newpage
\newpage

\pagenumbering{arabic}

\addcontentsline{toc}{chapter}{Introduction}  \markboth{Introduction}{Introduction}

\chapter*{Introduction\index{Introduction}}

\addcontentsline{toc}{section}{Quantum Chromodynamics and pure Yang-Mills theory}  \markboth{Quantum Chromodynamics and pure Yang-Mills theory}{Quantum Chromodynamics and pure Yang-Mills theory}
\section*{Quantum Chromodynamics and pure Yang-Mills theory\index{Quantum Chromodynamics and pure Yang-Mills theory}}

\noindent Quantum Chromodynamics (QCD) is the quantum theory of the strong interactions between the elementary constituents of the hadrons, the quarks and the gluons. It was formulated in the early 1970s by H. Fritzsch, M. Gell-Mann and H. Leutwyler \cite{fritz1,fritz2,fritz3} as an extension of the gauge theory of C.N. Yang and R. Mills \cite{yang} to the SU(3) color group with the goal of explaining why the gluons could not be observed as free particles.\\
The concept of the hadrons being composite particles dates back to as early as the 1950s, when the discovery of an ever-increasing number of particles subject to the nuclear interactions called for the need of an organizing principle to classify the observed spectrum of mesons and baryons. The first such principle -- termed the Eightfold Way -- was put forth by Gell-Mann \cite{gell1} and independently by Y. Ne'eman \cite{neem} in 1961, and was later developed by Gell-Mann himself \cite{gell2} and G. Zweig \cite{zweig1,zweig2} into what will come to be known as the quark model (1964). The quark model postulated that all the known hadrons could be considered as being made up of three kinds of spin-$1/2$ particles -- the $u$ quark, the $d$ quark and the $s$ quark -- bound by a yet unidentified interaction of nuclear type. The mesons would be bound states of a quark and an antiquark pair, whereas the baryons would be bound states of a triplet of quarks or antiquarks. The quark model succeeded in explaining the pattern of the hadron masses by organizing the mesons and baryons into multiplets of the flavor SU(3) group. However, it was soon realized that the existence of baryons such as the $\Delta^{++}$ or the $\Omega^{-}$ -- which in the quark model would be made up respectively of three $u$ quarks and three $s$ quarks in the same quantum state -- would violate the Pauli exclusion principle. This issue prompted O.W. Greenberg \cite{green} and M.Y. Han and Y. Nambu \cite{han} to postulate the existence of a new quantum number for the quarks, termed color charge. Each of the quarks would come in three varieties, known as colors; the mesons would be made up of quarks of opposite color, whereas the baryons would be made up of quarks of three different colors. In both cases, the quarks would be no longer in the same state -- so that the Pauli principle would not be violated -- and the resulting hadron would be color-neutral. The strong interactions were postulated to be symmetrical with respect to the continuous transformation of one color into the other, a feature that would formally imply that the laws of physics be globally invariant under the action of a SU(3) color group.\\
In the early stages of its formulation, the quark picture was though to be more of a mathematical device for organizing the spectrum of the observed hadrons, rather than a truly physical model for the internal structure of the mesons and baryons. Indeed, the existence of the quarks was challenged by the fact that such elementary components had never been observed as free particles. In 1969 R.P. Feynman had argued \cite{feyn} that the experimental data on hadron collisions was consistent with the picture of the hadrons being made up of more elementary point-like components, which he named partons, initially refraining from identifying them with Gell-Mann's quarks. However, the crucial breakthrough came only later on in the same year, when experiments on the deep-inelastic scattering of electrons from protons performed at the Stanford Linear Accelerator Center (SLAC) \cite{bloom,breid} revealed that the electron differential cross section exhibited a scaling behavior which had already been studied by J.D. Bjorken \cite{bjork1}. Bjorken himself, together with E.A. Paschos \cite{bjork2}, employed Feynman's parton picture to show that the electrons' behavior could be explained by assuming that during each inelastic collision the electron interacted electromagnetically with just one of the partons contained in the proton.\\
As the evidence for the compositeness of the hadrons accumulated, it remained to be explained why the quarks had never been observed individually. To this end, it was postulated that the particles subject to the strong interactions -- be they elementary or composite -- could only exist as free particles in color-neutral states. The quarks, being colored, would be among the particles that could not be observed if not in combination with one another, forming hadrons. This feature of the strong interactions came to be known as confinement. The precise mechanism by which the interactions between the quarks resulted in their confinement was (and is still to date) largely unknown.\\
As early as the mid 1960s it had been suggested that, in analogy with Quantum Electrodynamics, the interactions between the quarks could be mediated by the exchange of vector bosons, named gluons. In order to explain why such bosons were not observed in the experiments, in 1973 Fritzsch, Gell-Mann and Leutwyler \cite{fritz3} proposed that, just like the quarks, the gluons too might carry a color charge, so that they would only be observable in combination with other colored particles. The gluons themselves would be responsible for the exchange of the quarks' color charge, implying that from a mathematical point of view they would form an octet transforming under the adjoint representation of the SU(3) color group. The concepts of color as the charge associated to the strong interactions and of gluons as a color octet had already been suggested by Han and Nambu in their article of 1965 \cite{han}; the merit of Fritzsch, Gell-Mann and Leutwyler was in managing to formulate these ideas in terms of a gauge theory of Yang-Mills type, thus giving birth to the theory of strong interactions that will later be known as QCD. In what follows we will give a brief description of the mathematical formalism and fundamental features of both Yang-Mills theory and QCD.\\
\\
Pure Yang-Mills theory \cite{yang} is a quantum field theory of interacting vector bosons subject to a local SU(N) invariance. Its fundamental degrees of freedom are expressed in terms of an $N_{A}$-tuple of vector fields $A_{\mu}^{a}(x)$ ($a=1,\,\dots,\,N_{A}$), where $N_{A}=N^{2}-1$ is the dimension of the Lie group SU(N). At the classical level, it is defined by the Lagrangian\\
\[
\mc{L}_{\tx{YM}}=-\frac{1}{4}\ F^{a}_{\mu\nu}\,F^{a\,\mu\nu}
\]\\
Here $F_{\mu\nu}^{a}$ is the field-strength (or curvature) tensor associated to the fields $A_{\mu}^{a}$,\\
\[
F^{a}_{\mu\nu}=\partial_{\mu}A^{a}_{\nu}-\partial_{\nu}A^{a}_{\mu}+g\,f^{a}_{bc}\,A^{b}_{\mu}\,A^{c}_{\nu}
\]\\
where $g$ is a coupling constant and the structure constants $f^{a}_{bc}$ are defined by the commutation relations of the $\mf{su}$(N) algebra -- i.e. the Lie algebra associated to SU(N) --\\
\[
[T_{a},T_{b}]=i\,f_{ab}^{c}\,T_{c}\qquad\qquad\qquad (f_{ab}^{c}=-f_{ba}^{c})
\]\\
The generators of $\mf{su}$(N) are usually chosen in such a way as to satisfy the trace relation
\clearpage
\[
\text{Tr}\{T_{a}T_{b}\}=\frac{1}{2}\ \delta_{ab}
\]
\\
It can then be shown that the structure constants satisfy the antisymmetry relations\\
\[
f_{abc}=-f_{acb}=f_{cab}\qquad\quad f_{abc}f_{abd}=N\, \delta_{cd}\qquad\qquad\quad (f_{abc}=f_{ab}^{c})
\]\\
By expanding the field-strength tensor in the Yang-Mills Lagrangian, one finds that\\
\[
\mc{L}_{\tx{YM}}=-\frac{1}{4}\ (\partial_{\mu}A^{a}_{\nu}-\partial_{\nu}A^{a}_{\mu})(\partial^{\mu}A^{a\,\nu}-\partial^{\nu}A^{a\,\mu})-\frac{g}{2}\ f^{a}_{bc}\ \partial_{\mu}A_{\nu}^{a}\,A^{b\,\mu}A^{c\,\nu}-\frac{g^{2}}{4}\ f^{a}_{bc}f^{a}_{de}\ A^{b}_{\mu}A^{c}_{\nu}A^{d\,\mu}A^{e\,\nu}
\]\\
The first term in the above equation can be easily recognized as a generalization of the Maxwell Lagrangian to our $N_{A}$-tuple of vector fields: in the limit of vanishing coupling, Yang-Mills theory describes a set of $N_{A}$ massless vector bosons. If $g\neq 0$ the second and third term cause the bosons to interact: at variance with the photons of electrodynamics, the degrees of freedom of Yang-Mills theory do interact with each other. This crucial feature is ultimately responsible for the richness of both pure Yang-Mills theory and its extension to the quarks, Quantum Chromodynamics.\\
The Lagrangian of Yang-Mills theory is invariant with respect to the following local SU(N) transformation, parametrized by arbitrary functions $\chi^{a}(x)$:\\
\[
A_{\mu}(x)\to \widetilde{A_{\mu}}(x)=U(x)\left(A_{\mu}(x)+\frac{i}{g}\,\partial_{\mu}\right)U^{\dagger}(x)\qquad\qquad U(x)=\exp\Big(i\chi^{a}(x)\,T_{a}\Big)
\]\\
Here $A_{\mu}$ is defined as $A_{\mu}=A_{\mu}^{a}\,T_{a}$ and $U^{\dagger}$ is the hermitian conjugate of $U$. The invariance of $\mc{L}_{\tx{YM}}$ can be easily seen to follow from the corresponding transformation law for the field-strength tensor,\\
\[
F_{\mu\nu}(x)\to \widetilde{F_{\mu}}(x)=U(x)\,F_{\mu\nu}(x)\,U^{\dagger}(x)
\]\\
where $F_{\mu\nu}=F_{\mu\nu}^{a}\,T_{a}$. Since $F^{2}=F_{\mu\nu}^{a}F^{a\,\mu\nu}=2\,\text{Tr}\{F_{\mu\nu}F^{\mu\nu}\}$, the invariance of $F^{2}$ is a simple consequence of the cyclic property of the trace. Being invariant under an infinite set of local transformations, Yang-Mills theory is a gauge theory. The implications of this are twofold. First of all, if $A_{\mu}^{a}$ solves the equations of motion derived from the Yang-Mills Lagrangian, namely,\\
\[
\partial^{\mu}F_{\mu\nu}^{a}+g\,f^{a}_{bc}\ A^{b\,\mu}F_{\mu\nu}^{c}=0
\]\\
then, for any choice of the parameter functions $\chi^{a}$, its transform under SU(N) $\widetilde{A^{a}_{\mu}}$ also does. Since the transformation acts locally rather than globally, the pointwise values of the vector fields cannot have a genuine physical meaning: some of the local degrees of freedom of the theory are redundant. When passing to the quantum theory, this redundancy will cause problems with the definition of the quantum partition function, which will have to be dealt with by employing the so called Faddeev-Popov quantization procedure. Second of all, if we insist that gauge invariance be preserved at the level of the classical action, then Yang-Mills theory cannot be generalized to account for a classical mass for the bosons. Such a mass would be incorporated in the theory by adding to its Lagrangian a term of the form
\newpage
\[
\mc{L}_{\tx{m}}=\frac{1}{2}\ m^{2}\, A_{\mu}^{a}\,A^{a\,\mu}
\]\\
where $m$ is the bosons' mass. This term, however, is not invariant under local SU(N) transformations: it can be shown that under a gauge transformation\\
\[
\delta\left(\int d^{d}x\ \mc{L}_{\tx{m}}\right)= \frac{2i}{g}\ m^{2}\int d^{d}x\ \text{Tr}\{UA_{\mu}\partial^{\mu}U^{\dagger}\}\neq 0
\]\\
Therefore, gauge invariance constrains the vector bosons of Yang-Mills theory to be massless at the classical level\footnote{\ Whether this is still true at the quantum level will be discussed later on in this Introduction.}.\\
The quantum dynamics of Yang-Mills theory is defined by the partition function $Z[J]$\\
\[
Z[J]=\int\mc{D}A\ \exp\left\{i\int d^{d}x\ \ \mc{L}_{\tx{YM}}+J^{\mu}_{a}A_{\mu}^{a}\right\}
\]\\
where $J^{\mu}_{a}$ is an external source for the gauge field $A_{\mu}^{a}$. From a mathematical point of view, this functional integral is ill-defined: in integrating over all the possible configurations of the fields $A_{\mu}^{a}$, we are not taking into account that different configurations may actually be equivalent modulo gauge transformations; since the gauge group of $\mc{L}_{\tx{YM}}$ is infinite-dimensional\footnote{\ Although SU(N) is finite dimensional as a Lie group, its local action on $\mc{L}_{\tx{YM}}$ is parametrized by functions $\chi^{a}(x)$, rather than by constant parameters. Therefore, the symmetry group of Yang-Mills theory is actually infinite-dimensional.}, an infinite number of such equivalent configurations exists, resulting in the integrand being constant over an infinite volume of the configuration space. Therefore, in this form, $Z[J]$ is singular for every $J$. In order to solve this issue, one can adopt a quantization procedure due to L.D. Faddeev and V. Popov \cite{fadd}. In the Faddeev-Popov approach, the redundant gauge degrees of freedom of the theory are integrated over in such a way as to insulate the singularity of the partition function into an infinite multiplicative constant $\mc{C}$. Since the quantum behavior of the theory is dictated by the derivatives of the logarithm of $Z$ with respect to the source $J$, such a constant plays no role in the definition of the theory and can thus be ignored. However, as a result of the integration, the integrand of the partition function gets modified as follows:\\
\[
Z[J]=\mc{C}\int\mc{D}A\ \det\big(-\partial^{\mu}D_{\mu}\big)\ \exp\left\{i\int d^{d}x\ \ \mc{L}_{\tx{YM}}-\frac{1}{2\xi}\ (\partial^{\mu}A_{\mu}^{a})^{2}+J^{\mu}_{a}A_{\mu}^{a}\right\}
\]\\
Here $\xi$ is a gauge parameter that can take on any value from zero to infinity and $\det(-\partial^{\mu}D_{\mu})$ -- known as a Faddeev-Popov determinant -- depends on the vector fields $A_{\mu}^{a}$ through the covariant derivative $D_{\mu}$, which acts on the fields in the adjoint representation of SU(N) as\\
\[
D_{\mu}f^{a}=\partial_{\mu}f^{a}+g\,f^{a}_{bc}\,A^{b}_{\mu}\,f^{c}
\]\\
The Faddeev-Popov determinant is usually expressed in terms of an integral over the configurations of a pair of anticommuting fields -- known as ghost fields -- with values in the adjoint representation,\\
\[
\det(-\partial^{\mu}D_{\mu})=\int\mc{D}\overline{c}\,\mc{D}c\ \exp\left\{-i\int d^{d}x\ \cbar^{\,a}\partial^{\mu}D_{\mu}c^{a}\right\}=\int\mc{D}\overline{c}\,\mc{D}c\ \exp\left\{i\int d^{d}x\ \partial^{\mu}\cbar^{\,a}D_{\mu}c^{a}\right\}
\]
\newpage
\noindent The ghost fields $(c^{a},\,\cbar^{\,a})$ are a mathematical tool for keeping under control the gauge redundancy built into Yang-Mills theory, and should not be interpreted as being associated to any physical particle. As a matter of fact, they do not even obey the spin-statistic theorem, in that they are scalar (i.e. spin-0) fields with fermionic statistics.\\
The Faddeev-Popov quantization procedure leaves us with an effective Lagrangian $\mc{L}$ for Yang-Mills theory, in terms of which its partition function can be expressed as\footnote{\ Here we have suppressed the uninfluential constant $\mc{C}$ and added sources $j_{a}$ and $\overline{j}_{a}$ for the ghosts.}\\
\[
Z[J,j,\overline{j}]=\int\mc{D}A\,\mc{D}\cbar\,\mc{D}c\ \exp\left\{i\int d^{d}x\ \mc{L}+J^{\mu}_{a}A^{a}_{\mu}+\overline{j}_{a}c^{a}+\cbar^{\,a}j_{a}\right\}
\]
Explicitly,\\
\[
\mc{L}=\mc{L}_{\tx{YM}}+\mc{L}_{\tx{g.f.}}+\mc{L}_{\tx{ghost}}=-\frac{1}{4}\ F_{\mu\nu}^{a}\,F^{a\,\mu\nu}-\frac{1}{2\xi}\ \partial^{\mu}A_{\mu}^{a}\,\partial^{\nu}A_{\nu}^{a}+\partial^{\mu}\cbar^{\,a}D_{\mu}c^{a}
\]\\
Observe that this Lagrangian contains a term that causes the ghosts to interact with the vector bosons: the ghost Lagrangian can be expanded as\\
\[
\mc{L}_{\tx{ghost}}=\partial^{\mu}\cbar^{\,a}D_{\mu}c^{a}=\partial^{\mu}\cbar^{\,a}\partial_{\mu}c^{a}+g\,f^{a}_{bc}\,\partial^{\mu}\cbar^{\,a}\,c^{c}\, A_{\mu}^{b}
\]\\
where the first term is just the Klein-Gordon Lagrangian for a massless $N_{A}$-tuple of complex scalar bosons, whereas the second one is an interaction term for the ghosts. This was to be expected from the fact that the vector bosons interact with one another: if the ghosts are to cancel the unphysical content of the theory due to the gauge redundancy, then they should be coupled with the vector bosons in order to counterbalance the effects of their mutual interaction\footnote{\ For instance, the ghosts of Quantum Electrodynamics do not interact with the photons, since the photons themselves are not subject to mutual interactions. This is ultimately the reason why the ghosts are not an essential part of QED.}.\\
Moreover, it should be noticed that, as a result of the Faddeev-Popov procedure, $\mc{L}$ is not gauge-invariant anymore: neither the ghost Lagrangian $\mc{L}_{\tx{ghost}}$ nor the gauge-fixing term $\mc{L}_{\tx{g.f.}}=-(\partial^{\mu}A_{\mu}^{a})^{2}/2\xi$ are invariant under SU(N) local transformations. If on the one hand this is a necessary condition in order for the quantum partition function to be well-defined, on the other hand it was shown by I.V. Tyutin \cite{tyutin} and C. Becchi, A. Rouet and R. Stora \cite{becchi} that gauge invariance is not completely lost at the level of the Faddeev-Popov action: it has only been replaced by a kind of global supersymmetry called BRST symmetry. In order to see this, re-write the Faddeev-Popov action as\\
\[
\mc{L}=-\frac{1}{4}\ F_{\mu\nu}^{a}\,F^{a\,\mu\nu}+\frac{\xi}{2}\ B^{a}B^{a}+B^{a}\partial^{\mu}A_{\mu}^{a}+\partial^{\mu}\cbar^{\,a}D_{\mu}c^{a}
\]\\
where $B^{a}$ is an auxiliary field -- known as the Nakanishi-Lautrup field \cite{nak,laut} -- whose equations of motion are
\[
B^{a}=-\frac{1}{\xi}\ \partial^{\mu}A_{\mu}^{a}
\]
Since on shell\\
\[
\frac{\xi}{2}\ B^{a}B^{a}+B^{a}\partial^{\mu}A_{\mu}^{a}=-\frac{1}{2\xi}\ \partial^{\mu}A_{\mu}^{a}\,\partial^{\nu}A_{\nu}^{a}
\]
\newpage
\noindent our two expressions for $\mc{L}$ indeed coincide. However the new Lagrangian is invariant under the following global BRST transformation, parametrized by an anticommuting number $\epsilon$:\\
\begin{align*}
\delta A_{\mu}^{a}&=\epsilon\,D_{\mu}c^{a}\\
\delta c^{a}&=-\frac{1}{2}\ g\,\epsilon\,f^{a}_{bc}\ c^{b}c^{c}\\
\delta\cbar^{\,a}&=\epsilon\,B^{a}\\
\delta B^{a}&=0
\end{align*}\\
Observe that, as far as the vector fields are concerned, this is just the infinitesimal version of a local SU(N) transformation parametrized by the ghost fields themselves ($\chi^{a}=\epsilon\, c^{a}$).\\
BRST symmetry has been a fundamental tool for proving many crucial features of Yang-Mills theory. Among them, we cite the derivation of the non-abelian analogue of the Ward identities -- i.e. the Slavnov-Taylor identites \cite{taylor,slavnov} -- and the proof of the perturbative renormalizability of the theory by G. t' Hooft \cite{thooft1}.\\
\\
For $N=3$ (so that $N_{A}=8$), Yang-Mills theory describes the dynamics of an octet of vector fields that can be readily identified with the gluons of Quantum Chromodynamics. QCD, however, also comprises the quarks. Let us see how the full theory of Quantum Chromodynamics is defined.\\
Quantum Chromodynamics \cite{fritz3} is a gauge theory of Dirac fields in the fundamental representation of SU(3) -- the quark fields -- minimally coupled to an octet of Yang-Mills vector fields -- the gluon fields. Its Lagrangian $\mc{L}_{\tx{QCD}}$ is defined as\footnote{\ For simplicity we will be considering only one flavor of quark. Actual QCD contains six flavors of quark ($u,\,d,\,s,\,c,\,b,\,t$) with non-diagonal mass couplings given by the CKM matrix \cite{cabibbo,koba}.}\\
\[
\mc{L}_{\tx{QCD}}=\mc{L}_{\tx{YM}}+\mc{L}_{\tx{quark}}=-\frac{1}{4}\ F_{\mu\nu}^{a}\,F^{a\,\mu\nu}+i\psibar\gamma^{\mu}D_{\mu}\psi-m\,\psibar \psi
\]\\
In the above equation, $m$ is the quark mass and $D_{\mu}$ is the covariant derivative acting on the fundamental representation,
\[
D_{\mu}=\partial_{\mu}-ig\,A_{\mu}^{a}\,T_{a}
\]\\
where the $T_{a}$'s are the generators of $\mf{su}(3)$. By expanding the covariant derivative in $\mc{L}_{\tx{QCD}}$, one finds that
\[
\mc{L}_{\tx{quark}}=i\psibar\gamma^{\mu}\partial_{\mu}\psi-m\,\psibar \psi+g\,\psibar\gamma^{\mu}T_{a}\psi\,A_{\mu}^{a}
\]\\
The first two terms are just the Dirac Lagrangian for the quark fields. The third term is an interaction between the quarks and the gluon octet, generated by an octet of color currents $j^{\mu}_{a}$ defined as
\[
j^{\mu}_{a}=\psibar\gamma^{\mu}T_{a}\psi
\]\\
The classical equations of motion of QCD can be readily derived from $\mc{L}_{\tx{QCD}}$ and read
\begin{align*}
\partial^{\mu}F_{\mu\nu}^{a}+g\,f_{bc}^{a}\,A^{b\,\mu}\,F_{\mu\nu}^{c}&=-g\,j^{a}_{\nu}\\
(i\gamma^{\mu}D_{\mu}-m)\psi&=0
\end{align*}
\newpage
\noindent The QCD Lagrangian is invariant with respect to the following local SU(3) transformations, parametrized by arbitrary functions $\chi^{a}(x)$:\\
\begin{align*}
\psi(x)&\to\widetilde{\psi}(x)=U(x)\,\psi(x)\\
A_{\mu}(x)&\to \widetilde{A_{\mu}}(x)= U(x)\left(A_{\mu}(x)+\frac{i}{g}\, \partial_{\mu}\right)U^{\dagger}(x)
\end{align*}\\
where as usual $U(x)=\exp\big(i\chi^{a}(x)\,T_{a}\big)$. It follows that Quantum Chromodynamics is a gauge theory with gauge group SU(3).\\
At the quantum level, QCD is defined by the partition function\\
\[
Z[J_{A},J_{\psi},\overline{J}_{\psi}]=\int\mc{D}A\,\mc{D}\psibar\,\mc{D}\psi\ \exp\left\{i\int d^{d}x\ \ \mc{L}_{\tx{QCD}}+J^{\mu}_{A\,a}A_{\mu}^{a}+\overline{J}_{\psi}\psi+\psibar J_{\psi}\right\}
\]\\
Since the boson sector of QCD is that of pure Yang-Mills theory (modulo the interaction with the quarks), everything we previously said for Yang-Mills theory still applies to QCD. In particular, due to the gauge redundancy, the above partition function is ill-defined. By applying the Faddeev-Popov procedure to $Z$, we obtain the following gauge-dependent partition function for QCD:\\
\[
Z[\{J\}]=\int\mc{D}A\,\mc{D}\psibar\,\mc{D}\psi\,\mc{D}\cbar\,\mc{D}c\ \exp\left\{i\int d^{d}x\ \ \mc{L}+J^{\mu}_{A\,a}A_{\mu}^{a}+\overline{J}_{\psi}\psi+\psibar J_{\psi}+\overline{J}_{c}\,c+\cbar\, J_{c}\right\}
\]
where
\begin{align*}
\mc{L}&=\mc{L}_{\tx{YM}}+\mc{L}_{\tx{g.f.}}+\mc{L}_{\tx{quark}}+\mc{L}_{\tx{ghost}}=\\
&=-\frac{1}{4}\ F_{\mu\nu}^{a}\,F^{a\,\mu\nu}-\frac{1}{2\xi}\ \partial^{\mu}A_{\mu}^{a}\,\partial^{\nu}A_{\nu}^{a}+i\psibar\gamma^{\mu}D_{\mu}\psi-m\,\psibar \psi+\partial^{\mu}\cbar^{\,a}D_{\mu}c^{a}
\end{align*}\\
When re-written in terms of the Nakanishi-Lautrup field $B^{a}$, this Lagrangian can be shown to be invariant with respect to the global BRST transformation\\
\begin{align*}
\delta A_{\mu}^{a}&=\epsilon\,D_{\mu}c^{a}\\
\delta\psi&=ig\,\epsilon\, c^{a}T_{a}\,\psi\\
\delta c^{a}&=-\frac{1}{2}\ g\,\epsilon\,f^{a}_{bc}\ c^{b}c^{c}\\
\delta\cbar^{\,a}&=\epsilon\,B^{a}\\
\delta B^{a}&=0
\end{align*}\\
parametrized by an anticommuting number $\epsilon$. By exploiting the BRST symmetry of $\mc{L}$ one is able to derive the appropriate Slavnov-Taylor identities for QCD and prove its perturbative renormalizability.\\
Although the primary concern of Quantum Chromodynamics is explaining the dynamics and interactions between the quarks, pure Yang-Mills theory is still able to account for many of the essential features of the strong interactions by attributing them to the behavior of the gluons alone. For this reason, over the last fifty years Yang-Mills theory has been a very active field of research. In what follows we will discuss three different approaches to Yang-Mills theory: standard perturbation theory, lattice gauge theory and massive perturbation theory.
\clearpage

\addcontentsline{toc}{section}{Standard perturbation theory for Yang-Mills theory: the non-perturbative nature of the IR regime}  \markboth{Standard perturbation theory for Yang-Mills theory: the non-perturbative nature of the IR regime}{Standard perturbation theory for Yang-Mills theory: the non-perturbative nature of the IR regime}
\section*{Standard perturbation theory for Yang-Mills theory: the non-perturbative nature of the IR regime\index{Standard perturbation theory for Yang-Mills theory: the non-perturbative nature of the IR regime}}

Since its inception in the 1950s, the primary tool for making calculations in Yang-Mills theory has been perturbation theory. In the (modern) standard perturbative approach, the Faddeev-Popov gauge-fixed action $\mc{S}=\int d^{d}x\ \mc{L}$ is split into two terms,\\
\[
\mc{S}=\mc{S}_{0}+\mc{S}_{\tx{int}}
\]\\
where $\mc{S}_{0}$ is defined as the zero-coupling limit of $\mc{S}$,\\
\[
\mc{S}_{0}=\lim_{g\to 0}\ \mc{S}
\]\\
and $\mc{S}_{\tx{int}}=\mc{S}-\mc{S}_{0}$. In terms of $\mc{S}_{0}$ and $\mc{S}_{\tx{int}}$, the Faddeev-Popov partition function can be expressed as\footnote{\ For simplicity we set the gluon and ghost sources to zero.}
\[
Z=\int\mc{D}A\,\mc{D}\cbar\,\mc{D}c\ e^{i\mc{S}_{0}+i\mc{S}_{\tx{int}}}
\]\\
$Z$ is then expanded in powers of the interaction action $\mc{S}_{\tx{int}}$ to yield\\
\[
Z=\int\mc{D}A\,\mc{D}\cbar\,\mc{D}c\ e^{i\mc{S}_{0}}\ \sum_{n=0}^{+\infty}\ \frac{i^{n}}{n!}\ \mc{S}_{\tx{int}}^{n}=\sum_{n=0}^{+\infty}\ \frac{i^{n}}{n!}\ \int\mc{D}A\,\mc{D}\cbar\,\mc{D}c\ e^{i\mc{S}_{0}}\ \mc{S}_{\tx{int}}^{n}
\]\\
If an average operation $\avg{\cdot}_{0}$ is defined with respect to the zero-order action $\mc{S}_{0}$ as\\
\[
\avg{\mathcal{O}}=\frac{\int\mc{D}A\,\mc{D}\cbar\,\mc{D}c\ e^{i\mc{S}_{0}}\ \mc{O}}{\int\mc{D}A\,\mc{D}\cbar\,\mc{D}c\ e^{i\mc{S}_{0}}}
\]\\
where $\mc{O}$ is an arbitrary functional of the fields $A$, $c$ and $\cbar$, then the partition function can be further re-written as\\
\[
Z=\left(\int\mc{D}A\,\mc{D}\cbar\,\mc{D}c\ e^{i\mc{S}_{0}}\right)\ \left(\sum_{n=0}^{+\infty}\ \frac{i^{n}}{n!}\, \avg{\mc{S}_{\tx{int}}^{n}}_{0}\right)
\]\\
The averages $\avg{\mc{S}_{\tx{int}}^{n}}_{0}$ are usually computed in terms of Feynman diagrams, amongst which the connected diagrams play a fundamental role: it can be shown that\\
\[
\left(\sum_{n=0}^{+\infty}\, \frac{i^{n}}{n!}\, \avg{\mc{S}_{\tx{int}}^{n}}_{0}\right)=\exp\left(\sum_{n=1}^{+\infty}\ \frac{i^{n}}{n!}\, \avg{\mc{S}_{\tx{int}}^{n}}_{0,\tx{conn}}\right)
\]\\
where $\avg{\mc{S}_{\tx{int}}^{n}}_{0,\tx{conn}}$ is the restriction of $\avg{\mc{S}_{\tx{int}}^{n}}_{0}$ to its connected diagrams. In terms of the connected diagrams, the logarithm of the partition function reads\\
\[
\ln Z=\ln Z_{0}+\sum_{n=1}^{+\infty}\ \frac{i^{n}}{n!}\, \avg{\mc{S}_{\tx{int}}^{n}}_{0,\tx{conn}}
\]\\
where $Z_{0}=\lim_{g\to 0}\ Z$, yielding an expansion of $\ln Z$ in powers of the interaction action, i.e. -- since $\mc{S}_{\tx{int}}$ is proportional to the coupling constant -- in powers of $g$.
\clearpage
\noindent The zero-order action $\mc{S}_{0}$ describes the dynamics of an $N_{A}$-tuple of non-interacting massless vector bosons, together with two $N_{A}$-tuples of non-interacting, anticommuting scalar bosons with the wrong statistics. In momentum space, it reads\\
\[
\mc{S}_{0}=\int\frac{d^{d}p}{(2\pi)^{d}}\ \bigg\{-\frac{1}{2}\ A_{\mu}^{a}\ \delta_{ab}\ p^{2}\, \left[t^{\mu\nu}(p)+\frac{1}{\xi}\ \ell^{\mu\nu}(p)\right]\, A_{\nu}^{b}+\cbar^{\,a}\ \delta_{ab}\,p^{2}\ c^{b}\bigg\}
\]\\
where $t^{\mu\nu}(p)$ and $\ell^{\mu\nu}(p)$ are transverse and longitudinal projection tensors,\\
\BE
t^{\mu\nu}(p)=\eta^{\mu\nu}-\frac{p^{\mu}p^{\nu}}{p^{2}}\qquad\qquad \ell^{\mu\nu}(p)=\frac{p^{\mu}p^{\nu}}{p^{2}}
\EE
\\
The bare propagators $\mc{D}^{ab}_{0\,\mu\nu}(p)$ and $\mc{G}^{ab}_{0}(p)$ associated respectively to the vector and scalar bosons are defined by\\
\[
\mc{S}_{0}=i\ \int\frac{d^{d}p}{(2\pi)^{d}}\ \left\{\frac{1}{2}\ A_{\mu}^{a}\ \mc{D}_{0\,ab}^{\mu\nu}(p)^{-1}\, A_{\nu}^{b}+\cbar^{\,a}\ \mc{G}_{0\,ab}(p)^{-1}\, c^{b}\right\}
\]\\
Therefore\\
\[
\mc{D}_{0\,\mu\nu}^{ab}(p)=\delta^{ab}\ \bigg\{\frac{-i\, t_{\mu\nu}(p)}{p^{2}+i\epsilon}+\xi\ \frac{-i\, \ell_{\mu\nu}(p)}{p^{2}+i\epsilon}\bigg\}\qquad\qquad \mc{G}^{ab}_{0}(p)=\delta^{ab}\ \frac{i}{p^{2}+i\epsilon}
\]\\
where the $+i\epsilon$ term ($\epsilon>0$) is introduced in the denominators in order to select the correct integration contours for the loop integrals of the Feynman diagrams.\\
The interaction action $\mc{S}_{\tx{int}}$, on the other hand, can be expanded to yield\\
\[
\mc{S}_{\tx{int}}=\int d^{d}x\ \left\{-\frac{g}{2}\ f^{a}_{bc}\ \partial_{\mu}A_{\nu}^{a}\,A^{b\,\mu}A^{c\,\mu}-\frac{g^{2}}{4}\ f^{a}_{bc}f^{a}_{de}\ A_{\mu}^{b}\,A_{\nu}^{c}\,A^{d\,\mu}A^{e\,\nu}+g\,f^{a}_{bc}\ \partial^{\mu}\cbar^{\,a}A_{\mu}^{b}c^{c}\right\}
\]\\
Each of the three terms in $\mc{S}_{\tx{int}}$ gives rise to an interaction vertex involving the gluons and ghosts. The first one corresponds to a 3-gluon vertex, the second one corresponds to a 4-gluon interaction vertex, and the third one corresponds to a ghost-ghost-gluon vertex.\\
\\
The power series that defines $\ln Z$ perturbatively is plagued with infinities that arise from the divergent loop integrals in the diagrammatic expansion. In the context of Yang-Mills theory and QCD, these divergences are usually cured by a combination of dimensional regularization \cite{thooft2} and renormalization group (RG) methods \cite{callan,syman1,syman2}. In order to absorb the infinities into finite, renormalized parameters, one is forced to define a scale-dependent running coupling constant $g(\mu)$ whose behavior is determined by the equation\\
\[
\mu\,\frac{dg}{d\mu}(\mu)=\beta(g(\mu))
\]\\
The function $\beta(g)$ -- called the beta function -- can be computed perturbatively to any desired loop order. To one-loop order in standard perturbation theory it reads\footnote{\ The same result holds in full QCD with $\beta_{0}=11-2 n_{f}/3$, $n_{f}$ being the number of quark flavors.} \cite{peskin}\\
\[
\beta(g)=-\beta_{0}\,\frac{g^{3}}{16\pi^{2}}\qquad\qquad\qquad \beta_{0}=\frac{11}{3}\,N
\]
By defining the strong interaction analogue $\alpha_{s}$ of the electromagnetic fine structure constant $\alpha$ as
\[
\alpha_{s}=\frac{g^{2}}{4\pi}
\]\\
the one-loop solution to the equation of the running coupling constant can then be put in the form
\[
\alpha_{s}(\mu)=\frac{\alpha_{s}(\mu_{0})}{1+\frac{\beta_{0}\alpha_{s}(\mu_{0})}{4\pi}\,\ln(\mu^{2}/\mu^{2}_{0})}
\]\\
where $\alpha_{s}(\mu_{0})$ is the value of the running coupling $\alpha_{s}$ at some fixed renormalization scale $\mu_{0}$. $\alpha_{s}(\mu)$ is to replace the ordinary coupling constant in RG-improved expressions that describe processes occurring at energy scales of order $\mu$.\\
An alternative expression for $\alpha_{s}(\mu)$ is obtained by defining an energy scale $\Lambda_{\tx{YM}}$ as\\
\[
\Lambda_{\tx{YM}}=\mu_{0}\,e^{-2\pi/\beta_{0}\alpha_{s}(\mu_{0})}
\]
so that
\[
\alpha_{s}(\mu)=\frac{2\pi}{\beta_{0}\ln(\mu/\Lambda_{\tx{YM}})}
\]\\
This expression is interesting in two respects. First of all, observe that -- since $\ln(\mu/\Lambda_{\tx{YM}})\to +\infty$ as $\mu\to \infty$ -- in the high energy limit the running coupling constant goes to zero\footnote{\ The same behavior is shown by the coupling constant of full QCD -- unless the number of quark flavors is greater than 16.}. This result, known as asymptotic freedom, was discovered in 1973 by D.J. Gross and F. Wilczek \cite{gross} and by H.D. Politzer \cite{politz} and is able to explain why, for instance, in deep-inelastic scattering experiments at high momentum transfer the quarks and gluons contained in the hadrons can be approximately treated as free particles.\\
Second of all -- since $\ln(\mu/\Lambda_{\tx{YM}})\to 0$ as $\mu\to\Lambda_{\tx{YM}}$ -- at $\mu=\Lambda_{\tx{YM}}$ the running coupling becomes infinite\footnote{\ It can be shown that this behavior is not modified by higher order corrections to the beta function -- see for instance ref. \cite{deur}, where results for $\beta(g)$ are reported to order $g^{9}$.}. In the literature, the scale at which a coupling constant diverges is known as a Landau pole \cite{land1,land2,land3}. Because of the Landau pole, at energy scales $\mu>\Lambda_{\tx{YM}}$ of the same order of $\Lambda_{\tx{YM}}$ the coupling constant becomes so large that the ordinary perturbative approach loses its validity\footnote{\ It should be noted that since in the perturbative approach the beta function is itself computed perturbatively, the Landau poles of Yang-Mills theory and QCD may well be artifacts of ordinary perturbation theory. What the existence of a Landau pole actually tells us is that ordinary perturbation theory becomes inconsistent at energy scales of order $\Lambda_{\tx{YM}}$.}. At scales $\mu<\Lambda_{\tx{YM}}$ the coupling $\alpha_{s}(\mu)$ becomes negative -- i.e. $g(\mu)$ becomes imaginary -- and ordinary perturbation theory is manifestly ill-defined.\\
In full QCD the Landau pole $\Lambda_{\tx{QCD}}$ -- also known as the QCD scale -- is located at around $300$-$400$ MeV. Since this is a quite small scale compared to the energies involved in modern particle physics experiments\footnote{\ As long as they involve processes in which the momentum transfer is not too low.}, the ordinary approach to perturbative QCD has proved successful in explaining much of the experimental data gathered in the last fifty years at the high-energy colliders. This success was crucial for establishing that Quantum Chromodynamics is indeed the correct theory of the strong interactions. Critical nuclear phenomena such as the binding of the quarks inside the hadrons, or the onset of residual nuclear forces between the nucleons, however, occur at energies that are comparable to the QCD scale. With respect to the description of these phenomena, ordinary perturbation theory is utterly ineffective. In order to be able to make predictions about the low-energy behavior of the quarks and the gluons, one has to resort to non-perturbative computational methods such as lattice gauge theory (to be reviewed in the next section) or the numerical resolution of an infinite set of Schwinger-Dyson equations (SDEs) \cite{dyson,schwing}. Albeit successful in their own respect, these methods have the shortcoming of being non-analytical, thus providing numerical results with no control over the intermediate steps of the calculations.\\
\\
To conclude this brief review of ordinary perturbation theory, we wish to address the topic that will be the main subject of this thesis, namely, the issue of the mass of the gluons. As we saw in the previous section, by forbidding the inclusion of a mass term for the gluon fields in the Yang-Mills Lagrangian, gauge invariance constrains the gluons to be massless at the classical level. However, at the quantum level, the interactions could still be responsible for the generation of a dynamical gluon mass \cite{corn}. This mass would manifest itself in the finiteness of the transverse component of the dressed gluon propagator evaluated at zero momentum.\\
In ordinary perturbation theory, the dressed gluon propagator $\widetilde{\mc{D}}_{0\,\mu\nu}^{ab}$ can be expressed as \cite{peskin}\\
\[
\widetilde{\mc{D}}_{0\,\mu\nu}^{ab}(p)=\delta^{ab}\ \bigg\{\frac{-i\, t_{\mu\nu}(p)}{p^{2}-\Pi(p)}+\xi\ \frac{-i\, \ell_{\mu\nu}(p)}{p^{2}}\bigg\}
\]\\
where $\Pi(p)$ is the one-particle-irreducible gluon polarization. In the limit of vanishing momentum, its transverse component $\widetilde{\mc{D}}_{0,T}$ reads\\
\[
\widetilde{\mc{D}}_{0,T}(0)=\frac{-i}{-\Pi(0)}
\]\\
If $\Pi(0)=0$, as the momentum goes to zero the transverse dressed propagator grows to infinity. This behavior is typical of massless propagators and is displayed by the bare gluon propagator $\mc{D}_{0\,\mu\nu}^{ab}$ itself. On the other hand, if $\Pi(0)\neq 0$, the transverse dressed propagator remains finite at zero momentum. This is the limiting behavior that characterizes the massive propagators, as exemplified by the ordinary free propagator of a massive particle,
\[
\Delta(p)=\frac{i}{p^{2}-M^{2}}
\]\\
Therefore whether a mass is generated or not for the gluons depends on the zero momentum limit of the gluon polarization.\\
Now, it can be shown that to any finite loop order the gluon polarization of ordinary perturbation theory vanishes at zero momentum. In the context of pure Yang-Mills theory or full QCD with massless quarks this is clearly the case, since ordinary perturbation theory has no intrinsic mass scales and by dimensional analysis $\Pi(p)$ must be proportional to $p^{2}$~\footnote{\ In good regularization schemes the renormalization scale is contained in logarithmic corrections to the propagator that cannot modify this behavior.}. In full QCD with massive quarks more elaborate arguments based on gauge invariance (or rather BRST invariance) and the structure of the quark-gluon interaction are required to prove this claim \cite{itzyk}. In any case, ordinary perturbation theory is unable to describe the phenomenon of mass generation: the gluon is constrained to remain massless to any finite perturbative order. Of course, it could be argued that since mass generation is a low energy phenomenon, no conclusive evidence for its occurrence (or lack thereof) can be gathered through ordinary perturbation theory. In the next section we will see what a non-perturbative approach like lattice gauge theory has to say with respect to this issue.
\clearpage

\addcontentsline{toc}{section}{Yang-Mills theory on the lattice: dynamical mass generation}  \markboth{Yang-Mills theory on the lattice: dynamical mass generation}{Yang-Mills theory on the lattice: dynamical mass generation}
\section*{Yang-Mills theory on the lattice: dynamical mass generation\index{Yang-Mills theory on the lattice: dynamical mass generation}}

\begin{figure}[H]
\centering
\includegraphics[width=0.62\textwidth]{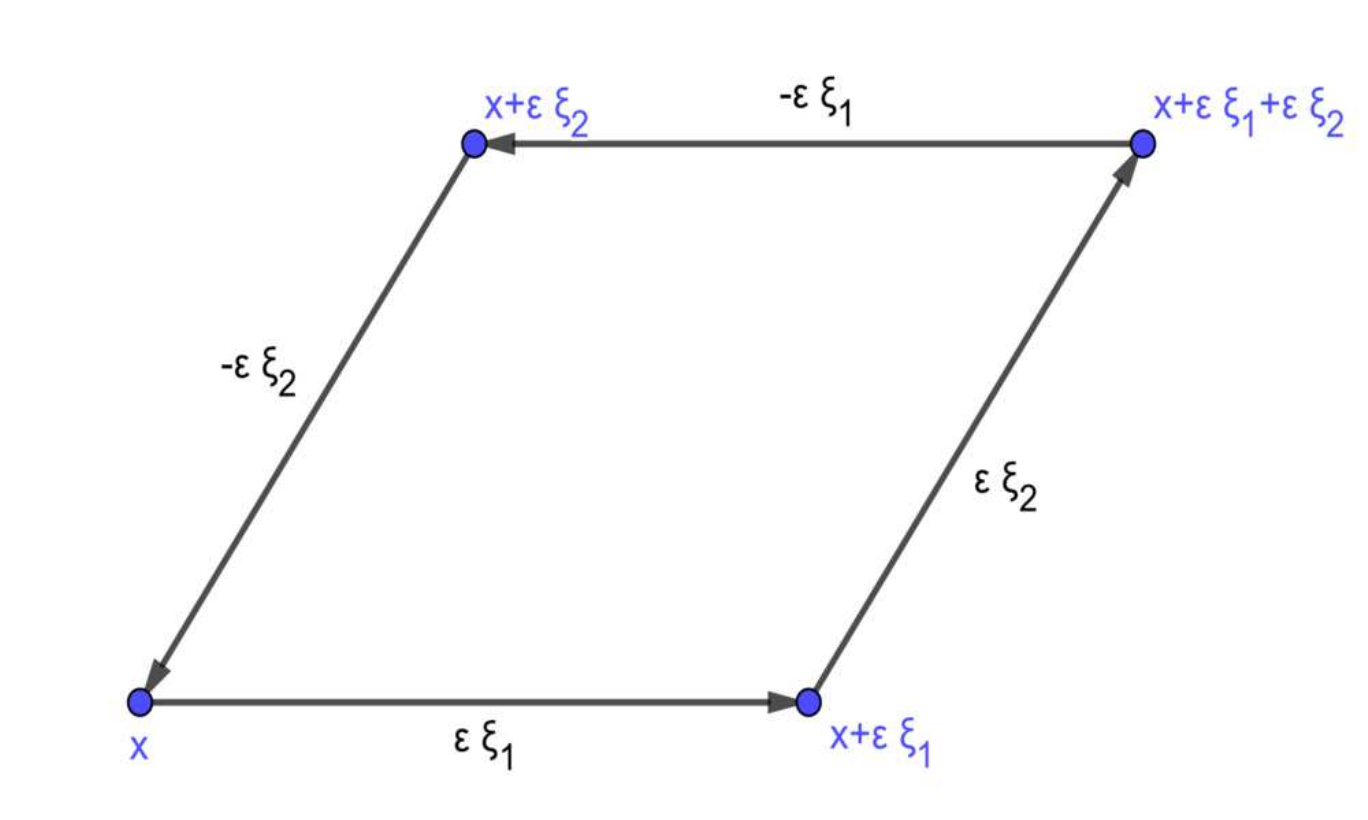}
\\
\caption{Integration path for recovering the field-strength tensor from the holonomies.}\label{wilsloop}
\end{figure}

\begin{figure}[H]
\centering
\vskip-20pt
\includegraphics[width=0.70\textwidth]{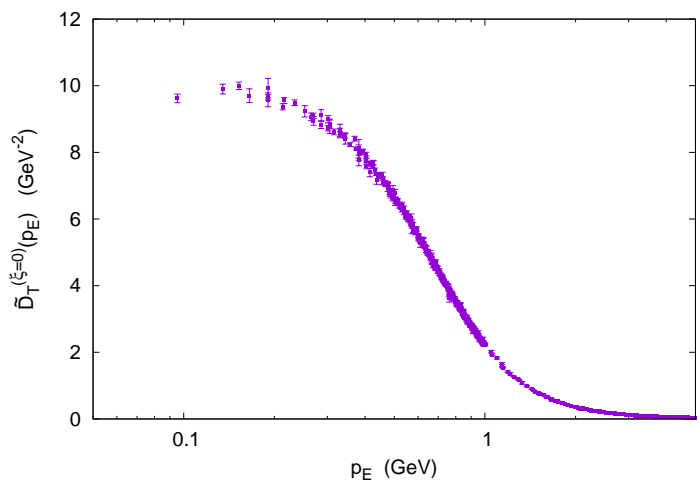}
\vskip-20pt
\caption{Transverse component of the Euclidean dressed gluon propagator computed on the lattice in the Landau gauge ($\xi=0$). Lattice data points from ref. \cite{duarte}.}\label{gluint}
\end{figure}

\begin{figure}[H]
\centering
\vskip-20pt
\includegraphics[width=0.70\textwidth]{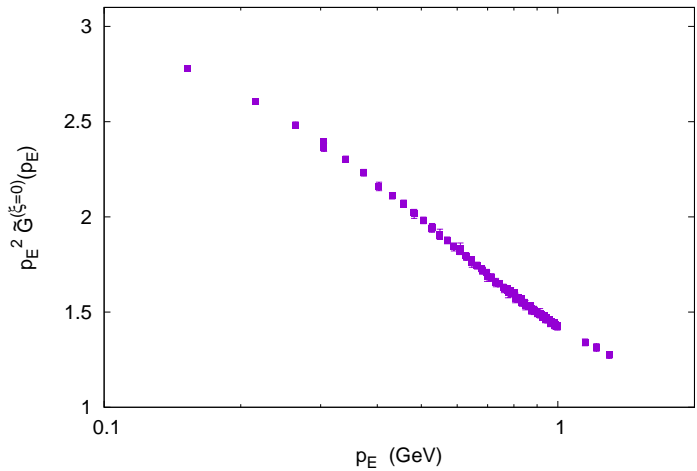}
\vskip-20pt
\caption{Lattice ghost dressing function in the Landau gauge ($\xi=0$). Data from ref. \cite{duarte}.}\label{ghostint}
\end{figure}

Lattice gauge theory \cite{rothe,pardata} is a non-perturbative numerical approach to quantum field theories with a gauge group based on the discretization of spacetime on a finite lattice. In what follows we will briefly review the definition of Yang-Mills theory on the lattice and discuss some crucial results which have recently been obtained by the lattice calculations.\\
\\
As a preliminary step in the definition of the lattice approach, we recall that a (finite four-dimensional cubic) lattice is a set of points of the form $a\,(n_{0},\,n_{1},\,n_{2},\,n_{3})\in\Bbb{R}^{4}$, where $a$ is the lattice spacing and the $n_{\mu}$'s are integers that span from zero to a finite number $L$. As we will see, the dynamical variables of lattice Yang-Mills theory are defined on the links that connect the neighboring sites of the lattice.\\
In order to formulate the lattice-equivalent of Yang-Mills theory, one starts by rewriting the quantum partition function $Z$ of the theory in terms of fields which are defined in Euclidean space rather than in Minkowski space. The Euclidean partition function is obtained from $Z$ by replacing everywhere the real time variable $t$ by an imaginary time variable $\tau$ defined as $\tau=it$. Since the $A_{t}^{a}$'s -- i.e. the time-components of the Yang-Mills fields -- are defined with respect to the real time $t$, in the Euclidean formulation the latter need to be replaced by analogous components in imaginary time; this is achieved by substituting $A_{t}^{a}\to iA_{\tau}^{a}$ in the Yang-Mills action. The derivatives with respect to real time too need to be exchanged with derivatives with respect to imaginary time: in the action we will have to replace $\partial_{t}\to i\partial_{\tau}$. These redefinitions leave us with a partition function that can be put in the form
\[
Z=\int\mc{D}A\ e^{-\mc{S}_{\tx{E}}}
\]\\
where $\mc{S}_{\tx{E}}$ is a Euclidean action defined as\\
\[
\mc{S}_{\tx{E}}=\int d\tau d^{3}x\ \ \frac{1}{4}\ F_{\mu\nu}^{a}\,F^{a\,\mu\nu}\Big|_{\partial_{t}\to i\partial_{\tau},A_{t}^{a}\to iA_{\tau}^{a}}
\]\\
The imaginary units are easily seen to drop out from the above equation if we replace the Minkowski metric by the Euclidean metric. The Euclidean Lagrangian $\mc{L}_{\tx{E}}$ then reads\\
\[
\mc{L}_{\tx{E}}=\frac{1}{4}\ F_{E\mu\nu}^{a}F_{E}^{a\mu\nu}=\frac{1}{4}\ \delta^{\mu\sigma}\delta^{\nu\lambda}\ F_{E\mu\nu}^{a}F_{E\sigma\lambda}^{a}
\]\\
where $F_{E}$ is the field-strength tensor associated to the Euclidean vector fields $(A_{\tau}^{a},A_{i}^{a})$. From now on we will drop the subscripts $_{E}$ and imply that the Yang-Mills fields, field-strength tensor, metric and action are all defined in Euclidean four-dimensional space.\\
The second step for formulating lattice Yang-Mills theory is to find dynamical variables that are appropriate to the discrete structure of the lattice. The hint as to how to do this comes from the geometrical structure of the gauge fields themselves. Observe that, since the Yang-Mills fields are actually covector fields (i.e. 1-forms) with values in $\mf{su}$(N), they can be meaningfully integrated along curves in spacetime to yield elements of the Lie algebra of SU(N). If $\gamma:[0,1]\to \Bbb{R}^{4}$ is such a curve, then we can define\\
\[
\int_{\gamma}A=\int_{0}^{1}ds\ \frac{d\gamma^{\mu}}{ds}(s)\,A_{\mu}^{a}(\gamma(s))\,T_{a}
\]\\
By exponentiating this Lie algebra element in a path-ordered fashion, we obtain an element $U_{\gamma}(A)$ of the group SU(N) that is functionally dependent on $A$, namely\\
\[
U_{\gamma}(A)=\mc{P}\exp\left\{ig\int_{\gamma}A\right\}
\]\\
Therefore any gauge field $A$ establishes a correspondence between curves in spacetime and group elements of SU(N) by associating a $U_{\gamma}(A)$ to each curve $\gamma$. In particular, the gauge field associates SU(N) group elements to each of the links $\ell$ that connect the neighboring lattice points in our discretized spacetime; we shall denote these group elements by $U_{\ell}(A)$. Any arbitrary $\ell$ has the form $x+as\,e_{\mu}$ for $s\in [0,1]$, where $x$ is the initial point of the link and $e_{\mu}$ -- the direction of the link -- can be one of $e_{0}=(1,0,0,0)$, $e_{1}=(0,1,0,0)$, $e_{2}=(0,0,1,0)$ or $e_{3}=(0,0,0,1)$. It follows from our general expression for $U_{\gamma}(A)$ that\\
\[
U_{\ell}(A)=\mathds{1}+iga\,A_{\mu}(x)+O(a^{2})
\]\\
where $A_{\mu}=A_{\mu}^{a}\,T_{a}$. In particular, in the limit of vanishing lattice spacing,\\
\[
A_{\mu}(x)=\lim_{a\to 0}\ \frac{U_{\ell}-\mathds{1}}{iga}
\]\\
where $\ell$ is the link in the direction $e_{\mu}$ originating from $x$. The above expression teaches us how to recover the gauge field starting from arbitrary group elements defined on the lattice. With such a procedure in our hands, we can seek for a formulation of lattice Yang-Mills theory that has the $U_{\ell}$'s, rather than the gauge field, as its dynamical variables. In order to do so, we take one step back and study the group elements associated by the gauge fields to the closed curves.\\
If $\gamma$ is a loop -- i.e. continuous closed curve -- then $U_{\gamma}(A)$ is called the holonomy of $\gamma$ with respect to the gauge field $A$. The holonomies can be related to the field-strength tensor as follows. Suppose that -- as shown in Fig.\ref{wilsloop} -- $\gamma$ is a loop composed by four rectilinear curves that join in succession the points $x$, $x+\epsilon\,\xi_{1}$, $x+\epsilon\,\xi_{1}+\epsilon\,\xi_{2}$, $x+\epsilon\,\xi_{2}$ and $x$, where $\epsilon$ is a small positive number. Then, by expanding $U_{\gamma}(A)$ in powers of $\epsilon$, one finds that \cite{rothe}\\
\[
U_{\gamma}(A)= \mathds{1}+ig\epsilon^{2}\ F_{\mu\nu}(x)\ \xi^{\mu}_{1}\,\xi^{\nu}_{2}-\frac{g^{2}\epsilon^{4}}{2}\ F_{\mu\nu}(x)\,F_{\sigma\tau}(x)\ \xi^{\mu}_{1}\xi^{\nu}_{2}\,\xi^{\sigma}_{1}\xi^{\tau}_{2}+O(\epsilon^{6})
\]
\newpage
\noindent where $F_{\mu\nu}=F_{\mu\nu}^{a}\,T_{a}$ is the field-strength tensor associated to $A_{\mu}$. In particular, if we define $U_{\mu\nu}(A)=U_{\gamma}(A)$ with $\xi_{1}=e_{\mu}$ and $\xi_{2}=e_{\nu}$ unitary vectors in the directions $\mu$ and $\nu$, from the above expression we obtain\\
\[
U_{\mu\nu}(A)= \mathds{1}+ig\epsilon^{2}\ F_{\mu\nu}(x)-\frac{g^{2}\epsilon^{4}}{2}\ F_{\mu\nu}(x)\,F_{\mu\nu}(x)+O(\epsilon^{6})\qquad\quad\text{(no sum over repeated indices)}
\]\\
Therefore the holonomy $U_{\mu\nu}(A)$ contains both the component $F_{\mu\nu}$ of the field-strength tensor and its square $(F_{\mu\nu})^{2}$.\\
Now, suppose that $\epsilon=a$, the lattice spacing. Then in the limit of vanishing lattice spacing the components of $F$ can be extracted from the holonomy as\\
\[
F_{\mu\nu}(x)=\lim_{a\to 0}\ \frac{U_{\mu\nu}-\mathds{1}}{iga^{2}}
\]\\
Moreover, recalling that -- since $T_{a}\in\mf{su}$(N) -- $\text{Tr}\{T_{a}\}=0$ and $\text{Tr}\{T_{a}T_{b}\}=\delta_{ab}/2$, by taking the trace of $U_{\mu\nu}(A)$ we find that\\
\[
\text{Tr}\left\{U_{\mu\nu}(A)\right\}= N-\frac{g^{2}a^{4}}{4}\ F_{\mu\nu}^{a}(x)\,F_{\mu\nu}^{a}(x)+O(a^{6})\qquad\ \text{(no sum over $\mu,\nu$ indices)}
\]\\
This expression brings us to the final step of the definition of the lattice approach, namely, the choice of a discrete action for the group elements $U_{\ell}$. By summing up both sides of the equation with respect to all the possible directions $\mu$ and $\nu$, subject to the constraint that $\mu<\nu$ in order to avoid the double count of the holonomies, we find that\\
\[
\frac{1}{4}\ F_{\mu\nu}^{a}\,F^{a\,\mu\nu}=\frac{2N}{g^{2}a^{4}}\ \sum_{\mu<\nu}\ \left(1-\frac{1}{N}\,\text{Tr}\left\{U_{\mu\nu}(A)\right\}\right)+O(a^{2})
\]\\
where this time the indices on the left-hand side are summed over. This result suggests the following definition for the lattice action:\\
\[
\mc{S}_{\tx{W}}=\frac{2N}{g^{2}}\ \sum_{x,\,\mu<\nu}\ \left(1-\frac{1}{N}\,\text{Tr}\left\{U_{\mu\nu}\right\}\right)
\]\\
$\mc{S}_{\tx{W}}$ is known as the Wilson action \cite{wils}; in deriving it from the holonomies we have proved that it reduces to the Yang-Mills action in the limit of vanishing lattice spacing (and infinite size of the lattice). In terms of the Wilson action, the lattice partition function $Z_{\tx{lat}}$ reads
\[
Z_{\tx{lat}}=\int\prod_{\ell}dU_{\ell}\ \,e^{-\mc{S}_{\tx{W}}}
\]\\
where the holonomies $U_{\mu\nu}$ and group elements $U_{\ell}$ are related by $U_{\mu\nu}=U_{\ell_{1}}U_{\ell_{2}}U_{\ell_{3}}U_{\ell_{4}}$, with $\ell_{1}$, $\ell_{2}$, $\ell_{3}$, and $\ell_{4}$ the links that make up the loop on which the holonomy is defined. Observe that since the number of links in a finite lattice is itself finite, $Z_{\tx{lat}}$ is an integral over the configurations of a finite number of degrees of freedom, and is thus mathematically well-defined.\\
\\
For $N=3$, upon introducing appropriate quark variables at the sites of the lattice, the construction given above defines the non-perturbative discrete approximation to Quantum Chromodynamics known as lattice QCD. Lattice QCD has proven successful in describing both qualitatively and quantitatively many of the low-energy features of the interactions between the quarks and the gluons. Among the notable results of the lattice approach we cite the derivation of the masses and quantum numbers of the hadrons from the dynamics of their elementary constituents \cite{lang}, the description of confinement in terms of gluonic flux tubes \cite{card} and the prediction of the crossover temperature between the confined phase and deconfined phase of quark-gluon matter \cite{petre}.\\
Of particular relevance for the purposes of this thesis is the take of the lattice on the issue of mass generation. As we saw in the last section, the generation of a dynamical mass for the gluons manifests itself in the finiteness of the transverse dressed gluon propagator in the limit of zero momentum. Through the lattice approach one is able to compute the gluon propagator non-perturbatively, albeit as a function of the Euclidean momentum $p_{E}$ rather than of the Minkowski momentum $p$. Nonetheless, since $p_{E}^{2}=(-ip^{0})^{2}+|{\bf p}|^{2}=-p^{2}$, where $p^{0}$ is the time-component of the Minkowski momentum, $p=(p^{0},{\bf p})$, we have that the limits $p^{2}\to 0$ and $p_{E}^{2}\to 0$ of the Minkowski and Euclidean propagators actually coincide. Therefore the question of whether or not the gluons acquire a dynamical mass can be answered as well by investigating the low momentum behavior of the Euclidean propagator computed on the lattice. In what follows we report the results of ref. \cite{duarte} (Duarte et al.) for the transverse component of the dressed gluon propagator computed on the lattice in the Landau gauge\footnote{\ For the problem of gauge fixing in lattice gauge theories see for instance \cite{giusti}.} in the framework of pure Yang-Mills SU(3) theory.\\
The lattice data of ref. \cite{duarte} for the gluon propagator is shown as a function of the Euclidean momentum in Fig.\ref{gluint}. As we can see, as the Euclidean momentum goes to zero the gluon propagator first changes concavity and then saturates to a finite value of order $(300\ \tx{MeV})^{-2}$. This result is of crucial importance, in that it proves that the gluons indeed acquire a dynamical mass in the infrared. This possibility was not unforeseen, as it had been anticipated already in the 1980s by SDE analyses of the Green functions of QCD \cite{corn}; nonetheless, the lattice calculations were the first approach to give reliable evidence of the occurrence of mass generation in QCD by making its low-energy regime accessible to the computations.\\
\\
\clearpage
\noindent To end this introduction we report the lattice data of ref. \cite{duarte} for the dressed ghost propagator\footnote{\ For the definition of the ghosts in the lattice approach see for instance \cite{cucc0}.} $\widetilde{G}(p_{E})$ in the Landau gauge. Rather than the propagator itself, in Fig.\ref{ghostint} we show the data points for the ghost dressing function $p_{E}^{2}\,\widetilde{G}(p_{E})$. As we can see, in the limit $p_{E}\to 0$ the dressing function approaches a finite value, implying that the ghost propagator becomes infinite at zero momentum. As discussed in the previous section, this behavior is typical of massless propagators. Therefore the lattice results inform us that -- at variance with the gluons -- the ghosts remain massless in the infrared as well as in the ultraviolet regime.\\
\\
\

\newpage


\stepcounter{count}
\addcontentsline{toc}{chapter}{1 Dynamical mass generation: the perturbative vacuum of Yang-Mills theory from a variational perspective}  \markboth{1 Dynamical mass generation: the perturbative vacuum of Yang-Mills theory from a variational perspective}{1 Dynamical mass generation: the perturbative vacuum of Yang-Mills theory from a variational perspective}
\chapter*{1\protect \\
\medskip{}
Dynamical mass generation: the perturbative vacuum of Yang-Mills theory from a variational perspective\index{Dynamical mass generation: the perturbative vacuum of Yang-Mills theory from a variational perspective}}

In recent years, lattice calculations \cite{cucc1,cucc2,bogol,dudal,oliv1,ayala,oliv2,burgio,duarte} have shown that the gluon develops an infrared dynamical mass that prevents its propagator from diverging in the limit of vanishing momentum.  In ordinary Yang-Mills and QCD perturbation theory, the phenomenon of dynamical mass generation cannot be described at any finite perturbative order. The reason for this is two-fold. On the one hand, the energy scale for the scaleless pure Yang-Mills theory (or even full QCD with chiral quarks) is set by the spontaneous breaking of scale invariance. In perturbation theory the breaking of scale invariance manifests itself with the appearance of corrections to the Green functions which depend logarithmically on the renormalization scale used to define the theory \cite{peskin}; in the absence of other energy scales, these terms are not able to generate a dynamical mass for the gluon at any finite order, so that pure Yang-Mills theory and chiral QCD in the ordinary perturbative approach remain massless even after the breaking has occurred. On the other hand, even for full QCD with non-chiral quarks (whose energy scale is set by the interplay between the renormalization scale and the masses of the quarks), in the absence of explicit gauge-symmetry breaking terms in the original Lagrangian, the Slavnov-Taylor identities \cite{taylor,slavnov} constrain the effective action to remain BRST-invariant at any finite order. This implies that the gluon mass does not get renormalized by the interactions, hence, again, that the gluon cannot acquire a mass at any finite perturbative order \cite{wein2}.\\
The inability to describe the phenomenon of mass generation is a limitation of ordinary perturbation theory, rather than of Yang-Mills theory itself. As a matter of fact, if we assume that the discretization on the lattice does not fundamentally spoil the symmetries of the theory, the lattice results lead to the conclusion that either BRST invariance is spontaneously broken at low energies -- so that the gluon can freely acquire a dynamical mass -- or that BRST invariance protects the gluon mass from radiative corrections only perturbatively -- so that non-perturbative approaches (or non-ordinary perturbative expansions) could still be able to describe the phenomenon of dynamical mass generation in Yang-Mills theory and QCD.\\
\\
Since dynamical mass generation is an intrinsically low-energy phenomenon, we should expect it to leave traces on the vacuum structure of the theory. As a matter of fact, it has been proposed that the mechanism for gluon mass generation relies on the presence of non-perturbative condensates that populate the rich vacuum of Yang-Mills theory and QCD \cite{corn}. Since our objective is to develop a non-standard perturbative expansion for Yang-Mills theory, the question we wish to ask in this chapter is the following: from a perturbative perspective, which free-particle vacuum state best approximates the true vacuum of the theory?\\
In order to answer this question, we pursue a simple variational approach that goes under the name of Gaussian Effective Potential (GEP) \cite{stev1,consoli,stev2,stev3,stev4,stev5,namg,ibanez,sir1}. The GEP is, roughly speaking, the energy density of a Gaussian state computed to first order in the interactions. Since the vacuum states of free field theories are Gaussian states \cite{wein1}, one can easily interpret these as the free-particle, unperturbed states starting from which one sets up perturbation theory. Since the width of the Gaussian depends on the mass of the particle, the GEP itself is a function of mass. For an ordinary bosonic field theory, the Jensen-Feynman inequality \cite{feyn2} states that the GEP is bounded from below by the exact vacuum energy of the theory. Therefore, by minimizing the GEP with respect to the mass of the particle, one obtains the best perturbative approximation to the true vacuum of the theory.\\
As we will see in what follows, it turns out that the GEP of Yang-Mills theory is minimized by a non-zero value of the mass of the gluon. This fact 1. may be interpreted as evidence of mass generation in Yang-Mills theory, 2. implies that the best perturbative approximation to the vacuum of Yang-Mills theory is attained by massive -- rather than massless -- gluons, foreshadowing the fact that a non-ordinary perturbation theory which treats the gluons as massive already at tree-level could lead to predictions which are in better agreement with the exact, non-perturbative, results.\\
\\
This chapter is organized as follows. In Sec.1.1 we will introduce the concept of optimized perturbation theory, define the Gaussian Effective Potential for a general quantum field theory and perform the GEP analysis of $\lambda\phi^{4}$ theory as a toy model for the phenomenon of mass generation. In Sec.1.2 we will apply the GEP approach to pure Yang-Mills theory and show that its perturbative vacuum is indeed massive, rather than massless, at far as the transverse gluons are concerned. In doing so, we will have to deal with subtleties arising due to the anticommuting nature of the ghost fields \cite{ibanez}; as we will see, these subtleties can be explicitly addressed and a variational statement can still be made regarding the lower bound set on the GEP by the exact vacuum energy of the theory.\\
\\
The results of this chapter were presented for the first time in ref. \cite{com1} and published in ref. \cite{com2}.

\newpage


\addcontentsline{toc}{section}{1.1 Optimized perturbation theory, the Gaussian Effective Potential and $\lambda\phi^{4}$ theory as a toy model for mass generation}  \markboth{1.1 Optimized perturbation theory, the Gaussian Effective Potential and $\lambda\phi^{4}$ theory as a toy model for mass generation}{1.1 Optimized perturbation theory, the Gaussian Effective Potential and $\lambda\phi^{4}$ theory as a toy model for mass generation}
\section*{1.1 Optimized perturbation theory, the Gaussian Effective Potential and $\boldsymbol{\lambda\phi^{4}}$ theory as a toy model for mass generation\index{Optimized perturbation theory, the Gaussian Effective Potential and $\lambda\phi^{4}$ theory as a toy model for mass generation}}

\addcontentsline{toc}{subsection}{1.1.1 Optimized perturbation theory and perturbative ground states}  \markboth{1.1.1 Optimized perturbation theory and perturbative ground states}{1.1.1 Optimized perturbation theory and perturbative ground states}
\subsection*{1.1.1 Optimized perturbation theory and perturbative ground states\index{Optimized perturbation theory and perturbative ground states}}

Consider an harmonic oscillator of frequency $\omega_{0}$ perturbed by a quartic potential,\\
\BE
H=\frac{p^{2}}{2}+\frac{\omega_{0}^{2}x^{2}}{2}+\lambda x^{4}
\EE
\\
with $\lambda$ a small parameter. In ordinary perturbation theory one splits $H$ as $H_{0}(\omega_{0})+V$, where
\BE
H_{0}(\omega_{0})=\frac{p^{2}}{2}+\frac{\omega_{0}^{2}x^{2}}{2}
\EE
\\
is the Hamiltonian of the unperturbed oscillator of frequency $\omega_{0}$. The ground state $\ket{\omega_{0}}$ of $H_{0}(\omega_{0})$ has wavefunction
\BE
\psi_{\omega_{0}}(x)=\bigg(\frac{\omega_{0}}{\pi}\bigg)^{1/4}\, \exp\bigg(-\frac{\omega_{0}x^{2}}{2}\bigg)
\EE
\\
and unperturbed energy $E_{g}^{(0)}=\omega_{0}/2$. To first order in ordinary perturbation theory, the energy $E_{g}^{(1)}(\omega_{0})$ of the ground state of $H$ is given by\\
\BE
E_{g}^{(1)}(\omega_{0})=\frac{\omega_{0}}{2}+\bra{\omega_{0}}V\ket{\omega_{0}}=\frac{\omega_{0}}{2}+\frac{3\lambda}{4\omega_{0}^{2}}
\EE
\\
In a non-ordinary formulation of perturbation theory, we may split $H$ as $H=H_{0}(\omega)+V'$, where
\BE
V'=\frac{(\omega_{0}^{2}-\omega^{2})\,x^{2}}{2}+\lambda x^{4}
\EE
\\
and $H_{0}(\omega)$ is the Hamiltonian of an unperturbed harmonic oscillator of frequency $\omega$. The ground state $\ket{\omega}$ of $H_{0}(\omega)$ has wavefunction\\
\BE
\psi_{\omega}(x)=\bigg(\frac{\omega}{\pi}\bigg)^{1/4}\, \exp\bigg(-\frac{\omega x^{2}}{2}\bigg)
\EE
\\
and unperturbed energy $\omega/2$. If we treat $V'$ as a perturbation to $H_{0}(\omega)$, then the energy $E_{g}^{(1)}(\omega)$ of the ground state of $H$ to first order in perturbation theory is given by\\
\BE
E_{g}^{(1)}(\omega)=\frac{\omega}{2}+\bra{\omega}V'\ket{\omega}=\frac{\omega}{4}+\frac{\omega_{0}^{2}}{4\omega}+\frac{3\lambda}{4\omega^{2}}
\EE
\\
Notice that since $E_{g}^{(1)}(\omega)=\bra{\omega}H\ket{\omega}$, the latter is precisely the energy obtained by applying the variational method to the test function $\psi_{\omega}$. Therefore we know that the exact ground state energy of the perturbed oscillator is less than or equal to $E_{g}^{(1)}(\omega)$, and we can minimize $E_{g}^{(1)}(\omega)$ with respect to $\omega$ to obtain the best estimate of the ground state energy:\\
\BE
0=\frac{\partial E_{g}^{(1)}}{\partial\omega}(\omega_{g})=\frac{1}{4}-\frac{\omega_{0}^{2}}{4\omega^{2}_{g}}-\frac{6\lambda}{4\omega^{3}_{g}}\qquad\Longleftrightarrow\qquad \omega_{g}^{3}-\omega_{0}^{2}\,\omega_{g}-6\lambda=0
\EE
where we have defined $\omega_{g}$ as the frequency that minimizes $E_{g}^{(1)}(\omega)$. We see that $\omega_{g}=\omega_{0}$ if and only if $\lambda=0$, i.e. if the harmonic oscillator is unperturbed. For $\lambda\neq 0$, we have $\omega_{g}>\omega_{0}$, so that the best approximation to the ground state energy is given by a Gaussian with a variance smaller than that of the unperturbed oscillator. This was to be expected, since at high $x$'s the quartic potential increases more rapidly than the harmonic potential, thus producing eigenstates which are bound around $x=0$ more tightly than those of the unperturbed oscillator. Moreover, as long as $\lambda$ is sufficiently small,\\
\BE
\omega_{g}=\omega_{0}+\frac{3\lambda}{\omega_{0}^{2}}+O(\lambda^{2})
\EE
\\
and $\omega_{g}$ is approximately equal to $\omega_{0}$. Therefore we can still regard $V'$ as a small perturbation to $H_{0}(\omega_{g})$ and our non-ordinary formulation of perturbation theory is still valid with $H_{0}(\omega_{g})$ as the zero-order Hamiltonian. Since $\ket{\omega_{g}}$ is closer than $\ket{\omega_{0}}$ to the true ground state of the perturbed oscillator, we expect the ground state average $\avg{\mathcal{O}}_{g}$ of an arbitrary operator $\mathcal{O}$ to be better approximated in perturbation theory if we compute it by expanding perturbatively around the eigenstates of $H_{0}(\omega_{g})$, rather than around those of $H_{0}(\omega_{0})$. For this reason, we call $\ket{\omega_{g}}$ the \textit{perturbative ground state} of the theory.\\
\\
This method for computing quantities in quantum mechanics is known as \textit{optimized perturbation theory} \cite{stev1}, and can be readily generalized to any quantum system. In a general setting, let $H$ be the exact Hamiltonian of some quantum system. $H$ can be arbitrarily split as $H_{0}+H_{\text{int}}$, where $H_{0}$ is an Hamiltonian of which we know the exact eigenstates and eigenenergies and $H_{\text{int}}=H-H_{0}$. If we choose an $H_{0}$ such that its eigenstates well approximate those of $H$, then the perturbative series having $H_{0}$ as the zero-order Hamiltonian will lead to more accurate predictions than those obtained by using a different $H_{0}$. If the optimal $H_{0}$ is chosen through a variational ansatz, then the perturbative series arising from the split $H=H_{0}+H_{\intr}$ is said to be \textit{optimized}. The ground state of the optimized $H_{0}$ is again called a perturbative ground state or -- if the quantum theory is a field theory -- a \textit{perturbative vacuum}.
\\
\\
\addcontentsline{toc}{subsection}{1.1.2 The perturbative vacuum of a quantum field theory and the Gaussian Effective Potential}  \markboth{1.1.2 The perturbative vacuum of a quantum field theory and the Gaussian Effective Potential}{1.1.2 The perturbative vacuum of a quantum field theory and the Gaussian Effective Potential}
\subsection*{1.1.2 The perturbative vacuum of a quantum field theory and the Gaussian Effective Potential\index{The perturbative vacuum of a quantum field theory and the Gaussian Effective Potential}}

In quantum field theory, ordinary perturbation theory follows from choosing as the zero-order Hamiltonian $H_{0}$ the free field Hamiltonian obtained in the limit of vanishing (renormalized) couplings. For instance, in $\lambda\phi^{4}$ theory, which has Hamiltonian\footnote{\ Here $H_{\text{c.t.}}$ contains all the relevant renormalization counterterms and vanishes at any perturbative order as $\lambda$ -- the renormalized coupling -- goes to zero.}\\
\BE
H=\int d^{D}x\ \left\{\frac{1}{2}\ \pi^{2}+\frac{1}{2}\,|{\bf \nabla}\phi|^{2}+\frac{m^{2}}{2}\, \phi^{2}+\frac{\lambda}{4!}\, \phi^{4}\right\}+H_{\text{c.t.}}
\EE
\\
with $m$ the pole mass of the scalar propagator and $D$ the number of spatial dimensions, the ordinary choice for $H_{0}$ is\\
\BE
H_{0}=\int d^{D}x\ \left\{\frac{1}{2}\ \pi^{2}+\frac{1}{2}\,|{\bf \nabla}\phi|^{2}+\frac{m^{2}}{2}\, \phi^{2}\right\}
\EE
The vacuum states of free field Hamiltonians are Gaussian functionals of the field configurations \cite{wein1}. For fixed spin, the only free parameter of these functionals is the mass of the particle. For instance, the vacuum wavefunctional of a free real scalar field of mass $m$ in $D$ spatial dimensions is given by\cite{wein1}\\
\BE
\psi[\Phi]=N\ \exp\bigg(-\frac{1}{2}\int d^{D}x\,d^{D}y\ \Phi({\bf x})\,\mathcal{E}_{m}({\bf x}-{\bf y})\,\Phi({\bf y})\bigg)
\EE
\\
where $N$ is a normalization constant and\\
\BE
\mathcal{E}_{m}({\bf x}-{\bf y})=\int\frac{d^{D}k}{(2\pi)^{D}}\ e^{i{\bf k}\cdot({\bf x}-{\bf y})}\ \sqrt{m^{2}+|{\bf k}|^{2}}
\EE
\\
is the Fourier transform of the energy of the scalar particle. Starting from $H_{0}$ and the corresponding vacuum wavefunctional, we can compute all the relevant quantities in ordinary perturbation theory by treating $H_{\text{int}}=H-H_{0}$ as a perturbation.\\ As long as we limit ourselves to free field Hamiltonians and Gaussian wavefunctionals, since, as we said, in this case the only free parameter is the mass of the functional, the obvious field-theoretic generalization of ordinary perturbation theory is obtained by choosing as the zero-order Hamiltonian $H_{0}$ a free field Hamiltonian with a mass different from that contained in the (renormalized) Lagrangian. Then one can optimize the value of the mass by requiring it to minimize the vacuum energy of the theory to first order in perturbation theory -- a procedure which is equivalent to applying the variational method to the ground state of $H_{0}$ --, thus obtaining an optimized perturbation theory with a Gaussian perturbative vacuum. With some abuse of language, the energy density of the vacuum state of a quantum field theory, computed to first order in its interactions by using as the zero-order vacuum wavefunctional a Gaussian with free parameters the masses of the particles, is called the Gaussian Effective Potential (GEP) \cite{stev1,consoli,stev2,stev3,stev4,stev5,namg,ibanez,sir1}.\\
The GEP is arguably the simplest tool for determining the perturbative ground state of a field theory. In principle, it may have as additional free parameters the vacuum expectation values of the fields; for example, for a scalar particle one may take as the vacuum wavefunctional for computing the GEP\\
\BE
\psi[\Phi]=N\ \exp\bigg(-\frac{1}{2}\int d^{D}x\,d^{D}y\ \big(\Phi({\bf x})-\avg{\Phi}\big)\,\mathcal{E}_{\mu}({\bf x}-{\bf y})\,\big(\Phi({\bf y})-\avg{\Phi}\big)\bigg)
\EE
\\
with an arbitrary mass $\mu$ and vacuum expectation value $\avg{\Phi}$, and compute its GEP. However, since we are only interested in theories whose fields have vanishing vacuum expectation values\footnote{\ Since the gluon field is a vector field, a non-zero vacuum expectation value for $A_{\mu}^{a}$ would lead to the spontaneous breaking of Lorentz symmetry, which we assume not to occur in any sensible relativistic field theory.}, in what follows we will limit ourselves to GEP's whose only free parameters are the masses of the particles.\\
\\
Of particular interest is the case in which the bare masses in the original Hamiltonian are zero. Then the interactions may or may not generate a dynamical mass for the excitations of the fields; likewise, the perturbative vacuum of the theory may or may not be the Gaussian vacuum of a massive particle. By applying the GEP approach to such theories, one is able to address the issue of mass generation both from a perturbative and from a non-perturbative perspective. If the GEP is found to be minimized by a non-zero value of the mass parameter, implying \cite{consoli,stev2} that the massless vacuum of the theory is unstable towards a massive vacuum, then one 1. has strong indications of the occurrence of the phenomenon of mass generation (non-perturbative aspect of the GEP analysis) and 2. has an even stronger indication that, since the massless perturbative vacuum is farther away from the true vacuum than the massive one, a non-ordinary perturbation theory which treats the excitations of the fields as massive already at tree-level may lead to more accurate predictions than those obtained by ordinary (massless) perturbation theory (perturbative aspect of the GEP analysis).\\
\\
We now proceed to give a formal definition of the GEP in the Lagrangian framework. The vacuum energy density $\mathcal{E}$ of a quantum field theory defined by the action $\mc{S}[F]$, describing a set of fields which we collectively denote by $F$, is given by\\
\BE
e^{-i\mathcal{E}\mathcal{V}_{d}}=\int \mathcal{D}F\ e^{i\mathcal{S}[F]}
\EE
\\
where $\mathcal{V}_{d}$ is the $d$-dimensional volume of spacetime. If $\mathcal{S}$ is polynomial in the fields and its derivatives, we know how to compute $\mathcal{E}$ perturbatively. We set $\mathcal{S}=\mathcal{S}_{0}+\mc{S}_{\text{int}}$, where $\mc{S}_{0}$ is an action term quadratic in the fields, and expand\\
\BE\label{abc}
\int \mathcal{D}F\ e^{i\mathcal{S}[F]}=\int \mathcal{D}F\ e^{i\mathcal{S}_{0}[F]}\ \sum_{n=0}^{+\infty}\ \frac{i^{n}\mc{S}^{n}_{\text{int}}[F]}{n!}=\left(\int \mathcal{D}F\ e^{i\mathcal{S}_{0}[F]}\right)\ \left(\sum_{n=0}^{+\infty}\ \frac{i^{n}\avg{\mc{S}_{\text{int}}^{n}}_{0}}{n!}\right)
\EE
\\
so that
\begin{align}\label{abd}
-i\mc{E}\mc{V}_{d}&=\ln\int \mathcal{D}F\ e^{i\mathcal{S}_{0}[F]}+\ln\left(1+\sum_{n=1}^{+\infty}\ \frac{i^{n}\avg{\mc{S}_{\text{int}}^{n}}_{0}}{n!}\right)
\end{align}
\\
In both \eqref{abc} and \eqref{abd}, the quantum average $\avg{\,\cdot\,}_{0}$ is defined with respect to the zero-order action $\mc{S}_{0}$. Since $\exp(i\mc{S}_{0})$ is Gaussian in the field configurations, in order to compute the averages $\avg{\mc{S}_{\text{int}}^{n}}_{0}$ one only needs to evaluate polynomial functional integrals with Gaussian kernels; this is usually done by making use of appropriate Feynman rules.\\
In ordinary perturbation theory, one chooses as the zero-order $\mc{S}_{0}$ the free action associated to the set of fields $F$, obtained, for instance, by taking the limit of vanishing renormalized couplings of the full action $\mc{S}$. In a more general setting, we may still define $\mc{S}_{0}$ to be the free action associated to the fields $F$, but with arbitrary -- rather than on-shell -- particle masses, which we collectively denote by $m$. With this choice, both $\mathcal{S}_{0}$ and $\mc{S}_{\tx{int}}=\mc{S}-\mc{S}_{0}$ are functions of the \textit{mass parameters} $m$. Going back to eq.~\eqref{abd} and expanding $\ln(1+x)=x+O(x^{2})$, we find\\
\BE
-i\mc{E}\mc{V}_{d}=\ln\int \mathcal{D}F\ e^{i\mathcal{S}_{0}(m)}+i\avg{\mc{S}_{\text{int}}(m)}_{0}+O(\avg{\mc{S}_{\text{int}}^{2}}_{0})
\EE
\\
The quantity $V_{G}(m)$, defined by\\
\BE\label{acz}
-iV_{G}(m)\,\mc{V}_{d}=\ln\int \mathcal{D}F\ e^{i\mathcal{S}_{0}(m)}+i\avg{\mc{S}_{\text{int}}(m)}_{0}
\EE
\\
is called the \textit{Gaussian Effective Potential} (GEP). It is the vacuum energy density of the field theory, computed to first order in its interactions as a function of the tree-level mass parameters $m$. Since the GEP is obtained by expanding the vacuum energy density to first order in $\mc{S}_{\tx{int}}$ rather than to first order in the coupling, $V_{G}$ is an essentially non-perturbative object. Since the GEP assumes the zero order action to be Gaussian in the fields, the GEP analysis addresses the issues of stability and mass generation from a perturbative perspective. If the fields $F$ are $c$-fields (i.e. if they are not Grassmann-valued), the Jensen-Feynman inequality \cite{feyn2} can be exploited to show that the exact vacuum energy of the system $\mc{E}$ sets an upper bound for the GEP $V_{G}(m)$ evaluated at any value of $m$:\\
\BE
V_{G}(m)\geq \mathcal{E}\qquad\qquad\forall\ m
\EE
\\
This implies that $V_{G}(m)$ computed at its minimum is the variational estimate of the vacuum energy density of the theory. By minimizing the GEP with respect to the mass parameters, one obtains the best Gaussian (i.e. free particle-) approximation to the vacuum of the system, that is, the perturbative vacuum of the theory. Once the perturbative vacuum is known, one can compute the quantities of interest in optimized perturbation theory by formulating the perturbative series so that $\mc{S}_{0}(m_{0})$ -- where $m_{0}$ is the value that realizes the minimum of the GEP -- is the zero-order action of the expansion.
\\
\\
\addcontentsline{toc}{subsection}{1.1.3 The Gaussian Effective Potential of $\lambda\phi^{4}$ theory: a toy model for mass generation}  \markboth{1.1.3 The Gaussian Effective Potential of $\lambda\phi^{4}$ theory: a toy model for mass generation}{1.1.3 The Gaussian Effective Potential of $\lambda\phi^{4}$ theory: a toy model for mass generation}
\subsection*{1.1.3 The Gaussian Effective Potential of $\boldsymbol{\lambda\phi^{4}}$ theory: a toy model for mass generation\index{The Gaussian Effective Potential of $\lambda\phi^{4}$ theory: a toy model for mass generation}}
\begin{figure}[H]
\centering
\includegraphics[width=0.74\textwidth]{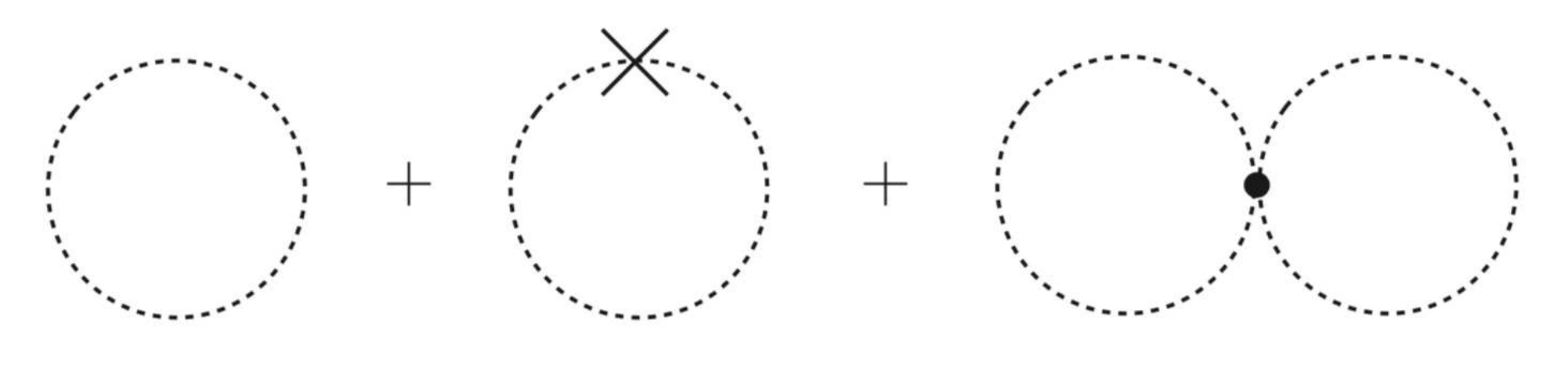}
\\
\caption{Diagrams which contribute to the GEP of $\lambda\phi^{4}$ theory. From left to right: the logarithmic contribution, the quadratic contribution, the quartic contribution.}\label{FA}
\end{figure}

\begin{figure}[H]
\centering
\includegraphics[width=0.3\textwidth]{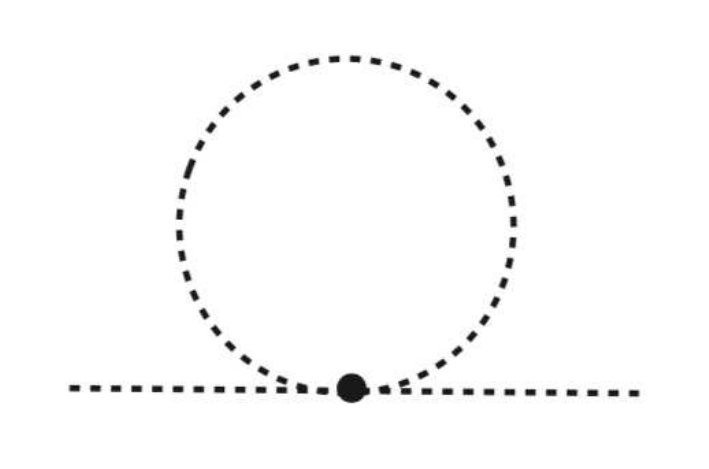}
\\
\caption{The tadpole diagram.}\label{FB}
\end{figure}

\begin{figure}[H]
\centering
\vskip-20pt
\includegraphics[width=0.70\textwidth]{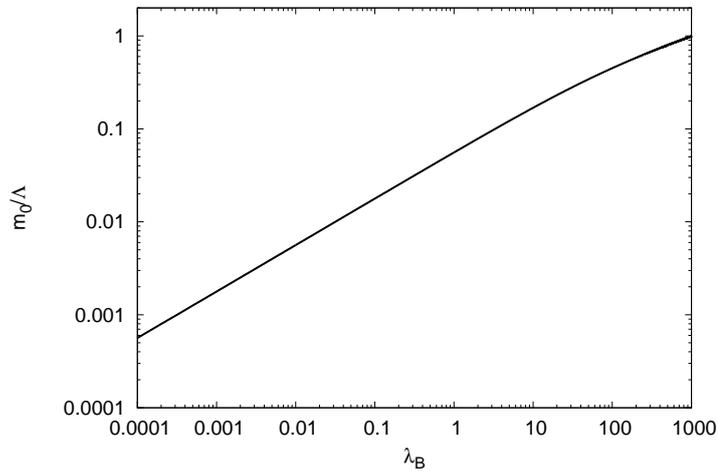}
\vskip-20pt
\caption{Numerical solution to the gap equation of the GEP in the sharp cutoff renormalization scheme.}\label{FC}
\end{figure}

\begin{figure}[H]
\centering
\vskip-20pt
\includegraphics[width=0.70\textwidth]{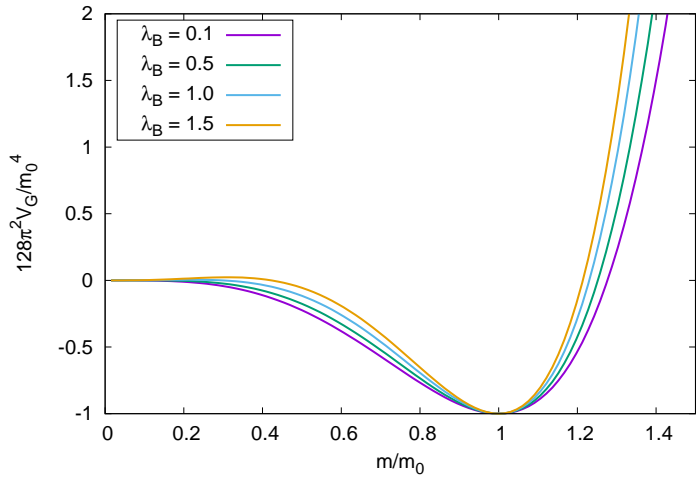}
\vskip-20pt
\caption{A normalized version of the dimensionally regularized ($d>4$) GEP as a function of $m/m_{0}$ for different values of the bare coupling $\lambda_{B}$. The relative minimum at $m=0$, relative maximum at $m=m_{0}\,e^{-1/\lambda_{B}-1/2}$ and absolute minimum at $m=m_{0}$ are clearly visible. The value at the minimum does not depend on the coupling.}\label{FD}
\end{figure}

\
\\
Before moving on to Yang-Mills theory, in order to get acquainted with the formalism, the basic features of the Gaussian Effective Potential and their connection to the issue of mass generation, let us define and compute the GEP of $\lambda\phi^{4}$ theory. The action of $\lambda\phi^{4}$ theory is given by
\BE
\mc{S}=\int d^{d}x\ \left\{\frac{1}{2}\,\partial_{\mu}\phi\,\partial^{\mu}\phi-\frac{m^{2}_{p}}{2}\,\phi^{2}-\frac{\lambda}{4!}\,\phi^{4}+\mathcal{L}_{\text{c.t}}\right\}
\EE
\\
where $m_{p}$ is the pole mass of the scalar particle and $\mathcal{L}_{\tx{c.t.}}$ contains the appropriate renormalization counterterms. The $\mathcal{S}_{0}$ and $\mc{S}_{\tx{int}}$ of ordinary perturbation theory are taken to be\\
\BE
\mc{S}_{0}=\int d^{d}x\ \left\{\frac{1}{2}\,\partial_{\mu}\phi\,\partial^{\mu}\phi-\frac{m^{2}_{p}}{2}\,\phi^{2}\right\}\qquad \mc{S}_{\tx{int}}=\mc{S}-\mc{S}_{0}=\int d^{d}x\ \left\{-\frac{\lambda}{4!}\,\phi^{4}+\mathcal{L}_{\text{c.t}}\right\}
\EE
\\
In order to define the GEP, we must allow for arbitrary tree-level masses. Hence we choose as $\mc{S}_{0}$
\BE
\mc{S}_{0}(m)=\int d^{d}x\ \left\{\frac{1}{2}\,\partial_{\mu}\phi\,\partial^{\mu}\phi-\frac{m^{2}}{2}\,\phi^{2}\right\}
\EE
\\
where $m$ is a mass parameter; it follows that
\BE
\mc{S}_{\tx{int}}(m)=\mc{S}-\mc{S}_{0}(m)=\int d^{d}x\ \left\{-\frac{m^{2}_{p}-m^{2}}{2}\,\phi^{2}-\frac{\lambda}{4!}\,\phi^{4}+\mathcal{L}_{\text{c.t}}\right\}
\EE
\\
From now on, we will work with bare -- rather than with renormalized -- masses and coupling constants: in order to define the renormalized mass and coupling, we are required to choose a renormalization scheme from the very start; we decide not to do so and rather to renormalize the GEP a posteriori, according to what divergences may arise from its computation. In terms of the bare mass $m_{B}$ and bare coupling $\lambda_{B}$, the GEP is given by\\
\BE\label{abf}
V_{G}(m)=\frac{i}{\mc{V}_{d}}\,\ln\int \mathcal{D}\phi\ e^{i\mathcal{S}_{0}(m)}+\frac{1}{\mc{V}_{d}}\ \int d^{d}x\ \left\{\frac{m^{2}_{B}-m^{2}}{2}\,\avg{\phi^{2}}_{0}+\frac{\lambda_{B}}{4!}\,\avg{\phi^{4}}_{0}\right\}
\EE
\\
\\
\textit{1.1.3.1 Computation of the GEP}\\
\\
To each term in $V_{G}$ we associate a Feynman diagram. The first, logarithmic term in eq.~\eqref{abf} is usually represented as a closed loop with no vertices (first diagram in Fig.\ref{FA}); the quadratic term, being proportional to the spacetime integral of the propagator, is represented as a closed loop with a two-point vertex (proportional to $m_{B}^{2}-m^{2}$, second diagram in Fig.\ref{FA}); the quartic term can be interpreted as the integral of the tadpole diagram (Fig.\ref{FB}), and as such it is represented by a double loop with a four-point vertex (the usual four-point coupling vertex, proportional to $\lambda_{B}$, third diagram in Fig.\ref{FA}). These diagrams may be computed by using appropriate Feynman rules. For better clarity, however, let us do the computation explicitly in coordinate space. We have\\
\BE
\avg{\phi^{2}(x)}_{0}=\lim_{y\to x}\avg{T\{\phi(x)\phi(y)\}}_{0}=\lim_{y\to x}\ \mathcal{D}^{m}_{F}(x-y)=\int\frac{d^{d}k}{(2\pi)^{d}}\frac{i}{k^{2}-m^{2}+i\epsilon}=J_{m}
\EE
\begin{align}\label{acf}
&\avg{\phi^{4}(x)}_{0}=\lim_{y_{1},y_{2},y_{3}\to x}\avg{T\{\phi(x)\phi(y_{1})\phi(y_{2})\phi(y_{3})\}}=\\
\notag&=\lim_{\{y_{i}\}\to x}\left\{\mathcal{D}^{m}_{F}(x-y_{1})\mathcal{D}^{m}_{F}(y_{2}-y_{3})+\mathcal{D}^{m}_{F}(x-y_{2})\mathcal{D}^{m}_{F}(y_{3}-y_{1})+\mathcal{D}^{m}_{F}(x-y_{3})\mathcal{D}^{m}_{F}(y_{1}-y_{2})\right\}=\\
\notag&=3[\mathcal{D}^{m}_{F}(0)]^{2}=3J_{m}^{2}
\end{align}
\\
where $\mathcal{D}_{F}^{m}(x)$ is the Feynman propagator (in coordinate space) of the free scalar field of mass $m$,
\BE
\mathcal{D}_{F}^{m}(x)=\int\frac{d^{d}k}{(2\pi)^{d}}\ e^{-ik\cdot x}\ \frac{i}{k^{2}-m^{2}+i\epsilon}
\EE
\\
\\
\\
\newpage
\
\
\\
\\
and $J_{m}$ is the Euclidean integral defined by\\
\BE\label{abp}
J_{m}=\int\frac{d^{d}k_{E}}{(2\pi)^{d}}\ \frac{1}{k^{2}_{E}+m^{2}}
\EE
\\
As for the first term in \eqref{abf}, an explicit computation of the Gaussian functional integral leads to\\
\BE
\ln \int \mathcal{D}F\ e^{i\mathcal{S}_{0}(m)}=-\frac{i\mathcal{V}_{d}}{2}\ \int\frac{d^{d}k_{E}}{(2\pi)^{d}}\ \ln(k^{2}_{E}+m^{2})=-i\mathcal{V}_{d}\,K_{m}
\EE
\\
where we have defined the Euclidean integral $K_{m}$ as\\
\BE
K_{m}=\frac{1}{2}\int\frac{d^{d}k_{E}}{(2\pi)^{d}}\ \ln(k^{2}_{E}+m^{2})
\EE
\\
By summing up the three contributions with the appropriate coefficients, we find that the GEP of $\lambda\phi^{4}$ theory is given by\\
\BE\label{abq}
V_{G}(m^{2})=K_{m}-\frac{1}{2}\ (m^{2}-m_{B}^{2})\,J_{m}+\frac{\lambda_{B}}{8}\, J_{m}^{2}
\EE
\\
\\
\textit{1.1.3.2 Minimization of the GEP and the gap equation for the massless theory}\\
\\
As it stands, the expression \eqref{abq} for $V_{G}(m^{2})$ is ill-defined: both $K_{m}$ and $J_{m}$ are divergent integrals which need to be regularized. Let us suppose for the moment that this has been done. Then, by taking the derivative of $V_{G}(m^{2})$ with respect to $m^{2}$, we obtain the stationarity condition for the GEP: $V_{G}(m^{2})$ is extremized by the values $m_{0}^{2}$ such that\\
\BE
\frac{\partial V_{G}}{\partial m^{2}}(m^{2}_{0})=-\frac{1}{2}\ \frac{\partial J_{m}}{\partial m^{2}}\bigg|_{m_{0}^{2}}\ \bigg\{m^{2}_{0}-m_{B}^{2}-\frac{\lambda_{B}}{2}\ J_{m_{0}}\bigg\}=0
\EE
\\
where we have used the formal identity\\
\BE\label{abh}
\frac{\partial K_{m}}{\partial m^{2}}=\frac{1}{2}\ J_{m}
\EE
\newpage
\noindent Since formally the derivative of $J_{m}$ with respect to $m^{2}$ is negative definite,\\
\BE\label{abw}
\frac{\partial J_{m}}{\partial m^{2}}=-\int\frac{d^{d}k_{E}}{(2\pi)^{d}}\ \frac{1}{(k^{2}_{E}+m^{2})^{2}}<0
\EE
\\
the derivative of $V_{G}$ is positive for $m^{2}>m_{B}^{2}+\lambda_{B}J_{m}/2$ and negative for $m^{2}<m_{B}^{2}+\lambda_{B}J_{m}/2$: if it exists, the value $m=m_{0}$ defined by\\
\BE\label{massgap0}
m_{0}^{2}=m_{B}^{2}+\frac{\lambda_{B}}{2}\ J_{m_{0}}
\EE
\\
is a minimum for the GEP. Eq.~\eqref{massgap0} is not new at all: provided that $m_{B}\neq 0$, when the dressed scalar propagator of $\lambda\phi^{4}$ theory is computed to one loop in ordinary perturbation theory, one finds that the relation between the bare mass $m_{B}$ and the pole mass $m_{p}$ of the scalar particle is\footnote{\ Recall that in $\lambda\phi^{4}$ theory $\lambda=\lambda_{B}$ to one loop order.} \cite{peskin}
\BE
m_{B}^{2}=m_{p}^{2}-\frac{\lambda}{2}\ J_{m_{p}}
\EE
\\
Therefore, the GEP approach predicts that the vacuum energy density of $\lambda\phi^{4}$ theory is minimized precisely by the pole mass of the scalar particle computed to one loop order: $m_{0}=m_{p}$.\\
\\
On the other hand, consider what happens in the case of a vanishing bare mass. For $m_{B}=0$, eq.~\eqref{massgap0} reads
\BE\label{massgap}
m_{0}^{2}=\frac{\lambda_{B}}{2}\ J_{m_{0}}
\EE
\\
Eq.~\eqref{massgap} is known as the \textit{gap equation} of the GEP. Assuming that it admits a non-zero solution, by fixing the value of the mass parameter that minimizes $V_{G}$, the gap equation predicts that the perturbative vacuum of $\lambda\phi^{4}$ theory is massive, even if the theory by itself was massless. This is at variance with ordinary perturbation theory, which in turn predicts that the propagator of massless $\lambda\phi^{4}$ theory remains massless even after the quantum corrections are included\footnote{\ This prediction is actually renormalization-scheme-dependent.}.\\
\\
In conclusion, not only through the GEP one is able to derive the perturbative one-loop relation between the bare mass and the pole mass of the massive theory, but the approach also sheds light on the non-perturbative issue of mass generation in the massless theory. Since we are only interested in the latter case, from now on we will set $m_{B}=0$ and study the behavior of the GEP of $\lambda\phi^{4}$ theory at vanishing bare mass.\\
\\
\\
\textit{1.1.3.3 Renormalization of the GEP and its solutions}\\
\\
Let us now turn to the issue of renormalization in $d=4$. As we will see, perhaps counterintuitively, different renormalization procedures lead to different conclusions with respect to the issue of mass generation.\\
To begin with, suppose that massless $\lambda\phi^{4}$ theory is defined with an intrinsic sharp cutoff $\Lambda$, so that all the integrals in Euclidean momentum space are convergent and $J_{m}$ and $\partial J_{m}/\partial m^{2}$ are respectively positive and negative definite. An explicit computation shows that\footnote{\ We have added a term proportional to $\Lambda^{4}\ln\Lambda^{2}$ to $K_{m}$ in order to adimensionalize the argument of the first logarithm. Such a modification does not spoil our computation, since it amounts to adding an arbitrary, $m$-independent, constant to the vacuum energy density of the system.}
\begin{align}\label{abz}
\notag K_{m}&=\frac{1}{64\pi^{2}}\ \left\{\Lambda^{4}\, \ln\left(1+\frac{m^{2}}{\Lambda^{2}}\right)-\frac{\Lambda^{4}}{2}+\Lambda^{2}m^{2}-m^{4}\,\ln\left(\frac{\Lambda^{2}}{m^{2}}+1\right)\right\}\\
J_{m}&=\frac{1}{16\pi^{2}}\left\{\Lambda^{2}-m^{2}\,\ln\left(\frac{\Lambda^{2}}{m^{2}}+1\right)\right\}\\
\notag \frac{\partial J_{m}}{\partial m^{2}}&=\frac{1}{16\pi^{2}}\left\{1-\ln\left(\frac{\Lambda^{2}}{m^{2}}+1\right)\right\}
\end{align}
\\
where we have not yet taken the limit $\Lambda\to \infty$ in order for the GEP to be defined for all $m$'s. According to our calculations, the gap equation reads\\
\BE
m^{2}_{0}=\frac{\lambda_{B}}{32\pi^{2}}\ \left\{\Lambda^{2}-m^{2}_{0}\,\ln\left(\frac{\Lambda^{2}}{m^{2}_{0}}+1\right)\right\}
\EE
\\
For arbitrarily large $\lambda_{B}$'s, the solution to this equation may be of order $\Lambda$ or greater (Fig.\ref{FC}). Since $\Lambda$ is a cutoff, if $m_{0}$ is to have any physical meaning at all it must be much smaller than $\Lambda$. This is verified if and only if $\lambda_{B}$ is sufficiently small, in which case the solution to the gap equation can be approximated as\\
\BE
m^{2}_{0}\approx\frac{\lambda_{B}\Lambda^{2}}{32\pi^{2}}
\EE
\\
\\
\newpage
\noindent Therefore, if we regularize the GEP through a cutoff, we find that 1. the solution to the gap equation is physically acceptable only if the bare coupling is sufficiently small, 2. if this is the case, then the optimal mass scale is roughly proportional (albeit through a small proportionality constant) to the cutoff, i.e. $m_{0}^{2}$ is proportional to the quadratic divergence of the tadpole diagram.\\
This last feature, in particular, is due to the fact that in $\lambda\phi^{4}$ theory -- at variance with gauge theories -- no special symmetry protects the mass of the scalar particle from receiving large quantum corrections from the quadratic divergences. If we are to interpret $\lambda\phi^{4}$ theory as a toy model for mass generation in Yang-Mills theory, the above solution cannot then be deemed satisfactory: one the one hand, it is well known that the quadratic divergences spoil the renormalizability of gauge theories by contributing with non-renormalizable terms to the masses of the gauge bosons, so that we should prevent them from appearing in our renormalized expressions; on the other hand, it is not even clear whether a mass generated through a quadratic divergence can be interpreted as a truly dynamically generated mass.\\
For future reference, we report the leading behavior of the expressions in eq.~\eqref{abz} in the limit $m\ll\Lambda$\footnote{\ Again, $K_{m}$ is defined modulo an $m$-independent additive constant with the dimensions of $\Lambda^{4}$.}:
\begin{align}\label{abv}
\notag K_{m}&=\frac{1}{64\pi^{2}}\ \left(-\frac{\Lambda^{4}+m^{4}}{2}+2\Lambda^{2}m^{2}+m^{4}\,\ln\frac{m^{2}}{\Lambda^{2}}\right)\\
J_{m}&=\frac{1}{16\pi^{2}}\left(\Lambda^{2}+m^{2}\,\ln\frac{m^{2}}{\Lambda^{2}}\right)\\
\notag \frac{\partial J_{m}}{\partial m^{2}}&=\frac{1}{16\pi^{2}}\left(\ln\frac{m^{2}}{\Lambda^{2}}+1\right)
\end{align}\\
\\
The considerations of the last paragraph lead us to turn to other renormalization schemes for the GEP of $\lambda \phi^{4}$ theory as a model of mass generation. With an eye to Yang-Mills theory, we examine a renormalization scheme known to prevent the gauge bosons from acquiring a mass due to the quadratic divergences, i.e. dimensional regularization (henceforth referred to also as \textit{dimreg}). Setting $\epsilon=4-d$, in dimreg we find that\\
\BE
J_{m}=-\frac{m^{2}}{16\pi^{2}}\ \left(\frac{2}{\epsilon}-\ln\frac{m^{2}}{\overline{\mu}^{2}}+1\right)\qquad\qquad \frac{\partial J_{m}}{\partial m^{2}}=-\frac{1}{16\pi^{2}}\ \left(\frac{2}{\epsilon}-\ln\frac{m^{2}}{\overline{\mu}^{2}}\right)
\EE
\\
where $\overline{\mu}^{2}=4\pi\mu^{2}e^{-\gamma_{E}}$ is the rationalized mass scale that results from defining the theory in $d\neq 4$. As for $K_{m}$, since this integral does not converge even in $d=1$, it is not clear at all what its dimensionally regularized expression should be. However, if we assume eq.~\eqref{abh} to hold also in dimreg, then -- modulo an irrelevant $m$-independent constant -- we are naturally lead to define $K_{m}$ as\\
\BE
K_{m}=\frac{1}{2}\int^{m^{2}}_{0}d\overline{m}^{2}\ J_{\overline{m}}=-\frac{m^{4}}{64\pi^{2}}\ \left(\frac{2}{\epsilon}-\ln\frac{m^{2}}{\overline{\mu}^{2}}+\frac{1}{2}\right)
\EE
\\
If we now introduce an $\epsilon$-dependent mass scale $\Lambda_{\epsilon}$ (not to be confused with the cutoff of the previous renormalization scheme), defined so that\\
\BE\label{aby}
\frac{2}{\epsilon}-\ln\frac{\Lambda^{2}_{\epsilon}}{\overline{\mu}^{2}}+1=0\qquad\qquad(\Lambda_{\epsilon}=\overline{\mu}\, e^{1/\epsilon+1/2})
\EE
then we can re-express our three divergent integrals in the form
\begin{align}
\notag K_{m}=\frac{m^{4}}{64\pi^{2}}\ \left(\ln\frac{m^{2}}{\Lambda_{\epsilon}^{2}}-\frac{1}{2}\right)\qquad\qquad
J_{m}=\frac{m^{2}}{16\pi^{2}}\ \ln\frac{m^{2}}{\Lambda_{\epsilon}^{2}}
\end{align}
\BE\label{abx}
\frac{\partial J_{m}}{\partial m^{2}}=\frac{1}{16\pi^{2}}\ \left(\ln\frac{m^{2}}{\Lambda_{\epsilon}^{2}}+1\right)
\EE
\\
Observe how radically different these results are from those given by eq.~\eqref{abv}. First of all, the quadratic divergence of $J_{m}$ has disappeared. This is a well known feature of dimensional regularization, and ultimately the main reason why dimreg is adopted for regularizing the gauge theories. Second of all, while the $\Lambda$ of eq.~\eqref{abv} is a cutoff -- hence a very large mass scale --, the $\Lambda_{\epsilon}$ of eq.~\eqref{abx}, defined by eq.~\eqref{aby}, is either a very large scale for $\epsilon>0$ (i.e. $d<4$), or a very small scale for $\epsilon<0$ (i.e. $d>4$). Correspondingly, we have
\BE
\ln\frac{m^{2}}{\Lambda_{\epsilon}^{2}}\ \begin{cases}<0&\quad d<4\\>0&\quad d>4\end{cases}
\EE
\\
in the regions of the GEP in which the mass parameter $m$ has a physical meaning. It follows that in dimreg $J_{m}$ and $\partial J_{m}/\partial m^{2}$ are not always respectively positive and negative definite, as implied by the formal definitions \eqref{abp} and \eqref{abw}. For this reason, we find ourselves in the following interesting situation.\\
\\
\underline{Case 1: $d<4$}\ \ If $d<4$, then $J_{m}<0$, at variance with the formal definition given in \eqref{abp}. In particular, the gap equation does not admit non-zero solutions (provided that $\lambda_{B}>0$, as it should be). On the other hand, since $\partial J_{m}/\partial m^{2}$ is not a priori negative, the full equation $\partial V_{G}/\partial m^{2}=0$ admits the solution\\
\BE
\frac{\partial J_{m}}{\partial m^{2}}=0\qquad\Longrightarrow\qquad m=\Lambda_{\epsilon}/\sqrt{e}
\EE
\\
This $m$ does not depend on the coupling, is of order $\Lambda_{\epsilon}$ and is actually a maximum for the GEP. Therefore we must conclude that for $d<4$ the GEP does not admit non-zero minima.\\
\\
\underline{Case 2: $d>4$}\ \ If $d>4$, then $J_{m}>0$ and the gap equation admits the non-zero solution
\BE
m_{0}=\Lambda_{\epsilon}\,e^{1/\alpha_{B}}\qquad\qquad\qquad\alpha_{B}=\frac{\lambda_{B}}{16\pi^{2}}
\EE
\\
Since again $\partial J_{m}/\partial m^{2}$ is not a priori negative, the GEP has a stationary point due to the vanishing of $\partial J_{m}/\partial m^{2}$; for the same reason, we must check explicitly whether the $m_{0}$ given above is a minimum or a maximum. In order to do so, we replace the mass scale $\Lambda_{\epsilon}$ in the GEP with the inverse solution $\Lambda_{\epsilon}=m_{0}\,e^{-1/\alpha_{B}}$ and express $V_{G}(m^{2})$ as a function of $m^{2}$ and $m_{0}^{2}$. An explicit computation shows that
\BE\label{aqd}
V_{G}(m^{2})=\frac{m^{4}}{128\pi^{2}}\ \left(\alpha_{B}\,\ln^{2}\frac{m^{2}}{m_{0}^{2}}+2\, \ln\frac{m^{2}}{m_{0}^{2}}-1\right)
\EE
\\
A plot of a normalized version of $V_{G}$ versus $m/m_{0}$ is shown in Fig.\ref{FD} for different values of $\alpha_{B}$. The third extremum (due to $\partial J_{m}/\partial m^{2}=0$, the other two being $m=0$ and $m=m_{0}$) is given by $m=m_{0}\,e^{-1/\alpha_{B}-1/2}<m_{0}$. Here the GEP has value\\
\BE\label{abn}
V_{G}(m_{0}^{2}\,e^{-2/\alpha_{B}-1})=\frac{m_{0}^{4}\,e^{-4/\alpha_{B}-2}}{128\pi^{2}}(1+\alpha_{B})>0
\EE
\\
On the other hand, for $m=m_{0}$ and $m=0$ the GEP has values\\
\BE\label{ajq}
V_{G}(0)=0\qquad\qquad\qquad V_{G}(m_{0}^{2})=-\frac{m_{0}^{4}}{128\pi^{2}}<0
\EE
\\
Therefore we conclude that for $d>4$ the value $m=m_{0}\,e^{-1/\alpha_{B}-1/2}$ is an absolute maximum, the values $m=0$ and $m=m_{0}$ are relative minima and in particular $m=m_{0}$ is an absolute minimum.\\
The GEP approach clearly predicts the existence of a non-zero minimum for the vacuum energy density of massless $\lambda\phi^{4}$ theory in $d=4+|\epsilon|$. By renormalization, this feature is inherited by the $d\to 4^{+}$ theory: the perturbative vacuum of massless $\lambda\phi^{4}$ theory, defined in dimreg by letting $d\to 4^{+}$, is indeed massive, and the scale of the theory is set precisely by the finite value of $m_{0}$. Of course, since the original classical theory in $d=4$ was invariant under scale transformations, the mass scale of the model comes from the quantum mechanical breaking of scale invariance and the actual value of $m_{0}$ cannot be predicted from first principles: $m_{0}$ must be determined a posteriori as a free parameter of the theory.\\
Finally, observe that the value of the GEP at its minimum -- i.e. at the only point at which it has a physical meaning -- does not depend on the bare coupling $\alpha_{B}$ and is completely determined by the value of $m_{0}$. Therefore, by fixing $m_{0}$, we obtain a fully renormalized value for the first order vacuum energy density of the theory, independent of the regulators and bare parameters, as it should be.\\
\\
\\
\newpage
\noindent \textit{1.1.3.4 Discussion and conclusions}\\
\\
In the previous paragraphs we have discovered that, depending on the renormalization scheme used to define the model, the GEP analysis of massless $\lambda\phi^{4}$ theory leads to very different conclusions with respect to the issue of mass generation.\\
In the presence of a sharp cutoff, the quadratic divergence of the tadpole diagram is found to generate a non-zero mass for the scalar particle. This is not satisfactory in two respects. First of all, it is not clear whether such a mass should be interpreted as a genuinely dynamical mass. Second of all, the quadratic divergence is known to break gauge invariance, so that if we are to regard massless $\lambda\phi^{4}$ theory as a toy model for mass generation in Yang-Mills theory, then we must discard results which depend on such a divergence.\\
In dimensional regularization, on the other hand, it seems that taking the limit $d\to 4$ from above or from below leads to two very different theories: whereas mass generation is predicted to occur in the $d\to 4^{+}$ regularized theory, the same is not true of the $d\to 4^{-}$ regularized one. This state of affairs is known in the literature and was first pointed out by Stevenson in \cite{stev5}, where he showed that mass generation in the dimensionally regularized theory with $d<4$ occurs if and only if the $\Bbb{Z}_{2}$ symmetry $\phi\to -\phi$ of the original Lagrangian is spontaneously broken, causing the scalar field $\phi$ to acquire a non-vanishing vacuum expectation value $\avg{\phi}$. Such a breaking is indeed predicted by the GEP equations themselves, provided that $\avg{\phi}$ is treated as a free parameter, rather than set to zero from the beginning as we did in our analysis. By minimizing the GEP with respect to $\avg{\phi}$ as well as $m$, one is able to prove that the massless, symmetric vacuum which we found in our analysis is unstable towards a massive, non-symmetric vacuum. Since we assume the expectation value of a gauge field to vanish in the vacuum, we must discard this solution too, and conclude that the dimensionally regularized theory in $d<4$ is not a suitable model for mass generation in Yang-Mills theory. Therefore, of the three proposed regularization schemes, only dimreg in $d>4$ leads to a viable model for our analysis.\\
One may ask how is it that different renormalization schemes lead to different results. With respect to this issue, we take the view that the choice of a renormalization scheme is part of the definition of the theory, rather than a formal procedure adopted to regularize the divergences in order to incorporate them into renormalized parameters. As a matter of fact, it is common knowledge that -- especially when gauge symmetries are involved -- not all renormalization schemes are equivalent from a physical point of view, or even from the point of view of mathematical consistency. Dimensional regularization has the great advantage of eliminating the symmetry-breaking quadratic divergence of the tadpole diagram from the very beginning, thus leading to a perturbative series which can be renormalized while preserving the symmetries of the theory. In the process of doing so, it modifies some of the formal aspects of the expansion. In the GEP approach this is exemplified by the fact that the divergent integrals $J_{m}$ and $\partial J_{m}/\partial m^{2}$ are not positive and negative definite as they should formally be, leading to physical consequences which, as we saw, include mass generation.

\newpage


\addcontentsline{toc}{section}{1.2 The Gaussian Effective Potential of pure Yang-Mills theory}  \markboth{1.2 The Gaussian Effective Potential of pure Yang-Mills theory}{1.2 The Gaussian Effective Potential of pure Yang-Mills theory}
\section*{1.2 The Gaussian Effective Potential of pure Yang-Mills theory\index{The Gaussian Effective Potential of pure Yang-Mills theory}}

In this section the machinery developed in Sec.1.1.2-3 will be applied to the GEP analysis of Yang-Mills theory. We will start by defining and computing the GEP (Sec.1.2.1-2) and then we will address the issue of renormalization and gauge invariance (Sec.1.2.3). In Sec.1.2.4 the variational status of the Yang-Mills GEP will be investigated in connection to the anticommuting nature of the ghost fields. In Sec.1.2.5 we will show that the purely gluonic contribution to the GEP is formally identical to the GEP of $\lambda\phi^{4}$ theory, so that the analysis of Sec.1.1.3 can be carried over verbatim to Yang-Mills theory. The massless perturbative vacuum of the transverse gluons employed in ordinary perturbation theory is found to be unstable towards a massive vacuum, motivating the massive perturbative expansion of Chapter 2.

\addcontentsline{toc}{subsection}{1.2.1 Mass parameters and the definition of the GEP of Yang-Mills theory}  \markboth{1.2.1 Mass parameters and the definition of the GEP of Yang-Mills theory}{1.2.1 Mass parameters and the definition of the GEP of Yang-Mills theory}
\subsection*{1.2.1 Mass parameters and the definition of the GEP of Yang-Mills theory\index{Mass parameters and the definition of the GEP of Yang-Mills theory}}

In a general covariant gauge, the Faddeev-Popov gauge fixed action of pure Yang-Mills theory is given by
\begin{align}
\mc{S}&=\int d^{d}x\ \bigg\{-\frac{1}{2}\ \partial_{\mu}A_{\nu}^{a}\,(\partial^{\mu}A^{a\,\nu}-\partial^{\nu}A^{a\,\mu})-\frac{1}{2\xi}\ \partial^{\mu} A^{a}_{\mu}\,\partial^{\nu} A^{a}_{\nu}+\partial^{\mu}\cbar^{\,a}\,\partial_{\mu}c^{a}+\\
\notag&\qquad-g\ f_{abc}\ \partial_{\mu}A^{a}_{\nu}\,A^{b\,\mu}A^{c\,\nu}-\frac{g^{2}}{4}\ f_{abc}\,f_{ade}\ A_{\mu}^{b}\,A_{\nu}^{c}\ A^{d\,\mu}\,A^{e\,\nu}-g\ f_{abc}\ \partial^{\mu}\cbar^{\,a}c^{b}A_{\mu}^{c}+\mathcal{L}_{\tx{c.t.}}\bigg\}
\end{align}
\\
where $\mathcal{L}_{\tx{c.t.}}$ contains the appropriate renormalization counterterms. In ordinary perturbation theory, one chooses as the zero-order action $\mc{S}_{0}$\\
\begin{align}\label{bab}
\mc{S}_{0}&=\int d^{d}x\ \bigg\{-\frac{1}{2}\ \partial_{\mu}A_{\nu}^{a}\,(\partial^{\mu}A^{a\,\nu}-\partial^{\nu}A^{a\,\mu})-\frac{1}{2\xi}\ \partial^{\mu} A^{a}_{\mu}\,\partial^{\nu} A^{a}_{\nu}+\partial^{\mu}\cbar^{\,a}\,\partial_{\mu}c^{a}\bigg\}=\\
\notag&=\int\frac{d^{d}p}{(2\pi)^{d}}\ \bigg\{-\frac{1}{2}\ A_{\mu}^{a}\ \delta_{ab}\ p^{2}\, \left[t^{\mu\nu}(p)+\frac{1}{\xi}\ \ell^{\mu\nu}(p)\right]\, A_{\nu}^{b}+\cbar^{\,a}\ \delta_{ab}\,p^{2}\ c^{b}\bigg\}
\end{align}
\\
where $t^{\mu\nu}(p)$ and $\ell^{\mu\nu}(p)$ are the transverse and longitudinal projection tensors,\\
\BE
t^{\mu\nu}(p)=\eta^{\mu\nu}-\frac{p^{\mu}p^{\nu}}{p^{2}}\qquad\qquad \ell^{\mu\nu}(p)=\frac{p^{\mu}p^{\nu}}{p^{2}}
\EE
\\
The corresponding gluon and ghost bare propagators $\mc{D}_{0\,\mu\nu}^{ab}$ and $\mc{G}^{ab}_{0}$ are readily determined to be
\BE
\mc{D}_{0\,\mu\nu}^{ab}(p)=\delta^{ab}\ \bigg\{\frac{-i\, t_{\mu\nu}(p)}{p^{2}+i\epsilon}+\xi\ \frac{-i\, \ell_{\mu\nu}(p)}{p^{2}+i\epsilon}\bigg\}\qquad\qquad \mc{G}^{ab}_{0}(p)=\delta^{ab}\ \frac{i}{p^{2}+i\epsilon}
\EE
\\
$\mc{D}_{0\,\mu\nu}^{ab}$ and $\mc{G}^{ab}_{0}$ are massless free particle propagators.\\
In order to define the GEP of Yang-Mills theory, we must add to eq.~\eqref{bab} appropriate mass terms for the gluon and ghost fields. Since the gluon propagator has a transverse and a longitudinal component, there is no reason to define a unique mass parameter for the transverse and longitudinal gluons. Indeed, in momentum space, the most general action term for the masses of the gluons and ghosts has the form\\
\BE
\Delta \mc{S}=\int\frac{d^{d}p}{(2\pi)^{d}}\ \bigg\{\frac{1}{2}\ A_{\mu}^{a}\ \delta_{ab}\ \left[m^{2}\ t^{\mu\nu}(p)+\frac{1}{\xi}\ m_{L}^{2}\ \ell^{\mu\nu}(p)\right]\, A_{\nu}^{b}-\cbar^{\,a}\ \delta_{ab}\,M^{2}\ c^{b}\bigg\}
\EE
where $M$ is the mass parameter for the ghosts, whereas $m$ and $m_{L}$ are the mass parameters for the transverse and longitudinal gluons respectively. In principle, we may compute the GEP by using as the zero-order action the sum $\mc{S}_{0}+\Delta\mc{S}$, with $\mc{S}_{0}$ given by eq.~\eqref{bab}. However, non-perturbatively, we know that due to gauge invariance the longitudinal part of the gluon propagator does not get corrected by the interactions \cite{peskin,itzyk}, so that in particular the longitudinal gluons cannot develop a mass. By setting $m_{L}=0$ from the very start, we obtain the exact, non-perturbative result for the longitudinal gluons. Therefore we will limit ourselves to study the GEP as a function of the transverse gluon and ghost mass at zero longitudinal gluon mass, and define as the zero-order action $\mc{S}_{0}$ for the computation of the GEP the quantity\\
\BE
\mc{S}_{0}=i\ \int\frac{d^{d}p}{(2\pi)^{d}}\ \left\{\frac{1}{2}\ A_{\mu}^{a}\ \mc{D}_{ab}^{\mu\nu}(p)^{-1}\, A_{\nu}^{b}+\cbar^{\,a}\ \mc{G}_{ab}(p)^{-1}\, c^{b}\right\}
\EE
where
\begin{align}
\mc{D}_{\mu\nu}^{ab}(p)^{-1}&=i\ \delta^{ab}\ \bigg\{(p^{2}-m^{2}+i\epsilon)\ t_{\mu\nu}(p)+\frac{1}{\xi}\ (p^{2}+i\epsilon)\ \ell_{\mu\nu}(p)\bigg\}\\
\notag \mc{G}^{ab}(p)^{-1}&=-i\ \delta^{ab}\ (p^{2}-M^{2}+i\epsilon)
\end{align}
\\
$\mc{D}_{\mu\nu}^{ab}$ and $\mc{G}^{ab}$ being the modified, massive gluon and ghost bare propagators\\
\begin{align}
\mc{D}_{\mu\nu}^{ab}(p)&=\delta^{ab}\ \bigg\{\frac{-i\, t_{\mu\nu}(p)}{p^{2}-m^{2}+i\epsilon}+\xi\ \frac{-i\, \ell_{\mu\nu}(p)}{p^{2}+i\epsilon}\bigg\}\\
\notag \mc{G}^{ab}(p)&=\delta^{ab}\ \frac{i}{p^{2}-M^{2}+i\epsilon}
\end{align}
\\
Accordingly, the interaction action $\mc{S}_{\tx{int}}=\mc{S}-\mc{S}_{0}$ reads\\
\begin{align}
&\mc{S}_{\tx{int}}=-\int\frac{d^{d}p}{(2\pi)^{d}}\ \bigg\{\frac{1}{2}\ \delta_{ab}\ m^{2}\ t^{\mu\nu}(p)\, A_{\mu}^{a}(p)\,A_{\nu}^{b}(-p)-\delta_{ab}\,M^{2}\ \cbar^{\,a}(p)\, c^{b}(p)\bigg\}+\\
\notag&-\int d^{d}x\ \bigg\{g_{B}\, f_{abc}\ \partial_{\mu}A^{a}_{\nu}A^{b\,\mu}\,A^{c\,\nu}+\frac{g^{2}_{B}}{4}\ f_{abc}\,f_{ade}\ A_{\mu}^{b}A_{\nu}^{c}A^{d\,\mu}A^{e\,\nu}+g_{B}\, f_{abc}\ \partial^{\mu}\cbar^{\,a}c^{b}A_{\mu}^{c}\bigg\}
\end{align}
\\
where the first line comes from the additional mass terms in $\mc{S}_{0}$ and we have expressed the second line in function of the bare strong coupling constant $g_{B}$, rather than its renormalized value $g$, just as we did in Sec.1.1.3 for $\lambda\phi^{4}$ theory.\\
Since both $\mc{S}_{0}$ and $\mc{S}_{\tx{int}}$ depend on $m$ and $M$, the GEP of Yang-Mills theory is a function of two mass parameters. Its defining expression is obtained by specializing eq.~\eqref{acz} to our choice of $\mc{S}_{0}$ and $\mc{S}_{\tx{int}}$ and reads\\
\BE\label{baq}
V_{G}(m,M)=\frac{i}{\mc{V}_{d}}\ \ln\int \mathcal{D}A\,\mc{D}\cbar\,\mc{D}c\ e^{i\mathcal{S}_{0}(m,M)}-\frac{1}{\mc{V}_{d}}\ \avg{\mc{S}_{\text{int}}(m,M)}_{0}
\EE
\newpage{}
\addcontentsline{toc}{subsection}{1.2.2 Computation of the GEP}  \markboth{1.2.2 Computation of the GEP}{1.2.2 Computation of the GEP}
\subsection*{1.2.2 Computation of the GEP\index{Computation of the GEP}}

\begin{figure}[H]
\centering
\includegraphics[width=0.95\textwidth]{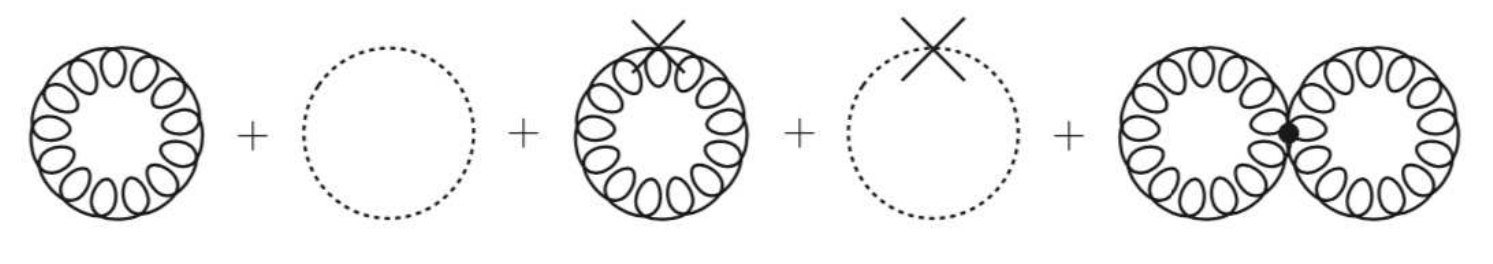}
\
\caption{Diagrams which contribute to the GEP of Yang-Mills theory.}\label{FE}
\end{figure}

Let us move on to the explicit computation of $V_{G}$. Since the vacuum expectation value of an odd number of field operators with respect to the action of a free theory is zero, the average of $\mc{S}_{\tx{int}}$ in eq.~\eqref{baq} reduces to\\
\begin{align}
\avg{\mc{S}_{\tx{int}}}_{0}&=-\int\frac{d^{d}p}{(2\pi)^{d}}\ \bigg\{\frac{1}{2}\ \delta_{ab}\ m^{2}\ t^{\mu\nu}(p)\, \avg{A_{\mu}^{a}(p)\,A_{\nu}^{b}(-p)}_{0}-\delta_{ab}\,M^{2}\ \avg{\cbar^{\,a}(p)\, c^{b}(p)}_{0}\bigg\}+\\
\notag&\qquad-\int d^{d}x\ \bigg\{\frac{g^{2}_{B}}{4}\ f_{abc}\,f_{ade}\ \avg{A_{\mu}^{b}A_{\nu}^{c}A^{d\,\mu}A^{e\,\nu}}_{0}\bigg\}
\end{align}
\\
As for the zero-order term of eq.~\eqref{baq}, we observe that the functional integral of $e^{i\mc{S}_{0}}$ can be factorized into the product of two integrals,\\
\BE
\int \mathcal{D}A\,\mc{D}\cbar\,\mc{D}c\ e^{i\mathcal{S}_{0}(m,M)}=\bigg(\int \mathcal{D}A\ e^{i\mathcal{S}_{0}^{(A)}(m)}\bigg)\bigg(\int\mc{D}\cbar\,\mc{D}c\ e^{i\mathcal{S}_{0}^{(c)}(M)}\bigg)
\EE
\\
where $\mc{S}^{(A)}_{0}$ and $\mc{S}^{(c)}_{0}$ are the $A$-dependent and $c$-dependent contributions to $\mc{S}_{0}$. Therefore, a preliminary expression for the GEP of Yang-Mills theory is given by\\
\begin{align}
V_{G}&=\frac{i}{\mathcal{V}_{d}}\ \ln\int \mathcal{D}A\ e^{i\mathcal{S}_{0}^{(A)}(m)}+\frac{i}{\mathcal{V}_{d}}\ \ln\int\mc{D}\cbar\,\mc{D}c\ e^{i\mathcal{S}_{0}^{(c)}(M)}+\\
\notag&+\frac{1}{\mathcal{V}_{g}}\int\frac{d^{d}p}{(2\pi)^{d}}\ \bigg\{\frac{1}{2}\ \delta_{ab}\ m^{2}\ t^{\mu\nu}(p)\, \avg{A_{\mu}^{a}(p)\,A_{\nu}^{b}(-p)}_{0}+\delta_{ab}\,M^{2}\ \avg{c^{a}(p)\,\cbar^{\,b}(p)}_{0}\bigg\}\\
\notag&+\frac{1}{\mathcal{V}_{d}}\ \frac{g_{B}^{2}}{4}\ f_{abc}\,f_{ade}\int d^{d}x\ \avg{A_{\mu}^{b}(x)\,A_{\nu}^{c}(x)\,A^{d\,\mu}(x)\,A^{e\,\nu}(x)}_{0}
\end{align}
\\
where inside the ghost quadratic average we have exchanged the order of the Grassmann fields. Again, to each of the terms in the equation we may associate a diagram. To the logarithmic terms we associate closed loops with no vertices (first and second diagram in Fig.\ref{FE}), with a wiggly line for the gluon and a dotted line for the ghosts; to the quadratic terms we associate closed loops with a single two-point vertex, proportional to the respective mass parameters squared (third and fourth diagram in Fig.\ref{FE}); to the quartic term we associate a double loop with a four-point vertex, proportional to $g_{B}^{2}$ (last diagram in Fig.\ref{FE}). We will now proceed to evaluate these diagrams.\\
\\
\\
\newpage
\noindent \textit{1.2.2.1 Logarithmic terms}\\
\\
Let us start from the logarithmic contributions to the GEP. In what follows we denote by $det$ the functional determinant and by $Tr$ the functional trace. Moreover, we define transverse and longitudinal bare gluon propagators $\mc{D}_{T}$ and $\mc{D}_{L}$ such that\\
\BE
\mc{D}_{\mu\nu}^{ab}(p)=\delta^{ab}\ [\mc{D}_{T}(p^{2})\, t_{\mu\nu}(p)+\xi\,\mc{D}_{L}(p^{2})\, \ell_{\mu\nu}(p)]
\EE
We have\\
\begin{align}\label{baw}
\ln\int \mathcal{D}A\ e^{i\mathcal{S}_{0}^{(A)}}&=\ln det[(\mc{D}_{\mu\nu}^{ab})^{1/2}]=\frac{1}{2}\, \ln det(\mc{D}_{\mu\nu}^{ab})=\\
\notag&=\frac{1}{2}\, \ln\left[det(\delta^{ab}\mc{D}_{T}t_{\mu\nu})\ det(\delta^{ab}\xi\mc{D}_{L}\ell_{\mu\nu})\right]=\\
\notag&=\frac{1}{2}\, \ln det(\delta^{ab}\mc{D}_{T}t_{\mu\nu})+\frac{1}{2}\, \ln det(\delta^{ab}\xi\mc{D}_{L}\ell_{\mu\nu})=\\
\notag&=\frac{1}{2}\ Tr\ln(\delta^{ab}\mc{D}_{T}t_{\mu\nu})+\frac{1}{2}\ Tr\ln(\delta^{ab}\mc{D}_{L}\ell_{\mu\nu})+\frac{1}{2}\ Tr\ln \xi=\\
\notag&=-iN_{A}(d-1)\,\mc{V}_{d}\, K_{m}-iN_{A}\,\mc{V}_{d}\,K_{0}+\frac{1}{2}\ Tr\ln\xi
\end{align}
\\
where we recall that
\BE
K_{m}=\frac{1}{2}\int\frac{d^{d}k_{E}}{(2\pi)^{d}}\ \ln(k_{E}^{2}+m^{2})
\EE
\\
and $N_{A}=N^{2}-1$, $N$ being the number of colors ($N=3$ for pure gauge QCD). The last term in eq.~\eqref{baw} is canceled \cite{com2} by the logarithm of the ($\xi$-dependent) constant $\mc{C}$ factored out from the partition function by the Faddeev-Popov gauge fixing procedure -- see the Introduction. By keeping only the relevant terms and multiplying eq.~\eqref{baw} by $i/\mc{V}_{d}$ we obtain the zero-order gluon contribution $V_{G}^{(0,A)}$ to the GEP,\\
\BE
V_{G}^{(0,A)}=N_{A}(d-1)\, K_{m}+N_{A}\,K_{0}
\EE
\\
As for the ghost loop, we simply have\\
\BE
\ln\int\mc{D}\cbar\,\mc{D}c\ e^{i\mathcal{S}_{0}^{(c)}}=\ln det[(\mc{G}^{ab})^{-1}]=-Tr\ln(\mc{G}^{ab})=2iN_{A}\,\mc{V}_{d}\,K_{M}
\EE
\\
so that the zero-order ghost contribution $V_{G}^{(0,c)}$ to the GEP is given by\\
\BE
V_{G}^{(0,c)}=-2N_{A}\,K_{M}
\EE
\\
\\
\textit{1.2.2.2 Quadratic terms}\\
\\
The quadratic averages are readily evaluated:\\
\BE
\avg{A_{\mu}^{a}(p)\,A_{\nu}^{b}(-p)}_{0}=\mc{V}_{d}\ \mc{D}_{\mu\nu}^{ab}(p)\qquad\qquad \avg{c^{a}(p)\,\cbar^{\,b}(p)}_{0}=\mc{V}_{d}\ \mc{G}^{ab}(p)
\EE
\newpage
\noindent If follows that the gluon loop with the two-point vertex contributes to the GEP with a term $V_{G}^{(2,A)}$ given by\\
\begin{align}
V_{G}^{(2,A)}&=\frac{1}{2}\ m^{2}\int\frac{d^{d}p}{(2\pi)^{d}}\ \ \delta_{ab}\ t^{\mu\nu}(p)\, \mc{D}_{\mu\nu}^{ab}(p)=-\frac{N_{A}(d-1)}{2}\ m^{2}\ J_{m}
\end{align}
\\
whereas the contribution $V_{G}^{(2,c)}$ due to the ghost loop with the two-point vertex is\\
\BE
V_{G}^{(2,c)}=M^{2}\int\frac{d^{d}p}{(2\pi)^{d}}\ \delta_{ab}\,\mc{G}^{ab}(p)=N_{A}\,M^{2}\,J_{M}
\EE
Recall that
\BE
J_{m}=\int\frac{d^{d}k_{E}}{(2\pi)^{d}}\frac{1}{k^{2}_{E}+m^{2}}
\EE
\\
\\
\textit{1.2.2.3 Quartic term}\\
\\
As for the quartic average, we have\\
\begin{align}
\avg{A_{\mu}^{b}(x)\,A_{\nu}^{c}(x)\,A^{d\,\mu}(x)\,A^{e\,\nu}(x)}_{0}=\lim_{\{y_{i}\}\to x}\ &\Big\{\mc{D}_{\mu\nu}^{bc}(x-y_{1})\,\mc{D}^{de\,\mu\nu}(y_{2}-y_{3})+\\
\notag&+\mc{D}_{\mu}^{bd\,\mu}(x-y_{2})\,\mc{D}^{ce\,\nu}_{\nu}(y_{1}-y_{3})+\\
\notag&+\mc{D}_{\mu\nu}^{be}(x-y_{3})\,\mc{D}^{cd\,\mu\nu}(y_{1}-y_{2})\Big\}
\end{align}
\\
where $\mc{D}_{\mu\nu}^{ab}(x)$ is the massive gluon propagator in coordinate space,
\\
\BE
\mc{D}_{\mu\nu}^{ab}(x)=\int\frac{d^{d}k}{(2\pi)^{d}}\ e^{-ik\cdot x}\ \mc{D}_{\mu\nu}^{ab}(k)
\EE
\\
When evaluated at $x=0$, the latter can be expressed as\\
\begin{align}
\lim_{x\to 0}\ \mc{D}_{\mu\nu}^{ab}(x)&=\delta^{ab}\int\frac{d^{d}k}{(2\pi)^{d}}\ \bigg\{\frac{-i\, t_{\mu\nu}(k)}{k^{2}-m^{2}+i\epsilon}+\xi\ \frac{-i\, \ell_{\mu\nu}(k)}{k^{2}+i\epsilon}\bigg\}=\\
\notag&=\delta^{ab}\ \eta_{\mu\nu}\int\frac{d^{d}k}{(2\pi)^{d}}\ \bigg\{\left(1-\frac{1}{d}\right)\ \frac{-i}{k^{2}-m^{2}+i\epsilon}+\frac{\xi}{d}\ \frac{-i}{k^{2}+i\epsilon}\bigg\}=\\
\notag&=-\delta^{ab}\ \eta_{\mu\nu}\ \bigg(\frac{d-1}{d}\ J_{m}+\frac{\xi}{d}\ J_{0}\bigg)
\end{align}
Therefore\\
\BE
\avg{A_{\mu}^{b}(x)\,A_{\nu}^{c}(x)\,A^{d\,\mu}(x)\,A^{e\,\nu}(x)}_{0}=d\,(\delta^{bc}\delta^{de}+\delta^{bd}\delta^{ce}\ d+\delta^{be}\delta^{cd})\ \bigg(\frac{d-1}{d}\ J_{m}+\frac{\xi}{d}\ J_{0}\bigg)^{2}
\EE
\\
and the gluon double loop contributes to the GEP with a term $V_{G}^{(4,A)}$ given by
\newpage
\begin{align}
V_{G}^{(4,A)}&=\frac{g_{B}^{2}}{4}\ f_{abc}\,f_{ade}\ d\,(\delta^{bc}\delta^{de}+\delta^{bd}\delta^{ce}\ d+\delta^{be}\delta^{cd})\ \bigg(\frac{d-1}{d}\ J_{m}+\frac{\xi}{d}\ J_{0}\bigg)^{2}=\\
\notag&=\frac{Ng_{B}^{2}}{4}\ N_{A}\ d(d-1)\,\bigg(\frac{d-1}{d}\ J_{m}+\frac{\xi}{d}\ J_{0}\bigg)^{2}
\end{align}
\\
where we have used $f_{abb}=0$, $f_{abc}f_{abd}=N\delta_{cd}$.\\
\\
By adding up the five contributions $V_{G}^{(0,A)}$, $V_{G}^{(0,c)}$, $V_{G}^{(2,A)}$, $V_{G}^{(2,c)}$ and $V_{G}^{(4,A)}$, we find our final expression for the GEP of Yang-Mills theory in an arbitrary covariant gauge and renormalization scheme:\\
\begin{align}\label{bhi}
\notag V_{G}(m,M)&=N_{A}(d-1)\,K_{m}+N_{A}\,K_{0}-2N_{A}\,K_{M}-N_{A}(d-1)\,\frac{m^{2}}{2}\,J_{m}+N_{A}\,M^{2}\,J_{M}+\\
&\qquad+\frac{Ng_{B}^{2}}{4}\ N_{A}\ d(d-1)\,\bigg(\frac{d-1}{d}\ J_{m}+\frac{\xi}{d}\ J_{0}\bigg)^{2}
\end{align}
\\
Observe that $V_{G}(m,M)$ is a sum of two terms, the first one depending only on the gluon mass parameter squared $m^{2}$ and the second one depending only on the ghost mass parameter squared $M^{2}$,
\BE
V_{G}(m^{2},M^{2})=V_{G}^{(A)}(m^{2})+V_{G}^{(c)}(M^{2})
\EE
where\\
\BE
\label{bhj}\frac{V_{G}^{(A)}(m^{2})}{N_{A}}=(d-1)\,K_{m}+K_{0}-(d-1)\,\frac{m^{2}}{2}\,J_{m}+\frac{Ng_{B}^{2}}{4}\ \frac{(d-1)^{3}}{d}\,\bigg(J_{m}+\frac{\xi}{d-1}\ J_{0}\bigg)^{2}
\EE
\BE
\frac{V_{G}^{(c)}(M^{2})}{2N_{A}}=-\left(K_{M}-\frac{M^{2}}{2}\, J_{M}\right)
\EE
\\
Therefore the stationary points of the GEP can be determined by separately extremizing $V_{G}^{(A)}$ with respect to $m^{2}$ and $V_{G}^{(c)}$ with respect to $M^{2}$.
\\
\\
\addcontentsline{toc}{subsection}{1.2.3 Renormalization and gauge invariance}  \markboth{1.2.3 Renormalization and gauge invariance}{1.2.3 Renormalization and gauge invariance}
\subsection*{1.2.3 Renormalization and gauge invariance\index{Renormalization and gauge invariance}}

In order to find the stationary points of the GEP, we must first of all regularize the integrals $K$ and $J$ by choosing a suitable renormalization scheme. In Sec.1.1.3 we have discussed the renormalization of the GEP of massless $\lambda\phi^{4}$ theory. There we saw that in cutoff regularization the squared mass generated for the scalar particle is proportional to the quadratic divergence of the tadpole diagram; we then discarded the scheme with the justification that quadratic divergences are known to spoil the gauge invariance of gauge theories. Let us see how this comes about in the GEP analysis.\\
Our computation lead us to the expression \eqref{bhi} for the GEP of Yang-Mills theory. The gauge dependence of the GEP comes entirely from the product $\xi J_{0}$ in the last term of eq.~\eqref{bhj}. In cutoff regularization,
\newpage
\BE
\xi J_{0}=\frac{\xi}{8\pi^{2}}\int^{\Lambda}_{0}dk_{E}\ k_{E}=\frac{\xi\Lambda^{2}}{16\pi^{2}}\neq 0
\EE
\\
Therefore in the presence of a sharp cutoff the GEP is explicitly gauge dependent, and the gauge dependence is caused precisely by the quadratic divergence of the tadpole diagram. Of course, a gauge-dependent GEP will have gauge-dependent minima which cannot be physically meaningful. We must then conclude that the sharp cutoff is not suitable for regularizing the GEP of Yang-Mills theory.\\
On the other hand, consider what happens in dimensional regularization. In dimreg, $J_{m}$ is given by eq.~\eqref{abv},
\BE
J_{m}=-\frac{m^{2}}{16\pi^{2}}\ \left(\frac{2}{\epsilon}-\ln\frac{m^{2}}{\overline{\mu}^{2}}+1\right)
\EE
\\
By taking the limit $m\to 0$, we find that $J_{0}=0$. It follows that in dimensional regularization the GEP, as well as its extrema, are gauge-independent.\\
In light of what we just saw, we take the following standpoint on the renormalization of the GEP of Yang-Mills theory (cf. Sec.1.1.3.4). We interpret dimensional regularization as the renormalization scheme which, by removing the quadratic divergence from the equations, preserves the gauge invariance of the theory and allows for the definition of a physically meaningful GEP. It is our scheme of choice for the regularization of the GEP of Yang-Mills theory, and the only one that we will consider in what follows. Since in dimreg $K_{m}\propto m^{4}$ -- eq.~\eqref{abx} --, modulo an inessential $m$-independent constant, $K_{0}$ as well as $J_{0}$ vanishes and we are left with the following regularized expressions for the gluonic and ghost contributions to the GEP:\\
\BE\label{bqq}
\frac{V_{G}^{(A)}(m^{2})}{N_{A}(d-1)}=K_{m}-\frac{m^{2}}{2}\,J_{m}+\frac{Ng_{B}^{2}}{4}\ \frac{(d-1)^{2}}{d}\ J_{m}^{2}
\EE
\BE\label{bqp}
\frac{V_{G}^{(c)}(M^{2})}{2N_{A}}=-\left(K_{M}-\frac{M^{2}}{2}\, J_{M}\right)
\EE
\\
In the above equations, $K$ and $J$ are given by the dimensionally regularized expressions in eq.~\eqref{abx}, where the mass scale $\Lambda_{\epsilon}$ is defined by eq.~\eqref{aby}.
\\
\addcontentsline{toc}{subsection}{1.2.4 The ghost mass parameter and the variational status of the GEP}  \markboth{1.2.4 The ghost mass parameter and the variational status of the GEP}{1.2.4 The ghost mass parameter and the variational status of the GEP}
\subsection*{1.2.4 The ghost mass parameter and the variational status of the GEP\index{The ghost mass parameter and the variational status of the GEP}}

As we saw, the GEP of Yang-Mills theory depends on the ghost mass $M$ through the ghost contribution $V_{G}^{(c)}(M^{2})$, eq.~\eqref{bqp}. Observe that $V_{G}^{(c)}$ is independent of the strong coupling constant $g_{B}$: as a matter of fact, $V_{G}^{(c)}$ is the same GEP that would be obtained from the free ghost Lagrangian alone. This feature of the Yang-Mills GEP challenges the applicability of the GEP analysis to the ghost sector, leading us to question whether it makes any sense at all to define a mass parameter for the ghost in the first place. After all -- as we showed in the Introduction -- through the non-perturbative lattice calculations the dressed ghost propagator was found to be massless in the infrared as well as in the ultraviolet regime. Since however -- to our knowledge -- the masslessness of the ghosts at low energies has not yet been proven analytically from first principles, it is still of some importance to take up this issue explicitly and ask whether it is possible to show through a GEP analysis that the ghosts do not acquire a dynamical mass in the infrared. The answer to this question turns out to be in the positive. In order to reach this conclusion, we start by studying the stationary points of the ghost GEP.\\
By taking the derivative of $V_{G}^{(c)}$ with respect to $M^{2}$ we find that\\
\BE
\frac{\partial V_{G}^{(c)}}{\partial M^{2}}=N_{A}\,M^{2}\ \frac{\partial J_{M}}{\partial M^{2}}
\EE
\\
so that the stationary ghost masses are given by $M=0$ and -- cf. eq.~\eqref{abx} -- $M=\Lambda_{\epsilon}/\sqrt{e}$. The second derivative of the ghost GEP, on the other hand, reads\\
\BE
\frac{\partial^{2} V_{G}^{(c)}}{\partial (M^{2})^{2}}=N_{A}\,\left(\frac{\partial J_{M}}{\partial M^{2}}+M^{2}\,\frac{\partial^{2} J_{M}}{\partial (M^{2})^{2}}\right)=\frac{N_{A}}{16\pi^{2}}\,\left(\ln\frac{M^{2}}{\Lambda_{\epsilon}^{2}}+2\right)
\EE
\\
Since the second derivative is negative at $M=0$ and positive at $M=\Lambda_{\epsilon}/\sqrt{e}$, $M=0$ is a maximum, whereas $M=\Lambda_{\epsilon}/\sqrt{e}$ is a minimum. At first sight this result may seem to imply that the ghosts actually acquire a mass; $M=\Lambda_{\epsilon}/\sqrt{e}$, however, does not depend on the coupling and as such cannot be interpreted as being due to the interactions. The maximum $M=0$, on the other hand, would be a far more acceptable result: not only its being zero would explain why the ghost GEP is independent of the coupling, but $M=0$ is also known to be the correct, non-perturbative result for the ghosts. So how is it that for the ghost sector it looks like maximizing the GEP, rather than minimizing it, is the correct procedure to follow in order to obtain physically meaningful results?\\
The answer to this question lies in the non-commuting nature of the ghost fields, as related to the variational status of the GEP in Yang-Mills theory. As discussed in Sec.1.1.2, the GEP approach is motivated by the Jensen-Feynman inequality, which shows that for a \textit{bosonic} field theory the Gaussian Effective Potential is bounded from below by the exact vacuum energy density of the system. However, once a gauge has been fixed through the Faddeev-Popov procedure, Yang-Mills theory becomes a theory of bosonic gluons on the one side and \textit{fermionic} ghosts -- yet with a bosonic kinetic action -- on the other side. For this reason, one must carefully check whether in the presence of such fields the GEP is still a consistent variational approximation to the exact vacuum energy density of the system, i.e. whether the latter still constitutes a lower bound for the GEP. We will now go through the computations needed to address this issue. A more detailed derivation may be found in \cite{com2}, where the variational statement is analyzed and extended to finite temperatures.\\
\\
In its Euclidean formulation, the exact vacuum energy density of Yang-Mills theory $\mathcal{E}$ is given by the equation\\
\BE
e^{-\mc{V}_{d}\,\mc{E}}=\int\mc{D}A\,\mc{D}\cbar\,\mc{D}c\ e^{-\mc{S}}
\EE
\\
where now $\mc{V}_{d}$ and $\mc{S}$ are the \textit{Euclidean} volume and action of the system. The presence of the anticommuting ghost fields $c$ and $\cbar$ spoils the direct applicability of the Jensen inequality, which states that for any positive (bosonic) probability measure with associated average $\avg{\cdot}$ and any function $f$ on the measure space $\avg{e^{f}}\geq e^{\avg{f}}$ \cite{feyn2}. One way to avoid this obstacle is to go one step backward in the derivation of the Faddeev-Popov Yang-Mills action and integrate out the ghost fields, so that the functional averages are taken only with respect to the gluon configurations. This integration leads to\\
\BE\label{bzz}
e^{-\mc{V}_{d}\,\mc{E}}=\int\mc{D}A\ e^{-\mc{S}'}\ \det(\mc{M}_{FP})
\EE
where $\mc{S}'$ is the pure gluonic (albeit gauge-fixed) action, obtained for instance from $\mc{S}$ by setting to zero the ghost fields, and $\det(\mc{M}_{FP}(A))=\det(-\partial\cdot D[A])$ -- $D[A]$ being the gauge covariant derivative associated to the gluon field -- is the Faddeev-Popov determinant. Rewriting the determinant as $\det(\mc{M}_{FP})=e^{\ln\det(\mc{M}_{FP})}$ and splitting the gluonic action $\mc{S}'$ as $\mc{S}_{0}'+\mc{S}_{\tx{int}}'$, where $\mc{S}_{0}'$ is the massive gluonic free action, we obtain\\
\BE
\mc{E}=V_{0}'-\frac{1}{\mc{V}_{d}}\ \ln\avg{e^{-\mc{S}_{\tx{int}}'+\ln\det(\mc{M}_{FP})}}_{0}'
\EE
where
\BE
V_{0}'=-\frac{1}{\mc{V}_{d}}\ \ln \int\mc{D}A\ e^{-\mc{S}'_{0}}
\EE
\\
is the gluonic kinetic vacuum energy density -- first two terms in eq.~\eqref{bqq} -- and $\avg{\cdot}_{0}'$ is the average with respect to the (bosonic) Euclidean action $\mc{S}_{0}'$. To the second term in eq.~\eqref{bzz} we can now apply the Jensen inequality in the form $\ln \avg{e^{f}}\geq \avg{f}$, to find that\\
\BE
\mc{E}\leq V_{0}'+\frac{1}{\mc{V}_{d}}\ \avg{\mc{S}_{\tx{int}}'}_{0}'-\frac{1}{\mc{V}_{d}}\ \avg{\ln\det(\mc{M}_{FP})}_{0}'
\EE
\\
Since the second term in this equation contains the purely gluonic interactions to first order, the first and second term sum to the gluonic contribution to the GEP, $V_{G}^{(A)}$. On the other hand, the third term contains the ghost kinetic vacuum energy density and the ghost loops to arbitrarily high order. Therefore, if we define an energy term $\delta \mc{E}$ as\\
\BE
\delta\mc{E}=-\frac{1}{\mc{V}_{d}}\ \avg{\ln\det(\mc{M}_{FP})}_{0}'-V_{G}^{(c)}
\EE
\\
where $V_{G}^{(c)}$ is the ghost contribution to the GEP, we find that\\
\BE
\mc{E}\leq V_{G}^{(A)}+V_{G}^{(c)}+\delta \mc{E}
\EE
or
\BE\label{bkq}
V_{G}=V_{G}^{(A)}+V_{G}^{(c)}\geq \mc{E}-\delta\mc{E}
\EE
\\
$\delta\mc{E}$ can actually be shown to be non-negative: rewrite the average in $\delta\mc{E}$ as\\
\BE
\avg{\ln\det(\mc{M}_{FP})}_{0}'=\text{Tr}\avg{\ln(\mc{M}_{FP})}_{0}'
\EE
\\
and apply the Jensen inequality in the form $\ln\avg{f}\geq\avg{\ln f}$ to obtain\\
\BE
-\frac{1}{\mc{V}_{d}}\, \avg{\ln\det(\mc{M}_{FP})}_{0}'=-\frac{1}{\mc{V}_{d}}\, \tx{Tr}\avg{\ln(\mc{M}_{FP})}_{0}'\geq -\frac{1}{\mc{V}_{d}}\,\tx{Tr}\ln \avg{\mc{M}_{FP}}_{0}'
\EE
\\
Since $\mc{M}_{FP}$ is the sum of minus the d'Alembert operator and of an operator which is linear in the gluon field, the latter averages to zero and\\
\BE
-\frac{1}{\mc{V}_{d}}\, \avg{\ln\det(\mc{M}_{FP})}_{0}'\geq -\frac{1}{\mc{V}_{d}}\,\tx{Tr}\ln \avg{\mc{M}_{FP}}_{0}'=-\frac{1}{\mc{V}_{d}}\,\text{Tr}\ln(-\partial^{2})=V_{G}^{(c)}\big|_{M=0}
\EE
\\
Hence, since $V_{G}^{(c)}$ attains its maximum at $M=0$,\\
\BE
\delta\mc{E}\geq V_{G}^{(c)}\big|_{M=0}-V_{G}^{(c)}\geq 0
\EE
Eq.~\eqref{bkq} with $\delta\mc{E}\geq 0$ tells us that the GEP of Yang-Mills theory is not actually bound from below by the exact vacuum energy density $\mc{E}$, but rather that the lower bound is set by an even lower energy density, $\mc{E}-\delta\mc{E}$. Therefore, by minimizing ``too much'' the GEP, we may drift away from the exact vacuum energy density $\mc{E}$ and approach $\mc{E}-\delta\mc{E}$, which is the farther away from $\mc{E}$ the larger is $\delta \mc{E}$. However, observe that by minimizing $\delta\mc{E}$ this potential error can be kept under control: if $\delta\mc{E}$ is very small, then $\mc{E}-\delta\mc{E}$ is not far away from $\mc{E}$ and by minimizing the GEP we still approach the exact vacuum energy density. Since as we saw $\delta\mc{E}\geq V_{G}^{(c)}\big|_{M=0}-V_{G}^{(c)}$, in order for $\delta\mc{E}$ to be the smallest the ghost GEP $V_{G}^{(c)}$ must approach its maximal value $V^{(c)}_{G}\big|_{M=0}$. Therefore, the Yang-Mills GEP can still be used as a variational tool provided that its ghost contribution is \textit{maximized} rather than minimized. In the next section we will give further evidence that the Yang-Mills GEP is a good variational estimate of the exact vacuum energy density of Yang-Mills theory by discussing some results which have been obtained in \cite{com1,com2} at non-zero temperatures.\\
\\
The modified Jensen-Feynman inequality given by eq.~\eqref{bkq} leads us to take $M=0$ -- i.e. the maximum of $V_{G}^{(c)}$ -- as the result of the GEP analysis of the ghost sector. $M=0$ is precisely the result found on the lattice. Since modulo an $M$-independent constant $V^{(c)}_{G}\big|_{M=0}=0$, in the presence of massless ghosts the GEP of Yang-Mills theory is given entirely by its gluonic contribution:\\
\BE
V_{G}(m^{2})=V_{G}^{(A)}(m^{2})
\EE
\\
\addcontentsline{toc}{subsection}{1.2.5 The perturbative vacuum of Yang-Mills theory: dynamical mass generation in the gluon sector}  \markboth{1.2.5 The perturbative vacuum of Yang-Mills theory: dynamical mass generation in the gluon sector}{1.2.5 The perturbative vacuum of Yang-Mills theory: dynamical mass generation in the gluon sector}
\subsection*{1.2.5 The perturbative vacuum of Yang-Mills theory: dynamical mass generation in the gluon sector\index{The perturbative vacuum of Yang-Mills theory: dynamical mass generation in the gluon sector}}

Having set the longitudinal gluon mass $m_{L}=0$ from the beginning and having found that by choosing a ghost mass $M=0$ the GEP is more closely bound to the exact vacuum energy density, we are left with the following expression for the Gaussian Effective Potential of Yang-Mills theory:\\
\BE\label{bql}
V_{G}(m^{2})=N_{A}(d-1)\,\left[K_{m}-\frac{m^{2}}{2}\,J_{m}+\frac{Ng_{B}^{2}}{4}\ \frac{(d-1)^{2}}{d}\ J_{m}^{2}\right]
\EE
\\
Following our discussions in Sec.1.1.3.3-4 and Sec.1.2.3, here the divergent integrals $K_{m}$ and $J_{m}$ must be understood as having been regularized in dimreg with $d>4$, so that their explicit expressions are given by eq.~\eqref{abx}.\\
We still need to understand how to treat the $d$'s that appear explicitly in eq.~\eqref{bql}. These come from the number of polarization degrees of freedom of the gluons in $d\neq 4$. Now, if we set $d=4-\epsilon$ in eq.~\eqref{bql}, the $O(\epsilon^{-1})$ and $O(\epsilon^{-2})$ terms in $K_{m}$, $J_{m}$ and $J_{m}^{2}$, when multiplied to the $O(\epsilon)$ and $O(\epsilon^{2})$ terms in $(d-1)$ and $(d-1)^{3}/d$, would give rise to finite terms that would spoil the structure of the gap equation. To be specific, the third term of the GEP could not be expressed as a coefficient times the square of the second term divided by $m^{4}$ -- as formally implied by eq.~\eqref{bql} --, leading to a first derivative of the GEP which is not proportional to $\partial J_{m}/\partial m^{2}$ times the gap equation. However, observe that we are still free to define dimensional regularization in such a way that the explicit $d$'s in the GEP are to be straightforwardly set to four. With this prescription, if we define a bare coupling constant $\alpha_{B}$ as
\BE\label{alph}
\alpha_{B}=\frac{9Ng_{B}^{2}}{32\pi^{2}}=\frac{9N\alpha_{sB}}{8\pi}
\EE
\\
then we can put the GEP in the final form\\
\BE\label{ymgep}
V_{G}(m^{2})=3N_{A}\,\left(K_{m}-\frac{m^{2}}{2}\,J_{m}+2\pi^{2}\alpha_{B}\, J_{m}^{2}\right)
\EE
\\
\\
Apart from a multiplicative factor of $3N_{A}$, eq.~\eqref{ymgep} is formally identical to the GEP of massless $\lambda\phi^{4}$ theory, given by eq.~\eqref{abq}. In eq.~\eqref{alph} we have defined $\alpha_{B}$ precisely so that the equations of Sec.1.1.3 can be carried over verbatim to Yang-Mills theory. Therefore we do not need to explicitly repeat the GEP analysis, and we can limit ourselves to just restating the results. From eq.~\eqref{ymgep}, the following gap equation of Yang-Mills theory can be derived:
\BE
m_{0}^{2}=8\pi^{2}\alpha_{B}\,J_{m_{0}}^{2}
\EE
\\
In $d\to4^{+}$ -- just as in massless $\lambda\phi^{4}$ theory --, the latter has a non-zero solution at $m_{0}=\Lambda_{\epsilon}\,e^{1/\alpha_{B}}$. In terms of $m_{0}$ rather than of the unrenormalized mass scale $\Lambda_{\epsilon}$ used to regularize the divergent integrals, the GEP reads\\
\BE
V_{G}(m^{2})=\frac{m^{4}}{128\pi^{2}}\ \left(\alpha_{B}\,\ln^{2}\frac{m^{2}}{m_{0}^{2}}+2\, \ln\frac{m^{2}}{m_{0}^{2}}-1\right)
\EE
\\
As shown in Fig.\ref{FD}, $V_{G}(m^{2})$ has a relative minimum at $m=0$, as well as a relative maximum at $m=m_{0}\,e^{-2/\alpha_{B}-1}<m_{0}$; since $V_{G}|_{m=0}=0$ while $V_{G}|_{m=m_{0}}<0$ -- cf. eq.~\eqref{ajq} --, $m_{0}$ is the absolute minimum of the GEP. As already discussed in the context of massless $\lambda\phi^{4}$ theory, since Yang-Mills theory is scale-free at the classical level, the actual value of $m_{0}$ cannot be predicted from first principles and must be fixed by experiment. The GEP analysis only informs us that $m_{0}$ is different from zero. Finally, the value of the GEP at its absolute minimum,
\BE
V_{G}(m_{0}^{2})=-\frac{m_{0}^{4}}{128\pi^{2}}
\EE
\\
does not depend on the bare coupling $\alpha_{B}$, but only on the position $m_{0}$ of the minimum. Since $V_{G}(m_{0}^{2})$ is the GEP approximation to the vacuum energy density of the system -- hence the only value of the GEP with a physical meaning --, from a physical point of view the bare coupling disappears, as it should, from the equations, having been absorbed into the definition of the renormalized mass scale $m_{0}$. On the other hand, the values of the GEP at $m\neq m_{0}$ are not physical, hence their dependence on $\alpha_{B}$ is uninfluential for the physics of the system.\\
\\
At this point of the analysis, one may wonder whether the non-zero mass derived through the GEP should be interpreted as a physically meaningful parameter, rather than an artifact of the variational approach. One way to address this issue is to ask whether the zero-temperature minimum found by the GEP is stable against thermal excitations, i.e. if it remains non-zero at small but finite temperatures, and whether its evolution with the temperature leads to physically meaningful predictions.
\newpage
\
\begin{figure}[H]
\centering
\vskip-20pt
\includegraphics[width=0.87\textwidth]{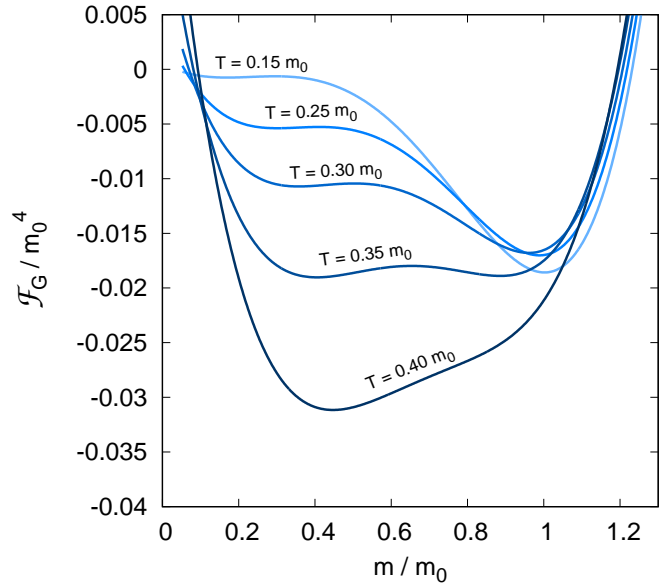}
\vskip-20pt
\caption{A normalized version of the finite-temperature Gaussian Effective Potential $\mc{F}_{G}$ as a function of $m/m_{0}$ for different values of the temperature $T$. The bare coupling constant is kept fixed at the value $\alpha_{B}=0.9$.}\label{FF}
\end{figure}
\
\\
\\
This aspect of the variational approach was explored in ref.\cite{com1,com2}. At finite temperatures, rather than by the exact vacuum energy density $\mc{E}$ of the system, the GEP is bound from below by the exact thermodynamical free energy density $\mc{F}$. The finite-temperature GEP is usually denoted by $\mc{F}_{G}$ and is a function of both the mass parameter $m$ and the temperature $T$ of the system. By minimizing $\mc{F}_{G}(m^{2},T)$ with respect to $m^{2}$ at fixed $T$, one obtains the optimal mass parameter at temperature $T$; as $T$ varies, these parameters define an optimal mass function $m(T)$. Since $\mc{F}_{G}|_{T=0}=V_{G}$, the zero-temperature minimum of $\mc{F}_{G}$ is precisely the $m_{0}$ that we found in this chapter. $m_{0}$ can then be used to set the scale of both the temperature $T$ and the finite-temperature minima $m(T)$, so that once $m_{0}$ has been fixed no other external parameters are needed to determine $m(T)$, which is thus is an actual prediction from first principles of the variational approach. Moreover, since the GEP itself is a variational approximation to the free energy density of the system, by evaluating $\mc{F}_{G}(m^{2},T)$ at its minimum $m(T)$ one obtains a variational estimate of the exact free energy density $\mc{F}(T)$. From this estimate the entropy and specific heat of the system can then be computed as functions of the temperature, allowing for a semi-quantitative characterization of the thermodynamics of the gluonic system.\\
In Fig.\ref{FF} the finite-temperature GEP $\mc{F}_{G}$ is shown as a function of the mass parameter for different values of the temperature\footnote{\ At variance with the zero-temperature $V_{G}$, at its minimum $\mc{F}_{G}$ still slightly depends on the bare coupling $\alpha_{B}$. The curve in Fig.\ref{FF} assumes an optimal value of $\alpha_{B}=0.9$, as discussed in \cite{com2}.}. As we can see, as the temperature grows from zero to finite values, the absolute minimum of the GEP remains quite close to $m_{0}$ until a critical temperature $T_{c}\approx 0.35\ m_{0}$ is reached. At $T=T_{c}$ the relative minimum which was once at $m=0$ becomes dominant, and the optimal mass parameter drops discontinuously to lower values. This discontinuity in the optimal mass causes the computed entropy of the system to be discontinuous as well. Therefore, when extended to finite temperatures, the GEP analysis gives clear evidence of the occurrence of a first order phase transition in the Yang-Mills system. Such a first order transition is indeed known to occur: it is the deconfinement phase transition of gluonic matter, connecting a low temperature phase in which the gluons cannot exist as free particles (confined gluons) to a high temperature phase in which the gluons are free to propagate (deconfined gluons).\\
The GEP analysis of the thermal behavior ofYang-Mills theory \cite{com1,com2} manages to reproduce in a semi-quantitative fashion results which had already been provided by lattice computations \cite{iwa,boyd,giusti2}. These results rest on the validity of the GEP approach at zero temperature -- especially in reference to its variational status, discussed in Sec.1.2.4 -- and on the physical significance of the non-zero vacuum minimum $m_{0}$. They lead us to interpret $m_{0}$ as a physically meaningful -- albeit a priori unfixed -- energy scale of Yang-Mills theory and to take seriously the conclusions of the GEP analysis already at zero temperature.\\
\\
The existence of a non-zero absolute minimum for the vacuum GEP implies that the massless gluon perturbative vacuum is unstable towards a massive vacuum. As discussed at length in Sec.1.1, this fact may be interpreted as evidence for the phenomenon of dynamical mass generation in the gluon sector.\\
Nonetheless, it must be kept in mind that, being of variational nature, the evidence gathered through the GEP approach cannot be understood as a rigorous proof that in the vacuum the gluons acquire a mass through the strong interactions. Furthermore, the GEP approach is not able to explain the actual mechanism by which the dynamical mass is generated. The only precise statement that we can make based on the GEP analysis is that the massless Gaussian (i.e. free-particle) state which is adopted as the zero-order vacuum state of ordinary perturbation theory is farther away from the true vacuum of the theory than a massive Gaussian state.\\
This statement leads us to believe that a perturbative formulation of Yang-Mills theory which treats the transverse gluons as massive already at tree-level may provide results which are closer to their exact counterpart than those obtained by ordinary perturbation theory. Since $m_{L}=0$ from first principles and $M=0$ by maximization of the ghost GEP, in such a formulation the ghosts and longitudinal gluons will need to be treated as massless.\\
The massive formulation of Yang-Mills perturbation theory will be presented in Chapter~2. There we will see that the massive expansion indeed reproduces features of the exact dressed gluon propagator -- including the generation of a truly dynamical gluon mass -- which cannot be described at any finite order by ordinary, massless perturbation theory, providing fully quantitative predictions in astonishing agreement with the lattice results. This fact reinforces the idea that massive perturbation theory actually may be the way to go when exploring the low energy dynamics of the strong interactions.

\newpage


\stepcounter{count}
\addcontentsline{toc}{chapter}{2 The massive perturbative expansion of Yang-Mills theory in an arbitrary covariant gauge}  \markboth{2 The massive perturbative expansion of Yang-Mills theory in an arbitrary covariant gauge}{2 The massive perturbative expansion of Yang-Mills theory in an arbitrary covariant gauge}
\chapter*{2\protect \\
\medskip{}
The massive perturbative expansion of Yang-Mills theory in an arbitrary covariant gauge\index{The massive perturbative expansion of Yang-Mills theory in an arbitrary covariant gauge}}

In Chapter 1 by a GEP analysis of Yang-Mills theory we were able to reach the conclusion that the massless perturbative vacuum of the transverse gluons is unstable towards a massive vacuum. We interpreted this fact as evidence for dynamical mass generation, and argued that because of it a non-ordinary perturbation theory which treats the transverse gluons as massive already at tree-level may lead to a perturbative series which more accurately captures the low energy dynamics of Yang-Mills theory. It is now time to leave the simplified realm of the Gaussian Effective Potential and confront ourselves with the richness of the full theory from the perspective of the massive perturbative expansion.\\
\\
The massive perturbative expansion is formally defined by the same procedure used to derive the GEP. One starts from the massless kinetic Lagrangian obtained by sending to zero the strong coupling $g$ in the full Faddeev-Popov gauge-fixed action and adds to it the relevant mass terms for the gluon and ghost fields. Since the ghosts and longitudinal gluons -- either by first principles or variationally -- are found to remain massless, from the standpoint of optimized perturbation theory a single mass term needs to be added to the ordinary kinetic Lagrangian of Yang-Mills theory, namely, the mass term for the transverse gluons. In order for the total action to remain unchanged, the very same term must then be subtracted from the interaction Lagrangian. As a result of this shift, the Feynman rules of the massive perturbative expansion are different from that of ordinary perturbation theory in two respects. First of all, the additional mass term in the kinetic Lagrangian causes the transverse component of the bare gluon propagator to be massive, rather than massless. Second of all, the subtraction of the mass term from the interaction Lagrangian gives rise to a new two-gluon vertex which is not present in the ordinary, massless expansion. The shift introduces a spurious mass parameter $m$ which cannot be fixed by first principles, since the original theory was scale-free at the classical level.\\
Equipped with the Feynman rules of the massive expansion, one can go on and compute the quantities of physical interest by using the usual machinery of perturbation theory. In this chapter we will be mainly concerned with the computation of the gluon and ghost two-point functions in momentum space, i.e. of the dressed gluon and ghost propagators, in an arbitrary covariant gauge. As we will see, the massive expansion is able to incorporate the phenomenon of dynamical mass generation for the transverse gluons in a non-trivial way, overcoming the limitations of ordinary perturbation theory. By contrast, in agreement with the lattice data, the ghost propagator will be shown to remain massless.\\
Massive perturbation theory provides explicit analytical expressions for the propagators which can then be continued to the whole complex plane and its Euclidean subdomain, where the lattice calculations are defined. The infrared results of the massive expansion can thus be tested against the predictions of the lattice in a quantitative as well as a qualitative fashion. In this chapter we will compare our expressions with the lattice data in the Landau gauge, where the data is the most reliable. Since the shift introduces a new free parameter -- namely the mass parameter -- in the equations, at this stage the comparison has the status of a fit, rather than of an actual comparison from first principles\footnote{\ This limitation will be overcome in the next chapter, where the alleged loss of predictivity will be investigated in connection to the gauge invariance of the massive expansion.}. As we will see, the Landau-gauge Euclidean ghost and gluon propagators turn out to be in astonishing agreement with the lattice data, reinforcing the idea that most of the non-perturbative content of Yang-Mills theory may actually be incorporated into the gluons' transverse mass.\\
\\
This chapter is organized as follows. In Sec.2.1 we will define the massive perturbative expansion and outline its general features both from a diagrammatic and from an analytic point of view. In Sec.2.2 we will define and compute the one-loop ghost propagator in an arbitrary covariant gauge, investigate its asymptotic limits and compare its low energy behavior with the lattice data in the Landau gauge. In Sec.2.3 we will do the same for the gluon propagator, with particular emphasis on the issue of mass generation. Due to their complexity, the calculations which lead to the final expression for the gluon propagator are described in much less detail than those presented for the ghost propagator. A thorough derivation of the one-loop gluon polarization may be found in the Appendix of ref.\cite{com3}.\\
\\
The results of this chapter were presented and published for the first time in ref. \cite{sir2,com3,sir3}.

\newpage


\addcontentsline{toc}{section}{2.1 Set-up of the massive expansion}  \markboth{2.1 Set-up of the massive expansion}{2.1 Set-up of the massive expansion}
\section*{2.1 Set-up of the massive expansion\index{Set-up of the massive expansion}}

\addcontentsline{toc}{subsection}{2.1.1 The Feynman rules of massive perturbation theory}  \markboth{2.1.1 The Feynman rules of massive perturbation theory}{2.1.1 The Feynman rules of massive perturbation theory}
\subsection*{2.1.1 The Feynman rules of massive perturbation theory\index{The Feynman rules of massive perturbation theory}}

\begin{figure}[H]
\centering
\includegraphics[width=0.92\textwidth]{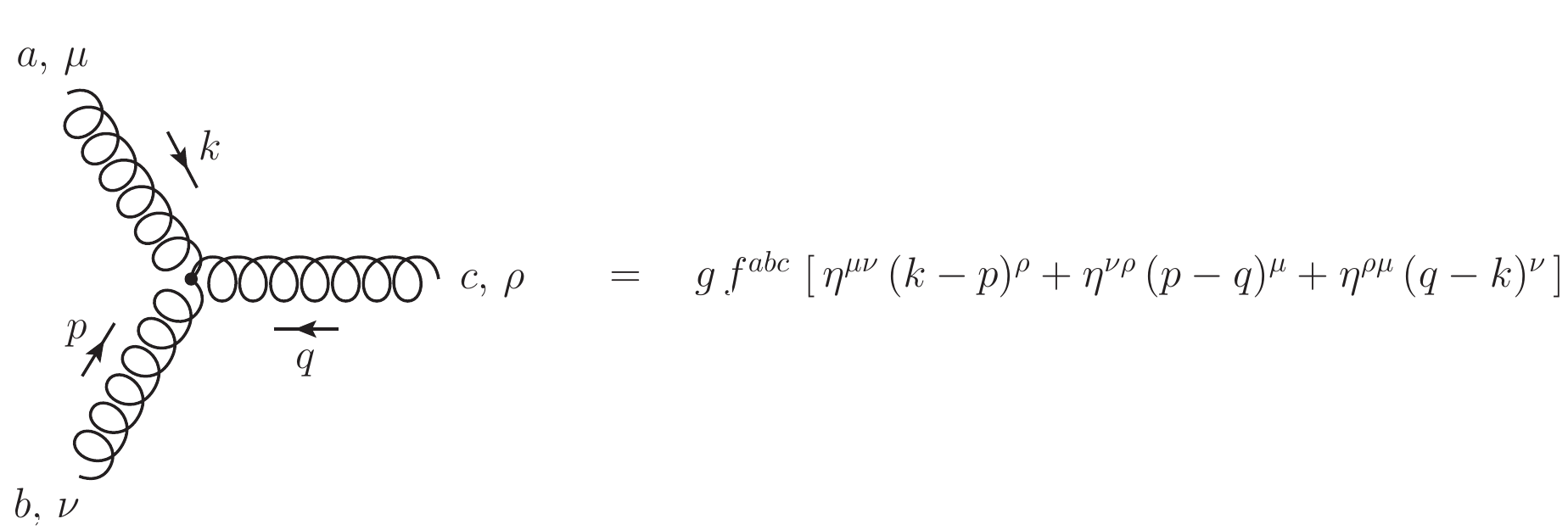}
\\
\caption{3-gluon vertex}\label{3gl}
\end{figure}

\begin{figure}[H]
\centering
\includegraphics[width=0.86\textwidth]{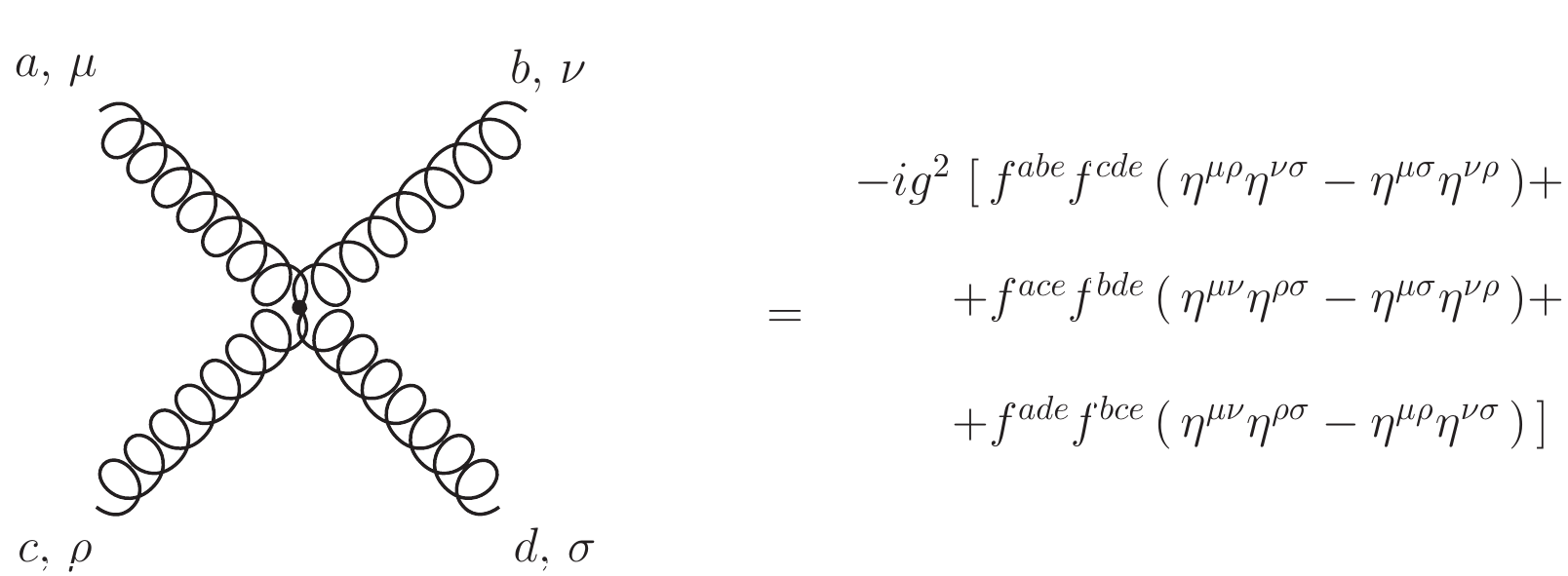}
\vspace{20pt}
\caption{4-gluon vertex}\label{4gl}
\end{figure}

\begin{figure}[H]
\centering
\includegraphics[width=0.55\textwidth]{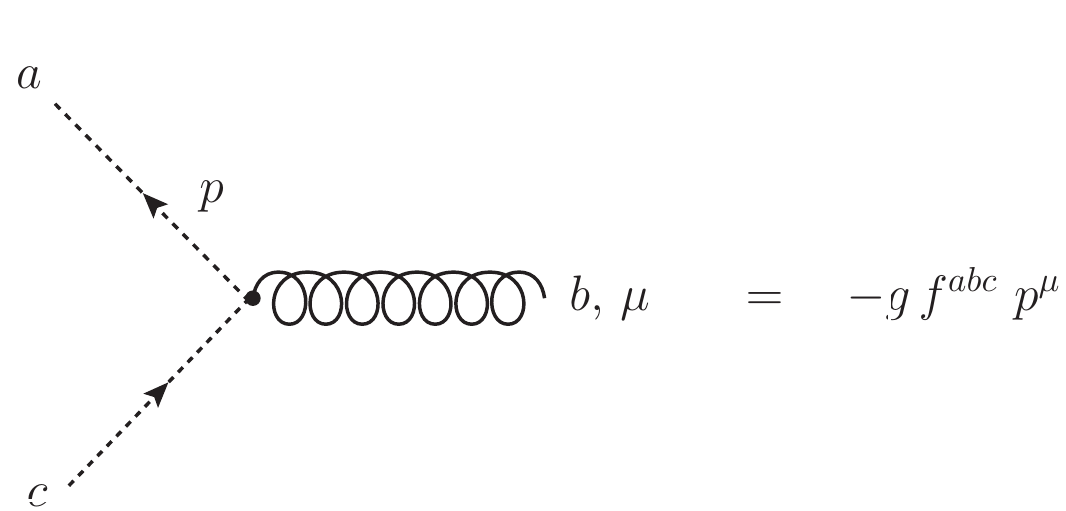}
\vspace{16pt}
\caption{Ghost-gluon vertex}\label{ghgl}
\end{figure}

\begin{figure}[H]
\centering
\includegraphics[width=0.63\textwidth]{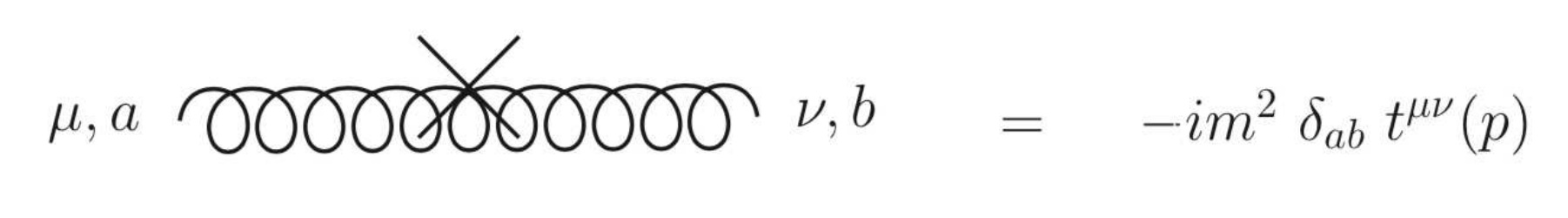}
\\
\caption{Mass counterterm}\label{mct}
\end{figure}

In this section we will proceed with the definition of the massive expansion and the derivation of its Feynman rules. We start again from the Faddeev-Popov gauge-fixed action of Yang-Mills theory in an arbitrary covariant gauge,\\
\begin{align}\label{sfp}
\mc{S}&=\int d^{d}x\ \bigg\{-\frac{1}{2}\ \partial_{\mu}A_{\nu}^{a}\,(\partial^{\mu}A^{a\,\nu}-\partial^{\nu}A^{a\,\mu})-\frac{1}{2\xi}\ \partial^{\mu} A^{a}_{\mu}\,\partial^{\nu} A^{a}_{\nu}+\partial^{\mu}\cbar^{\,a}\,\partial_{\mu}c^{a}+\\
\notag&\qquad-g\ f_{abc}\ \partial_{\mu}A^{a}_{\nu}\,A^{b\,\mu}A^{c\,\nu}-\frac{g^{2}}{4}\ f_{abc}\,f_{ade}\ A_{\mu}^{b}\,A_{\nu}^{c}\ A^{d\,\mu}\,A^{e\,\nu}-g\ f_{abc}\ \partial^{\mu}\cbar^{\,a}c^{b}A_{\mu}^{c}+\mathcal{L}_{\tx{c.t.}}\bigg\}
\end{align}
\\
were $\mathcal{L}_{\tx{c.t.}}$ contains the renormalization counterterms and $\xi\geq 0$ is the gauge parameter. In order to define a perturbation theory, we must fix a zero-order (kinetic) action $\mc{S}_{0}$ such that in momentum space the gluon and ghost bare propagators $\mc{D}_{\mu\nu}^{ab}$ and $\mc{G}^{ab}$ are given by\\
\BE\label{s00}
\mc{S}_{0}=i\ \int\frac{d^{d}p}{(2\pi)^{d}}\ \left\{\frac{1}{2}\ A_{\mu}^{a}\ \mc{D}_{ab}^{\mu\nu}(p)^{-1}\, A_{\nu}^{b}+\cbar^{\,a}\ \mc{G}_{ab}(p)^{-1}\, c^{b}\right\}
\EE
\\
In the massive perturbative expansion, the ghosts and longitudinal gluons are treated as massless at tree-level, whereas the transverse gluons are given a mass $m$; their bare propagators are therefore defined as\\
\BE\label{prop}
\mc{D}_{\mu\nu}^{ab}(p)=\delta^{ab}\ \bigg\{\frac{-i\, t_{\mu\nu}(p)}{p^{2}-m^{2}+i\epsilon}+\xi\ \frac{-i\, \ell_{\mu\nu}(p)}{p^{2}+i\epsilon}\bigg\}\qquad\qquad \mc{G}^{ab}(p)=\delta^{ab}\ \frac{i}{p^{2}+i\epsilon}
\EE
\\
and have inverses\\
\BE\label{propinv}
\mc{D}^{\mu\nu}_{ab}(p)^{-1}=i\ \delta_{ab}\ \bigg\{(p^{2}-m^{2})\,t^{\mu\nu}(p)+\frac{1}{\xi}\ p^{2}\ \ell^{\mu\nu}(p)\bigg\}\qquad\quad \mc{G}_{ab}(p)^{-1}=-i\,\delta_{ab}\ p^{2}
\EE
\\
where $t_{\mu\nu}$ and $\ell_{\mu\nu}$ are transverse and longitudinal projectors,\\
\BE
t_{\mu\nu}(p)=\eta_{\mu\nu}-\frac{p_{\mu}p_{\nu}}{p^{2}}\qquad\qquad \ell_{\mu\nu}(p)=\frac{p_{\mu}p_{\nu}}{p^{2}}
\EE
\\
Plugging eq.~\eqref{propinv} into eq.~\eqref{s00}, we obtain the following zero order action for the massive expansion:\\
\BE\label{s0}
\mc{S}_{0}=\int\frac{d^{d}p}{(2\pi)^{d}}\ \left\{-\frac{1}{2}\ A_{\mu}^{a}\ \delta_{ab}\left[(p^{2}-m^{2})\,t^{\mu\nu}(p)+\frac{1}{\xi}\ p^{2}\ \ell^{\mu\nu}(p)\right]\, A_{\nu}^{b}+\cbar^{\,a}\ \delta_{ab}\ p^{2}\, c^{b}\right\}
\EE
\\
The latter is identical to the kinetic action of ordinary perturbation theory -- first line of eq.~\eqref{sfp} in momentum space -- apart from a single term $\delta\mc{S}$ yielding the transverse gluon mass,
\BE
\delta\mc{S}=\int\frac{d^{d}p}{(2\pi)^{d}}\ \frac{1}{2}\ A_{\mu}^{a}\ m^{2}\,\delta_{ab}\,t^{\mu\nu}(p)\,A_{\nu}^{b}
\EE
\\
$\delta\mc{S}$ needs to be included in the interaction action $\mc{S}_{\tx{int}}$ in order for the total action $\mc{S}$ to remain unchanged, $\mc{S}=\mc{S}_{0}+\mc{S}_{\tx{int}}$. Setting $\mc{S}_{\tx{int}}=\mc{S}-\mc{S}_{0}$, we obtain\\
\begin{align}\label{sint}
\mc{S}_{\tx{int}}&=\int\frac{d^{d}p}{(2\pi)^{d}}\ \left\{-\frac{1}{2}\ A_{\mu}^{a}\ m^{2}\,\delta_{ab}\,t^{\mu\nu}(p)\,A_{\nu}^{b}\right\}+\int d^{d}x\ \bigg\{-g\ f_{abc}\ \partial_{\mu}A^{a}_{\nu}\,A^{b\,\mu}A^{c\,\nu}+\\
\notag&\qquad\qquad -\frac{g^{2}}{4}\ f_{abc}\,f_{ade}\ A_{\mu}^{b}\,A_{\nu}^{c}\ A^{d\,\mu}\,A^{e\,\nu}-g\ f_{abc}\ \partial^{\mu}\cbar^{\,a}c^{b}A_{\mu}^{c}+\mathcal{L}_{\tx{c.t.}}\bigg\}
\end{align}
\\
In eq.~\eqref{sint}, the coordinate space integral contains the ordinary 3-gluon, 4-gluon and ghost-gluon interactions and the ordinary renormalization counterterms; all of them are proportional either to the strong coupling constant $g$ or its square $g^{2}$. The momentum space integral, on the other hand, contains a new 2-gluon interaction, proportional to the gluon mass parameter squared $m^{2}$. This interaction is due to the shift induced by the term $\delta\mc{S}$ and is not present in ordinary perturbation theory.\\
Each of the interaction terms in $\mc{S}_{\tx{int}}$ yields a vertex for the diagrammatic computation of the quantities of physical interest in perturbation theory. The 3-gluon, 4-gluon and ghost-gluon vertices are left unchanged by the massive shift; their expressions are those of ordinary perturbation theory, Figg.\ref{3gl}-\ref{ghgl}. The 2-gluon vertex can be read out directly from $-i\delta\mc{S}$; it is given by Fig.\ref{mct}. Since it was obtained by the same addition/subtraction mechanism which usually defines the renormalization counterterms, we shall refer to it as the \textit{mass counterterm} and denote it with a cross.\\
\
\newpage
\
\
\\
\\
From the kinetic action $\mc{S}_{0}$, through eqq.~\eqref{s00} and \eqref{prop}, we read out the gluon and ghost bare propagators, to be associated to the internal lines of the Feynman diagrams. These are given in Fig.\ref{bglp}-\ref{bghp}.\\
\begin{figure}[H]
\centering
\includegraphics[width=0.79\textwidth]{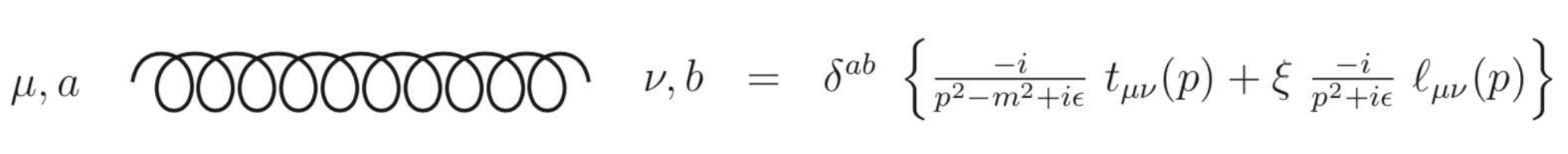}
\\
\caption{Bare gluon propagator}\label{bglp}
\end{figure}
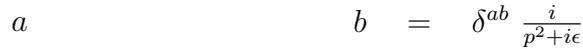
\begin{figure}[H]
\hspace{85pt}
\fcolorbox{white}{white}{
  \begin{picture}(160,22) (31,-11)
    \SetWidth{1.0}
    \SetColor{Black}
    \Line[dash,dashsize=2](42,-8)(148,-8)
    \Text(28,-10)[lb]{\large{\Black{$a$}}}
    \Text(156,-16)[lb]{\large{\Black{$b\ \ \ =\ \ \ \delta^{ab}\ \frac{i}{p^{2}+i\epsilon}$}}}
  \end{picture}
}
\\
\caption{Bare ghost propagator}\label{bghp}
\end{figure}
\
\\
The Feynman rules of Figg.\ref{3gl}-\ref{bghp} fully define the massive perturbative expansion. Since the shift induced by $\delta\mc{S}$ does not modify the total Faddeev-Popov action $\mc{S}$ -- hence the physical content of Yang-Mills theory --, the massive expansion is non-perturbatively equivalent to ordinary perturbation theory. However, due to the massiveness of its bare gluon propagator and the presence of the new mass counterterm, the truncation of the massive perturbative series to any finite order yields different results than those of standard, massless perturbation theory.
\\
\\

\addcontentsline{toc}{subsection}{2.1.2 The mass counterterm and the general features of the massive expansion}  \markboth{2.1.2 The mass counterterm and the general features of the massive expansion}{2.1.2 The mass counterterm and the general features of the massive expansion}
\subsection*{2.1.2 The mass counterterm and the general features of the massive expansion\index{The mass counterterm and the general features of the massive expansion}}

\begin{figure}[H]
\centering
\includegraphics[width=\textwidth]{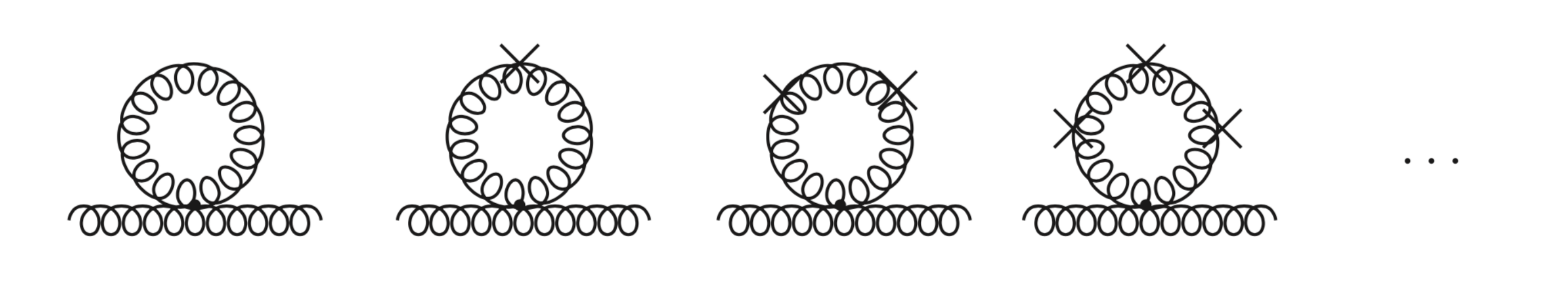}
\\
\caption{Tadpole diagrams of the massive perturbative expansion. The internal gluon line of the tadpole may contain an arbitrarily high number of mass counterterms.}\label{mtads}
\end{figure}

As we saw in the last section, the shift induced by the gluon mass term $\delta\mc{S}$ produces a new 2-gluon vertex, the mass counterterm, which is not part of the Feynman rules of ordinary perturbation theory. Due to the new vertex, the loop expansions of massive perturbation theory contain additional Feynman diagrams, called \textit{crossed diagrams}. These are diagrams which differ from those of ordinary perturbation theory by the presence of one or more mass counterterms in the internal gluon lines of the loops. As an example, consider the gluon tadpole diagram, Fig.\ref{mtads}. In ordinary perturbation theory there is a single tadpole: the first diagram on the left in the figure (no mass counterterms in the internal gluon line). In massive perturbation theory, on the other hand, there are an infinite number of tadpole diagrams, each with an increasing number of mass counterterms.\\
\
\\
Since $\delta\mc{S}$ -- unlike the gluon-gluon and ghost-gluon interactions -- is not proportional to the strong coupling constant, diagrams which only differ by the number of mass counterterms in any of their internal lines are of the same order in $g$. The non-perturbative nature of the massive expansion emerges precisely in this feature of the perturbative series: as far as the mass counterterms are concerned, in massive perturbation theory the correspondence between the number of vertices and the perturbative order in the coupling $g$ is lost.\\
\\
One of the tasks of the crossed diagrams is to counterbalance the effects of the tree-level gluon mass $m$. At the level of the full action $\mc{S}$, this is very clear: by definition, the mass counterterm in the interaction Lagrangian exactly cancels out the corresponding term in the kinetic Lagrangian. At the level of the perturbative series, the same cancellation would occur by the following mechanism. Consider some internal gluon line with a number $n\geq 1$ of mass counterterms in an arbitrary Feynman diagram. To such a line we would associate the expression
\BE\label{ctz}
\mc{D}(p)\cdot\left[-im^{2}\,t(p)\cdot\mc{D}(p)\right]^{n}
\EE
\\
where $p$ is the momentum of the line and we have suppressed the tensor indices of both the gluon propagator $\mathcal{D}(p)$ and the mass counterterm $-im^{2}\,t(p)$. Now, since\\
\BE
t(p)\cdot\mc{D}(p)=t(p)\cdot\left(\frac{-i}{p^{2}-m^{2}}\ t(p)+\frac{-i\xi}{p^{2}}\ \ell(p)\right)=\frac{-i}{p^{2}-m^{2}}\ t(p)
\EE
\\
as long as $n$ is non-zero, the $n$-th power of the quantity in parentheses in eq.~\eqref{ctz} reads\\
\BE
[-im^{2}t(p)\cdot\mc{D}(p)]^{n}=\left(\frac{-m^{2}}{p^{2}-m^{2}}\ t(p)\right)^{n}=\left(\frac{-m^{2}}{p^{2}-m^{2}}\right)^{n}\ t(p)
\EE
\\
Therefore, for $n\geq 1$,\\
\BE
\mathcal{D}(p)\cdot[-im^{2}t(p)\cdot\mc{D}(p)]^{n}=\frac{-i}{p^{2}-m^{2}}\ \left(\frac{-m^{2}}{p^{2}-m^{2}}\right)^{n}\ t(p)
\EE
\\
In the full perturbative series, there will be diagrams that are identical to that which we are considering, except for the number of mass counterterms in the very same internal line; this number goes from zero to infinity. By summing up these diagrams, we obtain a single resummed diagram whose internal gluon line is given by\\
\BE\label{resum}
\sum_{n=0}^{\infty}\ \mathcal{D}(p)\cdot[-im^{2}t(p)\cdot\mc{D}(p)]^{n}=\frac{-i}{p^{2}-m^{2}}\ \sum_{n=0}^{\infty}\ \left(\frac{-m^{2}}{p^{2}-m^{2}}\right)^{n}\ t(p)+\xi\ \frac{-i}{p^{2}}\ \ell(p)
\EE
\\
where the longitudinal term comes from the order zero summand. Since\\
\BE
\sum_{n=0}^{\infty}\ \left(\frac{-m^{2}}{p^{2}-m^{2}}\right)^{n}=\frac{1}{1+\frac{m^{2}}{p^{2}-m^{2}}}=\frac{p^{2}-m^{2}}{p^{2}}
\EE
the sum reduces to\\
\BE
\sum_{n=0}^{\infty}\ \mathcal{D}(p)\cdot[-im^{2}t(p)\cdot\mc{D}(p)]^{n}=\frac{-i}{p^{2}}\ t(p)+\xi\ \frac{-i}{p^{2}}\ \ell(p)
\EE
\\
which is precisely the bare propagator of a massless gluon. Therefore, by summing up all the crossed diagrams of the perturbative series, we obtain a new set of resummed diagrams with no mass counterterms in which the massive gluon internal lines have been replaced by massless lines. These are precisely the diagrams of ordinary perturbation theory. Of course, since this procedure restores the ordinary perturbative series, in massive perturbation theory we are not interested in the resummation of the whole set of crossed diagrams. Actually, this simple calculation teaches us an important lesson: if on the one hand, in the spirit of perturbation theory, a sufficient number of diagrams must be summed in order to approach the exact result, on the other hand the resummation of too many crossed diagrams brings us closer to massless perturbation theory and must thus be avoided.\\
\\
Clearly in the massive expansion some general prescription is needed in order to fix the number of crossed diagrams that are to be included in the perturbative series at any fixed loop order. Such a prescription, unfortunately, has not yet been established in full generality. Nonetheless, it turns out that the principle of renormalizability, together with the principle of minimality, are still sufficient to fix the number of mass counterterms to be included in expansions carried out to one loop. In order to motivate this statement, we need to take a preliminary step and describe a useful method for the computation of diagrams with an arbitrary number of mass counterterms. Then we will discuss the effect of the mass counterterm on the divergences of the massive perturbative expansion and draw our conclusions.\\
\\
Suppose that we want to sum some set of Feynman diagrams in which the number of mass counterterms goes from zero to a fixed integer $N$. Let us focus on a single diagram with $n\geq 1$ mass counterterms in one of its internal gluon lines. As we saw, to this line we may associate the expression\\
\BE\label{cnn}
\mathcal{D}(p)\cdot[-im^{2}t(p)\cdot\mc{D}(p)]^{n}=\frac{-i}{p^{2}-m^{2}}\ \left(\frac{-m^{2}}{p^{2}-m^{2}}\right)^{n}\ t(p)
\EE
Since
\BE
\frac{1}{(p^{2}-m^{2})^{n+1}}=\frac{1}{n!}\ \frac{\partial^{n}}{\partial (m^{2})^{n}}\frac{1}{p^{2}-m^{2}}
\EE
\\
the right hand side of eq.~\eqref{cnn} may be re-expressed as\\
\begin{align}
\frac{-i}{p^{2}-m^{2}}\ \left(\frac{-m^{2}}{p^{2}-m^{2}}\right)^{n}\ t(p)&=\frac{(-m^{2})^{n}}{n!}\frac{\partial^{n}}{\partial(m^{2})^{n}}\left(\frac{-i}{p^{2}-m^{2}}\ t(p)\right)=\\
\notag&=\frac{(-m^{2})^{n}}{n!}\frac{\partial^{n}}{\partial(m^{2})^{n}}\ \mc{D}(p)
\end{align}
Therefore
\BE\label{cmn}
\mathcal{D}(p)\cdot[-im^{2}t(p)\cdot\mc{D}(p)]^{n}=\frac{(-m^{2})^{n}}{n!}\frac{\partial^{n}}{\partial(m^{2})^{n}}\ \mc{D}(p)
\EE
\\
It follows that to any internal gluon line with $n\geq0$ mass counterterms\footnote{\ It is easy to see that the above equation holds also for $n=0$.} we can associate the $n$-th derivative of the massive propagator with respect to the mass parameter squared, multiplied by $(-m^{2})^{n}/n!$. Going back to our set of Feynman diagrams, let $\mc{R}^{0}$ be the sum of the subset of diagrams which have no mass counterterms in their internal gluon lines\footnote{\ These are precisely the diagrams of ordinary perturbation theory, albeit computed with massive rather than massless bare gluon propagators.}. From eq.~\eqref{cmn} it is easy to see that the quantity\\
\BE
\mc{R}^{n}=\frac{(-m^{2})^{n}}{n!}\frac{\partial^{n}\mc{R}^{0}}{\partial(m^{2})^{n}}
\EE
\\
is precisely the sum of the subset of diagrams with $n$ mass counterterms (not necessarily on the same gluon lines). Hence the sum $\mc{R}^{(N)}$ of the full set of diagrams, with zero to $N$ mass counterterms on any of their internal lines, may be expressed as\\
\BE\label{ccp}
\mc{R}^{(N)}=\sum_{n=0}^{N}\ \mc{R}^{n}=\sum_{n=0}^{N}\ \frac{(-m^{2})^{n}}{n!}\frac{\partial^{n}\mc{R}^{0}}{\partial(m^{2})^{n}}=\mc{R}^{0}-m^{2}\ \frac{\partial \mc{R}^{0}}{\partial m^{2}}+\frac{m^{4}}{2}\ \frac{\partial^{2} \mc{R}^{0}}{\partial (m^{2})^{2}}+\cdots
\EE
\\
Therefore, in order to compute $\mc{R}^{(N)}$, we only need to know $\mc{R}^{0}$ and its derivatives with respect to $m^{2}$ up to the order $N$. This is the procedure which we will follow in Sec.2.2-3 to obtain the ghost and gluon dressed propagators.\\
Of particular interest to us are the first two terms of the sum in eq.~\eqref{ccp}. Suppose that $\mc{R}^{0}$ contains a term which is linear in $m^{2}$, $\mc{R}^{0}=m^{2}L+\mc{Q}^{0}$, where $L$ is an $m$-independent coefficient and $\mc{Q}^{0}$ is a remainder. Then
\BE
\mc{R}^{1}=-m^{2}\,\frac{\partial \mc{R}^{0}}{\partial m^{2}}=-m^{2}L-m^{2}\,\frac{\partial\mc{Q}^{0}}{\partial m^{2}}
\EE
so that
\BE
\mc{R}^{0}+\mc{R}^{1}=\mc{R}^{0}-m^{2}\,\frac{\partial \mc{R}^{0}}{\partial m^{2}}=\mc{Q}^{0}-m^{2}\,\frac{\partial\mc{Q}^{0}}{\partial m^{2}}
\EE
\\
Therefore, even if $\mc{R}^{0}$ and $\mc{R}^{1}$ may separately contain linear terms in $m^{2}$, their sum does not: the linear term in $\mc{R}^{1}$ is precisely the same as that in $\mc{R}^{0}$ but with an opposite sign, leading to their reciprocal cancellation in the sum. This is especially important in two respects. First of all, as we will see in Sec.2.3, it turns out that because of this mechanism the mass squared $m^{2}$ in the denominator of the bare transverse gluon propagator gets cancelled by the quantum correction due to the simplest of the crossed diagrams, the single counterterm with no loops. It follows that the mass of the gluon, if any, must come from the loops of the perturbative series, instead of being a trivial consequence of the shift of the Fadeev-Popov action. For instance, the massive perturbative approach would not predict the generation of a mass for the photons of Quantum Electrodynamics. Second of all, the existence of the mass scale $m$ in the massive expansion may invalidate the perturbative series by giving rise to the so-called \textit{mass divergences}, i.e. divergences proportional to $m^{2}$, which cannot be removed from the series by making use of the renormalization counterterms of massless perturbation theory. However, thanks to the cancellation of the linear terms in $m^{2}$, it turns out that -- at least to one loop --, once the mass counterterms are taken into account, such divergences disappear from the series and the renormalizability of the theory is not spoiled.\\
\\
The potential appearance of mass divergences provides a criterion for fixing the minimum number of crossed diagrams that must be included in the perturbative series at a given loop order: such a number must be high enough for the cancellation of the mass divergences to occur. For the one-loop ghost and gluon dressed propagators, Sec.2.2-3, this criterion fixes the minimum number of crossed diagrams to be equal to one. The maximum number of crossed diagrams, on the other hand, is not constrained by the principle of renormalizability alone: since each mass counterterm multiplies the internal line on which it is attached by a factor of $(p^{2}-m^{2})^{-1}$, the loop integrals become less and less divergent as the number of mass counterterms in their internal lines is increased, and eventually become finite.\\
In order to fix the maximum number of crossed diagrams at a given loop order, one of the simplest criteria is given by the principle of minimality. Suppose that we want to compute some quantity in massive perturbation theory to a given order. Due to the mass scale $m$, the uncrossed loops in the expansion will give rise to non-renormalizable mass divergences and a minimum number of crossed diagrams will have to be included in order to preserve the renormalizability of the series. One or more of these diagrams will have a maximal total number of vertices, be they mass counterterms, gluon-gluon vertices or ghost-gluon vertices. Such diagrams arise from the average $\big\langle\mc{S}_{\tx{int}}^{N_{V}}\big\rangle_{0}$ in the perturbative series\footnote{\ Here $\avg{\cdot}_{0}$, as in Chapter 1, denotes the average with respect to the zero-order massive action.}, where $N_{V}$ is the maximal total number of vertices in the diagram. Therefore, in order to be consistent with the perturbative order in the powers of the interaction action, we may include all the diagrams with a number of mass counterterms such that the total number of vertices is less than or equal to $N_{V}$.\\
These criteria will be adopted for the computation of the one-loop ghost and gluon propagators in Sec.2.2-3. It turns out that the uncrossed diagrams of the ghost propagator do not contain mass divergences, whereas those of the gluon propagator do. In order to remove the mass divergences of the uncrossed gluon diagrams, we will need to include crossed diagrams with a total number of vertices up to three. Then, to be consistent with the order in the number of vertices, we will do the same for the ghost propagator.
\\
As a final remark, we wish to illustrate the general mechanism by which ordinary perturbation theory and the standard perturbative results are recovered in the high energy limit of the massive expansion. From eq.~\eqref{resum} we see that when we resum a finite number of mass counterterms in an internal gluon line, we obtain an effective line that propagates the gluons according to the expression\\
\BE
\sum_{n=0}^{N}\ \mathcal{D}(p)\cdot[-im^{2}t(p)\cdot\mc{D}(p)]^{n}=\frac{-i}{p^{2}-m^{2}}\ \sum_{n=0}^{N}\ \left(\frac{-m^{2}}{p^{2}-m^{2}}\right)^{n}\ t(p)+\xi\ \frac{-i}{p^{2}}\ \ell(p)
\EE
\\
where $N$ is the number of resummed mass counterterms. At high momenta ($|p^{2}|\gg m^{2}$) we have that $p^{2}-m^{2}\to p^{2}$ and for $n\geq 1$ the $n$-th power of the expression in parentheses becomes negligible with respect to the zero-order summand. Accordingly,\\
\BE
\lim_{|p^{2}|\to \infty}\ \sum_{n=0}^{N}\ \mathcal{D}(p)\cdot[-im^{2}t(p)\cdot\mc{D}(p)]^{n}=\frac{-i}{p^{2}}\ t(p)+\xi\ \frac{-i}{p^{2}}\ \ell(p)=\mathcal{D}(p)
\EE
\\
which implies that at high momenta the gluons effectively propagate through the ordinary, massless propagator.\\
As a result of this, the shift of the kinetic action does not modify the high energy behavior of the theory and in the UV, as we will explicitly see in Sec.2.2-3, the expressions derived in massive perturbation theory match the standard perturbative results. The mass parameter $m$ sets the scale at which the predictions of the massive expansion diverge from those of ordinary perturbation theory.
\\
\\
\clearpage

\addcontentsline{toc}{section}{2.2 The one-loop ghost propagator}  \markboth{2.2 The one-loop ghost propagator}{2.2 The one-loop ghost propagator}
\section*{2.2 The one-loop ghost propagator\index{The one-loop ghost propagator}}

\addcontentsline{toc}{subsection}{2.2.1 Definition and computation}  \markboth{2.2.1 Definition and computation}{2.2.1 Definition and computation}
\subsection*{2.2.1 Definition and computation\index{Definition and computation}}

\begin{figure}[H]
\centering
\includegraphics[width=0.68\textwidth]{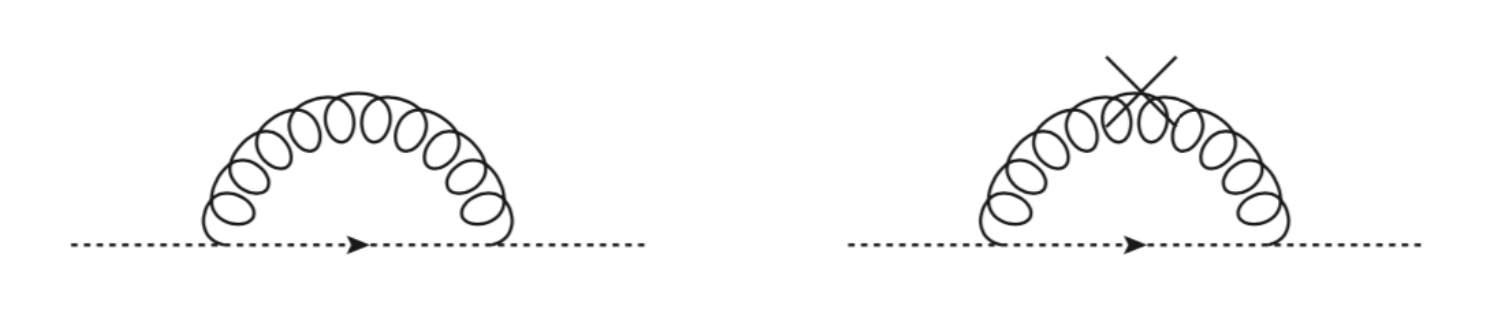}
\\
\caption{Diagrams which contribute to the one-loop ghost self-energy.}\label{ghosed}
\end{figure}

The dressed ghost propagator $\widetilde{\mc{G}}^{ab}(p)$ is defined as\\
\BE\label{ghprfor}
\widetilde{\mc{G}}^{ab}(p)=\int d^{d}x\ \ e^{ip\cdot x}\ \bra{\Omega}T\big\{c^{a}(x)\,\cbar^{b}(0)\big\}\ket{\Omega}
\EE
\\
where $T\{\cdot\}$ is the time ordering meta-operator and $\ket{\Omega}$ is the exact vacuum state of the theory. $\widetilde{\mc{G}}^{ab}(p)$ is related to the one-particle-irreducible ghost self-energy $\Sigma^{ab}(p)=\delta^{ab}\,\Sigma(p)$ through the equation
\BE
\widetilde{\mc{G}}^{ab}(p)=\delta^{ab}\ \frac{i}{p^{2}-\Sigma(p)}
\EE
\\
To one loop, a single irreducible uncrossed diagram contributes to the ghost self-energy: the first diagram in Fig.\ref{ghosed}. This diagram will turn out not to contain mass divergences, so that from the point of view of renormalizability no crossed diagrams need to be included in the ghost self-energy. However, as we will see, the uncrossed diagrams of the one-loop \textit{gluon} propagator do contain divergences proportional to $m^{2}$. To one-loop, a single mass counterterm is enough to remove such divergences, and by the principle of minimality we will need to include in the expansion crossed diagrams up to an order of three in the total number of vertices. Therefore, in order to be consistent with the maximum total number of vertices for both the gluon and the ghost propagator, the one-loop ghost self-energy will need to contain any crossed diagram with a total number of vertices up to three. The only single-loop diagram with this property is the second diagram in Fig.\ref{ghosed}. Thus the one-loop ghost self-energy, as computed in the massive expansion to order three in the interaction, is the sum of the two diagrams in figure.\\
\\
\
\\
\\
\textit{2.2.1.1 Computation of the ghost self-energy: the uncrossed diagram}\\
\\
Let us denote by $\Sigma^{0\,ab}(p)$ the uncrossed loop in Fig.\ref{ghosed}. Using the Feynman rules of massive perturbation theory, we find that\\
\begin{align}
\Sigma^{0\,ab}(p)&=ig^{2}f^{dca}f^{bcd}\int\frac{d^{d}q}{(2\pi)^{d}}\ \frac{(p-q)^{\mu}p^{\nu}}{(p-q)^{2}+i\epsilon}\ \left\{\frac{t_{\mu\nu}(q)}{q^{2}-m^{2}+i\epsilon}+\xi\ \frac{\ell_{\mu\nu}(q)}{q^{2}+i\epsilon}\right\}=\\
\notag&=-iNg^{2}\delta^{ab}\int\frac{d^{d}q}{(2\pi)^{d}}\ \frac{1}{(p-q)^{2}}\ \left\{\frac{q^{2}p^{2}-(p\cdot q)^{2}}{q^{2}(q^{2}-m^{2})}+\xi\ \frac{p\cdot q\,(p\cdot q-q^{2})}{q^{4}}\right\}=\\
\notag&=\delta^{ab}\,\left(\Sigma^{0}_{A}(p)+\Sigma^{0}_{B}(p)\right)
\end{align}
where $N$ is the number of colors ($N=3$ for pure gauge QCD) and\\
\BE
\Sigma^{0}_{A}(p)=-iNg^{2}\int\frac{d^{d}q}{(2\pi)^{d}}\ \frac{q^{2}p^{2}-(p\cdot q)^{2}}{q^{2}(p-q)^{2}(q^{2}-m^{2})}
\EE
\BE
\Sigma^{0}_{B}(p)=-iNg^{2}\, \xi\int\frac{d^{d}q}{(2\pi)^{d}}\ \frac{p\cdot q\,(p\cdot q-q^{2})}{q^{4}(p-q)^{2}}
\EE
\\
The triple denominator in the integral $\Sigma_{A}^{0}(p)$ may be transformed into a double denominator by using the identity\\
\BE
\frac{1}{q^{2}(q^{2}-m^{2})}=\frac{1}{m^{2}}\,\left(\frac{1}{q^{2}-m^{2}}-\frac{1}{q^{2}}\right)
\EE
\\
which allows us to put $\Sigma_{A}^{0}(p)$ in the form\\
\BE
\Sigma^{0}_{A}(p)=\frac{-iNg^{2}}{m^{2}}\int\frac{d^{d}q}{(2\pi)^{d}}\ \left\{\frac{q^{2}p^{2}-(p\cdot q)^{2}}{(p-q)^{2}(q^{2}-m^{2})}-\frac{q^{2}p^{2}-(p\cdot q)^{2}}{(p-q)^{2}q^{2}}\right\}
\EE
\\
By introducing a Feynman parameter $x$, changing variables of integration and discarding any term which is zero by Lorentz symmetry, we obtain\\
\begin{align}
\notag\Sigma^{0}_{A}(p)&=\frac{-iNg^{2}}{m^{2}}\int_{0}^{1}dx\,\int\frac{d^{d}q}{(2\pi)^{d}}\ \left\{\frac{q^{2}p^{2}-(p\cdot q)^{2}}{[q^{2}-xm^{2}+x(1-x)p^{2}]^{2}}-[m\to 0]\right\}=\\
&=-i\ \frac{N(d-1)g^{2}}{d}\ \frac{p^{2}}{m^{2}}\ \int_{0}^{1}dx\,\int\frac{d^{d}q}{(2\pi)^{d}}\ \left\{\frac{q^{2}}{[q^{2}-xm^{2}+x(1-x)p^{2}]^{2}}-[m\to 0]\right\}
\end{align}
\\
where $[m\to 0]$ denotes the limit $m\to 0$ of the first term in parentheses. In dimensional regularization,\\
\begin{align}
\int\frac{d^{d}q}{(2\pi)^{d}}\frac{q^{2}}{(q^{2}-\Delta)^{2}}&=-\frac{i\,\Delta}{(4\pi)^{2}}\,\frac{d}{2-d}\ \Gamma\left(2-\frac{d}{2}\right)\left(\frac{\Delta}{4\pi\mu^{2}}\right)^{d/2-2}\to\\
\notag&\to\frac{2i\,\Delta}{(4\pi)^{2}}\ \left[\frac{2}{\epsilon}-\ln\frac{\Delta}{\overline{\mu}^{2}}+\frac{1}{2}\right]
\end{align}
\\
where $\mu^{2}$ and $\overline{\mu}^{2}=4\pi\mu^{2}e^{-\gamma_{E}}$ are the $\tx{MS}$ and $\overline{\tx{MS}}$ mass scales resulting from the definition of the theory in $d\neq 4$ and we have set $d=4-\epsilon$. Therefore, with\\
\BE
\Delta=xm^{2}-x(1-x)p^{2}
\EE
\\
the dimensionally regularized expression for $\Sigma^{0}_{A}(p)$ reads\\
\begin{align}
\Sigma^{0}_{A}(p)&=\frac{3Ng^{2}}{2(4\pi)^{2}}\ \frac{p^{2}}{m^{2}}\ \int_{0}^{1}dx\,\left\{\Delta\ \left[\frac{2}{\epsilon}-\ln\frac{\Delta}{\overline{\mu}^{2}}+\frac{1}{3}\right]-[m\to 0]\right\}=\\
\notag&=\frac{3Ng^{2}}{4(4\pi)^{2}}\ p^{2}\ \left\{\frac{2}{\epsilon}-\ln\frac{-p^{2}}{\overline{\mu}^{2}}+\frac{1}{3}-\frac{2}{m^{2}}\int_{0}^{1}dx\ \left(\Delta\,\ln\frac{\Delta}{-p^{2}}-[m\to 0]\right)\right\}
\end{align}
Upon explicitly evaluating the integral in parentheses, we find that\\
\BE
\Sigma^{0}_{A}(p)=\frac{\alpha}{4}\ p^{2}\ \left(\frac{2}{\epsilon}-\ln\frac{m^{2}}{\overline{\mu}^{2}}-\frac{1}{3}\ g(s)+\frac{5}{3}\right)
\EE
\\
where $s=-p^{2}/m^{2}$ is an adimensionalized momentum variable, $g(s)$ is a function given by\\
\BE
g(s)=\frac{(1+s)^{3}}{s^{2}}\,\ln (1+s)-s\ln s-\frac{1}{s}
\EE
\\
and we have defined an effective coupling constant $\alpha$ by\\
\BE
\alpha=\frac{3N\alpha_{s}}{4\pi}\qquad\qquad \alpha_{s}=\frac{g^{2}}{4\pi}
\EE
\\
As for the integral $\Sigma_{B}^{0}(p)$, by introducing a Feynman parameter $x$, changing variables of integration and discarding any term which is zero by Lorentz symmetry, we obtain\\
\BE\label{ghsigb}
\Sigma^{0}_{B}(p)=-2iNg^{2}\, \xi\ p^{2}\int_{0}^{1}dx\ (1-x)\int\frac{d^{d}q}{(2\pi)^{d}}\ \frac{\big(1-(2+d)x\big)\,q^{2}/d+x^{2}(1-x)\,p^{2}}{(q^{2}+x(1-x)p^{2})^{3}}
\EE
\\
In dimensional regularization,\\
\begin{align}
\int\frac{d^{d}q}{(2\pi)^{d}}\ \frac{q^{2}}{(q^{2}-\Delta)^{3}}&=\frac{i}{(4\pi)^{2}}\,\frac{d}{4}\ \Gamma\left(2-\frac{d}{2}\right)\,\left(\frac{\Delta}{4\pi\mu^{2}}\right)^{d/2-2}\to\\
\notag&\to\frac{i}{(4\pi)^{2}}\,\frac{d}{4}\,\left(\frac{2}{\epsilon}-\ln\frac{\Delta}{\overline{\mu}^{2}}\right)
\end{align}
\\
The integrand term proportional to $p^{2}$, on the other hand, yields a finite integral, since\\
\begin{align}
\int\frac{d^{d}q}{(2\pi)^{d}}\ \frac{1}{(q^{2}-\Delta)^{3}}&= \frac{-i}{(4\pi)^{2}}\,\frac{1}{2\Delta}
\end{align}
\\
Therefore, the dimensionally regularized expression of $\Sigma^{0}_{B}(p)$ reads\\
\begin{align}
\Sigma^{0}_{B}(p)&=\frac{Ng^{2}}{2(4\pi)^{2}}\, \xi\ p^{2}\int_{0}^{1}dx\ (1-x)\ \bigg\{\big(1-(6-\epsilon)x\big)\,\left(\frac{2}{\epsilon}-\ln\frac{\Delta}{\overline{\mu}^{2}}\right)+2x\bigg\}=\\
\notag&=-\frac{Ng^{2}}{4(4\pi)^{2}}\, \xi\ p^{2}\ \left\{\frac{2}{\epsilon}-\ln\frac{m^{2}}{\overline{\mu}^{2}}-\frac{4}{3}+2\int_{0}^{1} dx\ (1-x)(1-6x)\,\ln\frac{\Delta}{m^{2}}\right\}
\end{align}
\\
where $\Delta=-x(1-x)\,p^{2}$. Upon explicitly computing the integral in the above equation, we find that
\BE
\Sigma^{0}_{B}(p)=-\frac{\alpha}{12}\, \xi\ p^{2}\ \left(\frac{2}{\epsilon}-\ln\frac{m^{2}}{\overline{\mu}^{2}}-\ln s\right)
\EE
\\
Finally, by summing up $\Sigma_{A}^{0}$ and $\Sigma_{B}^{0}$ we obtain the following expression for the uncrossed ghost self-energy:
\newpage
\BE
\Sigma^{0}(p)=\Sigma^{0,d}(p)+\Sigma^{0,f}(p)
\EE
where
\BE
\Sigma^{0,d}(p)=\frac{\alpha}{4}\ p^{2}\ \left(1-\frac{\xi}{3}\right)\ \left(\frac{2}{\epsilon}-\ln\frac{m^{2}}{\overline{\mu}^{2}}\right)
\EE
\\
is a divergent contribution, while\\
\BE
\Sigma^{0,f}(p)=-\frac{\alpha}{12}\ p^{2}\left(g(s)-5-\xi\,\ln s\right)
\EE
is a finite contribution.\\
\\
\\
\\
\textit{2.2.1.2 Computation of the ghost self-energy: the crossed diagram}\\
\\
Let $\Sigma^{1\,ab}(p)$ denote the crossed diagram in Fig.\ref{ghosed}. Following our discussion in Sec.2.1.2, $\Sigma^{1\,ab}(p)$  may be computed by taking a derivative of $\Sigma^{0\,ab}(p)$ with respect to the mass parameter squared $m^{2}$ according to the law\\
\BE
\Sigma^{1\,ab}(p)=-m^{2}\frac{\partial}{\partial m^{2}}\ \Sigma^{0\,ab}(p)=-\delta^{ab}\ m^{2}\,\frac{\partial}{\partial m^{2}}\ \Sigma^{0}(p)
\EE
\\
The derivative of the divergent part of $\Sigma^{0}(p)$ yields a finite term,\\
\BE
\frac{\partial}{\partial m^{2}}\ \Sigma^{0,d}(p)=-\frac{\alpha}{4}\ \frac{p^{2}}{m^{2}}\ \left(1-\frac{\xi}{3}\right)
\EE
\\
In order to compute the derivative of the finite part of $\Sigma^{0}(p)$, we may use chain differentiation to convert the derivative with respect to $m^{2}$ into a derivative with respect to the adimensional momentum variable $s$,\\
\BE
\frac{\partial}{\partial m^{2}}=\frac{\partial(-p^{2}/m^{2})}{\partial m^{2}}\frac{\partial}{\partial (-p^{2}/m^{2})}=-\frac{s}{m^{2}}\frac{\partial}{\partial s}
\EE
We then find that
\BE
\frac{\partial}{\partial m^{2}}\ \Sigma^{0,f}(p)=\frac{\alpha}{12}\ \frac{p^{2}}{m^{2}}\ \left(sg'(s)-\xi\right)
\EE
\\
Adding up the contributions due to the divergent and finite parts of $\Sigma^{0}(p)$, we obtain\\
\BE
\Sigma^{1}(p)=-\frac{\alpha}{12}\ p^{2}\left(sg'(s)-3\right)
\EE
\\
where the derivative of $g(s)$ reads\\
\BE
g'(s)=\frac{(1+s)^{2}(s-2)}{s^{3}}\,\ln(1+s)-\ln s+\frac{2}{s^{2}}+\frac{2}{s}
\EE
\newpage
\noindent\textit{2.2.1.3 Computation of the ghost self-energy: total self-energy}\\
\\
By adding up the crossed and uncrossed diagrams, we obtain the total ghost self-energy
\BE
\Sigma^{ab}(p)=\delta^{ab}\ \left(\Sigma^{d}(p)+\Sigma^{f}(p)\right)
\EE
where
\BE\label{ghodiv}
\Sigma^{d}(p)=\frac{\alpha}{4}\ p^{2}\ \left(1-\frac{\xi}{3}\right)\ \left(\frac{2}{\epsilon}-\ln\frac{m^{2}}{\overline{\mu}^{2}}\right)
\EE
\\
is a divergent contribution and\\
\BE
\Sigma^{f}(p)=-\frac{\alpha}{12}\ p^{2}\left(g(s)+sg'(s)-8-\xi\,\ln s\right)
\EE
\\
is a finite contribution. By explicitly evaluating the sum $g(s)+sg'(s)$, $\Sigma^{f}(p)$ can be put in the form
\BE
\Sigma^{f}(p)=-\alpha\ p^{2}\left(G(s)-\frac{2}{3}-\frac{\xi}{12}\,\ln s\right)
\EE
where
\BE\label{gfun}
G(s)=\frac{1}{12}\,[g(s)+sg'(s)]=\frac{1}{12}\left[\frac{(1+s)^{2}(2s-1)}{s^{2}}\,\ln(1+s)-2s\ln s+\frac{1}{s}+2\right]
\EE
\\
\\
\\
\textit{2.2.1.4 Renormalization of the ghost self-energy and a universal expression for the ghost propagator}\\
\\
As anticipated earlier in this section, neither of the two diagrams of the one-loop ghost self-energy contains mass divergences. The divergent part of the self-energy, given by eq.~\eqref{ghodiv}, is identical to its analogue in ordinary perturbation theory \cite{itzyk}, and may be absorbed in the standard ghost field-strength renormalization counterterm,\\
\BE
\Sigma^{ct}(p)=-\delta_{2}^{c}\ p^{2}
\EE
where $\delta_{2}^{c}=Z_{2}^{c}-1$. If we define the renormalization counterterm so that\\
\BE
\delta_{2}^{c}=\frac{\alpha}{4}\ \left\{\left(1-\frac{\xi}{3}\right)\ \left(\frac{2}{\epsilon}-\ln\frac{m^{2}}{\overline{\mu}^{2}}\right)+\frac{8}{3}+4g_{0}\right\}
\EE
\\
where $g_{0}$ is an arbitrary renormalization constant, then the sum $\Sigma^{d}(p)+\Sigma^{ct}(p)$ adds up to\\
\BE
\Sigma^{d}(p)+\Sigma^{ct}(p)=-\alpha\ p^{2}\ \left(\frac{2}{3}+g_{0}\right)
\EE
\\
and we are left with a finite renormalized self-energy given by\\
\BE\label{sigmaren}
\Sigma_{R}(p)=-\alpha\ p^{2}\left(G(s)+g_{0}-\frac{\xi}{12}\,\ln s\right)
\EE
\\
The constant $g_{0}$ depends on the renormalization conditions for the ghost propagator.
\newpage
\noindent Plugging $\Sigma(p)=\Sigma_{R}(p)$ into eq.~\eqref{ghprfor}, we obtain the following renormalized expression for the one-loop ghost propagator in an arbitrary covariant gauge:
\BE
\widetilde{\mc{G}}^{ab}(p)=\delta^{ab}\ \widetilde{\mc{G}}(p)
\EE
where
\BE\label{ghdrpr0}
\widetilde{\mc{G}}(p)=\frac{i}{p^{2}[1+\alpha\ (G(s)+g_{0}-\xi\,\ln s/12)]}
\EE
\\
A further simplification is achieved by absorbing the one in parentheses into a new arbitrary constant $G_{0}$, defined as
\BE
G_{0}=g_{0}+\frac{1}{\alpha}
\EE
in terms of which the ghost propagator reads\\
\BE\label{ghdrprs}
\widetilde{\mc{G}}(p)=\frac{iZ_{\mc{G}}}{p^{2}\,(G(s)+G_{0}-\xi\,\ln s/12)}\qquad\qquad\quad Z_{\mc{G}}=\frac{1}{\alpha}
\EE
\\
Since in general the dressed propagators are defined modulo a multiplicative normalization factor, $Z_{\mc{G}}$ may actually be given an arbitrary value.\\
\\
Through eq.~\eqref{ghdrprs} we managed to express the dressed ghost propagator in terms of two adimensional renormalization constants:  a multiplicative constant -- $Z_{\mc{G}}$ -- and an additive constant -- $G_{0}$. The coupling constant $g$ -- having been reabsorbed into the definition of $Z_{\mc{G}}$ and $G_{0}$ -- has completely disappeared from our equations and we are left with a propagator which depends only on the renormalization conditions and on the energy scale set by the mass parameter $m$.
\\
\\
\addcontentsline{toc}{subsection}{2.2.2 UV and IR behavior of the ghost propagator, the ghost mass and a comparison with the lattice data}  \markboth{2.2.2 UV and IR behavior of the ghost propagator, the ghost mass and a comparison with the lattice data}{2.2.2 UV and IR behavior of the ghost propagator, the ghost mass and a comparison with the lattice data}
\subsection*{2.2.2 UV and IR behavior of the ghost propagator, the ghost mass and a comparison with the lattice data\index{UV and IR behavior of the ghost propagator, the ghost mass and a comparison with the lattice data}}

\begin{figure}[H]
\centering
\vskip-20pt
\includegraphics[width=0.70\textwidth]{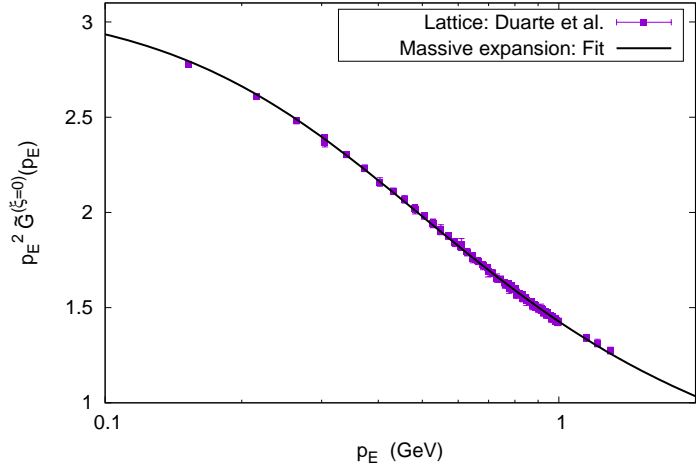}
\vskip-20pt
\caption{Ghost dressing function in the Landau gauge ($\xi=0$) as a function of the Euclidean momentum. Solid line: one-loop approximation in massive perturbation theory with the parameters of Tab.\ref{fitdatagh}. Squares: lattice data from ref.\cite{duarte}.}\label{ghproplandfit}
\end{figure}

Having computed the ghost propagator, we are now in a position to test our results against the predictions of both ordinary perturbation theory and the lattice calculations.\\
\\
Let us start from the high energy behavior of the propagator. In the UV, ordinary perturbation theory is effective in describing the dynamics of Yang-Mills theory and -- as discussed in Sec.2.1.2 -- the expressions derived in the massive expansion should match the standard results. By taking the high momentum limit ($|s|\gg 1$) of the function $G(s)$, we find that
\BE
\lim_{|s|\to\infty}\ G(s)=\frac{1}{4}\,\ln s+\frac{1}{3}
\EE
\\
Hence in the UV the renormalized ghost self-energy -- eq.~\eqref{sigmaren} -- reads\\
\BE\label{ghsehe}
\Sigma_{R}(p)=-\frac{\alpha}{4}\ p^{2}\left[\left(1-\frac{\xi}{3}\right)\ln\left(\frac{-p^{2}}{m^{2}}\right)+\frac{4}{3}+g_{0}\right]\qquad\qquad (|p^{2}|\gg m^{2})
\EE
\\
By inspection of eq.~\eqref{ghodiv}, we see that the logarithm in the above expression has the same coefficient as the logarithm in the divergent part $\Sigma^{d}$ of the unrenormalized self-energy. This was to be expected since, if in the high energy limit the massive expansion is to reduce to ordinary perturbation theory, then for sufficiently large momenta the mass parameter $m$ has to disappear from the equations. As a matter of fact, if we add back the divergent part of the self-energy to eq.~\eqref{ghsehe}, we find that\\
\BE\label{jkl}
\lim_{|p^{2}|\to \infty}\ \Sigma(p)=\frac{3N\alpha_{s}}{16\pi}\ p^{2}\ \left(1-\frac{\xi}{3}\right)\ \left(\frac{2}{\epsilon}-\ln\frac{-p^{2}}{\overline{\mu}^{2}}\right)+\alpha_{s}\ p^{2}\ \mc{C}
\EE
\\
where $\mc{C}$ is an additive renormalization constant. This is precisely the result of ordinary perturbation theory, as reported for instance in \cite{itzyk}. Eq.~\eqref{jkl} establishes that the ghost propagator, as computed in the massive expansion, has the correct high-energy behavior.\\
\\
In the infrared regime, most of our knowledge on the dynamics of Yang-Mills theory comes from the lattice calculations. These inform us that in the limit of vanishing momentum the exact ghost propagator grows to infinity, implying that the ghosts -- at variance with the gluons -- do not acquire a mass through the interactions. In order to investigate the $p\to 0$ limit of the ghost propagator, we need to send $s\to 0$ in the function $G(s)$,\\
\BE
\lim_{s\to0}\ G(s)=\frac{5}{24}
\EE
\\
and plug the result back into eq.~\eqref{ghdrprs}. By doing so, we find that the ghost propagator has the following zero-momentum limit:\\
\BE
\widetilde{\mc{G}}(p)=\frac{iZ_{\mc{G}}}{p^{2}\,[5/24+G_{0}-\xi\,\ln (-p^{2}/m^{2})/12]}\qquad\qquad(p^{2}\to 0)
\EE
\\
Since $p^{2}\ln(-p^{2})\to 0$ as $p^{2}\to 0$, in the limit of zero-momentum the denominator vanishes and the propagator grows to infinity. Therefore -- in accordance with the lattice results -- the massive expansion predicts that the ghosts remain massless even after the radiative corrections are included. In passing, notice that while in the Landau gauge ($\xi=0$) $p^{2}\widetilde{\mc{G}}(p)$ is finite and non-zero as $p\to 0$, in other gauges $p^{2}\widetilde{\mc{G}}(p)\to 0$ due to the logarithmic term.\\
\\
We can now go one step further and test the predictions of the massive expansion against the results of the lattice calculations. As discussed in the Introduction, the lattice data is defined in Euclidean space, obtained from Minkowski space by complexifying the time components of the Minkowski vectors and restricting them to imaginary values. In momentum space, starting from the Minkowski momentum $p=(p^{0},{\bf p})$, we assume that $p^{0}$ can take on any value in the complex plane and then ask that $p^{0}$ be imaginary, so that $p^{0}=ip^{4}$ for some real $p^{4}$. Then to the complexified Minkowski momentum $p=(ip^{4},{\bf p})$ we associate an ordinary Euclidean momentum $p_{E}=(p^{4},{\bf p})$ which has Euclidean norm $p_{E}^{2}=(p^{4})^{2}+|{\bf p}|^{2}=-p^{2}$, the latter being the Minkowski norm of $p$.\\
By complexifying the time component of the momentum variable of a Minkowski propagator, restricting to imaginary values and multiplying the propagator by $i$, we obtain the Euclidean version of the propagator, which can be computed non-perturbatively on the lattice. For the ghosts, this amounts to defining\\
\BE
\widetilde{G}(p_{E})=i\,\widetilde{\mc{G}}(p(p_{E}))
\EE
\\
$\widetilde{G}(p_{E})$ being the Euclidean ghost propagator. Going back to eq.~\eqref{ghdrprs}, by replacing $p^{2}$ with $-p_{E}^{2}$ and multiplying by $i$, we obtain the one-loop approximation to $\widetilde{G}(p_{E})$ in the form
\newpage
\BE\label{ghdrprseu}
\widetilde{G}(p_{E})=\frac{Z_{\mc{G}}}{p^{2}_{E}\,(G(s)+G_{0}-\xi\,\ln s/12)}
\EE
\\
where now $s=p_{E}^{2}/m^{2}$. The Euclidean ghost propagator is expressed in terms of the same renormalization constants $Z_{\mc{G}}$ and $G_{0}$ that define the propagator in Minkowski space.\\
Since -- especially for the ghosts -- the lattice calculations are the most reliable in the Landau gauge ($\xi=0$), in what follows we will limit ourself to this gauge and compare our results with the lattice data of ref.\cite{duarte} (Duarte et al.), which were already presented in the Introduction. In the Landau gauge,\\
\BE\label{ghdrprseuland}
\widetilde{G}^{(\xi=0)}(p_{E})=\frac{Z_{\mc{G}}}{p^{2}_{E}\,(G(s)+G_{0})}
\EE
\\
In the above expression, there are three parameters to be fixed, namely $Z_{\mc{G}}$, $G_{0}$ and the mass parameter $m$. The propagators of ordinary perturbation theory, by contrast, only depend on two parameters: the coupling constant and a multiplicative renormalization constant (together with some renormalization conditions). In Sec.2.2.1 we have absorbed the coupling constant into the renormalization constants, so that the number of adimensional free parameters of the ghost propagator is still equal to two; however, due to the shift of the kinetic action, we are still left with the spurious mass parameter, which cannot be fixed by first principles since classically Yang-Mills theory does not possess any mass scales. The presence of an extra free parameter in the equations, at face value, may seem to cause the massive expansion to lose its predictive power. Nonetheless, in the next chapter we will see that in order for gauge invariance to be preserved in the massive expansion, one of the three free parameters of the dressed \textit{gluon} propagator needs to be fixed to a specific value, resulting in a reduction in the number of adimensional free parameters which actually compensates for the arbitrariness of $m$.\\
Here and in the following section we do not concern ourselves with the issue of the predictivity of the massive expansion. Our only interest is determining whether our results agree with the lattice data for some choice of the values of the free parameters, a task which can be easily carried out by fitting the data to our functions and checking that the resulting fit well reproduces the data. Since the value of the mass parameter $m$ -- as we will see in the next section -- has a strong influence on the infrared behavior of the gluon propagator, it is best to fit it on the basis of the lattice data for the gluons, rather than on that of the ghosts. For the fit of the ghost propagator, we fix $m=0.654$ GeV, which is the value that we will find for the Landau gauge gluon propagator in Sec.2.3.2.\\
In Fig.\ref{ghproplandfit} we show the lattice data of ref.\cite{duarte} for the Landau gauge Euclidean ghost dressing function $p_{E}^{2}\,\widetilde{G}^{(\xi=0)}(p_{E})$ together with the one-loop results of massive perturbation theory, given by eq.~\eqref{ghdrprseuland} mutiplied by $p_{E}^{2}$. At fixed $m=0.654$ GeV, the values of $Z_{\mc{G}}$ and $G_{0}$ which were found to best fit the data are reported in Tab.\ref{fitdatagh}.\\
\
\\
\begin{table}[H]
\def\arraystretch{1.2}
\centering
\begin{tabular}{cc|c}
\hline
$G_{0}$&$Z_{\mc{G}}$&$m$ (GeV)\\
\hline
\hline
$0.1464$&$1.0994$&$0.654$\\
\hline
\end{tabular}
\vspace{3mm}
\caption{Parameters found from the fit of the lattice data of ref.\cite{duarte} for the Landau gauge ghost propagator in the range $0-2$ GeV at fixed $m=0.654$ GeV.}\label{fitdatagh}
\end{table}
\newpage
\
\
\\
\\
Setting $m=0.654$ GeV and $G_{0}=0.146$ in eq.~\eqref{ghdrprseuland} we obtain a dressing function $p_{E}^{2}\,\widetilde{G}^{(\xi=0)}(p_{E})$ -- hence a propagator $\widetilde{G}^{(\xi=0)}(p_{E})$ -- which is in excellent agreement with the lattice data for the ghosts in the infrared regime, showing that the massive perturbative expansion indeed is capable of accurately reproducing the low energy dynamics of the ghosts.

\clearpage

\addcontentsline{toc}{section}{2.3 The one-loop gluon propagator}  \markboth{2.3 The one-loop gluon propagator}{2.3 The one-loop gluon propagator}
\section*{2.3 The one-loop gluon propagator\index{The one-loop gluon propagator}}

\addcontentsline{toc}{subsection}{2.3.1 Definition and computation}  \markboth{2.3.1 Definition and computation}{2.3.1 Definition and computation}
\subsection*{2.3.1 Definition and computation\index{Definition and computation}}

\begin{figure}[H]
\centering
\includegraphics[width=\textwidth]{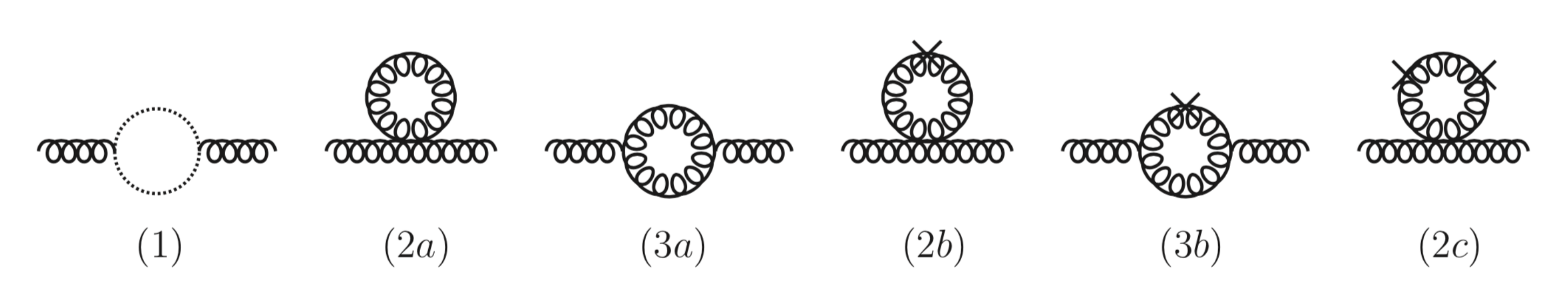}
\caption{Loop diagrams which contribute to the one-loop gluon polarization.}\label{glupoldiag}
\end{figure}

The dressed gluon propagator $\widetilde{\mc{D}}^{ab}_{\mu\nu}(p)$ is defined as\\
\BE\label{glprfor}
\widetilde{\mc{D}}_{\mu\nu}^{ab}(p)=\int d^{d}x\ \ e^{ip\cdot x}\ \bra{\Omega}T\big\{A_{\mu}^{a}(x)\,A_{\nu}^{b}(0)\big\}\ket{\Omega}
\EE
\\
$\widetilde{\mathcal{D}}^{ab}_{\mu\nu}(p)$ is related to the one-particle-irreducible gluon polarization $\Pi_{\mu\nu}^{ab}(p)$ through the equation
\BE\label{glprplfor}
\widetilde{\mathcal{D}}^{ab}_{\mu\nu}(p)=\delta^{ab}\ \left\{\frac{-i}{p^{2}-m^{2}-\Pi_{T}(p)}\ t_{\mu\nu}(p)+\xi\ \frac{-i}{p^{2}-\xi\,\Pi_{L}(p)}\ \ell_{\mu\nu}(p)\right\}
\EE
\\
where $\Pi_{T}$ and $\Pi_{L}$ are the transverse and longitudinal components of the polarization,\\
\BE
\Pi^{ab}_{\mu\nu}(p)=\delta^{ab}\ \left\{\Pi_{T}(p)\ t_{\mu\nu}(p)+\Pi_{L}(p)\ \ell_{\mu\nu}(p)\right\}
\EE
\\
Because of gauge invariance, the exact longitudinal component of the gluon polarization $\Pi_{L}$ is known to vanish \cite{peskin,itzyk}, so that the exact longitudinal gluon propagator $\widetilde{\mc{D}}_{L}$ reads\\
\BE\label{longprop}
\widetilde{\mc{D}}_{L}(p)=\xi\ \frac{-i}{p^{2}}
\EE
\\
If in ordinary perturbation theory this constraint can be shown to be fulfilled to any fixed loop order \cite{wein2,itzyk}, in the massive expansion the massiveness of the bare gluon propagator in general causes the approximate longitudinal polarization to be non-zero until the full set of crossed diagrams is resummed. In the Landau gauge ($\xi=0$) this poses no problem, since due to the $\xi$'s in eq.~\eqref{glprplfor} the dressed gluon propagator of the Landau gauge remains transverse anyway. In different gauges, however, $\Pi_{L}\neq 0$ may lead to an incorrect approximation to the exact longitudinal dressed propagator, eq.~\eqref{longprop}. With respect to this issue, we take the view that, since the exact longitudinal polarization is known to vanish, there is no actual need to compute it perturbatively: we may set it to zero from the very start. From a computational point of view, this may formally be achieved by resumming the infinite set of crossed diagrams limited to the longitudinal component of the polarization, thus reverting back to ordinary perturbation theory where we know the constraint $\Pi_{L}=0$ to hold at any fixed loop order. In what follows, we will assume that such a resummation has been performed and that $\Pi_{L}=0$. We will therefore limit ourselves to the computation of the transverse component of the gluon polarization $\Pi_{T}$, dropping the subscript $_T$, referring to it simply as the gluon polarization and setting\\
\BE\label{glprplfors}
\widetilde{\mathcal{D}}^{ab}_{\mu\nu}(p)=\delta^{ab}\ \left\{\frac{-i}{p^{2}-m^{2}-\Pi(p)}\ t_{\mu\nu}(p)+\xi\ \frac{-i}{p^{2}}\ \ell_{\mu\nu}(p)\right\}
\EE
\\
To one loop, the irreducible uncrossed diagrams which contribute to the gluon polarization are the first three in Fig.\ref{glupoldiag}. The gluon tadpole ($2a$) and the gluon loop ($3a$) will be shown to contain mass divergences which need to be eliminated by including in the polarization analogous crossed diagrams ($2b$ and $3b$). Since the crossed gluon loop (diagram $3b$) is of order three in the total number of vertices, we will also include in the polarization the doubly crossed tadpole ($3c$), which contains the same number of vertices.
\begin{figure}[H]
\centering
\includegraphics[width=0.25\textwidth]{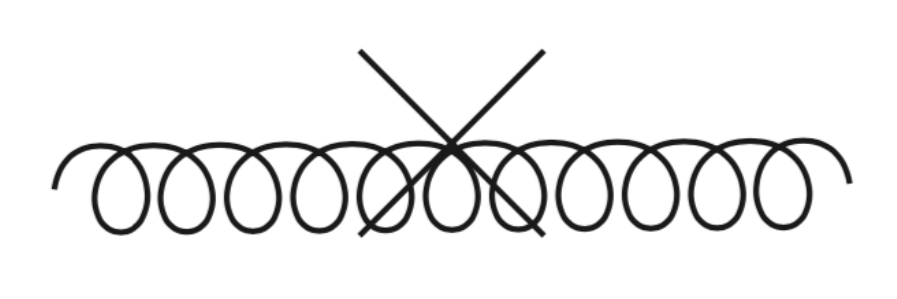}
\caption{Single-counterterm diagram.}\label{gluctdiag}
\end{figure}
\
\\
\\
Up to order three in the number of vertices, there is one last irreducible diagram, the loopless, single-counterterm diagram in Fig.\ref{gluctdiag}. From a theoretical point of view, this is a mostly important diagram. By multiplying the mass counterterm vertex by $-i$, we obtain its contribution $\Pi^{ab}_{ct\,\mu\nu}$ to the gluon polarization,\\
\BE
\Pi^{ab}_{ct\,\mu\nu}(p)=-\delta^{ab}\ m^{2}\ t_{\mu\nu}(p)\qquad\qquad\Pi_{ct}(p)=-m^{2}
\EE
\\
If we now split the one-loop gluon polarization $\Pi$ into the sum $\Pi_{ct}+\Pi_{\tx{loops}}$, where $\Pi_{\tx{loops}}$ contains the contributions due to the six diagrams in Fig.\ref{glupoldiag}, since $m^{2}+\Pi_{ct}(p)=0$, we find that
\BE\label{glprplforss}
\widetilde{\mathcal{D}}^{ab}_{\mu\nu}(p)=\delta^{ab}\ \left\{\frac{-i}{p^{2}-\Pi_{\tx{loops}}(p)}\ t_{\mu\nu}(p)+\xi\ \frac{-i}{p^{2}}\ \ell_{\mu\nu}(p)\right\}
\EE
\\
Therefore, once the single-counterterm diagram has been included in the polarization, the tree-level gluon mass $m^{2}$ produced by the shift of the kinetic action disappears from the propagator. Thus the mass of the gluon, if any, does not trivially result from the shift, but rather has to come from the loops of the polarization: the mass derived in the framework of massive perturbation theory is a truly dynamically generated mass. It is non-zero if in the limit of vanishing momentum $\Pi_{\tx{loops}}(p)$ approaches a finite, non-zero value, so that the propagator remains finite at $p=0$ instead of growing to infinity\footnote{\ It can be shown, for instance, that in the framework of massive QED perturbation theory the photon loop polarization vanishes at zero momentum, so that the photons do not acquire a mass.}. In order to reach a conclusion with respect to the issue of mass generation in Yang-Mills theory, we need to investigate the zero-momentum limit of $\Pi_{\tx{loops}}$.\\
\\
\\
\textit{2.3.1.1 Computation of the gluon polarization: the uncrossed diagrams}\\
\\
Let us denote with $\Pi^{ab}_{(1)\,\mu\nu}(p)$ the contribution to the gluon polarization due to the ghost loop -- diagram $1$ in Fig.\ref{glupoldiag}. $\Pi^{ab}_{(1)\,\mu\nu}(p)$ may be put in the form\\
\BE
\Pi^{ab}_{(1)\,\mu\nu}(p)=iNg^{2}\ \delta^{ab}\int\frac{d^{d}q}{(2\pi)^{d}}\ \frac{q^{\mu}\,(q-p)^{\nu}}{q^{2}(q-p)^{2}}
\EE
Since this diagram does not contain internal gluon lines nor mass counterterms, the above expression is identical to its analogue in ordinary perturbation theory. An explicit computation in dimensional regularization shows that the transverse component of $\Pi^{ab}_{(1)\,\mu\nu}(p)$, $\Pi_{(1)}(p)$, is given by
\BE
\Pi_{(1)}(p)=\frac{\alpha}{36}\ p^{2} \left(\frac{2}{\epsilon}-\ln\frac{m^{2}}{\overline{\mu}^{2}}-\ln s+2\right)
\EE
\\
where $\overline{\mu}$ is the $\overline{\tx{MS}}$ mass scale, $\epsilon=4-d$, $s=-p^{2}/m^{2}$ is an adimensionalized momentum variable and
\BE
\alpha=\frac{3N\alpha_{s}}{4\pi}\qquad\qquad\alpha_{s}=\frac{g^{2}}{4\pi}
\EE
as in Sec.2.2.\\
The gluon tadpole -- diagram $2a$ in Fig.\ref{glupoldiag} -- contributes to the self-energy with the momentum-independent term\\
\BE
\Pi^{ab}_{(2a)\,\mu\nu}(p)=\frac{iN(d-1)^{2}g^{2}}{d}\ \delta^{ab}\,\eta_{\mu\nu}\int\frac{d^{d}q}{(2\pi)^{d}}\ \frac{1}{q^{2}-m^{2}}
\EE
\\
which in dimensional regularization yields a transverse projection\\
\BE
\Pi_{(2a)}(p)=-\frac{3\alpha}{4}\ m^{2} \left(\frac{2}{\epsilon}-\ln\frac{m^{2}}{\overline{\mu}^{2}}+\frac{1}{6}\right)
\EE
\\
As anticipated earlier in this section, the tadpole diagram contains a non-renormalizable mass divergence which will be removed from the polarization by an opposite divergence in the singly-crossed tadpole -- diagram $2b$.\\
Finally, the contribution due to the gluon loop -- diagram $3a$ in Fig.\ref{glupoldiag} -- is given by\\
\begin{align}
\Pi^{ab}_{(3a)\,\mu\nu}(p)=\Pi^{0\,ab}_{(3a)\,\mu\nu}(p)+\xi\,\Pi^{\xi\,ab}_{(3a)\,\mu\nu}(p)+\xi^{2}\,\Pi^{\xi\xi\,ab}_{(3a)\,\mu\nu}(p)
\end{align}
\\
where the transverse components of the the three polarization terms can be put in the form\\
\begin{align}
\notag\Pi^{0}_{(3a)}(p)&=\frac{-iNg^{2}}{6}\int\frac{d^{d}q}{(2\pi)^{d}}\ \frac{q_{\perp}^{2}\mathcal{F}^{0}(q,p)}{(q^{2}-m^{2}[(q+p)^{2}-m^{2}]}\\
\frac{\mathcal{F}^{0}(q,p)}{12}&=\frac{q^{2}+p^{2}}{q^{2}}+\frac{p^{2}}{(q+p)^{2}}-\frac{p^{2}q_{\perp}^{2}}{3(q+p)^{2}q^{2}}\\
\notag q_{\perp}^{2}&=q^{2}-\frac{(q\cdot p)^{2}}{p^{2}}
\end{align}
\begin{align}
\notag\Pi^{\xi}_{(3a)}(p)&=\frac{-iNg^{2}}{6}\int\frac{d^{d}q}{(2\pi)^{d}}\ \left\{\frac{\mc{F}^{0\xi}(q,p)}{(q^{2}-m^{2})(q+p)^{2}}+\frac{\mc{F}^{\xi0}(q,p)}{q^{2}[(q+p)^{2}-m^{2}]}\right\}\\
\mc{F}^{0\xi}(q,p)&=\frac{(3q^{2}-q_{\perp}^{2})(q^{2}-p^{2})^{2}}{q^{2}(q+p)^{2}}\\
\notag \mc{F}^{\xi0}(q,p)&=3q^{2}+12\,p^{2}+12\,q\cdot p-q_{\perp}^{2}\left(\frac{11p^{2}+2q\cdot p}{q^{2}}+\frac{p^{4}}{(q+p)^{2}q^{2}}\right)
\end{align}
\clearpage
\begin{align}\label{xixi}
\Pi^{\xi\xi}_{(3a)}(p)&=\frac{-iNg^{2}}{6}\int\frac{d^{d}q}{(2\pi)^{d}}\ \frac{\mc{F}^{\xi\xi}(q,p)}{q^{2}(q+p)^{2}}\\
\notag \mc{F}^{\xi\xi}(q,p)&=\frac{p^{4}q_{\perp}^{2}}{q^{2}(q+p)^{2}}
\end{align}
\\
The integral in eq.~\eqref{xixi} is convergent and turns out to be equal to a finite constant times $\xi^{2}p^{2}$. Since these constants can be absorbed in the gluon field-strength renormalization counterterm (see ahead), we will not keep track of it. A straightforward albeit tedious calculation then leads to the following expression for the transverse component of the gluon loop:\\
\begin{align}
\Pi_{(3a)}(p)&=\alpha\left[\left(1+\frac{\xi}{4}\right)m^{2}+\left(\frac{25}{36}-\frac{\xi}{6}\right)\ p^{2}\right]\left(\frac{2}{\epsilon}-\ln\frac{m^{2}}{\overline{\mu}^{2}}\right)+\\
\notag&\quad\quad+\frac{\alpha}{72}\ p^{2}\left(\frac{2}{s^{2}}-\frac{135}{s}+\frac{226}{3}+s^{2}\ln s-l_{A}(s)-l_{B}(s)+\mc{C}_{0}\right)+\\
\notag&\quad\quad+\frac{\alpha}{12}\ \xi\ p^{2}\left(\frac{(1+s)(1-s)^{3}}{s^{3}}\ln(1+s)+s\ln s-\frac{1}{s^{2}}+\mc{C}_{\xi}\right)
\end{align}
\\
where $l_{A}(s)$ and $l_{B}(s)$ are logarithmic functions given by
\begin{align}
l_{A}(s)&=(s^{2}-20s+12)\left(\frac{4+s}{s}\right)^{3/2}\ln\left(\frac{\sqrt{4+s}-\sqrt{s}}{\sqrt{4+s}+\sqrt{s}}\right)\\
l_{B}(s)&=\frac{2(1+s)^{3}}{s^{3}}(s^{2}-10s+1)\ln(1+s)
\end{align}
\\
and $\mc{C}_{0}$ and $\mc{C}_{\xi}$ are irrelevant finite constants. Just as the tadpole, the gluon loop contains a non-renormalizable mass divergence which is removed from the polarization by an opposite divergence in the crossed gluon loop -- diagram $3b$.\\
\\
\\
\textit{2.3.1.2 Computation of the gluon polarization: the crossed diagrams}\\
\\
The three crossed diagrams in Fig.\ref{glupoldiag} may be computed by deriving the uncrossed diagrams $2a$ and $3a$ with respect to the mass parameter squared $m^{2}$ (cfr. Sec.2.1.2). To be specific, denoting with $\Pi_{(2b)},\,\Pi_{(2c)}$ and $\Pi_{(3b)}$ the transverse components of the polarization terms due to the respective diagrams, we have\\
\begin{align}
\notag\Pi_{(2b)}&=-m^{2}\,\frac{\partial}{\partial m^{2}}\ \Pi_{(2a)}\\
\Pi_{(2c)}&=\frac{m^{4}}{2}\,\frac{\partial^{2}}{\partial (m^{2})^{2}}\ \Pi_{(2a)}\\
\notag\Pi_{(3b)}&=-m^{2}\,\frac{\partial}{\partial m^{2}}\ \Pi_{(3a)}
\end{align}
\\
Again, when needed, we may replace the derivative with respect to $m^{2}$ by a derivative with respect to $s=-p^{2}/m^{2}$ through chain differentiation,\\
\BE
\frac{\partial}{\partial m^{2}}=-\frac{s}{m^{2}}\,\frac{\partial}{\partial s}
\EE
\\
An explicit calculation leads to the following expressions for the crossed polarization terms:\\
\begin{align}
\Pi_{(2b)}(p)&=\frac{3\alpha}{4}\ m^{2}\left(\frac{2}{\epsilon}-\ln\frac{m^{2}}{\overline{\mu}^{2}}-\frac{5}{6}\right)\\
\Pi_{(2c)}(p)&=\frac{3\alpha}{8}\ m^{2}\\
\Pi_{(3b)}(p)&=-\alpha\left(1+\frac{\xi}{4}\right)m^{2}\left(\frac{2}{\epsilon}-\ln\frac{m^{2}}{\overline{\mu}^{2}}\right)+\\
\notag&\quad+\frac{\alpha}{72}\ p^{2}\bigg(-\frac{4}{s^{2}}+\frac{63}{s}+50+s^{2}+2s^{2}\ln s-s\,l_{A}'(s)-s\,l_{B}'(s)\bigg)+\\
\notag&\quad-\frac{\alpha}{12}\ \xi\ p^{2}\left(\frac{(1-s)^{2}(s^{2}+2s+3)}{s^{3}}\ln(1+s)-s\ln s-\frac{3}{s^{2}}-\frac{3}{s}-1+3s\right)
\end{align}
\\
where the derivatives of the logarithmic functions read\\
\begin{align}
l'_{A}(s)&=\sqrt{\frac{4+s}{s}}\left\{\frac{2(s^{3}-9s^{2}+20s-36)}{s^{2}}\ln\left(\frac{\sqrt{4+s}-\sqrt{s}}{\sqrt{4+s}+\sqrt{s}}\right)-\frac{s^{2}-20s+12}{s}\right\}\\
l'_{B}(s)&=\frac{2(1+s)^{2}}{s^{4}}(2s^{3}-11s^{2}+20s-3)+\frac{2(1+s)^{2}}{s^{3}}(s^{2}-10s+1)
\end{align}
\\
\\
\textit{2.3.1.3 Computation of the gluon polarization: total polarization}\\
\\
By adding up the contributions due to the six loops in Fig.\ref{glupoldiag}, we find the following expression for the transverse component $\Pi_{\tx{loops}}$ of the one-loop gluon polarization:\\
\begin{align}
\Pi_{\tx{loops}}(p)=\Pi_{\tx{loops}}^{d}(p)+\Pi_{\tx{loops}}^{f}(p)
\end{align}
where
\BE\label{gluprdiv}
\Pi_{\tx{loops}}^{d}(p)=\frac{\alpha}{3}\ \left(\frac{13}{6}-\frac{\xi}{2}\right)\, p^{2}\left(\frac{2}{\epsilon}-\ln\frac{m^{2}}{\overline{\mu}^{2}}\right)
\EE
\\
is a divergent term and\\
\BE
\Pi_{\tx{loops}}^{f}(p)=-\alpha\ p^{2}\left(F(s)+\xi F_{\xi}(s)+\mc{C}\right)
\EE
\\
is a finite term, with $\mc{C}$ an irrelevant finite constant. In $\Pi_{\tx{loops}}^{f}$ the functions $F(s)$ and $F_{\xi}(s)$ are defined as\\
\begin{align}
\label{ffunct}F(s)&=\frac{5}{8s}+\frac{1}{72}\ \left[L_{a}(s)+L_{b}(s)+L_{c}(s)+R_{a}(s)+R_{b}(s)+R_{c}(s)\right]\\
F_{\xi}(s)&=\frac{1}{4s}-\frac{1}{12}\left[2s\ln s-\frac{2(1-s)(1-s^{3})}{s^{3}}\ln(1+s)+\frac{3s^{2}-3s+2}{s^{2}}\right]
\end{align}
where the functions $L_{a}(s),\,L_{b}(s),\,L_{c}(s),\,R_{a}(s),\,R_{b}(s)$ and $R_{c}(s)$ are given by\\
\begin{align}
\notag L_{a}(s)&=\frac{3s^{3}-34s^{2}-28s-24}{s}\ \sqrt{\frac{4+s}{s}}\ \ln\left(\frac{\sqrt{4+s}-\sqrt{s}}{\sqrt{4+s}+\sqrt{s}}\right)\\
L_{b}(s)&=\frac{2(1+s)^{2}}{s^{3}}\ (3s^{3}-20s^{2}+11s-2)\ \ln(1+s)\\
\notag L_{c}(s)&=(2-3s^{2})\ \ln s\\
\notag \\
\notag R_{a}(s)&=-\frac{4+s}{s}\ (s^{2}-20s+12)\\
\label{rfunct}R_{b}(s)&=\frac{2(1+s)^{2}}{s^{2}}\ (s^{2}-10s+1)\\
\notag R_{c}(s)&=\frac{2}{s^{2}}+2-s^{2}
\end{align}\\
\\
\\
\textit{2.3.1.4 Renormalization of the gluon polarization and a universal expression for the gluon propagator}\\
\\
Once the crossed diagrams are included in the polarization, the mass divergences arising from the gluon tadpole and the gluon loop cancel and we are left with a renormalizable divergence -- eq.~\eqref{gluprdiv} -- which is identical to its analogue in ordinary perturbation theory \cite{peskin,itzyk}. As usual, this divergence may be absorbed in the gluon field-strength renormalization contribution to the polarization,\\
\BE
\Pi_{ct\,\mu\nu}^{ab}(p)=-\delta_{3}\ p^{2}\ \delta^{ab}\,t_{\mu\nu}(p)
\EE
\\
by defining the field-strength renormalization counterterm $\delta_{3}=Z_{3}-1$ to be equal to\\
\BE
\delta_{3}=\frac{\alpha}{3}\ \left(\frac{13}{6}-\frac{\xi}{2}\right)\, p^{2}\left(\frac{2}{\epsilon}-\ln\frac{m^{2}}{\overline{\mu}^{2}}\right)+\alpha\ (f_{0}-\mc{C})
\EE
\\
where $f_{0}$ is an arbitrary renormalization constant. With $\delta_{3}$ as above, the renormalized transverse component of the gluon polarization reads\\
\BE\label{glupolff}
\Pi_{R}(p)=-\alpha\ p^{2}\left(F(s)+\xi F_{\xi}(s)+f_{0}\right)
\EE
\\
Therefore, setting $\Pi_{\tx{loops}}=\Pi_{R}$ in eq.\eqref{glprplforss}, we find that the renormalized expression for the transverse component of the gluon propagator is given by\\
\BE\label{gluprfin0}
\widetilde{\mc{D}}_{T}(p)=\frac{-i}{p^{2}[1+\alpha\,(F(s)+\xi F_{\xi}(s)+f_{0})]}
\EE
\\
Finally, by defining an arbitrary renormalization constant $F_{0}$ as\\
\BE
F_{0}=f_{0}+\frac{1}{\alpha}
\EE
we can put the propagator in the form\\
\BE\label{gluprfin}
\widetilde{\mc{D}}_{T}(p)=\frac{-iZ_{\mc{D}}}{p^{2}(F(s)+\xi F_{\xi}(s)+F_{0})}\qquad\qquad \quad Z_{\mc{D}}=\frac{1}{\alpha}
\EE
\\
where, since the propagator is defined modulo an arbitrary multiplicative factor, $Z_{\mc{D}}$ may be given an arbitrary value.\\
\\
Through eq.~\eqref{gluprfin} we have managed to express the gluon propagator in terms of two adimensional renormalization constants: a multiplicative constant -- $Z_{\mc{D}}$ -- and an additive constant -- $F_{0}$. The coupling constant $g$ has disappeared from the equation and we are left with a propagator which depends only on the renormalization conditions and on the energy scale set by the mass parameter $m$. 
\\
\\
\addcontentsline{toc}{subsection}{2.3.2 UV and IR behavior of the gluon propagator, the gluon mass and a comparison with the lattice data}  \markboth{2.3.2 UV and IR behavior of the gluon propagator, the gluon mass and a comparison with the lattice data}{2.3.2 UV and IR behavior of the gluon propagator, the gluon mass and a comparison with the lattice data}
\subsection*{2.3.2 UV and IR behavior of the gluon propagator, the gluon mass and a comparison with the lattice data\index{UV and IR behavior of the gluon propagator, the gluon mass and a comparison with the lattice data}}

\begin{figure}[H]
\centering
\vskip-20pt
\includegraphics[width=0.70\textwidth]{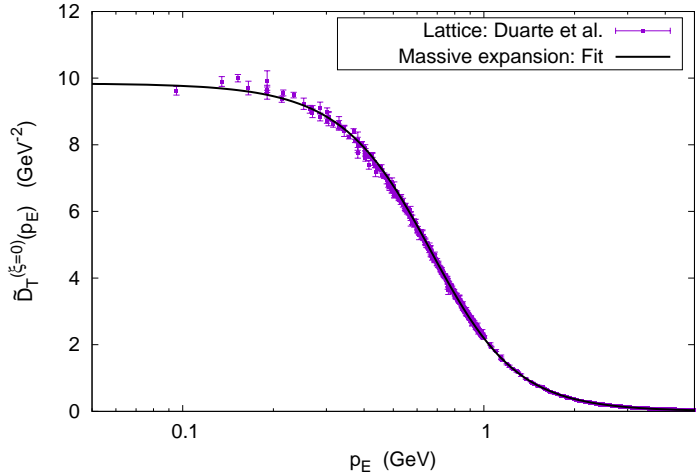}
\vskip-20pt
\caption{Transverse gluon propagator in the Landau gauge ($\xi=0$) as a function of the Euclidean momentum. Solid line: one-loop approximation in massive perturbation theory with the parameters of Tab.\ref{fitdata}. Squares: lattice data from ref.\cite{duarte}.}\label{gluproplandfit}
\end{figure}

Having computed the one-loop gluon propagator, we can now proceed to investigate its UV and IR behavior and compare our results with the predictions of ordinary perturbation theory and the lattice computations.\\
\\
The UV limit of the gluon propagator is obtained by sending $|s|\to \infty$ in the functions $F(s)$ and $F_{\xi}(s)$ defined by eqq.~\eqref{ffunct}-\eqref{rfunct}. An explicit calculation shows that\\
\BE
\lim_{|s|\to \infty}\ F(s)=\frac{17}{18}+\frac{13}{18}\ \ln s\qquad\qquad \lim_{|s|\to\infty}\ F_{\xi}(s)=-\frac{1}{6}\ \ln s -\frac{1}{12}
\EE
\\
Thus, going back to eq.~\eqref{glupolff}, we find that in the UV\\
\BE
\lim_{|p^{2}|\to \infty}\ \Pi_{R}(p)=-\frac{\alpha}{3}\ p^{2}\left(\frac{13}{6}-\frac{\xi}{2}\right)\, \ln \frac{-p^{2}}{m^{2}}+\alpha_{s}\ p^{2}\ \mc{C}
\EE
\\
where $\mc{C}$ is an irrelevant constant. The coefficient of the logarithmic term in $\Pi_{R}$ is the same as that of the divergence in $\Pi^{d}_{\tx{loops}}$ -- eq.~\eqref{gluprdiv} --, as needed to eliminate the mass parameter $m$ from the expressions in the high energy limit. By adding back the divergence to the renormalized polarization, we find that\\
\BE
\Pi_{\tx{loops}}(p)=\frac{N\alpha_{s}}{4\pi}\left(\frac{13}{6}-\frac{\xi}{2}\right)\, p^{2}\left(\frac{2}{\epsilon}-\ln\frac{-p^{2}}{\overline{\mu}^{2}}\right)+\alpha_{s}\ p^{2}\ \mc{C}\qquad\qquad(|p^{2}|\gg m^{2})
\EE
\\
which is the very same result of ordinary perturbation theory, as reported for instance in ref.\cite{itzyk}. Therefore, as anticipated by our discussion in Sec.2.1.2, the one-loop gluon propagator of the massive expansion has the correct, ordinary UV behavior.\\
\\
In the IR, as discussed at length in the Introduction and in Chapter 1, the transverse gluons are reported by the lattice calculations to acquire a dynamical mass. The gluon mass manifests itself in the finiteness of the gluon propagator at vanishing momentum, at variance with the typical behavior of massless propagators which grow to infinity as $p\to 0$. In the limit of vanishing momentum ($s\to 0$), the functions $F(s)$ and $F_{\xi}(s)$ both tend to infinity as $1/s$:
\clearpage
\BE
\lim_{s\to 0}\ F(s)=\frac{5}{8s}\qquad\qquad \lim_{s\to0}\ F_{\xi}(s)=\frac{1}{4s}
\EE
\\
Since in the propagator -- eq.~\eqref{gluprfin} --, the functions $F$ and $F_{\xi}$ are multiplied by a factor of $p^{2}$, the zero-momentum limit of the transverse component of the dressed gluon propagator is finite and reads
\BE
\widetilde{\mc{D}}_{T}(0)=\frac{-iZ_{\mc{D}}}{-M_{\xi}^{2}}
\EE
\\
where $M_{\xi}$ is a gauge-dependent mass scale defined by\\
\BE
M_{\xi}^{2}=\frac{5m^{2}}{8}\left(1+\frac{2\xi}{5}\right)
\EE
\\
Therefore, in accordance with the lattice data, the massive perturbative expansion indeed predicts that the transverse gluon propagator acquires a mass in the infrared regime.\\
As discussed in the introduction to this section, the mass derived in the framework of the massive expansion is dynamical in nature: it does not automatically result from the massive shift of the kinetic action, but rather is generated by the loops of the polarization. Let us see how this comes about. By taking the zero-momentum limit of the loop diagrams in Fig.\ref{glupoldiag} we find that
\begin{align}
\notag\Pi_{(1)}(0)&=0\\
\Pi_{(2a)}(0)+\Pi_{(2b)}(0)+\Pi_{(2c)}(0)&=-\frac{3\alpha}{8}\ m^{2}\\
\notag\Pi_{(3a)}(0)+\Pi_{(3b)}(0)&=\alpha\left(1+\frac{\xi}{4}\right) m^{2}
\end{align}
\\
where we have summed the diagrams which only differ by the number of mass counterterms in order to get rid of the spurious mass divergences. At zero momentum, the ghost loop -- diagram $1$ in Fig.\ref{glupoldiag} -- vanishes, hence it does not contribute to the gluon mass. The gluon tadpoles and the gluon loops -- diagrams $2a$-$2c$ and $3a$-$3b$ in Fig.\ref{glupoldiag} --, on the other hand, do not vanish and sum to $\alpha$ times $M_{\xi}^{2}$. Therefore we conclude that the mass of the gluons is generated both by the gluon tadpoles and by the gluon loops.\\
\\
The ability to predict a non-zero mass for the gluon propagator is a necessary condition for the validity of any result on the low energy dynamics of the gluons in Yang-Mills theory. Having shown that in massive perturbation theory the one-loop transverse gluon propagator is massive as it should be, we can now proceed to compare our predictions with the lattice data. For the comparison we will again use the data reported in ref.\cite{duarte} by Duarte et al., already presented in the Introduction. By analytically continuing the Minkowski momentum $p$ to the Euclidean momentum $p_{E}$ -- equivalently, by setting $p^{2}=-p_{E}^{2}$ -- and by multiplying the Minkowski propagator by $-i$\footnote{\ The minus sign in $-i$ is due to the vector nature of the gluon field: as we go from Minkowski space to the Euclidean space we must replace everywhere the Minkowski metric $\eta_{\mu\nu}$ with the Euclidean metric $-\delta_{\mu\nu}$, causing a change of sign in the tensor propagator $\mc{D}_{\mu\nu}(p)$.}, we obtain the transverse gluon propagator in Euclidean space $\widetilde{D}_{T}(p_{E})$:\\
\BE\label{gluprfineu}
\widetilde{D}_{T}(p_{E})=\frac{Z_{\mc{D}}}{p^{2}_{E}\,(F(s)+\xi F_{\xi}(s)+F_{0})}
\EE
where now $s=p_{E}^{2}/m^{2}$. The Euclidean propagator is expressed in terms of the same renormalization constants of the Minkowski propagator, namely $Z_{\mc{D}}$ and $F_{0}$. Again, since the lattice data is the most reliable in the Landau gauge ($\xi=0$), in this section we will limit our comparison to this gauge. Setting $\xi=0$ in eq.~\eqref{gluprfineu} we obtain\\
\BE\label{gluprfineuland}
\widetilde{D}_{T}^{(\xi=0)}(p_{E})=\frac{Z_{\mc{D}}}{p^{2}_{E}\,(F(s)+F_{0})}
\EE
\\
$Z_{\mc{D}}$ and $F_{0}$, together with the value of the mass parameter $m$, are free parameters which we need to fit starting from the lattice data. We observe that, like the ghost propagator -- see Sec.2.2.2 --, the gluon propagator too depends on a free parameter in excess of the two required for the perturbative results to be true predictions from first principles. What we will see in the next chapter is that gauge invariance constrains the renormalization constant $F_{0}$ to take a specific value, thus reducing the number of free parameters back to two (the constant $Z_{\mc{D}}$ and the mass $m$) and making our results as predictive as they can be in the context of perturbation theory. For the moment, we treat $F_{0}$ as a free parameter to be fitted to the lattice data, just as we did for the constant $G_{0}$ in the ghost propagator. The fitted value of $F_{0}$ will be shown to be in perfect agreement with the one dictated by gauge invariance in the next chapter.\\
In Fig.\ref{gluproplandfit} we show the lattice data of ref.\cite{duarte} for the Landau gauge transverse gluon propagator together with the one-loop results of massive perturbation theory, given by eq.~\eqref{gluprfineuland}. The parameters which were found to best fit the data in the momentum range $0-4$ GeV are reported in Tab.\ref{fitdata}.\\
\\
\\
\\
\newpage
\
\begin{table}[H]
\def\arraystretch{1.2}
\centering
\begin{tabular}{ccc}
\hline
$F_{0}$&$m$ (GeV)&$Z_{\mc{D}}$\\
\hline
\hline
$-0.8872$&$0.6541$&$2.6307$\\
\hline
\end{tabular}
\\
\caption{Parameters found from the fit of the lattice data of ref.\cite{duarte} for the Landau gauge Euclidean transverse gluon propagator in the range $0-4$ GeV.}\label{fitdata}
\end{table}
\
\\
As we can see, by choosing $m=0.654$ GeV and $F_{0}=-0.887$ we obtain a propagator which is in astonishing agreement with the lattice data already at one loop, proving that the massive expansion is able to describe the low energy dynamics of the gluons not only qualitatively through the generation of a dynamical mass for the gluons, but also quantitatively, to an unexpected degree of accuracy.\\
In the next chapter we will extend our comparison with the lattice to covariant gauges up to $\xi=0.5$, where some data -- albeit less accurate than that of the Landau gauge -- is already available.
\clearpage

\addcontentsline{toc}{section}{2.4 Summary and conclusions}  \markboth{2.4 Summary and conclusions}{2.4 Summary and conclusions}
\section*{2.4 Summary and conclusions\index{Summary and conclusions}}

In this chapter we have computed the one-loop ghost and gluon propagators in the framework of massive perturbation theory and compared our results with the predictions of ordinary perturbation theory (UV regime) and -- limited to the Landau gauge -- of the lattice calculations (IR regime).\\
In massive perturbation theory, the one-loop ghost and gluon propagators can be expressed in terms of five free parameters: two multiplicative renormalization constants -- $Z_{\mc{G}}$ and $Z_{\mc{D}}$~--, two additive renormalization constants -- $G_{0}$ and $F_{0}$ -- and a mass parameter -- $m$. The mass parameter is a spurious free parameter, in that it is artificially introduced in the perturbative series by the shift of the Faddeev-Popov action which defines the massive expansion. It cannot be fixed from first principles since Yang-Mills theory is scale-free at the classical level.\\
The one-loop ghost propagator was found to remain massless in any covariant gauge. In the high-energy limit, its expression was shown to match the results of ordinary perturbation theory. In the low-energy limit, an appropriate choice of the renormalization constants and of the mass parameter -- cf. Tab.\ref{fitdatagh} -- leads to a Euclidean propagator which accurately reproduces the lattice data of ref.\cite{duarte} (Duarte et al.). The value of the additive renormalization constant which was found to best fit the data at fixed $m^{2}=0.654$ GeV\footnote{\ This is the value of $m^{2}$ obtained from the fit of the gluon propagator, see ahead.} is $G_{0}=0.146$ -- cf. Tab.\ref{fitdatagh}.\\
The transverse component of the gluon propagator was found to non-trivially acquire a mass in any covariant gauge due to the non-vanishing zero-momentum limit of the gluon tadpoles and of the gluon loops. The high-energy limit of the propagator was shown to match the predictions of ordinary perturbation theory. In the low-energy limit, the propagator was found to reproduce the lattice data of ref.\cite{duarte} (Duarte et al.) to an astonishing degree of accuracy. The parameters which were found to best fit the data -- cf. Tab.\ref{fitdata} -- are $F_{0}=-0.887$ and $m=0.654$ GeV.\\
\\
The asymptotic analysis of the one-loop ghost and gluon propagators shows that massive perturbation theory is capable of describing both the UV and IR behavior of Yang-Mills theory by the use of elementary quantum field theoretic methods. In the framework of the massive expansion, the non-perturbative content which was responsible for the failure of ordinary perturbation theory is effectively incorporated in the dynamical mass generated for the gluons through the loops of the massive series. Our results suggest that the breakdown of the ordinary perturbative methods at low energies may be due to a bad choice of the expansion point of the perturbative series, rather than to an intrinsic limitation of the methods themselves. In the next chapter we will perfect the massive perturbative method and restore the predictivity of the expansion, which was weakened by the introduction of the spurious mass parameter $m$.

\clearpage


\stepcounter{count}
\addcontentsline{toc}{chapter}{3 The analytic structure of the gluon propagator and the Nielsen identities: gauge invariance and the predictivity of the massive expansion} \markboth{3 The analytic structure of the gluon propagator and the Nielsen identities: gauge invariance and the predictivity of the massive expansion}{3 The analytic structure of the gluon propagator and the Nielsen identities: gauge invariance and the predictivity of the massive expansion}
\chapter*{3\protect \\
\medskip{}
The analytic structure of the gluon propagator and the Nielsen identities: gauge invariance and the predictivity of the massive expansion\index{The analytic structure of the gluon propagator and the Nielsen identities: gauge invariance and the predictivity of the massive expansion}}

As we saw in Chapter 2, through massive perturbation theory one is able to derive one-loop ghost and gluon propagators which have the correct asymptotic behavior and -- by an appropriate choice of free parameters -- quantitatively reproduce the lattice data to a high degree of accuracy. Nonetheless, the theory presented in the previous chapter is incomplete in at least two respects. First of all, the shift of the kinetic action that defines the massive expansion introduces a spurious mass parameter which cannot be fixed from first principles and must be provided as an external input. The need for an external input, at face value, results in a loss of the predictive power of the theory, which is not desirable if one wants to interpret the massive perturbation theory as a true approach form first principles. Second of all, the fact that the massive expansion treats the gluons as massive at tree-level makes it unclear to which extent gauge invariance -- or rather BRST invariance -- is preserved at the level of its perturbative series. If non-perturbatively, by leaving the Faddeev-Popov action $\mc{S}$ unchanged, the massive expansion is guaranteed to preserve gauge symmetry, from the point of view of perturbation theory the massiveness of the bare gluon propagator and the presence of the gluon mass counterterm may lead to results in which BRST invariance is not automatically implemented to any finite perturbative order.\\
The aim of this chapter is to present a common solution to both of these issues. The key idea is that gauge invariance imposes strict constraints on the analytic structure of the dressed gluon propagator, which in massive perturbation theory are not automatically satisfied due to the massiveness of the propagator. By enforcing these constraints in the massive expansion, one is able to fix the value of the renormalization constant $F_{0}$ that defines the gluon propagator. Once $F_{0}$ has been fixed from first principles, we are left with a gluon propagator which complies with BRST invariance and that depends on two free parameters only -- a multiplicative renormalization constant $Z_{\mc{D}}$ and the value of the mass parameter $m$ --, two being the correct number of free parameters for a propagator computed from first principles.\\
Gauge invariance constrains the analytical structure of the gauge-dependent gluon propagator through the Nielsen identities. In the context of non-abelian gauge theories, the propagators associated to the gauge bosons -- and, more generally, any of the Green functions of the theory -- depend on the gauge in a non-trivial fashion. The Nielsen identities are equations that describe how the Green functions change as functions of the gauge parameter $\xi$. They are a consequence of the BRST invariance of the Faddeev-Popov action, and can be derived by ordinary functional methods. Through the Nielsen identities one can prove that, even though the gluon propagator depends on the gauge, the position of its poles in the complex plane is gauge-independent, i.e. it is fixed regardless of the value of $\xi$. In ordinary perturbation theory, this requirement is trivially met: the gluon propagator possesses a single pole at $p^{2}=0$, which is obviously a gauge-independent value. In massive perturbation theory, on the other hand, dynamical mass generation causes the analytic structure of the propagator to be much richer: as we will see, the gluon propagator turns out to have two non-zero complex-conjugated poles, whose position depends on the value of the renormalization constant $F_{0}$. In order to comply with gauge invariance, $F_{0}$ cannot be arbitrary, but must be given the value that fixes the gluon poles in the correct position. In addition, the Nielsen identities can be used to show that the residues of the gluon propagator at its poles also are gauge-invariant. Ultimately, this condition will allow us to completely determine the gluon propagator as a function of $\xi$ and to prove that the requirement of gauge invariance alone is able to restore the predictivity of the massive expansion by fixing the value of the renormalization constant $F_{0}$.\\
\\
This chapter is organized as follows. In Sec.3.1 we discuss the Nielsen identities and describe the method that will allow us to fix the value of $F_{0}$ by enforcing the gauge invariance of the pole structure of the gluon propagator. In Sec.3.2. we investigate the analytic structure of the gluon propagator in the Landau gauge as $F_{0}$ is tuned across the real numbers, apply the method laid out in the previous section and present our results. The  value of $F_{0}$ singled out by our requirements of gauge invariance will be shown to be very close to that which was obtained in Sec.2.3.2 by fitting the propagator to the lattice data. In Sec.3.3 we discuss a renormalization-scheme-dependent method presented in ref. \cite{sir4} for fixing the value of the ghost additive renormalization constant $G_{0}$ in the Landau gauge.\\
\\
The results of this chapter were presented and published for the first time in ref. \cite{com3}.

\newpage

\addcontentsline{toc}{section}{3.1 Optimizing the massive expansion by the requirement of gauge invariance}  \markboth{3.1 Optimizing the massive expansion by the requirement of gauge invariance}{3.1 Optimizing the massive expansion by the requirement of gauge invariance}
\section*{3.1 Optimizing the massive expansion by the requirement of gauge invariance\index{Optimizing the massive expansion by the requirement of gauge invariance}}

\addcontentsline{toc}{subsection}{3.1.1 The Nielsen identities}  \markboth{3.1.1 The Nielsen identities}{3.1.1 The Nielsen identities}
\subsection*{3.1.1 The Nielsen identities\index{The Nielsen identities}}

In Chapter 2 we saw that the gluon propagator computed in massive perturbation theory depends on the gauge chosen for the definition of the theory, in that it contains terms which are proportional to the gauge parameter $\xi$. This is a general, non-perturbative feature of the Green functions of non-abelian gauge theories: although the quantities which can be measured in the laboratory are gauge-invariant, the basic building blocks from which they are made of typically depend on the gauge. In order for the gauge dependence to ultimately factor out from the equations, the Green functions of any gauge theory must satisfy specific constraints, which are formulated in terms of the so-called Nielsen identities.\\
The Nielsen identities \cite{niels,piguet} are non-perturbative equations which relate the gauge dependence of a given Green function to the content of other Green functions. They are derived by ordinary functional methods starting from the BRST invariance of the Fadeev-Popov action. In this chapter we will not go through their explicit derivation, for which we refer to \cite{breck} (Breckenridge et al.). Rather, we will use the identities to prove two results which are of fundamental importance in the analysis of the massive perturbative expansion, namely, the gauge invariance of the position and residues of the poles of the gluon propagator.\\
\\
In a general covariant gauge parametrized by $\xi$, the gauge dependence of the transverse component of the dressed gluon propagator $\widetilde{\mathcal{D}}_{T}(p^{2},\xi)$ is described by the following Nielsen identity (cf. eq.(3.6d) in ref.\cite{breck}):\\
\BE\label{niels}
\frac{\partial \widetilde{\mc{D}}_{T}^{-1}}{\partial \xi}(p^{2},\xi)=H_{T}(p^{2},\xi)\,\widetilde{\mc{D}}_{T}^{-2}(p^{2},\xi)
\EE
\\
In this equation, $H_{T}$ is the transverse component of a Green function $H_{\mu\nu}^{ab}$ defined as\\
\BE\label{uglyh}
H_{\mu\nu}^{ab}(p,\xi)=\int d^{d}x\,d^{d}y\ e^{ip\cdot (x-y)}\ \avg{T\left\{D_{\mu}c^{a}(x)\,A_{\nu}^{b}(y)\,\overline{c}^{\,c}(0)\,B^{c}(0)\right\}}_{\xi}
\EE
\\
where $B^{a}$ is an auxiliary field -- the Nakanishi-Lautrup field\footnote{See for instance the Introduction to this thesis.} -- introduced to enforce the off-shell BRST invariance of the Faddeev-Popov action and $\avg{\cdot}_{\xi}$ denotes the vacuum expectation value computed in the gauge $\xi$. For our purposes, the explicit form of $H_{T}$ is irrelevant. What matters to us is that the position of the poles of $H_{T}$, in general, is different from that of the gluon propagators'. This is important for the following reason. Suppose that as $\xi$ varies the gluon propagator has a pole at $p^{2}=p_{0}^{2}(\xi)$, so that\\
\BE\label{mst}
\widetilde{\mc{D}}_{T}^{-1}(p^{2}_{0}(\xi),\xi)=0
\EE
\\
for every $\xi$. Then, by taking the total derivative of the above equation with respect to $\xi$, we find that
\BE\label{msw}
\frac{\partial\widetilde{\mc{D}}_{T}^{-1}}{\partial\xi}(p_{0}^{2}(\xi),\xi)+\frac{dp_{0}^{2}}{d\xi}(\xi)\,\frac{\partial\widetilde{\mc{D}}_{T}^{-1}}{\partial p^{2}}(p_{0}^{2}(\xi),\xi)=0
\EE
However, from eq.~\eqref{niels} we know that\\
\BE\label{mnt}
\frac{\partial \widetilde{\mc{D}}_{T}^{-1}}{\partial \xi}(p^{2}_{0}(\xi),\xi)=H_{T}(p^{2}_{0}(\xi),\xi)\,\widetilde{\mc{D}}_{T}^{-2}(p^{2}_{0}(\xi),\xi)=0
\EE
where the vanishing of the right-hand side is due to eq.~\eqref{mst} together with the fact that $H_{T}(p^{2},\xi)$ was assumed not to have poles -- hence to be finite -- at $p^{2}_{0}(\xi)$. Therefore the partial derivative with respect to $\xi$ in eq.~\eqref{msw} vanishes and, since the partial derivative of $\widetilde{D}_{T}^{-2}$ with respect to $p^{2}$ in general is different from zero, we are left with\\
\BE\label{poleinv}
\frac{dp_{0}^{2}}{d\xi}(\xi)=0
\EE
\\
Eq.~\eqref{poleinv} informs us that if the gluon propagator possesses a pole, then the position of this pole must be gauge-invariant: because of eq.~\eqref{niels} we must have $p_{0}^{2}(\xi_{1})=p_{0}^{2}(\xi_{2})$ for any gauges $\xi_{1}$ and $\xi_{2}$.\\
Actually, we can go one step further and show that the residues of the gluon propagator at its poles must also be gauge-invariant. In order to do this, we start by taking the partial derivative of eq.~\eqref{niels} with respect to $p^{2}$. By exchanging the order of the derivatives on the left-hand side, we obtain\\
\BE\label{kkx}
\frac{\partial^{2} \widetilde{\mc{D}}_{T}^{-1}}{\partial \xi\partial p^{2}}(p^{2},\xi)=\frac{\partial H_{T}}{\partial p^{2}}(p^{2},\xi)\,\widetilde{\mc{D}}_{T}^{-2}(p^{2},\xi)+2\,H_{T}(p^{2},\xi)\,\widetilde{\mc{D}}_{T}^{-1}(p^{2},\xi)\,\frac{\partial \widetilde{\mc{D}}_{T}^{-1}}{\partial p^{2}}(p^{2},\xi)
\EE
\\
Since $\widetilde{\mc{D}}_{T}^{-1}$ vanishes at the poles and since the other terms on the right-hand side are finite, when computed at $p^{2}=p^{2}_{0}(\xi)$ eq.~\eqref{kkx} simply reads\\
\BE\label{mjt}
\frac{\partial^{2} \widetilde{\mc{D}}_{T}^{-1}}{\partial \xi\partial p^{2}}(p^{2}_{0}(\xi),\xi)=0
\EE
\\
Now, recall the definition of the residue of a function $f$ at one of its poles $z_{0}$,\\
\BE
\mc{R}(f,z_{0})=\lim_{z\to z_{0}}\ f(z)(z-z_{0})=\lim_{z\to z_{0}}\ \frac{z-z_{0}}{\frac{1}{f(z)}-\frac{1}{f(z_{0})}}=\left[\frac{d}{dz}\frac{1}{f(z)}\right]_{z=z_{0}}^{-1}
\EE
\\
By applying this definition to $\widetilde{\mc{D}}_{T}$, we can express the inverse of the residue of the propagator at $p_{0}^{2}(\xi)$ as
\BE\label{polesder}
\mc{R}^{-1}(\widetilde{\mc{D}}_{T},p_{0}^{2}(\xi))=\frac{\partial \widetilde{\mc{D}}_{T}^{-1}}{\partial p^{2}}(p_{0}^{2}(\xi),\xi)
\EE
\\
By taking the total derivative of the above equation with respect to $\xi$ we find that\\
\BE
\frac{d}{d\xi}\ \mc{R}^{-1}(\widetilde{\mc{D}}_{T},p_{0}^{2}(\xi))=\frac{\partial^{2} \widetilde{\mc{D}}_{T}^{-1}}{\partial \xi\partial p^{2}}(p^{2}_{0}(\xi),\xi)+\frac{dp_{0}^{2}}{d\xi}(\xi)\,\frac{\partial^{2} \widetilde{\mc{D}}_{T}^{-1}}{\partial (p^{2})^{2}}(p^{2}_{0}(\xi),\xi)=0
\EE
\\
where the vanishing of the right-hand side is due to eqq.~\eqref{poleinv} and \eqref{mjt}. Therefore, the residues of the gluon propagator at its gauge-invariant poles are themselves gauge-invariant.\\
\\
Through the Nielsen identities one is able to show that -- despite the gluon propagator being gauge-dependent -- the position and the residues of its poles are gauge-invariant. In the absence of mass generation, this result -- at least as far as the position of the poles is concerned -- is trivial: a massless gluon propagator has a single pole at $p^{2}=0$, which is obviously a gauge-invariant position. If a dynamical mass is generated for the gluons, on the other hand, the poles may be found at any location in the complex plane. Hence, in the presence of mass generation, the Nielsen identities impose strict new constraints on the analytic structure of the gluon propagator.
\\
\\

\addcontentsline{toc}{subsection}{3.1.2 Enforcing the gauge invariance of the poles in the massive expansion}  \markboth{3.1.2 Enforcing the gauge invariance of the poles in the massive expansion}{3.1.2 Enforcing the gauge invariance of the poles in the massive expansion}
\subsection*{3.1.2 Enforcing the gauge invariance of the poles in the massive expansion\index{Enforcing the gauge invariance of the poles in the massive expansion}}

The gauge invariance of the position and residues of the poles of the gluon propagator is a non-perturbative feature of Yang-Mills theory which is not guaranteed to be retained at any finite order in massive perturbation theory. The aim of this section is to provide a method for reducing the number of free parameters of the massive expansion by exploiting this apparent weakness of the massive approach.\\
\\
Let us investigate under which conditions the poles of the gluon propagator derived in Chapter 2 have a gauge-invariant position. In Sec.2.3.1 we saw that the one-loop transverse gluon propagator of the massive expansion can be expressed as
\BE\label{gluprfin}
\widetilde{\mc{D}}_{T}(p)=\frac{-iZ_{\mc{D}}}{p^{2}(F(s)+\xi F_{\xi}(s)+F_{0})}
\EE
\\
where $Z_{\mc{D}}$ and $F_{0}$ are arbitrary renormalization constants, $s=-p^{2}/m^{2}$ and the functions $F(s)$ and $F_{\xi}(s)$ are defined in eqq.~\eqref{ffunct}-\eqref{rfunct}. In Sec.2.3.2 we showed that the propagator is finite at $p^{2}=0$. Hence $\widetilde{\mc{D}}_{T}$, as is obvious in the presence of mass generation, does not have a massless pole. Since the value $p^{2}=0$ is excluded, a necessary and sufficient condition for $\widetilde{\mc{D}}_{T}$ to have a pole at $p^{2}=p_{0}^{2}$ is that\\
\BE\label{posstu}
F(-p_{0}^{2}/m^{2})+\xi\, F_{\xi}(-p_{0}^{2}/m^{2})+F_{0}=0
\EE
\\
We remark that what we are actually looking for are the poles of the analytic continuation of $\widetilde{\mc{D}}_{T}$ to the whole complex plane, so that $p_{0}^{2}$ in general is complex and the above equation is a complex equation. Now, the Nielsen identities inform us that eq.~\eqref{posstu} should be verified for every $\xi$, since $p_{0}^{2}$ is a gauge-independent value. It is easy to see that this condition cannot be satisfied unless $F_{0}$, and perhaps even the mass parameter $m$, are functions of the gauge. $F_{0}$ -- being a renormalization constant -- can actually be given different values in different gauges. This is because, first of all, the renormalization conditions may be chosen to be different in different gauges and, second of all, because the propagator itself is gauge-dependent, so that equal renormalization conditions for different gauges can be only implemented by choosing different $F_{0}$'s. Therefore, in general, we should regard $F_{0}$ as being a gauge-dependent constant: $F_{0}=F_{0}(\xi)$. As for the mass parameter, we observe that the role of $m$ in the framework of the massive expansion is to absorb the non-perturbative content of the theory by introducing a mass scale in the perturbative series; for a gauge-dependent propagator, this task may require the scale itself to be different in different gauges. Therefore, in general, $m$ also may also be regarded as being a gauge-dependent parameter: $m^{2}=m^{2}(\xi)$.\\
With $F_{0}$ and $m^{2}$ dependent on the gauge, the condition for the gauge invariance of the pole $p_{0}^{2}$ reads
\BE\label{masterpos}
F(-p_{0}^{2}/m^{2}(\xi))+\xi\, F_{\xi}(-p_{0}^{2}/m^{2}(\xi))+F_{0}(\xi)=0\qquad\quad \forall\ \xi
\EE
For arbitrary $F_{0}(\xi)$ and $m^{2}(\xi)$, this condition will not be met. Therefore, in general, the one-loop gluon propagator computed in massive perturbation theory does not comply with the Nielsen identities. However, observe that if we knew in advance the position of the pole, then the above equation could be solved to find the functions $F_{0}(\xi)$ and $m^{2}(\xi)$ which realize the gauge invariance of its position. Indeed, since eq.~\eqref{masterpos} is a complex equation, the vanishing of its real and imaginary parts at fixed $p_{0}^{2}$ is in principle sufficient to determine the two real functions. Knowing $F_{0}(\xi)$ and $m^{2}(\xi)$ would then be equivalent to knowing the gluon propagator in any gauge, with no dependence left on any free parameter\footnote{\ Except for the multiplicative renormalization constant $Z_{\mc{D}}$.}. The problem is, of course, that we do not actually know $p_{0}^{2}$. Therefore what we need to ask is: is it possible to determine the position of the poles starting from the requirement of the gauge invariance alone? The answer turns out to be yes, as long as we are willing to retain the value of $m$ in some fixed gauge as a free parameter.\\
To prove our claim, we start by observing that, in order to determine the position of the poles of the propagator, we only need to know the value of $F_{0}$ and $m^{2}$ in some specific gauge. For instance, we may pick the Landau gauge ($\xi=0$) and solve for $p_{0}^{2}$ the equation\\
\BE\label{mastland}
F(-p_{0}^{2}/m^{2}_{L})+F_{0}^{L}=0
\EE
\\
where $m_{L}^{2}=m^{2}(0)$\footnote{\ Not to be confused with the longitudinal mass parameter of Chapter 1.} and $F_{0}^{L}=F_{0}(0)$ are the mass parameter and additive renormalization constant in the Landau gauge. If $F_{0}^{L}$ is known, then $p_{0}^{2}=m^{2}_{L}\,z_{0}^{2}$, where $z_{0}^{2}$ solves the equation
\BE
F(-z_{0}^{2})+F_{0}^{L}=0
\EE
\\
Of course, since Yang-Mills theory is scale-free at the classical level, the value of the mass parameter $m_{L}^{2}$ cannot be determined from first principles. However, once $m_{L}^{2}$ has been provided as an external input, we are left with a solution $p_{0}^{2}$ that depends exclusively on the value of the additive renormalization constant $F_{0}^{L}$ in the Landau gauge: the solution to eq.~\eqref{mastland} takes the form of an $F_{0}^{L}$-dependent pole function $p_{0}^{2}(F_{0}^{L})$ such that\\
\BE
F(-p_{0}^{2}(F_{0}^{L})/m^{2}_{L})+F_{0}^{L}=0
\EE
\\
Once some value for $F_{0}^{L}$ has been fixed and eq.~\eqref{mastland} has been solved to find the corresponding pole, we can then plug the solution back into eq.~\eqref{masterpos} to determine the functions $F_{0}(\xi)$ and $m^{2}(\xi)$ for any value of $\xi$. This procedure leaves us with gauge-dependent functions that depend only on the value of $F_{0}^{L}$ which was used in the first place to compute the position of the pole, with $m_{L}^{2}$ providing the scale for the mass parameter in an arbitrary gauge. By construction, the functions $F_{0}(\xi)$ and $m^{2}(\xi)$ so obtained realize the gauge invariance of the position of the poles of the gluon propagator, which now depends on $m_{L}^{2}$ and $F_{0}^{L}$ only. Luckily, however, this is not the end of the story. As a matter of fact, as we saw in the last section, the Nielsen identities constrain not only the positions of the poles, but also their residues to be gauge-invariant. This constraint can be turned to our advantage as follows.\\
Once $F_{0}(\xi)$ and $m^{2}(\xi)$ have been obtained by picking some value for $F^{L}_{0}$, we can go on and compute the residue of the gluon propagator at the pole $p_{0}^{2}(F^{L}_{0})$ as a function of the gauge. By virtue of eq.~\eqref{polesder} and of the vanishing of the inverse propagator at the pole, the residue can be computed as
\newpage
\BE
\mc{R}_{p_{0}^{2}}(\xi)=\frac{m^{2}(\xi)}{p_{0}^{2}}\,\frac{iZ_{\mc{D}}}{F'(-p_{0}^{2}/m^{2}(\xi))+\xi\, F'_{\xi}(-p_{0}^{2}/m^{2}(\xi))}
\EE
\\
where $F'(s)$ and $F_{\xi}'(s)$ are the derivatives of $F(s)$ and $F_{\xi}(s)$. By the Nielsen identities, we should have $\mc{R}_{p_{0}^{2}}(\xi_{1})=\mc{R}_{p_{0}^{2}}(\xi_{2})$ for any two gauges $\xi_{1}$ and $\xi_{2}$. However, observe that the multiplicative renormalization constant $Z_{\mc{D}}$ can actually be given arbitrary values in arbitrary gauges: we could set $Z_{\mc{D}}=Z_{\mc{D}}(\xi)$ for any positive real function $Z_{\mc{D}}(\xi)$ without violating the principles of renormalization. It follows that the residues of the propagator can actually be defined only modulo a positive real function, so that the Nielsen identities cannot ensure that the modulus $|\mc{R}_{p_{0}^{2}}(\xi)|$ of the residue is the same in different gauges. Nonetheless, since the phase of the residue cannot be modified by a change of $Z_{\mc{D}}$, the Nielsen identities still apply to the \textit{phase} of the residue, constraining the latter to be gauge-invariant and the residue to be of the form\footnote{\ The factor of $-i$ comes from the definition of the gluon propagator in Minkowski space.}\\
\BE\label{wnz}
\mc{R}_{p_{0}^{2}}(\xi)=-i\,|\mc{R}_{p_{0}^{2}}(\xi)|\ e^{i\varphi}
\EE\\
for some real, gauge-independent phase $\varphi$.\\
In massive perturbation theory, for an arbitrary choice of the free parameters eq.~\eqref{wnz} is not satisfied. Indeed, if we denote by $\theta_{p_{0}^{2}}(\xi)$ the difference between the phases of the residues at $p_{0}^{2}$ in the gauge $\xi$ and in the Landau gauge, namely,\\
\BE
\theta_{p_{0}^{2}}(\xi)=\text{Arg}\left\{\frac{\mc{R}_{p_{0}^{2}}(\xi)}{\mc{R}_{p_{0}^{2}}(\xi=0)}\right\}=\text{Arg}\left\{\frac{F'(-p_{0}^{2}/m^{2}_{L})}{F'(-p_{0}^{2}/m^{2}(\xi))+\xi\, F'_{\xi}(-p_{0}^{2}/m^{2}(\xi))}\right\}
\EE
\\
where $m^{2}(\xi)$ is computed by the procedure described on the previous page starting from the value of $F_{0}^{L}$ that produced $p_{0}^{2}$, a numerical evaluation of the phases informs us that $\theta_{p_{0}^{2}}(\xi)$ in general is not equal to zero, as it should be as a consequence of the Nielsen identities. Therefore, in the framework of massive perturbation theory, the Nielsen identities impose a very strict condition on the value of $F_{0}^{L}$\footnote{\ We remark that in the context of this derivation the Landau gauge does not have any special status: it is just our gauge of choice for the definition of the mass parameter $m^{2}$ which needs to be supplied to the theory as an external input.}: $F_{0}^{L}$ must be such that, once the quantities $p_{0}^{2}$, $F_{0}(\xi)$ and $m^{2}(\xi)$ have been derived by the procedure of the previous page, the poles of the gauge-dependent propagator resulting from these quantities have residues with gauge-invariant phases. Equivalently, $F_{0}^{L}$ must realize the vanishing of the function $\theta_{p_{0}^{2}}(\xi)$ associated to the poles of the propagator. Observe that a value of $F_{0}^{L}$ with such a property is not by any means guaranteed to exists. Since we are dealing with an approximation to the full theory, we should not expect $\theta_{p_{0}^{2}}(\xi)$ to be exactly zero for any value of $F_{0}^{L}$. Nonetheless, in the next section we will see that there actually are values of $F_{0}^{L}$ for which $|\theta_{p_{0}^{2}}(\xi)|$ is as small as three parts in one thousand. For these values the phases of the residues of the gluon propagator are are gauge-invariant for all practical purpose.\\
\\
In summary, the strategy that we will adopt in the next section to obtain the optimal value of $F_{0}^{L}$ and, together with it, the position $p_{0}^{2}$ of the pole, the optimal gauge-dependent renormalization parameter $F_{0}(\xi)$ and the optimal gauge-dependent mass parameter $m^{2}(\xi)$ is the following: 1. we will choose some arbitrary value for $F_{0}^{L}$ and solve eq.~\eqref{mastland} for the gauge-invariant position $p_{0}^{2}$ of the pole, which we will express as an adimensional complex number $z_{0}^{2}$ times an external mass parameter $m_{L}^{2}$, 2. we will solve eq.~\eqref{masterpos} for different values of $\xi$ in order to obtain the functions $F_{0}(\xi)$ and $m^{2}(\xi)$ in an arbitrary gauge, 3. we will use these functions to compute $\theta_{p_{0}^{2}}(\xi)$. The optimal value of $F_{0}^{L}$ will be the one which results in a $\theta_{p_{0}^{2}}(\xi)$ that is as close as possible to zero, implying that for this value the gauge invariance of both the position and the phases of the residues of the poles of the gluon propagator is achieved.

\newpage

\addcontentsline{toc}{section}{3.2 The analytic structure of the gluon propagator and the optimal gauge-dependent parameters}  \markboth{3.2 The analytic structure of the gluon propagator and the optimal gauge-dependent parameters}{3.2 The analytic structure of the gluon propagator and the optimal gauge-dependent parameters}
\section*{3.2 The analytic structure of the gluon propagator and the optimal gauge-dependent parameters\index{The analytic structure of the gluon propagator and the optimal gauge-dependent parameters}}
\addcontentsline{toc}{subsection}{3.2.1 Analytic structure of the Landau gauge gluon propagator: gluonic poles as a function of $F_{0}^{L}$}  \markboth{3.2.1 Analytic structure of the Landau gauge gluon propagator: gluonic poles as a function of $F_{0}^{L}$}{3.2.1 Analytic structure of the Landau gauge gluon propagator: gluonic poles as a function of $F_{0}^{L}$}
\subsection*{3.2.1 Analytic structure of the Landau gauge gluon propagator: gluonic poles as a function of $\boldsymbol{F_{0}^{L}}$\index{Analytic structure of the Landau gauge gluon propagator: gluonic poles as a function of $F_{0}^{L}$}}

\begin{figure}[H]
\centering
\vskip-20pt
\includegraphics[width=0.78\textwidth]{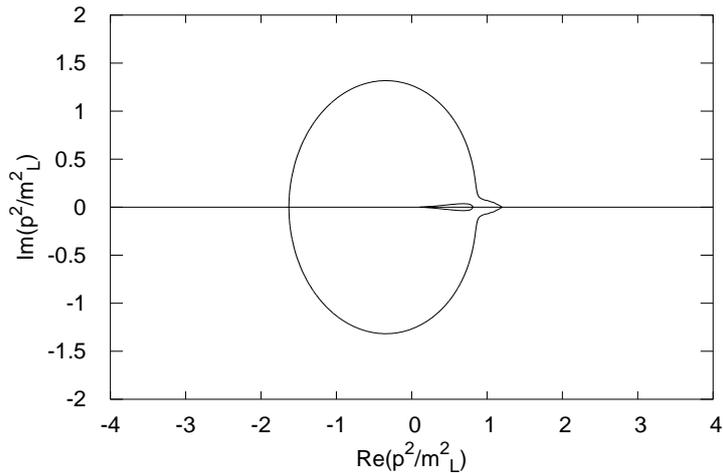}
\vskip-20pt
\caption{Zero set of the imaginary part of $F(-p^{2}/m_{L}^{2})$. The positive real axis is an artifact of the graphical algorithm and is not actually part of the zero set.}\label{imagzeros}
\end{figure}

\begin{figure}[H]
\centering
\vskip -20pt
\includegraphics[width=0.70\textwidth]{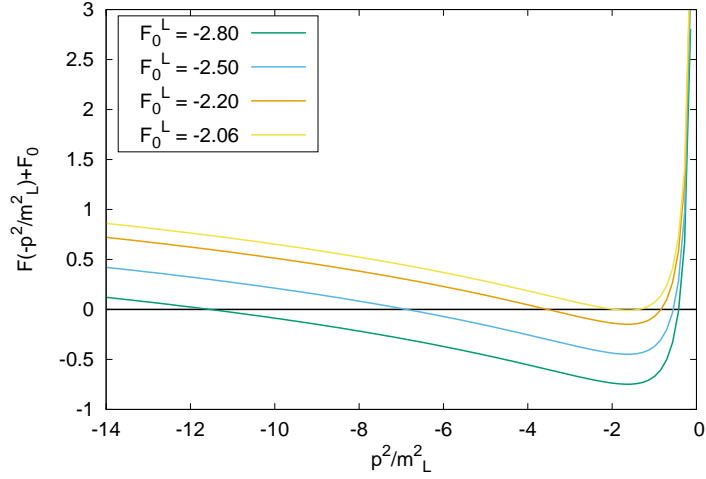}
\vskip -20pt
\caption{Case 1 -- graph of the function $F(-p^{2}/m_{L}^{2})+F^{L}_{0}$ for negative real $p^{2}$ at $F_{0}^{L}=-2.80,\,-2.50,\,-2.20,\,-2.06$.}\label{ffunctreg1}
\end{figure}

\begin{figure}[H]
\centering
\vspace{-20pt}
\begin{minipage}{.47\textwidth}
\centering
\includegraphics[width=1.1\linewidth]{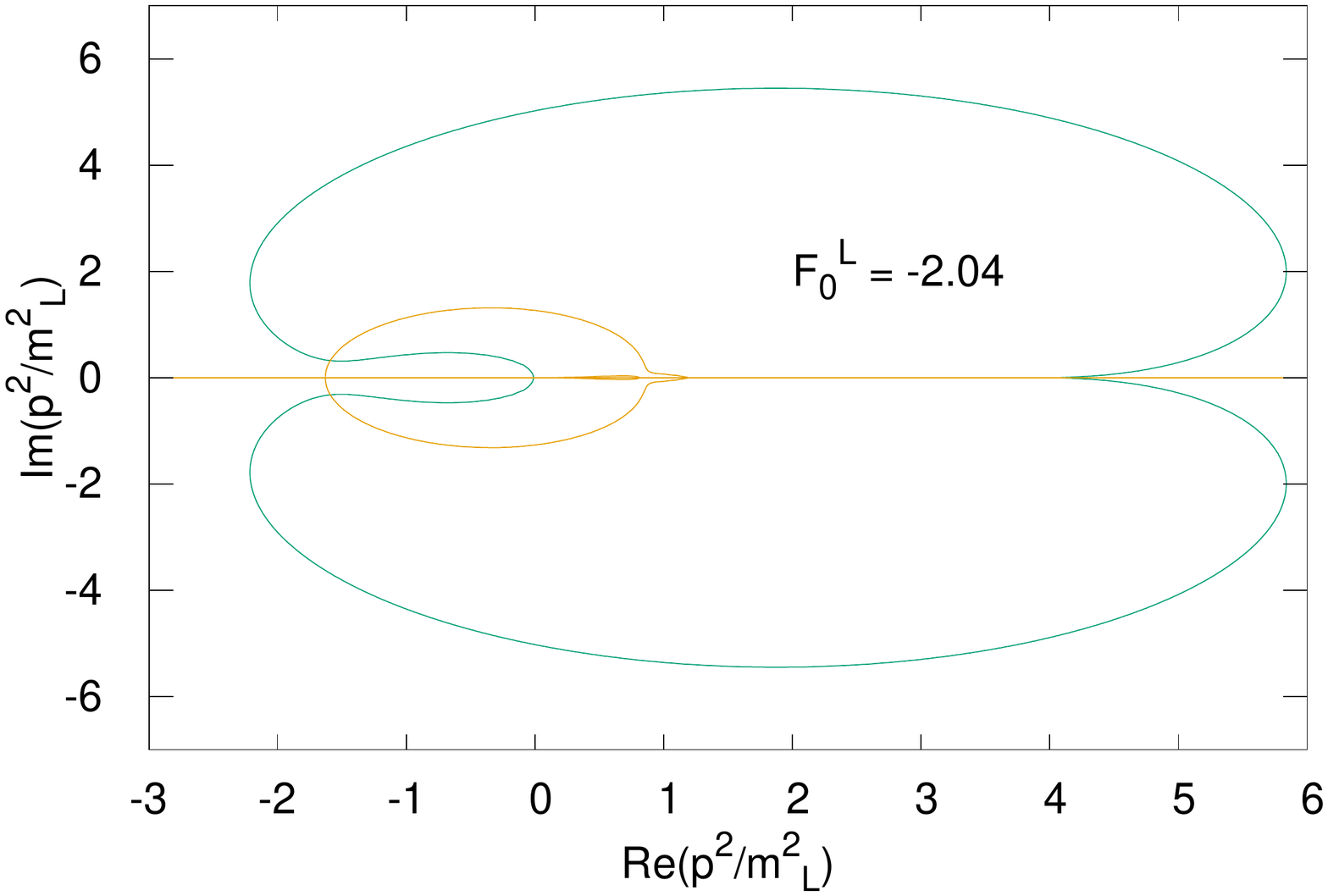}
\end{minipage}
\hspace{15pt}
\begin{minipage}{.47\textwidth}
\centering
\includegraphics[width=1.1\linewidth]{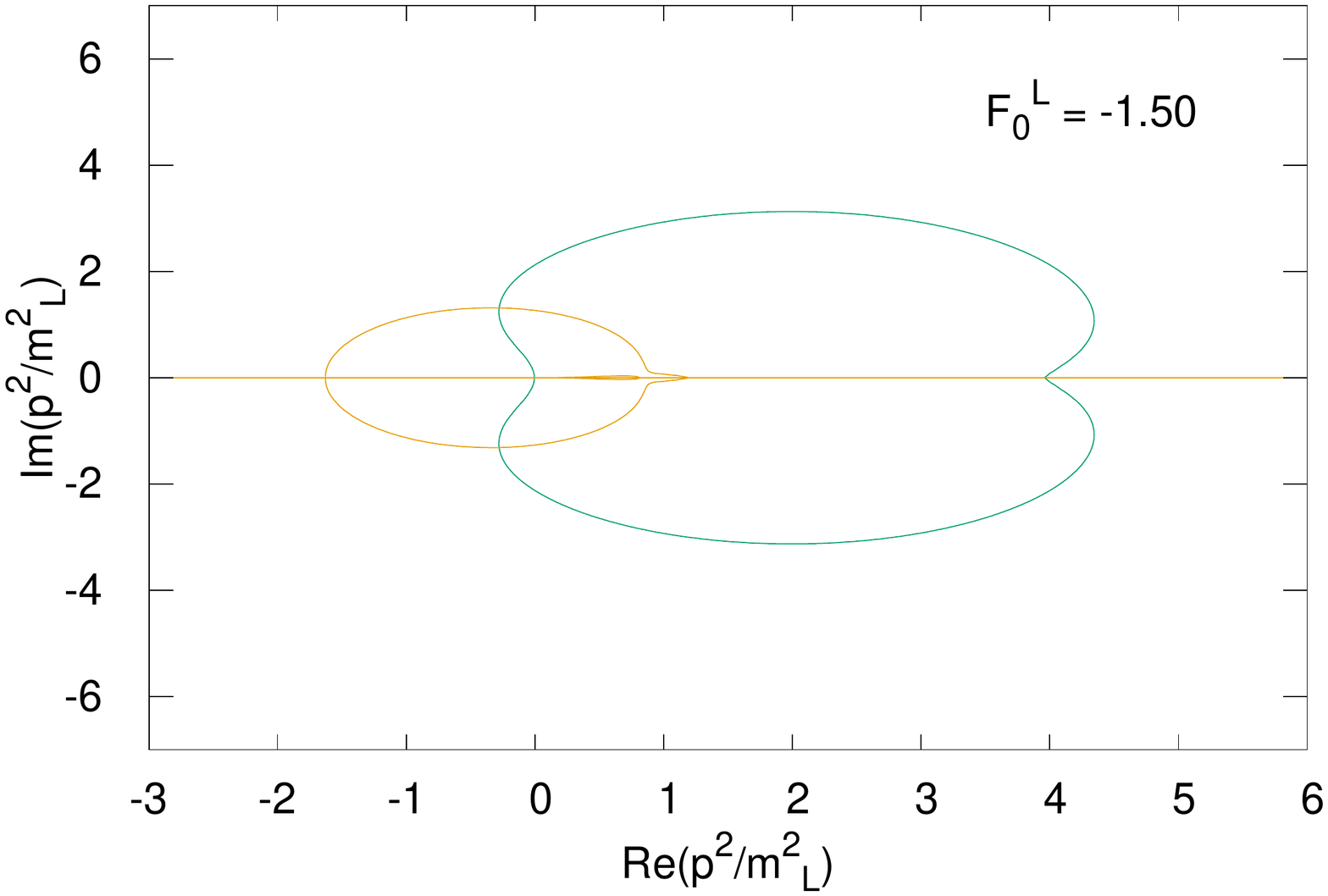}
\end{minipage}
\vspace{-20pt}
\begin{minipage}{.47\textwidth}
\centering
\includegraphics[width=1.1\linewidth]{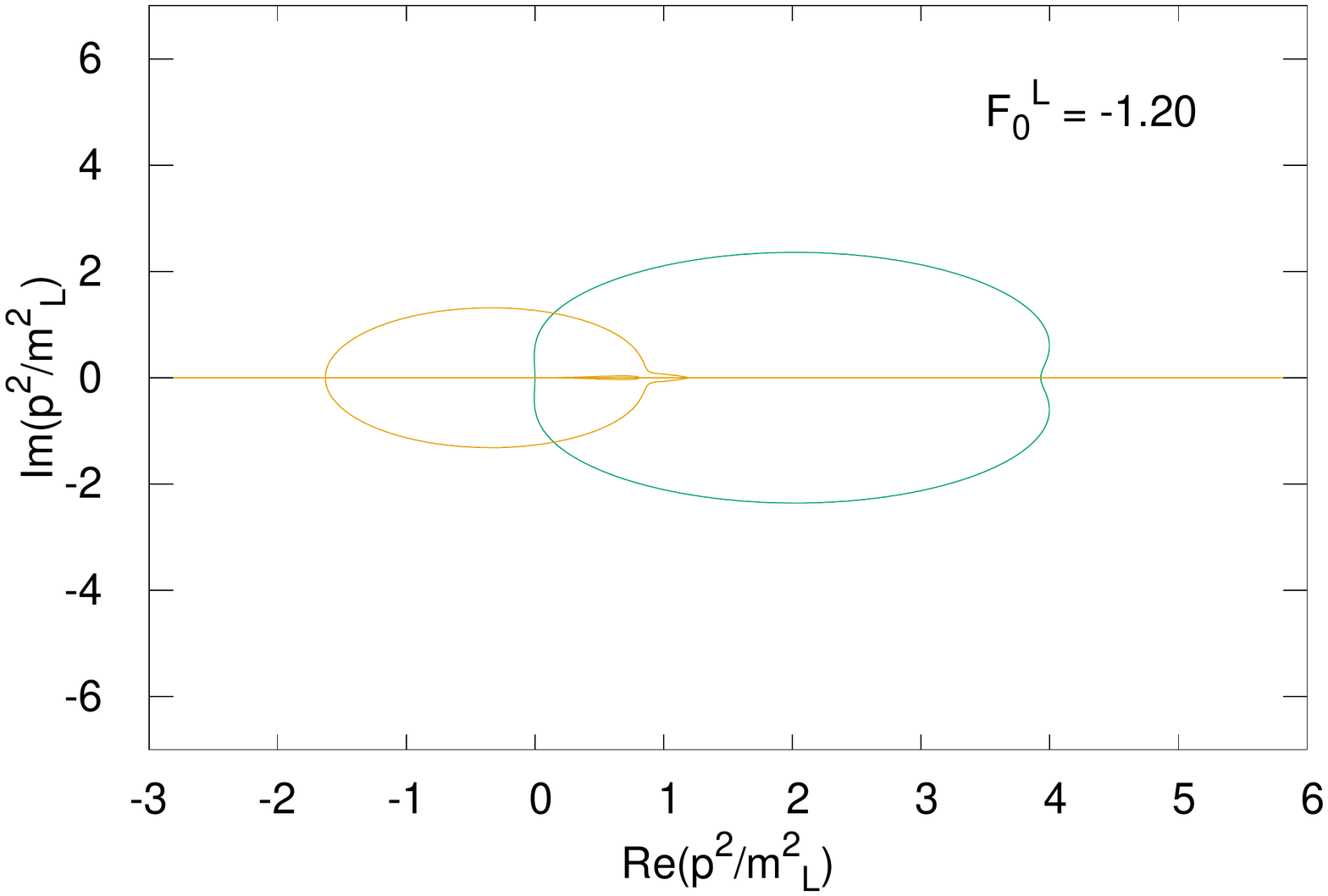}
\end{minipage}
\hspace{15pt}
\begin{minipage}{.47\textwidth}
\centering
\includegraphics[width=1.1\linewidth]{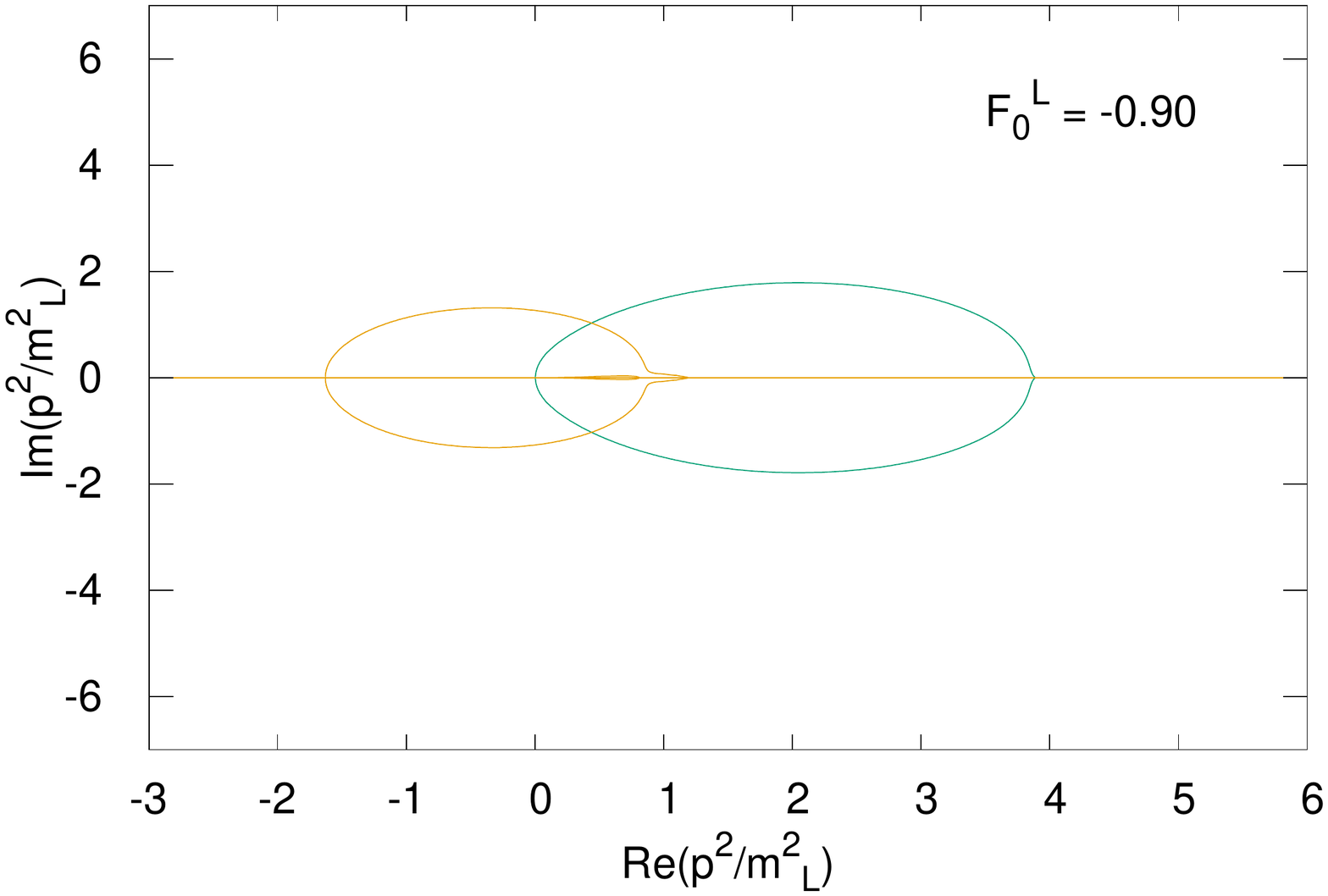}
\end{minipage}
\vspace{-20pt}
\begin{minipage}{.47\textwidth}
\centering
\includegraphics[width=1.1\linewidth]{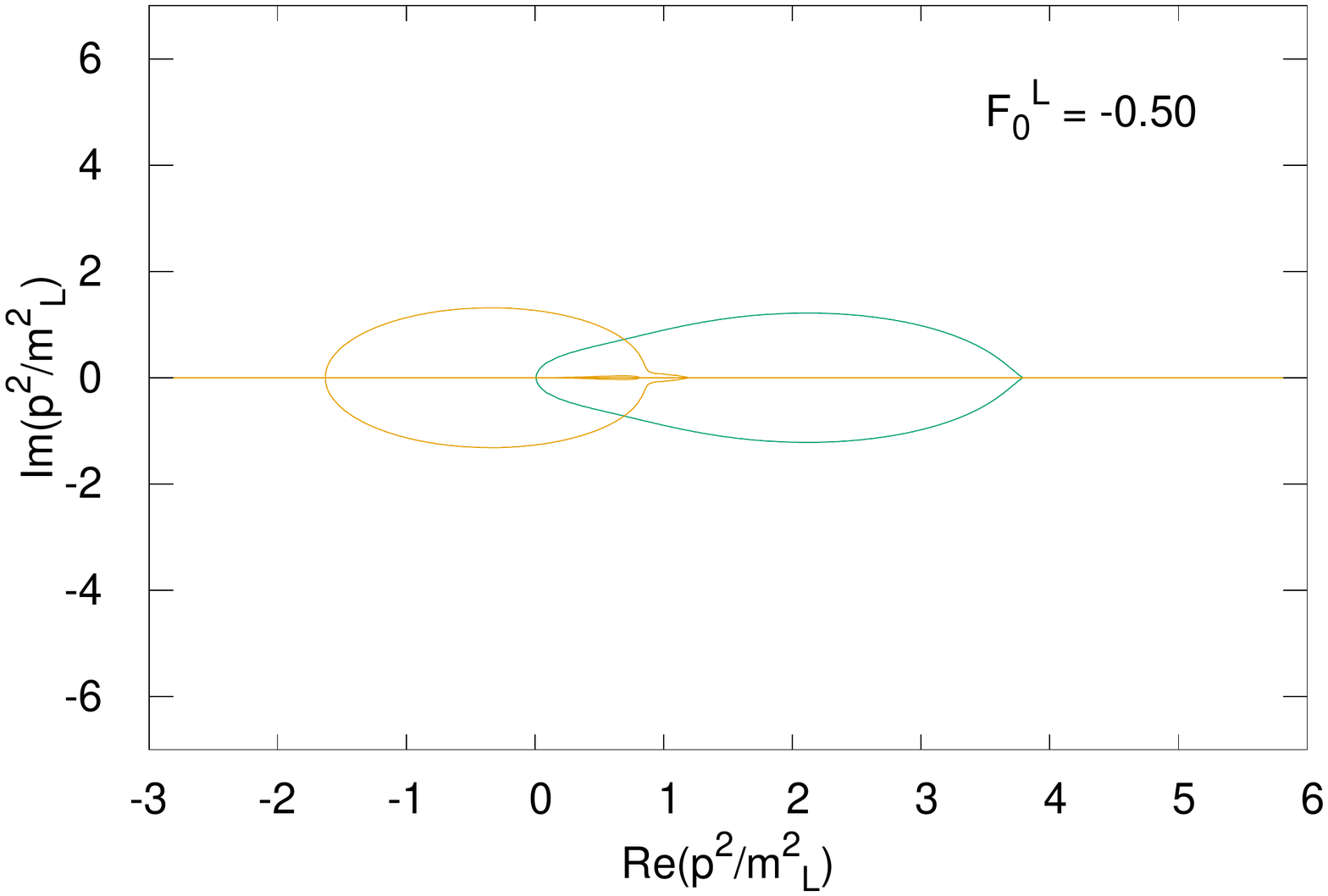}
\end{minipage}
\hspace{15pt}
\begin{minipage}{.47\textwidth}
\centering
\includegraphics[width=1.1\linewidth]{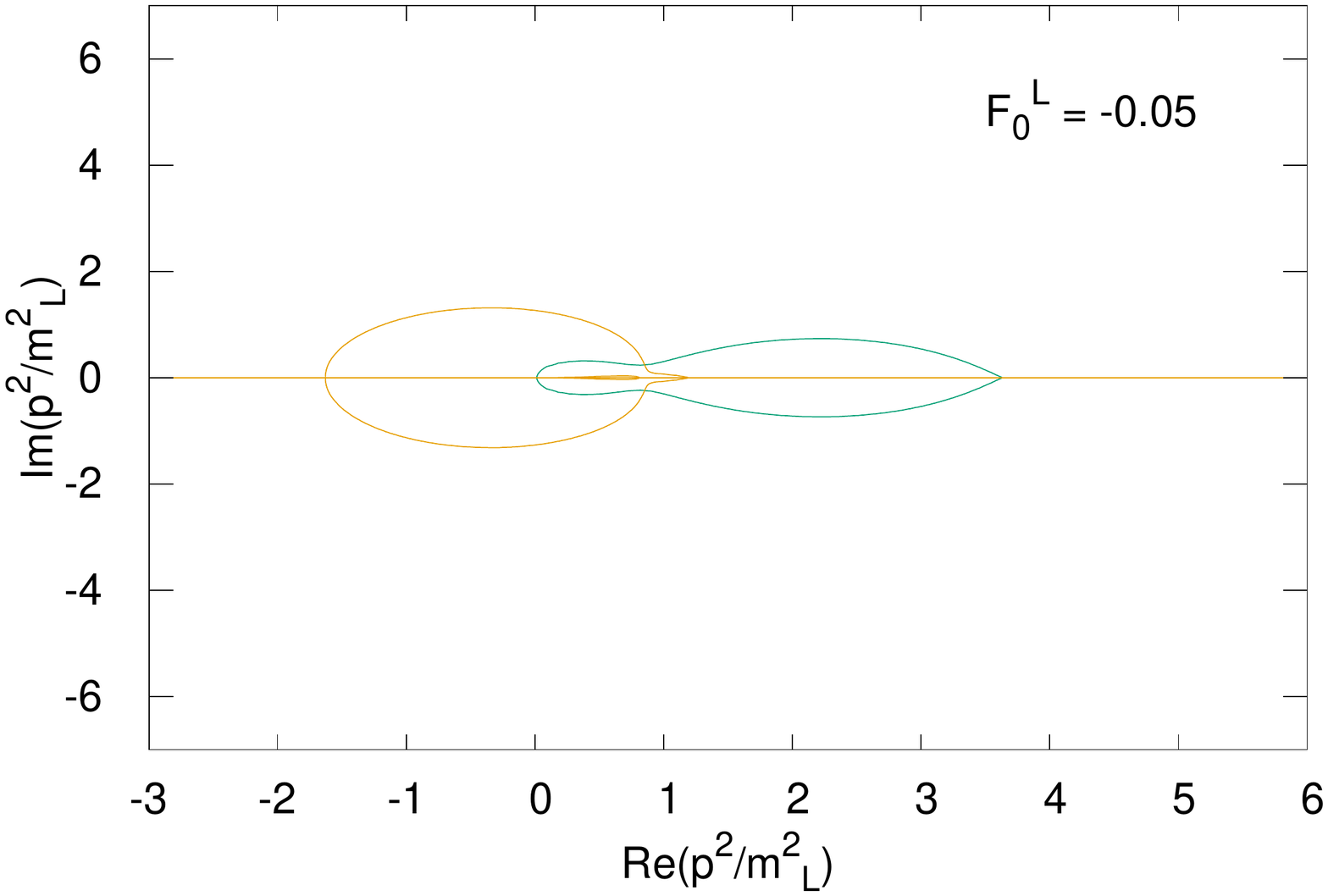}
\end{minipage}
\\
\caption{Case 2 -- zero sets of the real and imaginary part of $F(-p^{2}/m_{L}^{2})+F_{0}^{L}$ for $F_{0}^{L}=-2.04,\,-1.50,\,-1.20,\,-0.90,\,-0.50,\,-0.05$. The propagator has two complex conjugate poles.}\label{zeros2}
\end{figure}

\begin{figure}[H]
\centering
\begin{minipage}{.47\textwidth}
\centering
\includegraphics[width=1.1\linewidth]{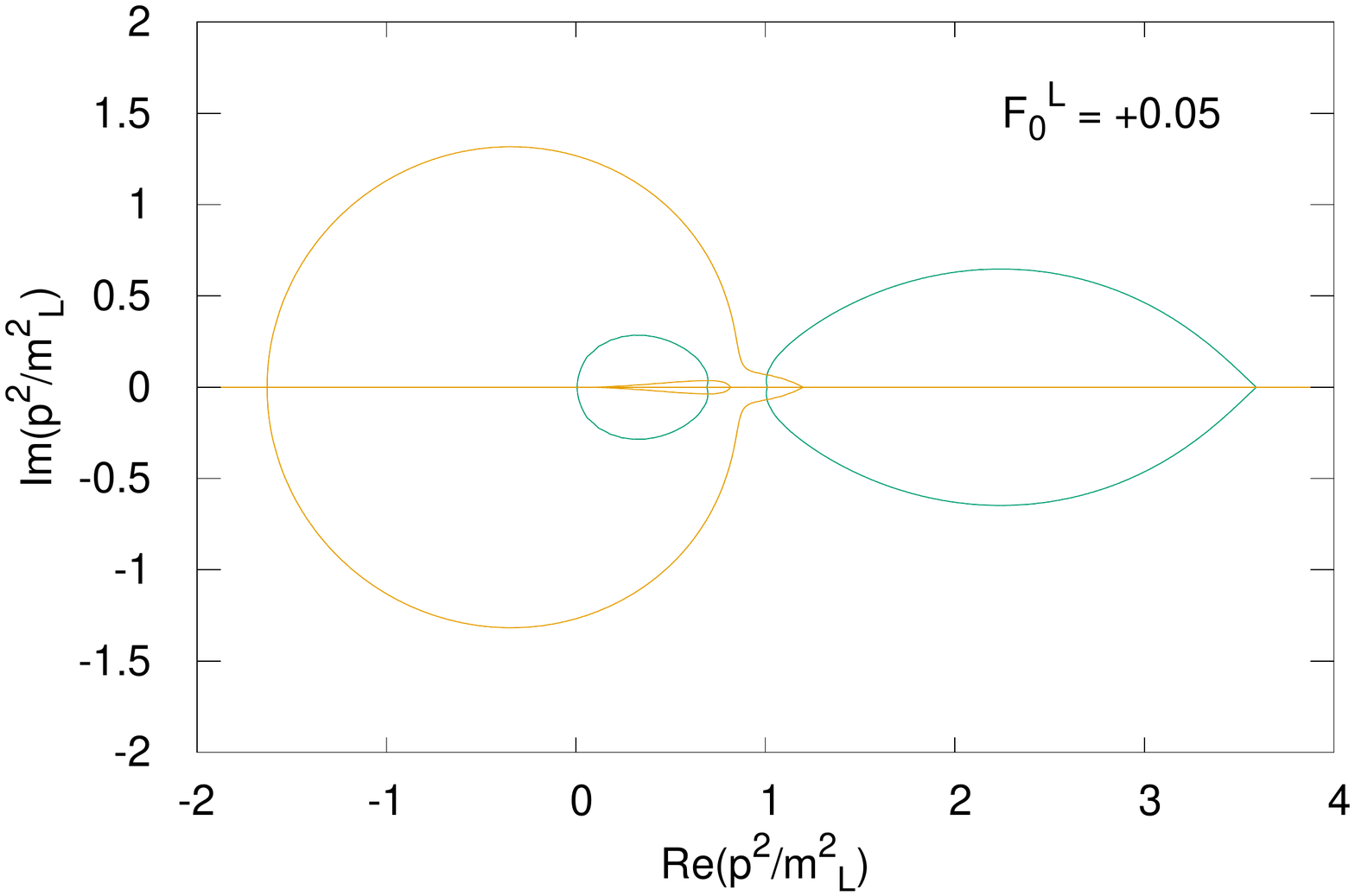}
\end{minipage}
\hspace{15pt}
\begin{minipage}{.47\textwidth}
\centering
\includegraphics[width=1.1\linewidth]{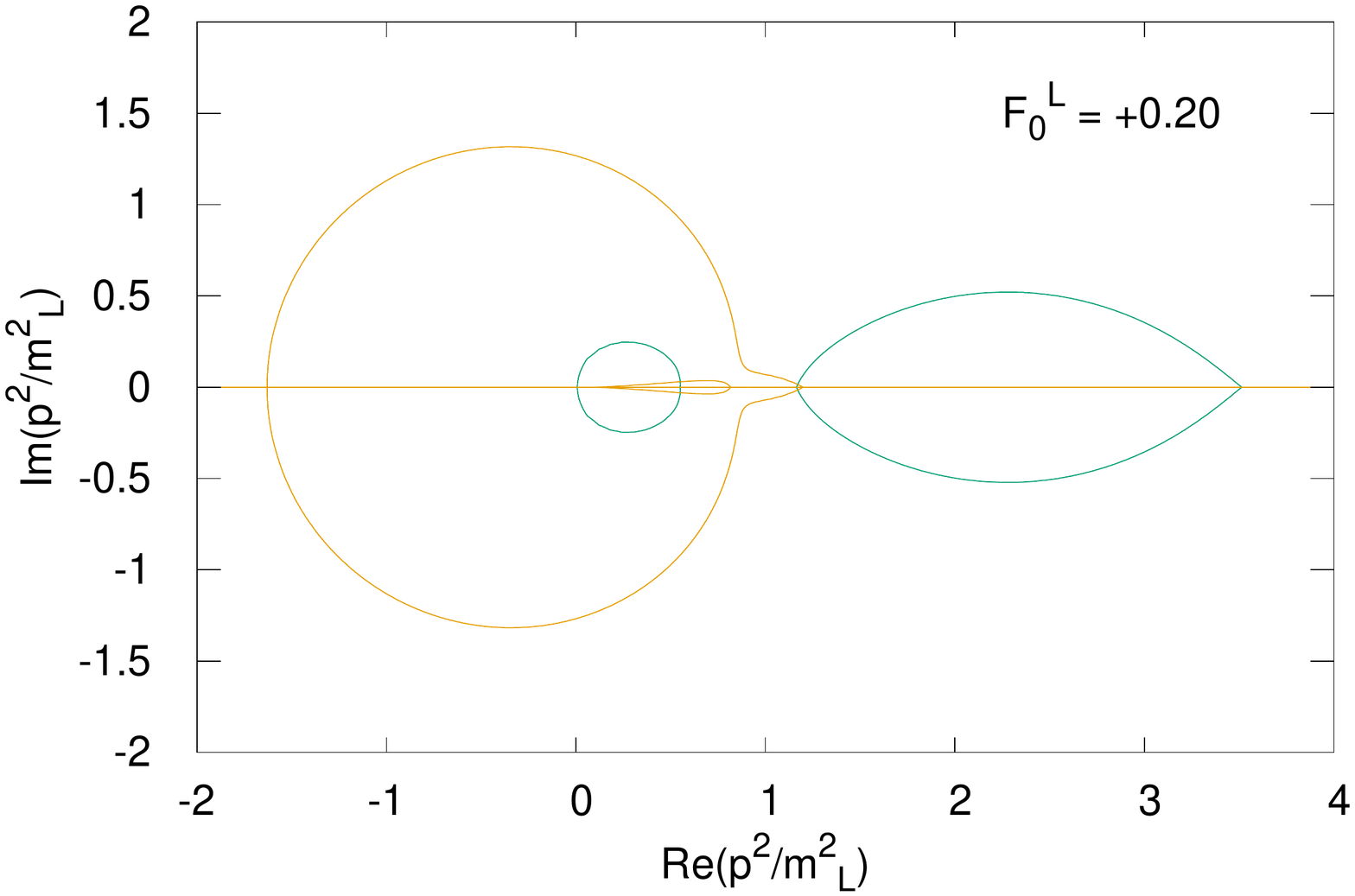}
\end{minipage}
\\
\caption{Case 3 -- zero sets of the real and imaginary part of $F(-p^{2}/m_{L}^{2})+F_{0}^{L}$ for $F_{0}^{L}=+0.05,\,+0.20$. The propagator has two pairs of complex conjugate poles.}\label{zeros3}
\end{figure}

\begin{figure}[H]
\centering
\vskip -20pt
\includegraphics[width=0.78\textwidth]{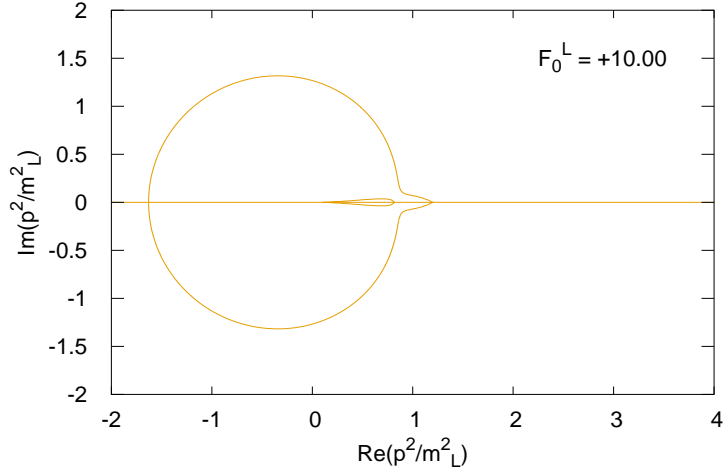}
\vskip -20pt
\caption{Case 5 -- zero sets of the real and imaginary part of $F(-p^{2}/m_{L}^{2})+F_{0}^{L}$ for $F_{0}^{L}=+10.00$. The zero set of $\Bbb{R}\tx{e}\{F(-p^{2}/m_{L}^{2})+F_{0}^{L}\}$ is empty, therefore the propagator has no poles.}\label{zeros5}
\end{figure}

In order to carry out the programme of Sec.3.1.2, the first thing we need to do is study the analytic structure of the gluon propagator in the Landau gauge as a function of $F_{0}^{L}$, the additive renormalization constant at $\xi=0$. What we are interested in is the number and position of the poles of the propagator as $F_{0}^{L}$ is tuned across the real numbers. Recall that in the Landau gauge the poles of the propagator are found by solving the equation\\
\BE\label{xet}
F(-p^{2}/m^{2}_{L})+F_{0}^{L}=0
\EE
\\
where $m_{L}^{2}$ is the mass parameter at $\xi=0$, to be provided as an external input. Since $F(s)$ is a complicated function of $s$, eq.~\eqref{xet} can only be solved numerically. However, in order to figure out the general structure of its solutions, we can start from a graphical analysis of the zero set of the sum on its left hand side.\\
\\
Eq.~\eqref{xet} can be expressed in terms of its real and imaginary parts as\\
\BE
\Bbb{R}\tx{e}\{F(-p^{2}/m^{2}_{L})\}+F_{0}^{L}=0\qquad\qquad \Bbb{I}\tx{m}\{F(-p^{2}/m^{2}_{L})\}=0
\EE
\\
The imaginary part of the equation does not depend on $F^{L}_{0}$; therefore, regardless of the value of $F^{L}_{0}$, the poles of the propagator are constrained to lie in the one-dimensional subset of the complex plane defined by $\Bbb{I}\tx{m}\{F\}=0$. This subset is shown in Fig.\ref{imagzeros}.\\
\\
\newpage
\noindent The zero set of $\Bbb{I}\tx{m}\{F(-p^{2}/m_{L}^{2})\}$ consists of two concentric subsets with the topology of a circle, together with the negative real axis ($F(s)$ is real for positive $s$). The positive real axis is not actually part of the set: since $F(s)$ contains a natural logarithm of $s$, $F(-p^{2}/m_{L}^{2})$ has a branch cut for positive real $p^{2}$'s; the small imaginary discontinuity of the function across the branch cut causes the graphical algorithm to erroneously include the positive real axis in the zero set of the function.\\
The zero set of $\Bbb{I}\tx{m}\{F\}$ is symmetric with respect to the real axis. This is a consequence of the identity $\overline{F(s)}=F(\overline{s})$ -- where the bar denotes complex conjugation -- which holds for $F$ since the latter is a sum of products of logarithms, square roots and rational functions, all of which have real coefficients. The same identity also tells us that if $p_{0}^{2}$ solves eq.~\eqref{xet}, then $\overline{p_{0}^{2}}$ does too: the gluon propagator has complex conjugate poles, which must thus be real or come in conjugate pairs.\\
In order to be poles of the propagator, the zeros of $\Bbb{I}\tx{m}\{F\}$ must also be zeros of $\Bbb{R}\tx{e}\{F\}+F_{0}^{L}$. The latter condition depends on $F^{L}_{0}$, whose value therefore influences the position of the poles. Actually, it turns out that the number of poles also depends on the value of $F^{L}_{0}$. Again, this can be shown through a graphical analysis.\\
\\
\vspace{-20pt}
\begin{figure}[H]
\centering
\begin{minipage}{.47\textwidth}
\centering
\includegraphics[width=1.1\linewidth]{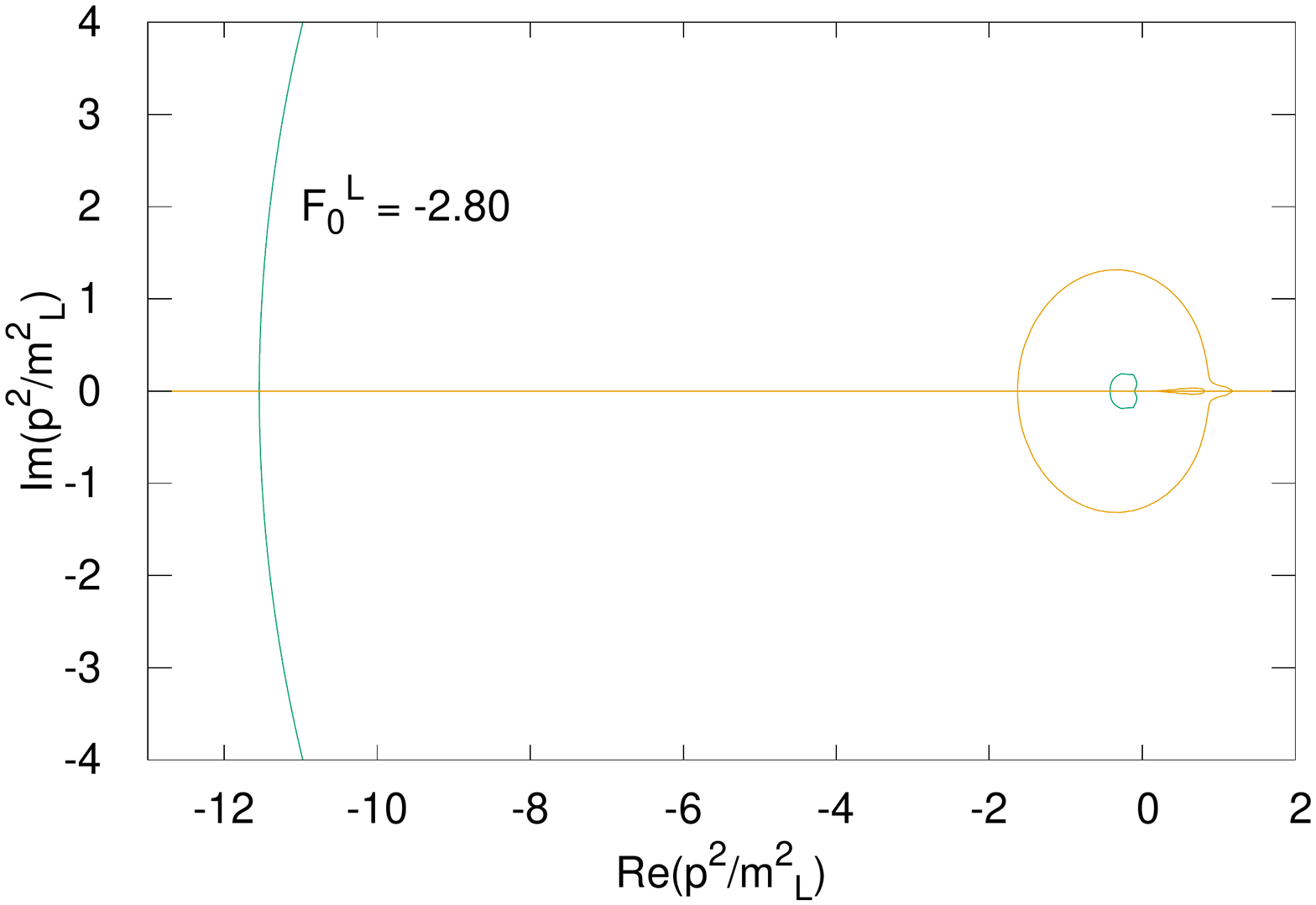}
\end{minipage}
\hspace{15pt}
\begin{minipage}{.47\textwidth}
\centering
\includegraphics[width=1.1\linewidth]{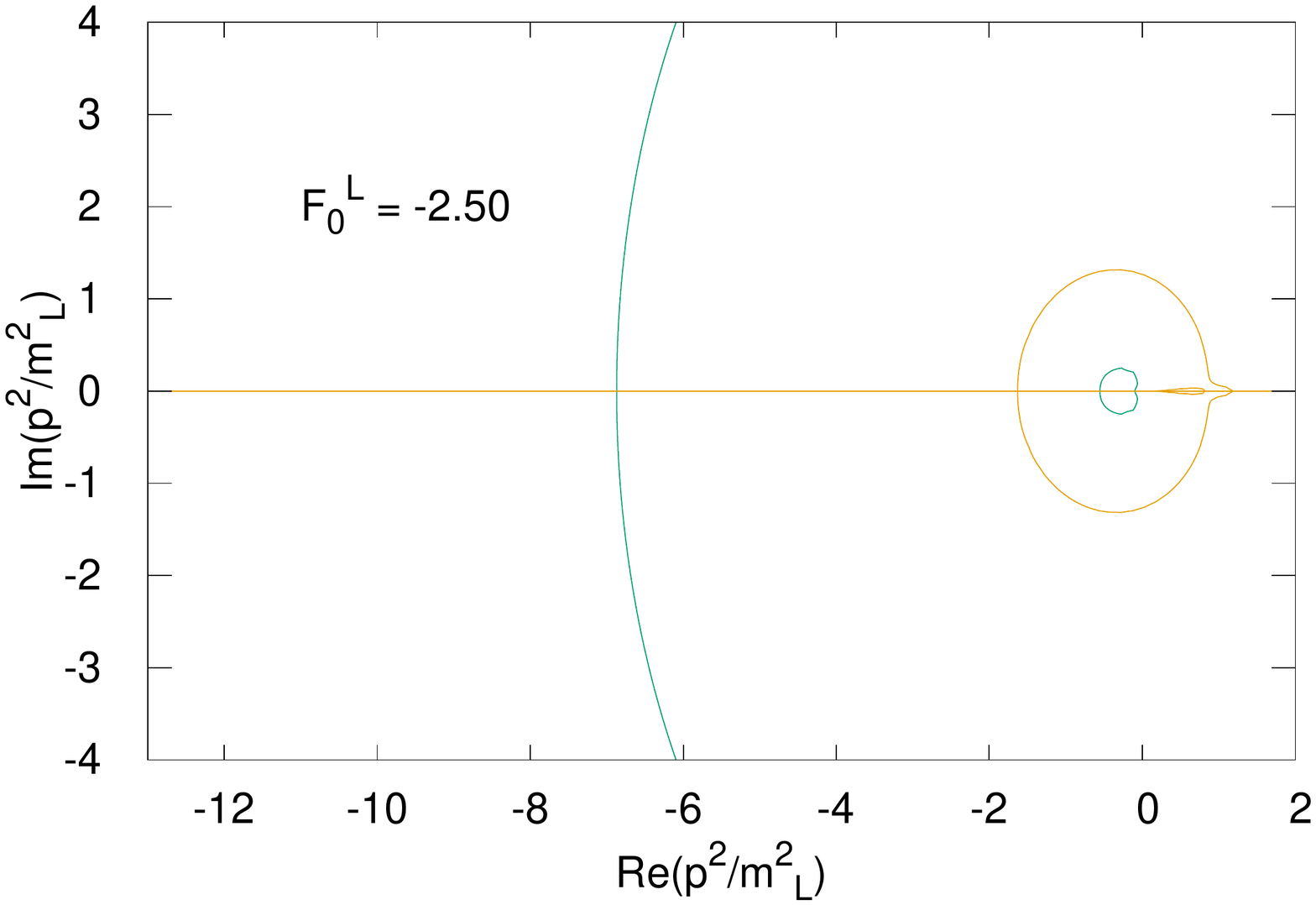}
\end{minipage}
\begin{minipage}{.47\textwidth}
\centering
\includegraphics[width=1.1\linewidth]{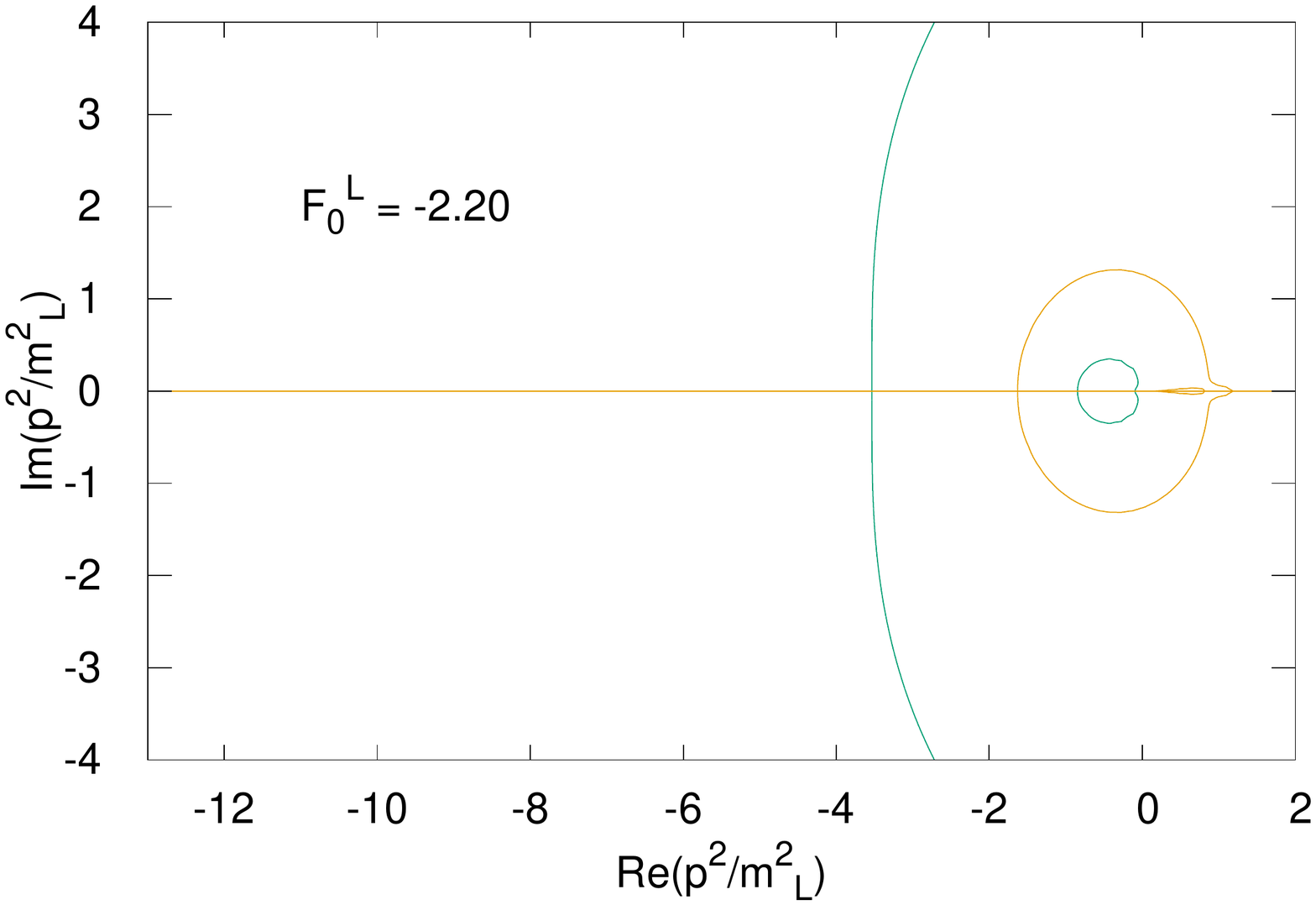}
\end{minipage}
\hspace{15pt}
\begin{minipage}{.47\textwidth}
\centering
\includegraphics[width=1.1\linewidth]{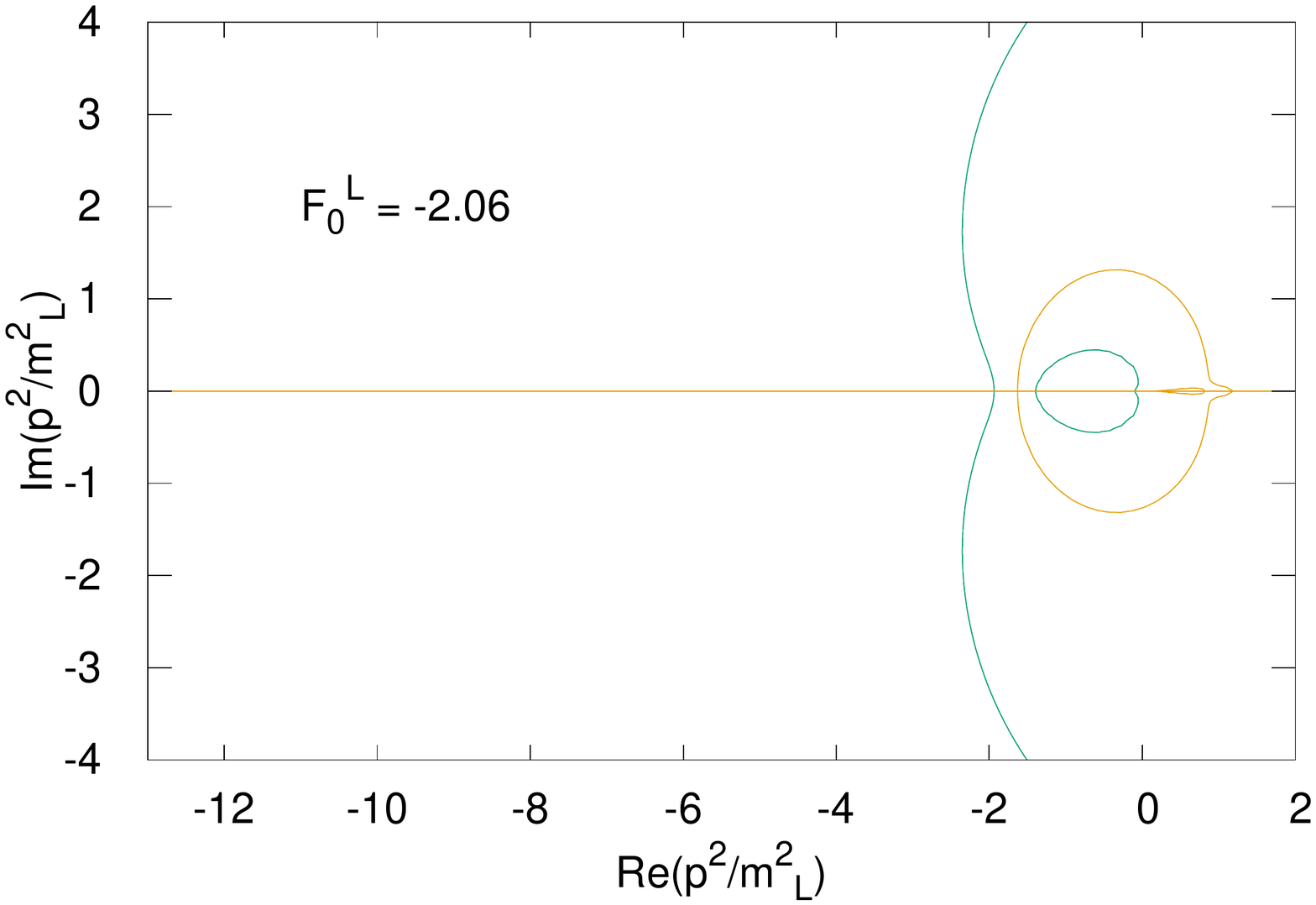}
\end{minipage}
\\
\caption{Case 1 -- zero sets of the real and imaginary part of $F(-p^{2}/m_{L}^{2})+F_{0}^{L}$ for $F_{0}^{L}=-2.80,\,-2.50,\,-2.20,\,-2.06$. The propagator has two negative real poles.}\label{zeros1}
\end{figure}
\newpage
\noindent\underline{Case 1}: $F_{0}^{L}\lessapprox-2.05$.\\
\\
The zero sets of the real and imaginary part of $F(-p^{2}/m_{L}^{2})+F_{0}^{L}$ are shown in Fig.\ref{zeros1} for different values of $F_{0}^{L}<-2.05$. The poles of the propagator are found at the intersection of the two sets.\\
For $F_{0}^{L}\lessapprox-2.05$, the propagator has two poles on the negative real axis. This is an immediate consequence of $F(s)$ being real and greater than approximately $2.05$ for positive real $s$: as $F^{L}_{0}$ is tuned below $-2.05$, the graph of $F(-p^{2}/m^{2}_{L})+F^{L}_{0}$ (for $p^{2}\in\Bbb{R}^{-}$) intersects the horizontal axis in two points, as shown in Fig.\ref{ffunctreg1}. Since $F(-p^{2}/m^{2}_{L})$ tends to $+\infty$ both in the $p^{2}\to 0$ and in the $p^{2}\to -\infty$ limit, this behavior is easily seen to be shared by propagators with arbitrarily negative $F^{L}_{0}$.\\
\\
\
\\
\\
\\
\underline{Case 2}: $F_{0}^{L}\gtrapprox -2.05$, $F_{0}^{L}<0$.\\
\\
The zero sets of the real and imaginary part of $F(-p^{2}/m_{L}^{2})+F_{0}^{L}$ are shown in Fig.\ref{zeros2} for different values of $F_{0}^{L}\in\,]\,-2.05,0\,[$. The intersections at $p^{2}=0$ and $\Bbb{I}\tx{m}(p^{2})=0$, $\Bbb{R}\tx{e}(p^{2})>0$ are not actual poles: we know $F(s)$ to be infinite at $s=0$, which excludes the solution $p^{2}=0$, and we know $F(s)$ to have a branch cut for negative real $s$, which excludes the positive real solutions.\\
As $F_{0}^{L}$ is tuned above approximately $-2.05$, the function $F(-p^{2}/m_{L}^{2})+F^{L}_{0}$ becomes strictly positive for negative real $p^{2}$, so that the negative real poles of Case 1 disappear. In their place, the propagator develops two complex conjugate poles. In this range of $F_{0}^{L}$, the real part of the conjugate poles can be either negative or positive, implying that the real part of $\sqrt{p^{2}}$ can be either less or greater than its imaginary part. The value of $F^{L}_{0}$ for which the real part of the poles is zero -- equivalently, for which $\Bbb{R}\tx{e}(\sqrt{p^{2}})=\Bbb{I}\tx{m}(\sqrt{p^{2}})$ -- is $F_{0}^{L}\approx -1.33$. As $F_{0}^{L}$ approaches zero, the zero set of $\Bbb{R}\tx{e}\{F(-p^{2}/m_{L}^{2})+F_{0}^{L}\}$ starts to shrink.
\newpage
\
\\
\\
\underline{Case 3}: $F_{0}^{L}>0$, $F_{0}^{L}\lessapprox +0.23$.\\
\\
The zero sets of the real and imaginary part of $F(-p^{2}/m_{L}^{2})+F_{0}^{L}$ are shown in Fig.\ref{zeros3} for different values of $F_{0}^{L}\in\,]\,0,0.23\,[$. Again, the intersections at $p^{2}=0$ and $\Bbb{I}\tx{m}(p^{2})=0$, $\Bbb{R}\tx{e}(p^{2})>0$ are not actual poles.\\
The $\Bbb{R}\tx{e}\{F+F_{0}\}=0$ set intersects both of the circular subsets of $\Bbb{I}\tx{m}\{F+F_{0}\}=0$. Therefore the propagator has four complex poles, conjugated in pairs.
\newpage
\
\
\\
\\
\begin{figure}[H]
\centering
\begin{minipage}{.47\textwidth}
\centering
\includegraphics[width=1.1\linewidth]{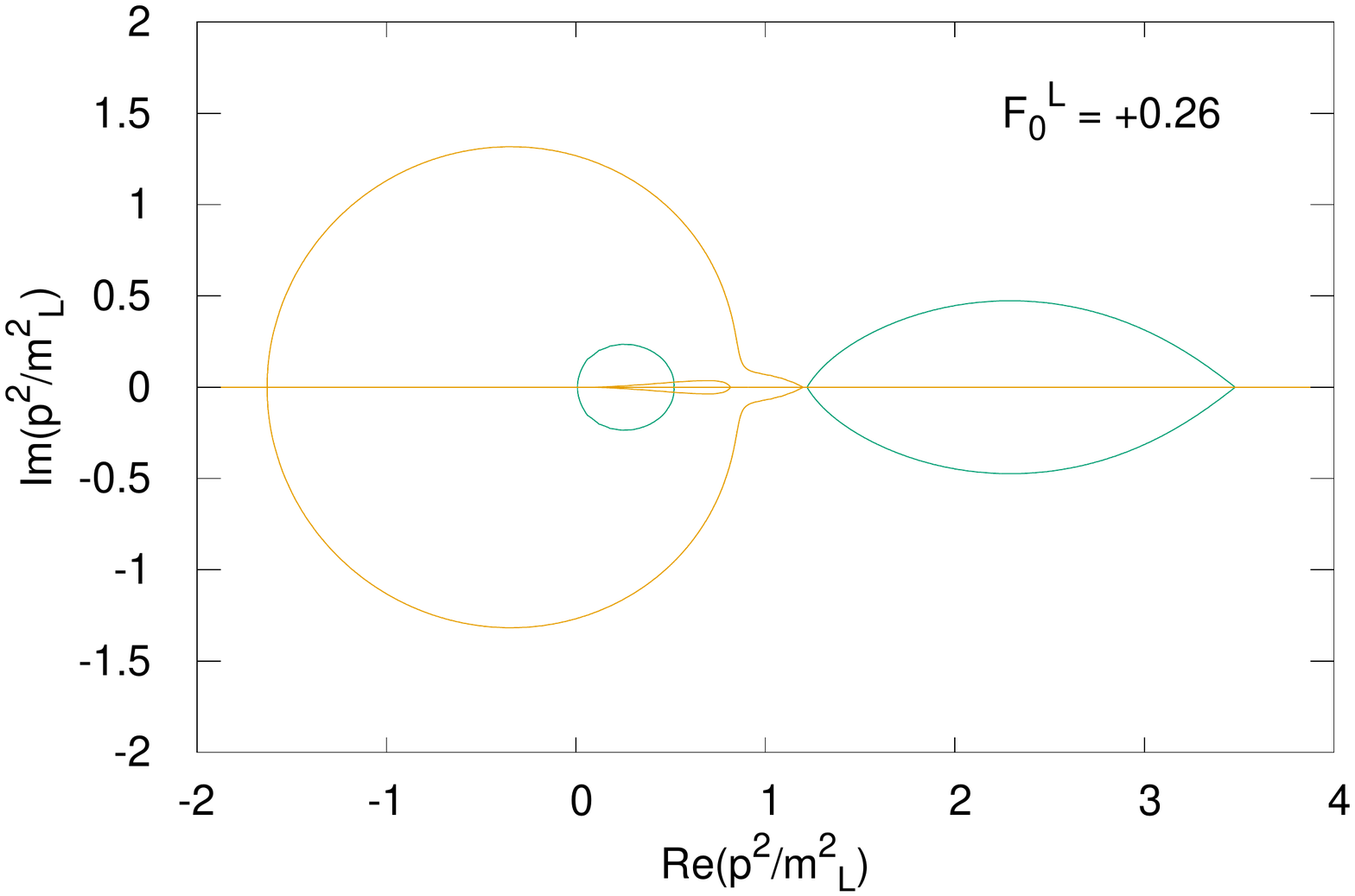}
\end{minipage}
\hspace{15pt}
\begin{minipage}{.47\textwidth}
\centering
\includegraphics[width=1.1\linewidth]{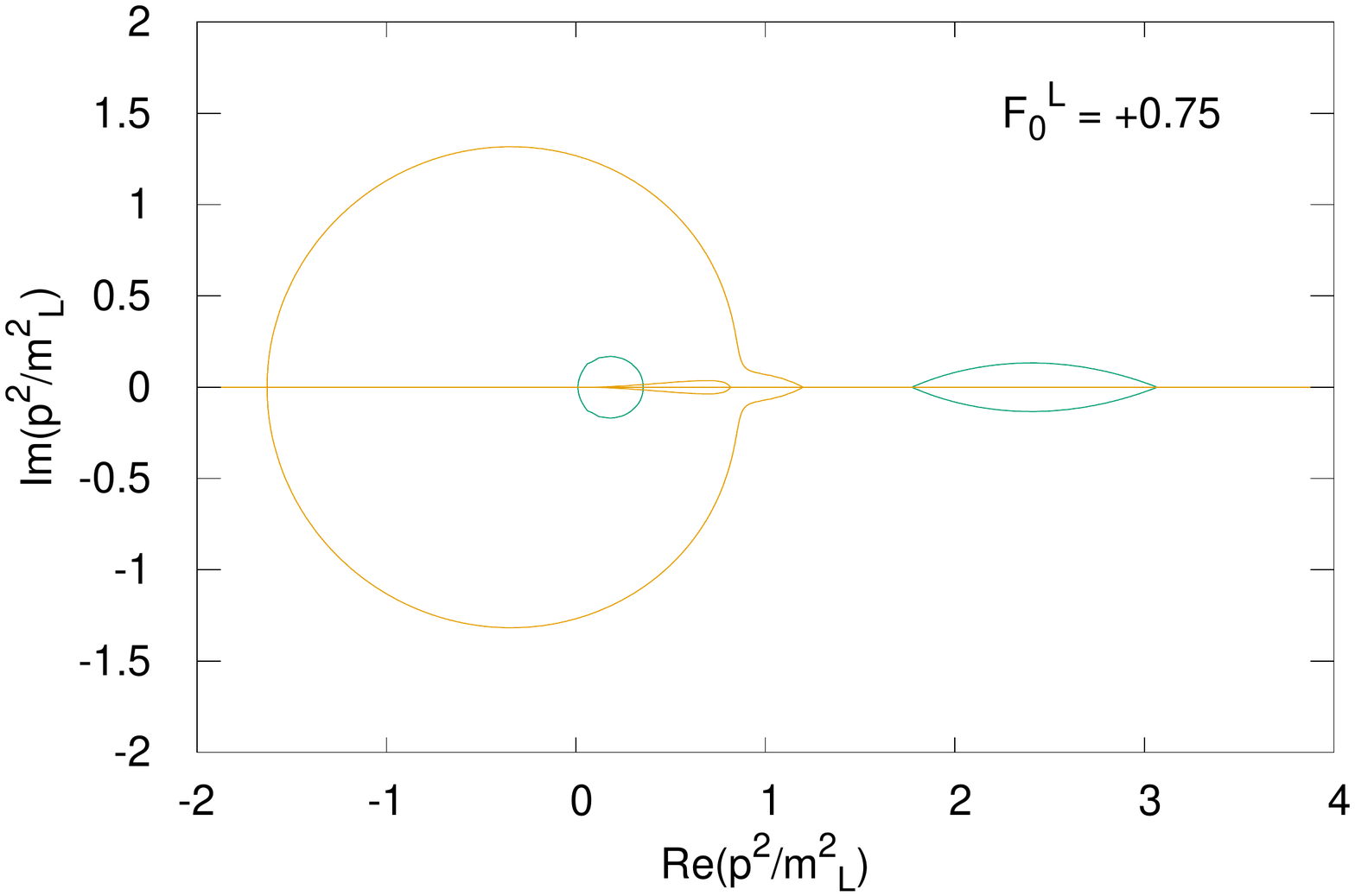}
\end{minipage}
\begin{minipage}{.47\textwidth}
\centering
\includegraphics[width=1.1\linewidth]{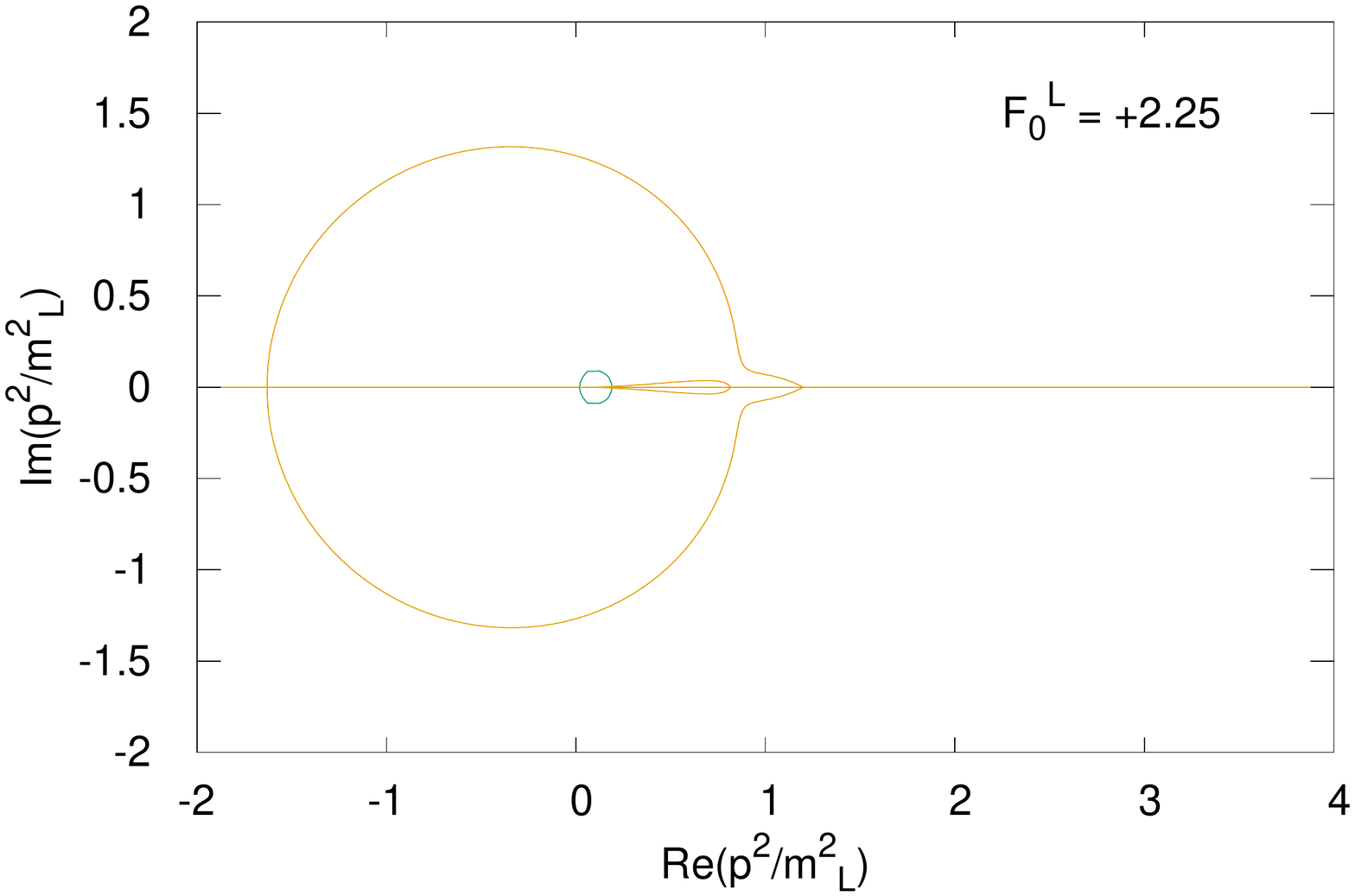}
\end{minipage}
\hspace{15pt}
\begin{minipage}{.47\textwidth}
\centering
\includegraphics[width=1.1\linewidth]{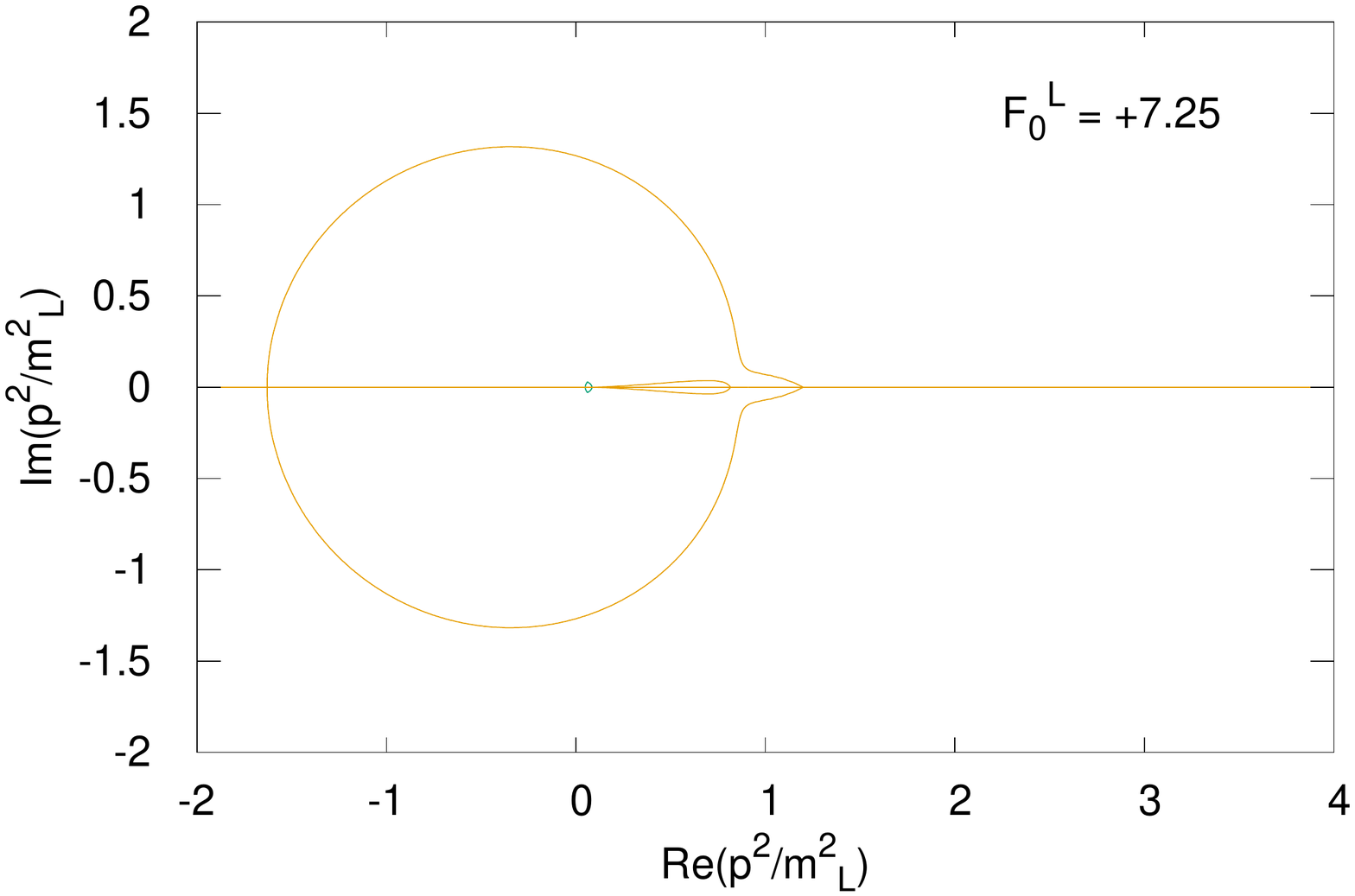}
\end{minipage}
\caption{Case 4 -- zero sets of the real and imaginary part of $F(-p^{2}/m_{L}^{2})+F_{0}^{L}$ for $F_{0}^{L}=+0.26,\,+0.75,\,+2.25,\,+7.25$. The propagators has two complex conjugate poles.}\label{zeros4}
\end{figure}
\newpage
\noindent\underline{Case 4}: $F_{0}^{L}\gtrapprox +0.23$, $F_{0}^{L}\lessapprox +9.20$.\\
\\
The zero sets of the real and imaginary part of $F(-p^{2}/m_{L}^{2})+F_{0}^{L}$ are shown in Fig.\ref{zeros4} for different values of $F_{0}^{L}\in\,]\,0.23,9.20\,[$. The intersections at $p^{2}=0$ and $\Bbb{I}\tx{m}(p^{2})=0$, $\Bbb{R}\tx{e}(p^{2})>0$ are not actual poles.\\
As $F_{0}^{L}$ is tuned above approximately 0.23, one of the two connected components of $\Bbb{R}\tx{e}\{F+F_{0}\}=0$ ceases to intersect the zero set of $\Bbb{I}\tx{m}\{F+F_{0}\}$, then shrinks to zero and disappears: in this range the propagator has two complex conjugate poles. As $F_{0}^{L}$ approaches approximately $9.20$ the second connected component of $\Bbb{R}\tx{e}\{F+F_{0}\}=0$ also shrinks to zero.\\
\\
\\
\underline{Case 5}: $F_{0}^{L}\gtrapprox +9.20$.\\
\\
The zero sets of the real and imaginary part of $F(-p^{2}/m_{L}^{2})+F_{0}^{L}$ are shown in Fig.\ref{zeros5} for a single value of $F_{0}^{L}\gtrapprox 9.20$. The equation $\Bbb{R}\tx{e}\{F(-p^{2}/m_{L}^{2})+F_{0}^{L}\}=0$ has no solutions. Therefore the propagator has no poles.\\
\
\\
\\
\\
The graphical analysis of the zero sets of $F(-p^{2}/m_{L}^{2})+F_{0}^{L}$ informs us that the topology of the poles of the gluon propagator is very much dependent on the value of $F_{0}^{L}$. As $F_{0}^{L}$ is tuned from arbitrarily negative values to arbitrarily positive values, the propagator goes from having two real negative poles (Euclidean poles), to having two complex conjugate poles, to having two pairs of complex conjugate poles, to having two complex conjugate poles again, to having no poles at all. In principle, any of these solutions might yield the true analytic structure of the gluon propagator. However, we have reasons to believe that only the values of Case 2, namely, $F_{0}^{L}\in\,]\,-2.05,0\,[\,$, are reasonable parameters for the propagator. The motivations for this are the following.\\
First of all, consider Case 1: $F_{0}^{L}\lessapprox -2.05$. The poles of Case 1 are real negative because for these values of $F_{0}^{L}$ the function $F(s)+F^{L}_{0}$ evaluated at positive real $s$ goes from being positive, to being negative, to being positive again as $s$ is tuned from $0$ to $+\infty$ -- cf. Fig.\ref{ffunctreg1}. Observe that $s\in\Bbb{R}^{+}_{0}$ is precisely the domain of definition of the Euclidean propagator, which therefore, in this range of $F_{0}^{L}$, not only becomes infinite at two finite values of the Euclidean momentum, but is also negative in some momentum interval. This is not a reasonable behavior for the Euclidean propagator. Therefore we must conclude that $\,]-\infty,-2.05\,[\,$ is not an acceptable range of values for $F_{0}^{L}$.\\
As for Cases 3 and 4, we observe that the poles resulting from $F_{0}^{L}>0$ lie in regions of $\Bbb{I}\tx{m}\{F+F_{0}\}=0$ which have a fairly bizarre shape and are quite flattened against the branch cut on the positive real axis. Now, since we are working with an approximation to the exact propagator, we might expect the true zero set of the imaginary part of the inverse propagator to be somewhat shifted from $\Bbb{I}\tx{m}\{F+F_{0}\}=0$. Due to the closeness of the aforementioned regions to the real axis, a small shift originating in higher order radiative corrections to the propagator may cause the $F_{0}^{L}>0$ poles to completely disappear: because of their position and of the shape of the set to which they belong, such poles cannot be guaranteed to be genuine, so much as an artifact of the massive expansion. In addition to this, we could also argue that the values $F_{0}^{L}>0$ fail to reproduce the Euclidean propagator computed on the lattice: in Sec.2.3.2 we found that the value of $F_{0}^{L}$ which best fits the lattice data in the Landau gauge is $F_{0}^{L}=-0.887$, which is quite far away from the positive values of Cases 3 to 5.\\
These considerations lead us to conclude that the most sensible range of values for the additive renormalization constant $F_{0}^{L}$ is that of Case 2, namely, $\,]\,-2.05,\,0\,[\,$. As we will see in the following section, this range contains values of $F_{0}^{L}$ for which the phase difference $\theta(\xi)$ is smaller than $0.003$ in absolute value.\\
\\
Having restricted the optimal range of $F_{0}^{L}$ to $\,]\,-2.05,\,0\,[\,$, we can now move on to the numerical solution of eq.~\eqref{mastland}. The position of the poles as a function of $F_{0}^{L}$ is reported in Tab.\ref{pospolestab} for different values of the renormalization constant in the aforementioned range.\\
\begin{table}[H]
\def\arraystretch{1.3}
\centering
\begin{tabular}{c|c}
$F_{0}^{L}$&$p^{2}_{0}/m_{L}^{2}$\\
\hline
$-2.00$&$-1.458\pm0.654\ i$\\
$-1.90$&$-1.159\pm1.016\ i$\\
$-1.75$&$-0.776\pm1.240\ i$\\
$-1.50$&$-0.279\pm1.316\ i$\\
$-1.25$&$0.085\pm1.240\ i$\\
$-1.00$&$0.352\pm1.097\ i$\\
$-0.75$&$0.549\pm0.921\ i$\\
$-0.50$&$0.691\pm0.722\ i$\\
$-0.25$&$0.791\pm0.494\ i$\\
$-0.10$&$0.836\pm0.319\ i$
\end{tabular}
\\
\caption{Poles of the gluon propagator as a function of $F_{0}^{L}\in \,]\,-2.05,\,0\,[\,$.}\label{pospolestab}
\end{table}
\
\begin{figure}[H]
\centering
\vskip -20pt
\includegraphics[width=0.70\textwidth]{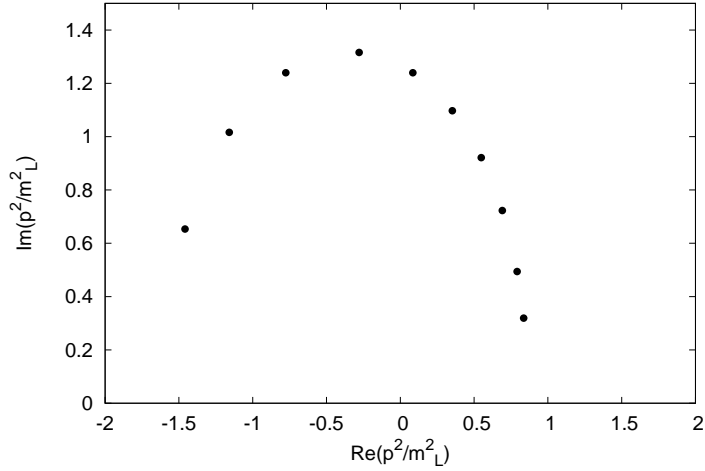}
\vskip -20pt
\caption{Poles of the gluon propagator as a function of $F_{0}^{L}\in\,]-2.05,\,0\,[\,$. Left to right: $F_{0}=-2.00,\,-1.90,\,-1.75,\,-1.50,\,-1.25,\,-1.00,\,-0.75,\,-0.50,\,-0.25,\,-0.10$.}\label{pospolesfig}
\end{figure}
\
\\
\\
In Fig.\ref{pospolesfig} we show the location of the poles of Tab.\ref{pospolestab} in the complex plane\footnote{\ Only the pole with a positive imaginary part is shown in the figure.}. As $F^{L}_{0}$ is tuned from $-2.00$ to $-0.10$, the real part of the complex conjugate poles grows from $-1.458$ to $0.836$ (in units of $m_{L}^{2}$); their imaginary part, on the other hand, grows in absolute value from $0.654$ to $1.316$ at $F_{0}^{L}=-1.50$ and then falls back to $0.319$. These poles are not guaranteed to have residues whose phases -- in compliance with the Nielsen identities -- are gauge-invariant. In order to be able to compute the phases of the residues as a function of the gauge parameter $\xi$, we first need to know the gauge-dependent parameter functions $F_{0}(\xi)$ and $m^{2}(\xi)$. These will be derived in the next section.
\\
\\

\addcontentsline{toc}{subsection}{3.2.2 The functions $F_{0}(\xi)$, $m^{2}(\xi)$ and $\theta(\xi)$}  \markboth{3.2.2 The functions $F_{0}(\xi)$, $m^{2}(\xi)$ and $\theta(\xi)$}{3.2.2 The functions $F_{0}(\xi)$, $m^{2}(\xi)$ and $\theta(\xi)$}
\subsection*{3.2.2 The functions $\boldsymbol{F_{0}(\xi)}$, $\boldsymbol{m^{2}(\xi)}$ and $\boldsymbol{\theta(\xi)}$\index{The functions $F_{0}(\xi)$, $m^{2}(\xi)$ and $\theta(\xi)$}}

\begin{figure}[H]
\centering
\vskip -20pt
\includegraphics[width=0.70\textwidth]{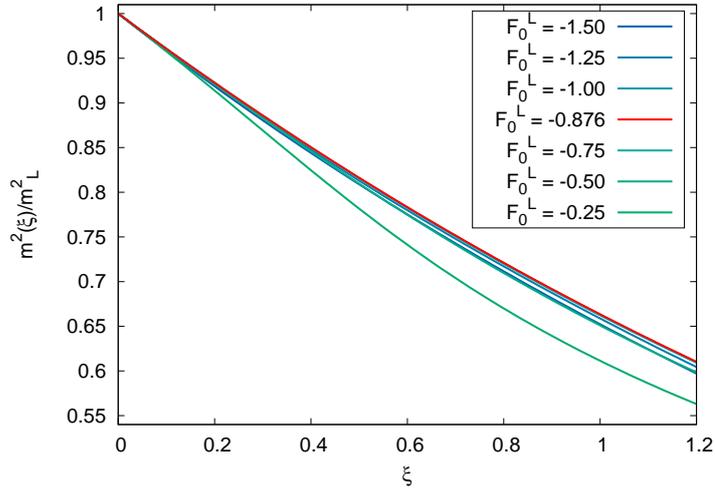}
\vskip -20pt
\caption{Gauge-dependent mass parameter $m^{2}(\xi)$ as a function of $\xi$ for different values of $F_{0}^{L}$ in the range $\,]-2.05,\,0\,[\,$.}\label{massfig}
\end{figure}

Having found the location of the poles of the gluon propagator as $F_{0}^{L}$ is tuned across the interval $\,]-2.05,\,0\,[\,$, we are now in a position to compute the gauge-dependent parameter functions $F_{0}(\xi)$ and $m^{2}(\xi)$ associated to such poles. Recall that the latter are defined as the renormalization and mass parameters which lead to the poles in the gauge $\xi$ being in the same position of those in the Landau gauge. Equivalently, $F_{0}(\xi)$ and $m^{2}(\xi)$ can be defined as the functions which solve the equation\\
\BE\label{fmeq}
F(-p_{0}^{2}/m^{2}(\xi))+\xi\, F_{\xi}(-p_{0}^{2}/m^{2}(\xi))+F_{0}(\xi)=0\qquad\quad \forall\ \xi
\EE
\\
where $p_{0}^{2}$ has been computed by fixing some value for the Landau gauge renormalization parameter $F_{0}^{L}$. Observe that since $\xi$, $F_{0}(\xi)$ and $m^{2}(\xi)$ are real, and since $\overline{F(s)}=F(\overline{s})$, $\overline{F_{\xi}(s)}=F_{\xi}(\overline{s})$, this equation yields the same solutions with both $p_{0}^{2}$ and $\overline{p_{0}^{2}}$ as an input. Therefore its solutions do not depend on the choice of the pole, as it should be\footnote{\ In the presence of two pairs of complex conjugate poles, the solutions of eq.~\eqref{fmeq} depend on the choice of one of the two pairs, so that both pairs cannot, in general, be simultaneously gauge-invariant. This gives us further confidence that Case 3 in Sec.3.2.1 must be discarded.}.\\
Eq.~\eqref{fmeq} can be solved by isolating its real and imaginary parts. Since $F_{0}(\xi)$ is real, the imaginary part of the equation depends on $m^{2}(\xi)$ only and reads\\
\BE\label{massgauge}
\Bbb{I}\tx{m}\{F(-p_{0}^{2}/m^{2}(\xi))+\xi\, F_{\xi}(-p_{0}^{2}/m^{2}(\xi))\}=0
\EE
\\
This equation can be solved for $m^{2}(\xi)$ alone, without having to worry about $F_{0}(\xi)$. Once $m^{2}(\xi)$ is known, one can compute $F_{0}(\xi)$ as\\
\BE\label{f0gauge}
F_{0}(\xi)=-\Bbb{R}\tx{e}\{F(-p_{0}^{2}/m^{2}(\xi))+\xi\, F_{\xi}(-p_{0}^{2}/m^{2}(\xi))\}
\EE
\\
Of course, both $m^{2}(\xi)$ and $F_{0}(\xi)$ will depend on the value of $F_{0}^{L}$ which has been used to compute $p_{0}^{2}$ in the first place.\\
\\
In Fig.\ref{massfig} we show the gauge-dependent mass parameter $m^{2}(\xi)$ -- as obtained by numerically solving eq.~\eqref{massgauge} -- as a function of $\xi$ for different values of $F_{0}^{L}$ in the range $\,]-2.05,\,0\,[\,$. As $F_{0}^{L}$ is increased from $-1.50$ to $-0.25$ at constant $\xi$, the value of $m^{2}(\xi)$ first increases and then decreases. For $F_{0}^{L}$ between approximately $-1.00$ and $-0.75$, the functions $m^{2}(\xi)$ computed at different $F_{0}^{L}$'s are almost indistinguishable from one another. $F_{0}^{L}=-0.876$ (red line in figure) is the value for which the mass parameter is both stationary with respect to $F_{0}^{L}$ at any fixed $\xi$ and closer to $m_{L}^{2}$. This feature can be shown \cite{com3} to be a direct consequence of the phase of the residue being gauge independent (or almost so) for the aforementioned value of $F_{0}^{L}$ (see ahead). $m^{2}(\xi)$ decreases with the gauge for every value of $F_{0}^{L}$.\\
\\
\newpage
\begin{figure}[H]
\centering
\vskip -20pt
\includegraphics[width=0.70\textwidth]{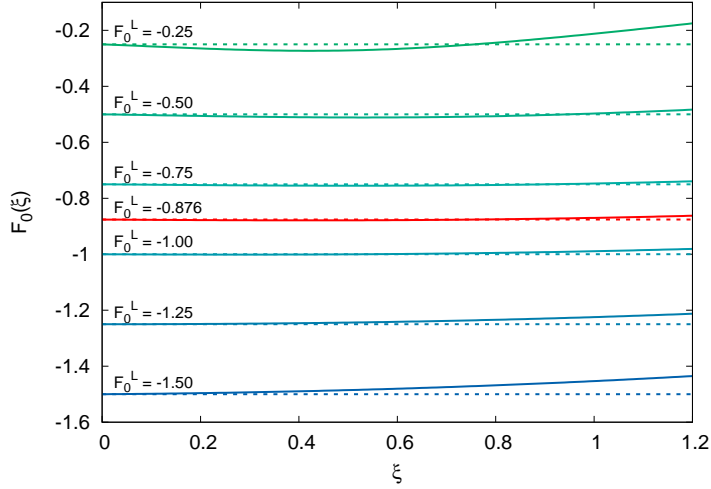}
\vskip -20pt
\caption{Gauge-dependent renormalization constant $F^{L}_{0}(\xi)$ as a function of $\xi$ for different values of $F_{0}^{L}$ in the range $\,]-2.05,\,0\,[\,$.}\label{f0fig}
\end{figure}
\
\\
\\
In Fig.\ref{f0fig} we show the gauge-dependent renormalization constant $F^{L}_{0}(\xi)$ -- as computed from eq.~\eqref{f0gauge} -- as a function of $\xi$ for different values of $F_{0}^{L}$ in the range $\,]-2.05,\,0\,[\,$. As $\xi$ increases from zero to higher values, the function $F_{0}(\xi)$ deviates from its value $F^{L}_{0}$ in the Landau gauge in a non-monotonic fashion. For $F_{0}^{L}$ in the range $ [\,-1.00,\,-0.75\,]$, $F_{0}(\xi)$ is nearly equal to $F_{0}^{L}$ up to and beyond the Feynman gauge ($\xi=1$).\\
The solutions presented in Figg.\ref{massfig}-\ref{f0fig} for the functions $F_{0}(\xi)$ and $m^{2}(\xi)$ show some form of stationarity for values of $F_{0}^{L}$ in the range around $-1.00$ to about $-0.75$, suggesting that something interesting might be happening in this interval. As we will see in a moment, this is indeed the case.\\
\\
Having computed $F_{0}(\xi)$ and $m^{2}(\xi)$ for different values of $F_{0}^{L}$, we are now in possession of a family of propagators which in any gauge -- modulo an arbitrary multiplicative gauge-dependent constant $Z_{\mc{D}}(\xi)$ -- depend only on the mass parameter and additive renormalization constant $m^{2}_{L}$ and $F_{0}^{L}$ in the Landau gauge:\\
\BE
\widetilde{\mc{D}}_{T}(p^{2},\xi)=\frac{-iZ_{\mc{D}}(\xi)}{p^{2}\big(F(-p^{2}/m^{2}(\xi))+\xi F_{\xi}(-p^{2}/m^{2}(\xi))+F_{0}(\xi)\big)}
\EE
\\
These propagators comply with the Nielsen identities in that their poles are by construction gauge-invariant. It is now time to see whether for any of the values of $F_{0}^{L}$ in the range $\,]-2.05,\,0\,[\,$ the phases of the residues at their poles also are gauge-invariant. Recall the definition of $\theta_{p_{0}^{2}}(\xi)$ from Sec.3.1.2:\\
\BE\label{phaseq}
\theta_{p_{0}^{2}}(\xi)=\text{Arg}\left\{\frac{F'(-p_{0}^{2}/m^{2}_{L})}{F'(-p_{0}^{2}/m^{2}(\xi))+\xi\, F'_{\xi}(-p_{0}^{2}/m^{2}(\xi))}\right\}
\EE
\\
$\theta_{p_{0}^{2}}(\xi)$ is simply the phase difference between the residue at $p_{0}^{2}$ in the gauge $\xi$ and that in the Landau gauge. From eq.~\eqref{phaseq} one can easily prove that $\theta_{\overline{p_{0}^{2}}}(\xi)=-\theta_{p_{0}^{2}}(\xi)$.
\newpage
\begin{figure}[H]
\centering
\vskip -15pt
\includegraphics[width=0.65\textwidth,angle=0]{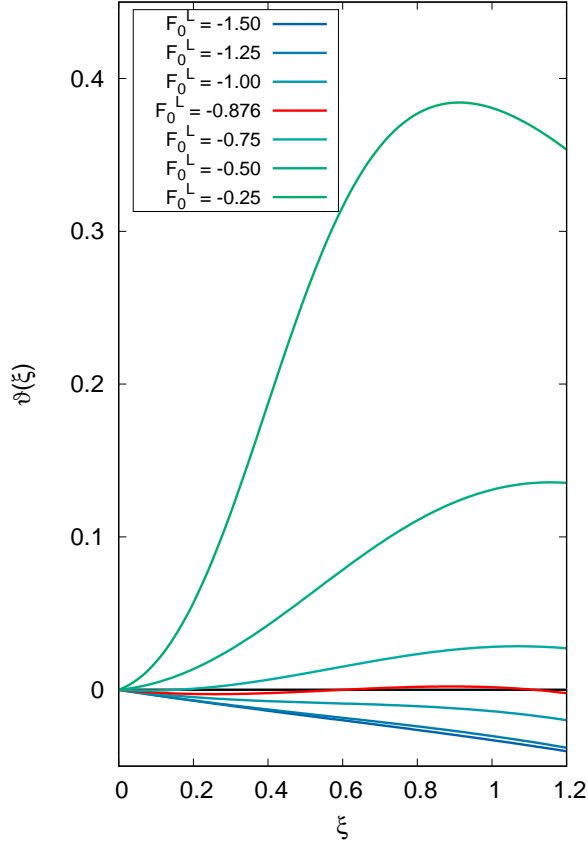}
\vskip -20pt
\caption{Phase difference of the residues $\theta(\xi)$ as a function of $\xi$ for different values of $F_{0}^{L}$ in the range $\,]-2.05,\,0\,[\,$.}\label{phasefig}
\end{figure}
\
\\
\\
Since the two complex conjugate poles have opposite phase differences, in what follows we will drop the label $p_{0}^{2}$ and denote by $\theta(\xi)$ the phase difference at the pole with a positive imaginary part. Explicit analytic expressions for the functions $F'(s)$ and $F'_{\xi}(s)$ that appear in the definition of $\theta(\xi)$ can be found in Appendix A.\\
In Fig.\ref{phasefig} we show the phase difference $\theta(\xi)$ -- as computed from eq.~\eqref{phaseq} -- as a function of $\xi$ for different values of $F_{0}^{L}$ in the range $\,]-2.05,\,0\,[\,$. As we can see, for arbitrary values of $F_{0}^{L}$ in said range, the phase of the residue in the gauge $\xi$ can be quite different from that of the Landau gauge. $|\theta(\xi)|$ can be as high as $20^{\circ}$ or more unless $F_{0}^{L}$ is chosen in the stationarity interval $\,]-1.00,\,-0.75\,[\,$, where it is less than about $2^{\circ}$. The stationarity interval contains a value for which the phase difference is exceedingly small up to and beyond the Feynman gauge ($\xi=1$): for $F_{0}^{L}=-0.876$ (red line in figure) one finds that\footnote{\ Numerical values for the function $\theta(\xi)$ computed at $F_{0}^{L}=-0.876$ can be found in Appendix B.} $|\theta(\xi)|<2.76\cdot 10^{-3}$. For all practical purposes, this difference is so small that we can safely state that the phases of the residues of the gluon propagator computed for this value of $F_{0}^{L}$ are gauge-invariant, in compliance with the Nielsen identities.\\
We conclude that from the perspective of gauge invariance $F_{0}^{L}=-0.876$ is the optimal value of the additive renormalization constant in the Landau gauge. Notice that this value is within about $1\%$ from that which in Chapter 2 was found to best fit the lattice data in the Euclidean space, namely $F_{0}=-0.887$.
\newpage

\addcontentsline{toc}{subsection}{3.2.3 Gauge-optimized parameters for the massive expansion and the gauge-independent gluon poles}  \markboth{3.2.3 Gauge-optimized parameters for the massive expansion and the gauge-independent gluon poles}{3.2.3 Gauge-optimized parameters for the massive expansion and the gauge-independent gluon poles}
\subsection*{3.2.3 Gauge-optimized parameters for the massive expansion and the gauge-independent gluon poles\index{Gauge-optimized parameters for the massive expansion and the gauge-independent gluon poles}}

In the previous section we have established that from the point of view of gauge invariance $F_{0}=-0.876$ is the optimal value of the gluon additive renormalization constant in the Landau gauge: starting from this value and employing the procedure laid out in Sec.3.1.2 one obtains gauge-dependent functions $F_{0}(\xi)$ and $m^{2}(\xi)$ which yield a gluon propagator that complies with the Nielsen identities. This section will be devoted to giving some quantitative details about the functions $F_{0}(\xi)$ and $m^{2}(\xi)$ and the position and phases of the residues of the poles computed by using the optimal value of $F_{0}^{L}$. In what follows $m_{L}^{2}$ is the value of the mass parameter in the Landau gauge, which must be provided as an external input.\\
Let us start from the poles of the propagator. In Tab.\ref{paramtab} we report the position $z_{0}^{2}=p^{2}_{0}/m_{L}^{2}$ and the phase $\varphi$ of the residue of the gluon poles computed at $F_{0}^{L}=-0.876$. As in Sec.3.1.2 the phase is defined modulo the $-i$ of the propagator in Minkowski space.\\
\\
\begin{table}[H]
\def\arraystretch{1.4}
\centering
\begin{tabular}{|c|ccc|}
\hline
\hline
$F_{0}^{L}=-0.876$&$z_{0}^{2}=0.4575\pm1.0130\ i$&$z_{0}=\pm0.8857\pm0.5718\ i$&$\varphi=\pm\,1.262$\\
\hline
\hline
\end{tabular}
\\
\caption{Position and phases of the residues of the gluon poles for $F_{0}^{L}=-0.876$.}\label{paramtab}
\end{table}
\
\\
The gauge-dependent mass parameter $m^{2}(\xi)$ computed at $F_{0}^{L}=-0.876$ is shown in Fig.\ref{massfitfig} as a function of $\xi$. The numerical data -- presented in Appendix B -- can be polynomially interpolated to yield the following approximate expression for $m^{2}(\xi)$:\\
\BE\label{polymass}
m^{2}(\xi)\approx m_{L}^{2}\, \left(1-0.39997\,\xi+0.064141\,\xi^{2}\right)
\EE
\\
This approximation is very precise up to and beyond the Feynman gauge ($\xi=1$).\\
\\
\begin{figure}[H]
\centering
\vskip -20pt
\includegraphics[width=0.70\textwidth]{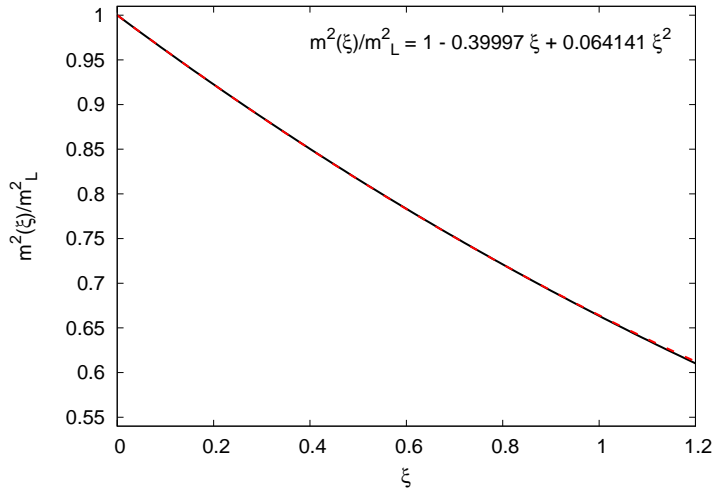}
\vskip -20pt
\caption{Optimal mass parameter $m^{2}(\xi)$ as a function of $\xi$ computed at $F_{0}^{L}=-0.876$. Dashed red line: polynomial interpolation given by eq.\eqref{polymass}.}\label{massfitfig}
\end{figure}
\newpage
\begin{figure}[H]
\centering
\vskip -20pt
\includegraphics[width=0.70\textwidth]{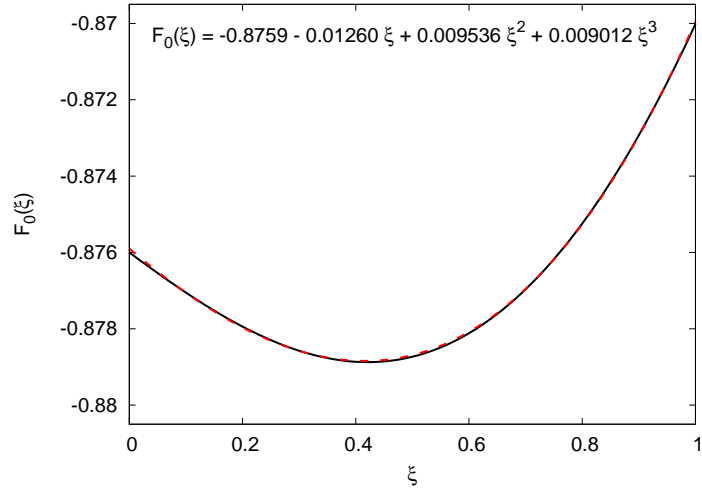}
\vskip -20pt
\caption{Optimal additive renormalization constant $F_{0}(\xi)$ as a function of $\xi$ computed at $F_{0}^{L}=-0.876$. Dashed red line: polynomial interpolation given by eq.\eqref{polyf0}.}\label{f0fitfig}
\end{figure}
\
\\
\\
The gauge-dependent renormalization constant $F_{0}(\xi)$ computed at $F_{0}^{L}=-0.876$ is shown in Fig.\ref{f0fitfig} as a function of $\xi$. The numerical data -- presented in Appendix B -- can be polynomially interpolated to yield the following approximate expression for $F_{0}(\xi)$:\\
\BE\label{polyf0}
F_{0}(\xi)\approx -0.8759-0.01260\,\xi+0.009536\,\xi^{2}+0.009012\,\xi^{3}
\EE
\\
Again, this approximation is very precise up to and beyond the Feynman gauge ($\xi=1$).\\
\\
In the next section we will use the values and functions presented above to investigate the behavior of the optimized gluon propagator in different gauges. Our results will be shown to be in good agreement with the available lattice data for the Euclidean gluon propagator in gauges different from the Landau gauge.

\newpage

\addcontentsline{toc}{section}{3.3 Gauge-optimized one-loop gluon and ghost propagators in an arbitrary covariant gauge}  \markboth{3.3 Gauge-optimized one-loop gluon and ghost propagators in an arbitrary covariant gauge}{3.3 Gauge-optimized one-loop gluon and ghost propagators in an arbitrary covariant gauge}
\section*{3.3 Gauge-optimized one-loop gluon and ghost propagators in an arbitrary covariant gauge\index{Gauge-optimized one-loop gluon and ghost propagators in an arbitrary covariant gauge}}
\addcontentsline{toc}{subsection}{3.3.1 Gluon propagator}  \markboth{3.3.1 Gluon propagator}{3.3.1 Gluon propagator}
\subsection*{3.3.1 Gluon propagator\index{Gluon propagator}}

\begin{figure}[H]
\centering
\vskip -20pt
\includegraphics[width=0.70\textwidth]{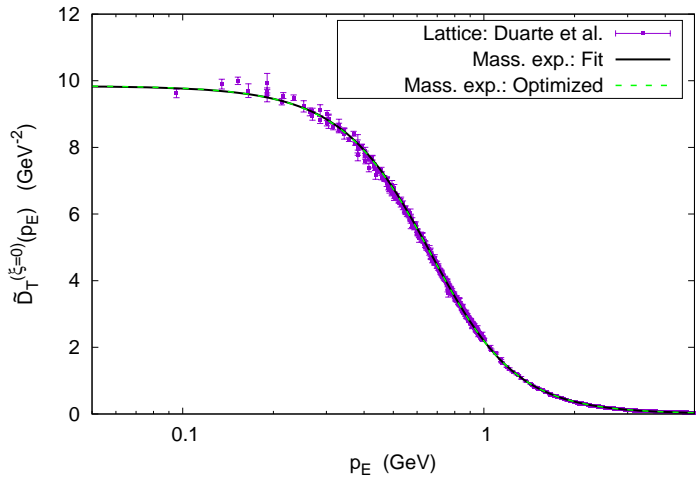}
\vskip -20pt
\caption{Gluon propagator in the Landau gauge. Black solid line: massive expansion with $F_{0}$ fitted from the lattice data. Green dashed line: massive expansion with $F_{0}$ optimized by gauge invariance ($F_{0}=-0.876$). Data points: lattice propagator from ref. \cite{duarte}.}.\label{proplandopt}
\end{figure}

\begin{figure}[H]
\centering
\vskip -20pt
\includegraphics[width=0.70\textwidth]{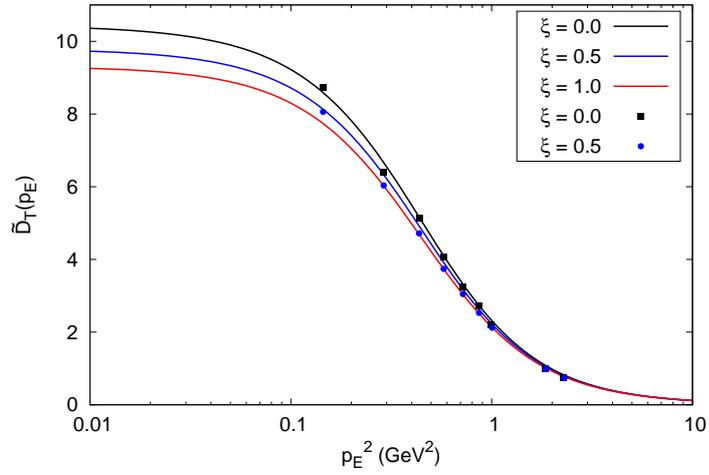}
\vskip -20pt
\caption{Euclidean gluon propagator in different gauges, renormalized at $\mu=4.317$ GeV. Data points from ref.\cite{bicudo} (Bicudo et al.).}.\label{propgauge}
\end{figure}

\begin{figure}[H]
\centering
\vskip -20pt
\includegraphics[width=0.70\textwidth]{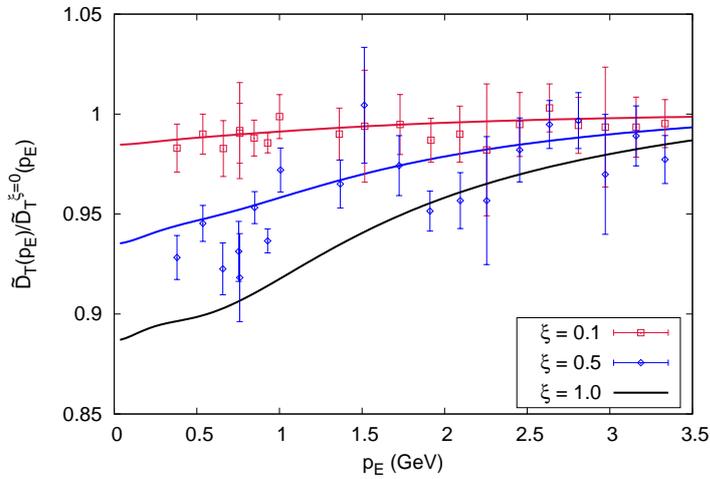}
\vskip -20pt
\caption{Ratio of the Euclidean gluon propagator in different gauges to the propagator in the Landau gauge, renormalized at $\mu=4.317$ GeV. Data points from ref.\cite{bicudo} (Bicudo et al.).}.\label{propgaugerat}
\end{figure}

Our analysis of the gauge invariance of the poles of the gluon propagator equipped us with precise values for the position and residues of the poles and with accurate approximations for the optimal gauge-dependent functions $F_{0}(\xi)$ and $m^{2}(\xi)$. It is now time to put to use our results and investigate the behavior of the optimized gluon propagator as a function of both the momentum and the gauge.\\
\\
Let us start from the principal part of the propagator. Recall that the principal part of an analytic function $f$ is the part of its series expansion that contains its poles. If $f$ has poles at $\{z_{k}\}_{k}$ with residues $\{\mc{R}_{k}\}_{k}$, then its principal part $f|_{\tx{PP}}$ is defined as\\
\BE
f(z)\Big|_{\tx{PP}}=\sum_{k}\ \frac{\mc{R}_{k}}{z-z_{k}}
\EE
\\
Due to the Nielsen identities, the principal part of the gluon propagator, namely\\
\BE
\widetilde{\mc{D}}_{T}(p^{2})\Big|_{\tx{PP}}=\frac{-i|\mc{R}|\,e^{i\varphi}}{p^{2}-p_{0}^{2}}+\frac{-i|\mc{R}|\,e^{-i\varphi}}{p^{2}-\overline{p_{0}^{2}}}
\EE
\\
is constrained to be gauge-invariant modulo an arbitrary gauge-dependent multiplicative factor. The invariance of $\widetilde{\mc{D}}_{T}|_{\tx{PP}}$ suggests that the latter may play an important role in the definition of physically meaningful quantities; therefore, it is worth to study its behavior and contrast it with that of the full propagator. Since the gluon propagator has a branch cut at positive $p^{2}$, in order to investigate the momentum-dependence of its principal part we switch to Euclidean space. Here $\widetilde{D}_{T}|_{\tx{PP}}$ reads\\
\begin{figure}[H]
\centering
\vskip -20pt
\includegraphics[width=0.70\textwidth]{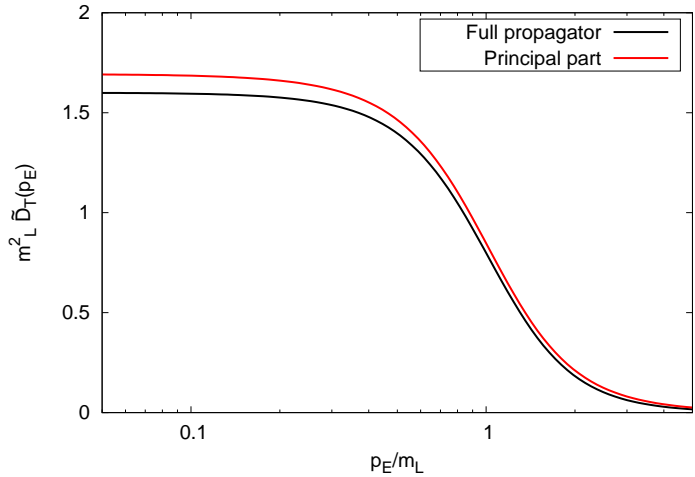}
\vskip -20pt
\caption{Optimized gluon propagator in the Landau gauge ($F_{0}=-0.876$). Black line: full propagator. Red line: principal part of the propagator, with poles and residues given by Tab.\ref{paramtab}. $|\mc{R}|=0.947$.}.\label{glupropprinc}
\end{figure}
\newpage
\BE\label{princpeu}
\widetilde{D}_{T}(p_{E}^{2})\Big|_{\tx{PP}}=\frac{|\mc{R}|\,e^{i\varphi}}{p_{E}^{2}+p_{0}^{2}}+\frac{|\mc{R}|\,e^{-i\varphi}}{p_{E}^{2}+\overline{p_{0}^{2}}}
\EE
\\
Of course, $|\mc{R}|$ depends on the value of the multiplicative renormalization constant $Z_{\mc{D}}$. A numerical evaluation shows that for $Z_{\mc{D}}=1$ the modulus of the residue is $|\mc{R}|=0.947$. In Fig.\ref{glupropprinc} we show the optimized gluon propagator in the Landau gauge together with its gauge-invariant principal part $\widetilde{D}_{T}|_{\tx{PP}}$, both computed at $Z_{\mc{D}}=1$. As we can see, at low momenta the full propagator is slightly suppressed with respect to its principal part. Nonetheless, $\widetilde{D}_{T}|_{\tx{PP}}$ still makes up for the largest part of the propagator. Actually, as shown in Fig.\ref{glupropprinc2}, most of the difference between $\widetilde{D}_{T}|_{\tx{PP}}$ and the full propagator can be absorbed into the normalization of the former, to the extent that in practical applications one may as well replace the propagator by a normalized version of its principal part.\\
By adding up the contributions due to the two poles in eq.\eqref{princpeu}, we find that $\widetilde{D}_{T}|_{\tx{PP}}$ can be put in the Gribov-Zwanziger form \cite{dudal2}\\
\BE
\widetilde{D}_{T}(p_{E}^{2})\Big|_{\tx{PP}}=Z_{\tx{GZ}}\ \frac{p_{E}^{2} +M_{1}^{2}}{p_{E}^{4}+M_{2}^{2}\,p_{E}^{2}+M_{3}^{4}}\qquad\quad Z_{\tx{GZ}}=2\Bbb{R}\tx{e}\{\mc{R}\}
\EE
\\
where, with
\BE
p_{0}=M+i\gamma\qquad\qquad t=\Bbb{I}\tx{m}\{\mc{R}\}/\Bbb{R}\tx{e}\{\mc{R}\}
\EE
\\
the mass scales $M_{1}$, $M_{2}$ and $M_{3}$ are defined as\\
\BE
M_{1}^{2}=M^{2}-\gamma^{2}+2M\gamma t\qquad M_{2}^{2}=2\,(M^{2}-\gamma^{2})\qquad\qquad M_{3}^{4}=(M^{2}+\gamma^{2})^{2}
\EE
\\
In Tab.\ref{gztab} we report the optimized values of the three scales. Observe that $t=\tan\varphi$, so that with $\varphi=1.262$ we have $t=3.132$.\\
\\
\\
\begin{figure}[H]
\centering
\vskip -20pt
\includegraphics[width=0.70\textwidth]{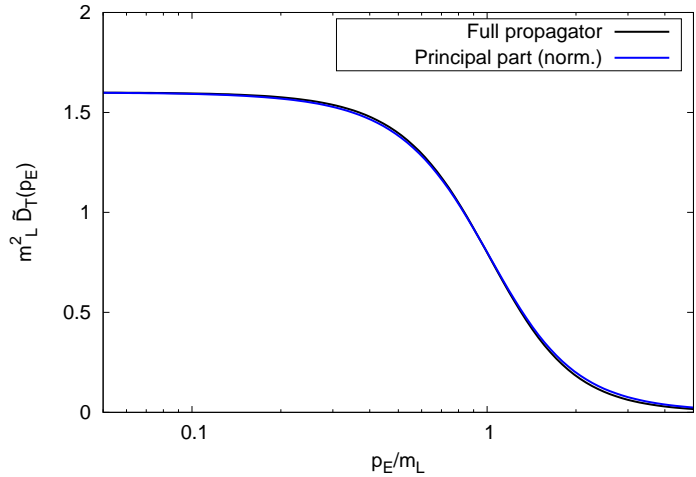}
\vskip -20pt
\caption{Same plot as Fig.\ref{glupropprinc}, with the principal part normalized by a factor of $0.945$.}.\label{glupropprinc2}
\end{figure}
\newpage
\
\begin{table}[H]
\def\arraystretch{1.4}
\centering
\begin{tabular}{ccc}
\hline
\hline
$M_{1}^{2}/m_{L}^{2}=3.630$&$M_{2}^{2}/m_{L}^{2}=0.915$&$M_{3}^{4}/m_{L}^{4}=1.235$\\
\hline
\hline
\end{tabular}
\\
\caption{Gribov-Zwanziger parameters for the principal part of the gluon propagator.}\label{gztab}
\end{table}
\
\\
The next thing we want to do is confront our results for the optimized gluon propagator with the predictions of the lattice. In Chapter 2 we saw that the Euclidean gluon propagator computed in the massive expansion well reproduces the lattice data in the Landau gauge provided that the parameters given in Tab.\ref{fitdata} (Sec.2.3.2) are used for its definition. Now, the value of $F_{0}$ which was obtained from the fit of the lattice data, namely $F_{0}=-0.887$, falls within $1\%$ from the optimized value obtained by enforcing the Nielsen identities, $F_{0}=-0.876$. Therefore we expect that by fixing $F_{0}=-0.876$ and fitting the other free parameters of the propagator -- $Z_{\mc{D}}$ and $m^{2}$ -- to the lattice data, we will obtain a curve which is in good agreement with the lattice results. This is indeed the case, as we show in Fig.\ref{proplandopt} by displaying the lattice data of ref.\cite{duarte} together with the results of massive perturbation theory both for the fitted value of $F_{0}$ and for its gauge-optimized value. The value of $m$ which is found to best fit the data at $F_{0}=-0.876$ is $m=0.656$ GeV -- see Tab.\ref{fitdataopt}. This is to be compared with the value obtained by freely fitting $F_{0}$ to the lattice data, namely $m=0.654$ GeV (Tab.\ref{fitdata}), which is just 0.3\% lower. Since the optimized parameters are extremely close to the fitted parameters, the optimized propagator is nearly undistinguishable from the fitted propagator. This major result validates the method of optimization by gauge invariance and illustrates the extent to which the massive expansion is able to make predictions from first principles about the infrared behavior of Yang-Mills theory. In Tab.\ref{paramtabfit} we report the dimensionful position of the optimized gluon poles obtained by using $m=0.656$ GeV as the value of the mass parameter in the Landau gauge.\\
\\
\newpage
\
\begin{table}[H]
\def\arraystretch{1.2}
\centering
\begin{tabular}{c|cc}
\hline
$F_{0}$&$m$ (GeV)&$Z_{\mc{D}}$\\
\hline
\hline
$-0.876$&$0.6557$&$2.6841$\\
\hline
\end{tabular}
\\
\caption{Parameters found from the fit of the lattice data of ref.\cite{duarte} to the gluon propagator of the massive expansion at fixed $F_{0}=-0.876$ in the range $0-4$ GeV.}\label{fitdataopt}
\end{table}
\
\begin{table}[H]
\def\arraystretch{1.4}
\centering
\begin{tabular}{|c|cc|}
\hline
\hline
$F_{0}^{L}=-0.876$&$p_{0}^{2}=(0.1969\pm0.4359\ i)$ GeV$^{2}$&$p_{0}=(\pm0.5810\pm0.3751\ i)$ GeV\\
\hline
\hline
\end{tabular}
\\
\caption{Position of the gluon poles for $F_{0}^{L}=-0.876$ and $m_{L}^{2}=0.656$. $m_{L}^{2}$ was fixed by fitting the optimized gluon propagator to the lattice data of ref. \cite{duarte}.}\label{paramtabfit}
\end{table}
\
\\
As for the gauge dependence of the optimized propagator, in order to test our results we will again resort to a comparison with the lattice data. It must be noted, however, that the lattice calculations for Yang-Mills theory are not yet very accurate outside the Landau gauge; therefore any comparison for $\xi\neq 0$ must be made with caution. In what follows we will employ the lattice data of ref.\cite{bicudo} (Bicudo et al.).\\
\noindent In Fig.\ref{propgauge} we show the lattice data of ref.\cite{bicudo} for the Euclidean gluon propagator in the covariant gauges $\xi=0.0,\,0.5$, together with the corresponding propagators computed in the massive expansion and optimized by gauge invariance. In addition, we also show our optimized results for the Feynman gauge ($\xi=1$) propagator, for which no lattice data is yet available. The comparison between different gauges is made by renormalizing the value of the propagator at the scale $\mu=4.317$ GeV. As we can see, our propagators follow the general trend of the lattice. As the gauge is increased at fixed, low Euclidean momenta the propagator is suppressed. At higher momenta, on the other hand, the propagator becomes less and less dependent from the gauge.\\
In Fig.\ref{propgaugerat} we show the lattice data of ref.\cite{bicudo} for the ratio of the Euclidean gluon propagator in the gauges $\xi=0.1,\,0.5$ to the propagator in the Landau gauge, together with the predictions of the optimized massive perturbation theory up to the Feynman gauge ($\xi=1$). The large errors to which the lattice calculations are still subject in gauges different from the Landau gauge are clearly displayed in the figure. Nonetheless, again, our predictions can be seen to follow the general trend of the lattice.
\newpage
\
\
\\
\\
\newpage

\addcontentsline{toc}{subsection}{3.3.2 Ghost propagator}  \markboth{3.3.2 Ghost propagator}{3.3.2 Ghost propagator}
\subsection*{3.3.2 Ghost propagator\index{Ghost propagator}}

\begin{figure}[H]
\centering
\vskip -20pt
\includegraphics[width=0.70\textwidth]{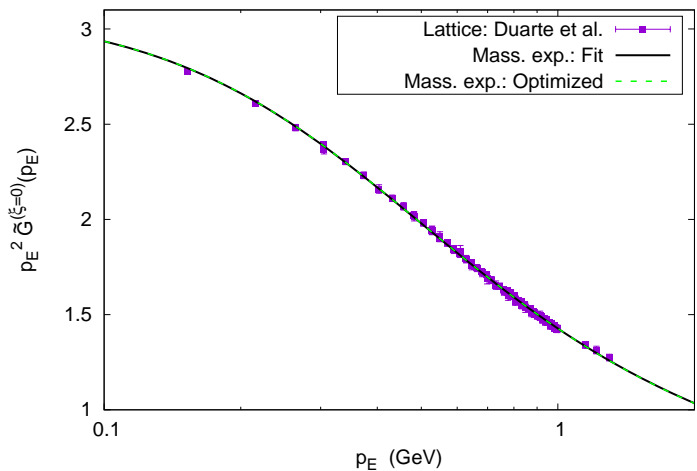}
\vskip -20pt
\caption{Ghost dressing function in the Landau gauge. Black solid line: massive expansion with $G_{0}$ and $m^{2}$ fitted from the lattice data. Green dashed line: massive expansion with $G_{0}$ optimized in the SMOM scheme by the principle of minimal sensitivity and $m^{2}$ optimized by gauge-invariance. Data points: lattice dressing function from ref. \cite{duarte}.}\label{ghproplandopt}
\end{figure}

In this section we will discuss a renormalization-scheme-dependent method for fixing the value of the ghost additive renormalization constant $G_{0}$ starting from the value of the gluon renormalization constant $F_{0}$. For simplicity, we will limit our discussion to the Landau gauge ($\xi=0$); our results can be straightforwardly generalized to an arbitrary covariant gauge. The contents of this section have been recently presented in ref. \cite{sir4}.\\
\\
In Sec.2.2.1 and 2.3.1 we saw that in the massive expansion the Landau gauge ghost and transverse gluon dressed propagators can be expressed as -- cf. eqq.\eqref{ghdrpr0} and \eqref{gluprfin0}~--\\
\BE
\widetilde{\mc{G}}(p^{2})=\frac{i}{p^{2}\,[1+\alpha(G(s)+g_{0})]}
\EE
\BE
\widetilde{\mc{D}}_{T}(p^{2})=\frac{-i}{p^{2}[1+\alpha (F(s)+f_{0})]}
\EE
\\
where $\alpha=3N\alpha_{s}/4\pi$ and $g_{0}$ and $f_{0}$ are renormalization-scheme-dependent additive renormalization constants related to the constants $G_{0}$ and $F_{0}$ by the equations\\
\BE\label{constrel}
G_{0}=\frac{1}{\alpha}+g_{0}\qquad\qquad\qquad F_{0}=\frac{1}{\alpha}+f_{0}
\EE
\\
In an arbitrary renormalization scheme, $g_{0}$ and $f_{0}$ are defined by the values $\widetilde{\mc{G}}(-\mu^{2})$ and $\widetilde{\mc{D}}_{T}(-\mu^{2})$ of the propagators at the renormalization scale $p^{2}=-\mu^{2}$:\\
\begin{align}
g_{0}(\mu)&=\frac{1}{\alpha(\mu)}\,\left[\big(i\mu^{2}\,\widetilde{\mc{G}}(-\mu^{2})\big)^{-1}-1\right]-G(\mu^{2}/m^{2})\label{g0mu}\\
f_{0}(\mu)&=\frac{1}{\alpha(\mu)}\,\left[\big(-i\mu^{2}\,\widetilde{\mc{D}}_{T}(-\mu^{2})\big)^{-1}-1\right]-F(\mu^{2}/m^{2})\label{f0mu}
\end{align}
\\
where $\alpha(\mu)$ is the coupling at the scale $\mu$ in the chosen renormalization scheme. From the above equations it follows that, as soon as we fix the renormalization conditions for the propagators -- i.e. the values of $\widetilde{\mc{G}}(-\mu^{2})$ and $\widetilde{\mc{D}}_{T}(-\mu^{2})$ --, $g_{0}(\mu)$ and $f_{0}(\mu)$ are not independent from one another, but rather are related to each other by the value of the renormalized coupling. Indeed, eq.~\eqref{f0mu} allows us to express the inverse coupling as\\
\BE\label{couplla}
\frac{1}{\alpha(\mu)}=\left[\big(-i\mu^{2}\,\widetilde{\mc{D}}_{T}(-\mu^{2})\big)^{-1}-1\right]^{-1}\left(F(\mu^{2}/m^{2})+f_{0}(\mu)\right)
\EE\\
which can be plugged back into eq.~\eqref{g0mu} to yield\\
\BE\label{constla}
g_{0}(\mu)=\frac{\big(i\mu^{2}\,\widetilde{\mc{G}}(-\mu^{2})\big)^{-1}-1}{\big(-i\mu^{2}\,\widetilde{\mc{D}}_{T}(-\mu^{2})\big)^{-1}-1}\left(F(\mu^{2}/m^{2})+f_{0}(\mu)\right)-G(\mu^{2}/m^{2})
\EE
\\
In what follows, in order to fix the value of $g_{0}$ starting from the value of $f_{0}$ -- hence $G_{0}$ starting from $F_{0}$, cf. eq.~\eqref{constrel} --, we will adopt a renormalization scheme termed Screened MOMentum subtraction scheme (SMOM) \cite{sir4}. The SMOM scheme is defined by setting\\
\BE
\widetilde{\mc{G}}(-\mu^{2})=\frac{i}{-\mu^{2}}\qquad\qquad\qquad\widetilde{\mc{D}}_{T}(-\mu^{2})=\frac{-i}{-\mu^{2}-m^{2}}
\EE
where $m$ is the gluon mass parameter, so that at the scale $\mu$ the ghost and gluon propagators are given respectively the values of a bare massless propagator and a bare massive propagator. The mass parameter in $\widetilde{\mc{D}}_{T}(-\mu^{2})$ prevents the latter from diverging at low renormalization scales, hence the name of the scheme. In the SMOM scheme we have\\
\BE\label{smoms}
[i\mu^{2}\,\widetilde{\mc{G}}(-\mu^{2})]^{-1}=1\qquad\qquad\qquad [-i\mu^{2}\,\widetilde{\mc{D}}_{T}(-\mu^{2})]^{-1}=1+\frac{m^{2}}{\mu^{2}}
\EE
\\
It follows from eqq.~\eqref{couplla}, \eqref{constla} and \eqref{smoms} that the SMOM coupling constant can be expressed in terms of $F(\mu^{2}/m^{2})$ and $f_{0}(\mu)$ as\\
\BE
\alpha_{\tx{SMOM}}(\mu)=\frac{m^{2}}{\mu^{2}}\ \left(F(\mu^{2}/m^{2})+f_{0}(\mu)\right)^{-1}
\EE
\\
whereas $g_{0}(\mu)$ is related to the value of the gluon function $G(s)$ at the renormalization scale $\mu$ by the simple relation
\BE
g_{0}(\mu)=-G(\mu^{2}/m^{2})
\EE
\\
In terms of the renormalization constants $G_{0}$ and $F_{0}$ at the scale $\mu$ -- cf. eq.~\eqref{constrel} --, the above equations read\\
\BE\label{asmom}
\alpha_{\tx{SMOM}}(\mu)=\left(1+\frac{m^{2}}{\mu^{2}}\right)\,[F(\mu^{2}/m^{2})+F_{0}(\mu)]^{-1}
\EE
\BE\label{g0smom}
G_{0}(\mu)=\left(1+\frac{m^{2}}{\mu^{2}}\right)^{-1}\,[F(\mu^{2}/m^{2})+F_{0}(\mu)]-G(\mu^{2}/m^{2})
\EE
\\
Eqq.~\eqref{asmom} and \eqref{g0smom} completely fix the value of the SMOM coupling $\alpha_{\tx{SMOM}}$ and ghost renormalization constant $G_{0}$ at the scale $\mu$ starting from the knowledge of the ghost renormalization constant $F_{0}$ at the scale $\mu$.\\
\\
In this chapter the value of $F_{0}$ in the Landau gauge was optimized by the requirement of the gauge invariance of the position and phases of the residues of the poles of the gluon propagator. Our derivation did not assume any specific renormalization condition for the gluon propagator and thus provided us with a renormalization-scheme-independent value for $F_{0}$, namely, $F_{0}=-0.876$. Now, from eq.~\eqref{g0smom} it is clear that if $F_{0}$ does not depend on the renormalization scale, then the constant $G_{0}$ does. Therefore, in the framework of the gauge-optimized massive perturbation theory renormalized in the SMOM scheme, the ghost constant $G_{0}$ is predicted to be dependent on the renormalization scale.\\
In ref. \cite{sir4} it was shown that a renormalization-scale-dependent $G_{0}$, in general, spoils the multiplicative renormalizability of the ghost propagator. This apparent drawback of the gauge-optimized SMOM framework was exploited to fix the value of $G_{0}$ according to Stevenson's principle of minimal sensitivity \cite{stev6}. The principle of minimal sensitivity states that the best approximation to a Green function renormalized in a momentum-subtraction-like scheme in which some parameter is required to be independent from the renormalization scale $\mu$ is obtained by choosing the renormalization scale in such a way that the aforementioned parameter is less sensitive to a variation of $\mu$. In other words, the renormalization scale should be chosen so that the parameter is stationary with respect to $\mu$. In particular, in the gauge-optimized SMOM framework, the principle of minimal sensitivity requires that we renormalize the propagators at the scale $\mu=\mu^{\star}$ such that $G_{0}'(\mu^{\star})=0$, where the prime denotes a derivative with respect to $\mu$.
\newpage
\begin{figure}[H]
\centering
\vskip -20pt
\includegraphics[width=0.70\textwidth]{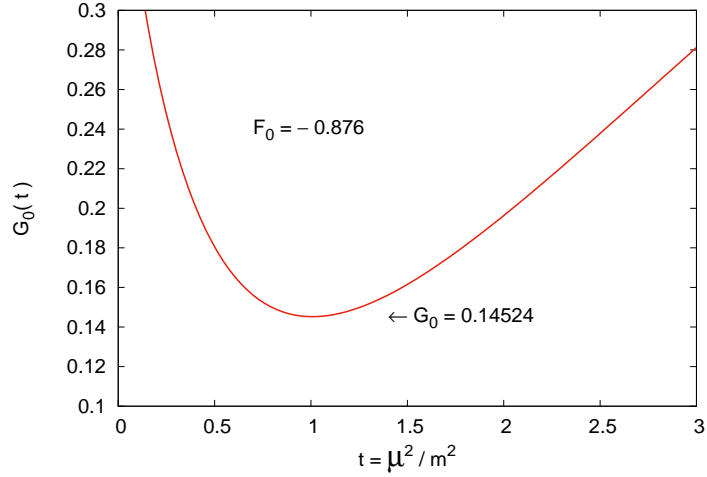}
\vskip -20pt
\caption{SMOM function $G_{0}(\mu)$ as a function of the adimensionalized renormalization scale squared $\mu^{2}/m^{2}$ for the optimal value $F_{0}=-0.876$. The minimum $G_{0}(\mu^{\star})=0.14524$ is found at the renormalization scale $\mu^{\star}=1.004\,m$.}\label{g0smomfig}
\end{figure}
\
\\
\\
\noindent In Fig.\ref{g0smomfig} we display the SMOM function $G_{0}(\mu)$ -- as defined by eq.~\eqref{g0smom} -- computed at fixed $F_{0}=-0.876$. $G_{0}(\mu)$ can be seen to have a pronounced minimum at a renormalization scale $\mu^{\star}\approx m$. A numerical evaluation shows that $\mu^{\star}=1.004\,m$ and $G_{0}(\mu^{\star})=0.1452$ \cite{sir4}. Observe that the optimal renormalization scale $\mu^{\star}$ derived by the principle of minimal sensitivity is remarkably close to the value of the gluon mass parameter. Moreover, the optimal value $G_{0}=G_{0}(\mu^{\star})$ is within 0.8\% from the value $G_{0}=0.1464$ which was found to best fit the lattice data in Chapter 2 (see Tab.\ref{fitdatagh} in Sec.2.2.2). Since the optimized mass $m$ was also found to be very close to the fitted mass -- cf. Sec.3.3.1 --, we expect the optimized ghost propagator to accurately reproduce the lattice data. This is confirmed by Fig.\ref{ghproplandopt}, where we plot the Euclidean Landau gauge ghost dressing function $p_{E}^{2}\widetilde{G}(p_{E})$ obtained by setting $G_{0}=G_{0}(\mu^{\star})=0.1452$ and $m=0.656$ GeV -- cf. Tab.\ref{fitdataghsmom} -- together with the lattice results of ref. \cite{duarte} and the fitted dressing function. As we can see, the fitted dressing function and the optimized dressing function are almost indistinguishable from one another. Therefore a comparison with the lattice data validates the method of optimization by minimal sensitivity.\\
\\
\\
\begin{table}[H]
\def\arraystretch{1.2}
\centering
\begin{tabular}{cc|c}
\hline
$m$ (GeV)&$G_{0}$&$Z_{\mc{G}}$\\
\hline
\hline
$0.656$&$0.1452$&$1.0959$\\
\hline
\end{tabular}
\vspace{3mm}
\caption{Parameter $Z_{\mc{G}}$ found from the fit of the lattice data of ref.\cite{duarte} for the Landau gauge ghost propagator in the range $0-2$ GeV at fixed $m=0.656$ GeV and $G_{0}=0.1452$.}\label{fitdataghsmom}
\end{table}
\newpage
\
\newpage
\thispagestyle{empty}
\newpage


\addcontentsline{toc}{chapter}{Conclusions}  \markboth{Conclusions}{Conclusions}

\chapter*{Conclusions\index{Conclusions}}

In this thesis we have addressed the issue of dynamical mass generation from different perspectives and presented a new perturbative framework for making computations in low-energy Yang-Mills theory, the massive perturbative expansion.\\
In Chapter 1 we performed a GEP analysis of the perturbative vacuum of Yang-Mills theory. The Yang-Mills GEP was computed at zero longitudinal gluon mass, in compliance with the non-perturbative transversality of the gluon one-particle-irreducible polarization. By analyzing the influence of the ghost contribution to the GEP, we have concluded that in order for the Gaussian Effective Potential at the gluon minimum to be a good variational estimate of the exact vacuum energy density of Yang-Mills theory the mass of the ghosts must be set to zero. By solving the gap equation of the GEP we have discovered that the massless perturbative vacuum of the transverse gluons is unstable towards a massive vacuum, giving evidence for the occurrence of dynamical mass generation and indicating that a non-standard perturbation theory that treats the transverse gluons as massive already at tree-level could be more suitable for making calculations in low-energy Yang-Mills theory. In such a framework, both the ghosts and the longitudinal gluons are to be treated as massless.\\
In Chapter 2 we formulated the massive perturbative framework and used it to compute the ghost and gluon dressed propagators to one loop. Massive perturbation theory was defined by a shift of the expansion point of the Yang-Mills perturbative series, achieved by adding a mass term for the transverse gluons to the kinetic Yang-Mills Lagrangian and subtracting the same term from the interaction Lagrangian. While leaving the total action and physical content of the theory unchanged, the shift was shown to modify the Feynman rules of the diagrammatic expansion by replacing the massless transverse bare gluon propagator of ordinary perturbation theory with a massive propagator and giving rise to a new two-gluon vertex, proportional to the gluon mass parameter squared $m^{2}$. The one-loop ghost and gluon propagators computed in the massive framework were shown to be free of spurious, non-renormalizable mass divergences and to have the correct UV behavior, matching the results of ordinary perturbation theory in the high-energy limit. At low energies, the gluon propagator was shown to develop a dynamical mass by way of a non-trivial mechanism that involves the gluon loops and the gluon tadpoles. The ghost propagator, on the other hand, was shown to remain massless. In the IR we compared our results with the Euclidean lattice data in the Landau gauge and found that the two agree to a high degree of accuracy. In Chapter 2 we also discussed the issue of the predictivity of the massive framework. We concluded that one of the free parameters of the massive expansion must be fixed in order to make its results as predictive as those obtained in standard perturbation theory.\\
In Chapter 3 we advanced a method for optimizing the massive expansion by the requirement of the gauge-invariance of the position and phases of the residues of the poles of the dressed gluon propagator, in compliance with the Nielsen identities. This requirement was shown to be sufficient to fix the value of the gluon additive renormalization constant in the Landau gauge to $F_{0}=-0.876$, thus reducing the number of free parameters of the gluon propagator computed in the massive expansion back to two and restoring the predictivity of the massive approach. The optimization procedure left us with an expression for the dressed gluon propagator in a general covariant gauge whose principal part -- modulo a gauge-dependent multiplicative renormalization factor -- is gauge-invariant. A comparison with the lattice data allowed us to fix the optimal value of the gluon mass parameter to $m=0.656$ GeV and the position of the complex conjugate poles of the gluon propagator to $p_{0}=(\pm0.581\pm 0.375\,i)$ GeV. The optimal value of the gluon renormalization constant $F_{0}$ was then used to determine the optimal value of the gluon additive renormalization constant $G_{0}$ in the SMOM renormalization scheme according to Stevenson's principle of minimal sensitivity. The optimal value of $G_{0}$ was found to be $G_{0}=0.145$. The optimized values of the gluon and ghost renormalization constants were shown to agree with those obtained from a fit to the lattice data to less than 1\%.\\
\\
The massive perturbative expansion is able to provide a clear picture of the infrared behavior of Yang-Mills theory from first principles and without the need of any external phenomenological parameter other than the mass scale of the theory. The main achievement of the massive framework is the prediction that at low energies the gluons acquire a dynamically generated transverse mass of the order of the QCD scale. Although this result had already been anticipated through SDE methods \cite{corn}, first proved numerically through lattice calculations \cite{cucc1,cucc2,bogol,dudal,oliv1,ayala,oliv2,burgio,duarte} and built into phenomenological massive models of QCD \cite{tissier1,tissier2,tissier3}, the massive approach to perturbation theory is the first analytical method to prove it from first principles, employing only elementary quantum-field-theoretic techniques.\\
The analytic expressions derived in the massive framework can be continued to the complex plane both to study the behavior of the propagators in the Euclidean space and to seek for complex poles which, being gauge invariant, might be directly related to physical observables \cite{dudal3}. As a matter of fact, a knowledge of the analytic structure of the gluon propagator alone is sufficient to gain significant insight on the low-energy dynamics of the gluons. For instance, the existence of two complex conjugate poles for the gluon propagator -- which is predicted by the gauge-optimized massive expansion -- was shown in \cite{com3} to result in a violation of the positivity of the gluon spectral function, a feature which in the literature has been linked to the problem of gluon confinement \cite{haya}.\\
The success of the massive perturbative expansion in accurately reproducing the lattice data in the Euclidean space gives us confidence in the general validity of the method and leads us to present it as a viable framework for doing calculations in low energy Yang-Mills theory.

\newpage
\thispagestyle{empty}
\newpage


\newcounter{count2}
\setcounter{count2}{1}
\setcounter{equation}{0}
\renewcommand{\theequation}{\Alph{count2}.\arabic{equation}}
\addcontentsline{toc}{chapter}{Appendix}  \markboth{Appendix}{Appendix}

\chapter*{Appendix\index{Appendix}}

\addcontentsline{toc}{section}{A. Residues of the gluon propagator: the functions $F'(s)$ and $F_{\xi}'(s)$}  \markboth{A. Residues of the gluon propagator: the functions $F'(s)$ and $F_{\xi}'(s)$}{A. Residues of the gluon propagator: the functions $F'(s)$ and $F_{\xi}'(s)$}
\section*{A. Residues of the gluon propagator: the functions $\boldsymbol{F'(s)}$ and $\boldsymbol{F_{\xi}'(s)}$\index{Residues of the gluon propagator: the functions $F'(s)$ and $F_{\xi}'(s)$}}
\
\\
In the framework of massive perturbation theory in order to compute the residues of the gluon propagator at its poles one needs to know the derivatives of the functions $F(s)$ and $F_{\xi}(s)$ which we have defined in Sec.2.3.1. These are given by\\
\begin{align}
F'(s)&=-\frac{5}{8s^{2}}+\frac{1}{72}\,[L'_{a}(s)+L'_{a}(s)+L'_{a}(s)+R'(s)]\\
F'_{\xi}(s)&=\frac{s^{4}+6s-3}{6s^{4}}\,\ln(1+s)-\frac{1}{6}\,\ln s+\frac{(1-s)(1-s^{3})}{6s^{3}(1+s)}+\frac{1}{3s^{3}}-\frac{1}{2s^{2}}-\frac{1}{6}
\end{align}
\\
where
\begin{align}
L_{a}'(s)&=\frac{6s^{4}-16s^{3}-68s^{2}+80s+144}{s^{2}(s+4)}\,\sqrt{\frac{s+4}{s}}\ \ln\left(\frac{\sqrt{s+4}-\sqrt{s}}{\sqrt{s+4}+\sqrt{s}}\right)\\
L_{b}'(s)&=\frac{4(1+s)}{s^{4}}\,(3s^{4}-10s^{3}+10s^{2}-10s+3)\,\ln(1+s)\\
L_{c}'(s)&=-6s\,\ln s\\
R'(s)&=\frac{12}{s}+\frac{106}{s^{2}}-\frac{12}{s^{3}}
\end{align}
\\
In terms of $F'(s)$ and $F_{\xi}'(s)$, the residue $\mc{R}_{p_{0}^{2}}(\xi)$ of the gluon propagator at $p_{0}^{2}$ in the gauge $\xi$ can be expressed as -- cf. Sec.3.1.2 --\\
\BE
\mc{R}_{p_{0}^{2}}(\xi)=\frac{m^{2}(\xi)}{p_{0}^{2}}\,\frac{iZ_{\mc{D}}}{F'(-p_{0}^{2}/m^{2}(\xi))+\xi\, F'_{\xi}(-p_{0}^{2}/m^{2}(\xi))}
\EE
\\
where $m^{2}(\xi)$ is the gauge-dependent mass parameter in the gauge $\xi$.
\newpage
\stepcounter{count2}
\setcounter{equation}{0}
\addcontentsline{toc}{section}{B. Gauge-optimized functions $F_{0}(\xi)$ and $m^{2}(\xi)$ for the gluon propagator: numerical data}  \markboth{B. Gauge-optimized functions $F_{0}(\xi)$ and $m^{2}(\xi)$ for the gluon propagator: numerical data}{B. Gauge-optimized functions $F_{0}(\xi)$ and $m^{2}(\xi)$ for the gluon propagator: numerical data}
\section*{B. Gauge-optimized functions $\boldsymbol{F_{0}(\xi)}$ and $\boldsymbol{m^{2}(\xi)}$ for the gluon propagator: numerical data\index{Gauge-optimized functions $F_{0}(\xi)$ and $m^{2}(\xi)$ for the gluon propagator: numerical data}}
\
\\
In this Appendix we report the numerical values of the gauge-dependent functions $F_{0}(\xi)$ and $m^{2}(\xi)$, as obtained by the optimization procedure described in Sec.3.1.2 for the value $F_{0}^{L}=-0.876$. The third column in the table contains the phase difference $\theta(\xi)$ between the residue at the pole in the gauge $\xi$ and that in the Landau gauge ($\xi=0$).\\
\\
\begin{table}[H]
\def\arraystretch{1.2}
\centering
\begin{tabular}{c|ccc}
\hline
\hline
$\xi$&$m^{2}(\xi)/m^{2}_{L}$&$F_{0}(\xi)$&$\theta(\xi)$\\
\hline
\hline
$0.00$&$1.00000$&$-0.87600$&$+0.00000$\\
$0.01$&$0.99800$&$-0.87611$&$-0.00023$\\
$0.02$&$0.99600$&$-0.87622$&$-0.00045$\\
$0.03$&$0.99400$&$-0.87633$&$-0.00066$\\
$0.04$&$0.99200$&$-0.87643$&$-0.00086$\\
$0.05$&$0.99001$&$-0.87654$&$-0.00105$\\
$0.06$&$0.98803$&$-0.87665$&$-0.00123$\\
$0.07$&$0.98604$&$-0.87675$&$-0.00140$\\
$0.08$&$0.98406$&$-0.87685$&$-0.00155$\\
$0.09$&$0.98208$&$-0.87695$&$-0.00170$\\
$0.10$&$0.98010$&$-0.87705$&$-0.00183$\\
$0.11$&$0.97813$&$-0.87715$&$-0.00196$\\
$0.12$&$0.97616$&$-0.87725$&$-0.00208$\\
$0.13$&$0.97419$&$-0.87734$&$-0.00218$\\
$0.14$&$0.97223$&$-0.87743$&$-0.00228$\\
$0.15$&$0.97027$&$-0.87752$&$-0.00237$\\
$0.16$&$0.96831$&$-0.87761$&$-0.00244$\\
$0.17$&$0.96635$&$-0.87770$&$-0.00251$\\
$0.18$&$0.96440$&$-0.87778$&$-0.00257$\\
$0.19$&$0.96245$&$-0.87786$&$-0.00262$\\
$0.20$&$0.96050$&$-0.87794$&$-0.00267$\\
$0.21$&$0.95856$&$-0.87802$&$-0.00270$\\
$0.22$&$0.95662$&$-0.87809$&$-0.00272$\\
$0.23$&$0.95468$&$-0.87816$&$-0.00274$\\
$0.24$&$0.95275$&$-0.87823$&$-0.00275$\\
$0.25$&$0.95082$&$-0.87830$&$-0.00275$\\
$0.26$&$0.94889$&$-0.87836$&$-0.00275$\\
$0.27$&$0.94696$&$-0.87842$&$-0.00274$\\
$0.28$&$0.94504$&$-0.87847$&$-0.00272$\\
$0.29$&$0.94312$&$-0.87853$&$-0.00269$
\end{tabular}
\end{table}
\newpage
\
\\
\begin{table}[H]
\def\arraystretch{1.2}
\centering
\begin{tabular}{c|ccc}
\hline
\hline
$\xi$&$m^{2}(\xi)/m^{2}_{L}$&$F_{0}(\xi)$&$\theta(\xi)$\\
\hline
$0.30$&$0.94120$&$-0.87858$&$-0.00266$\\
$0.31$&$0.93928$&$-0.87862$&$-0.00262$\\
$0.32$&$0.93737$&$-0.87866$&$-0.00257$\\
$0.33$&$0.93546$&$-0.87870$&$-0.00252$\\
$0.34$&$0.93356$&$-0.87874$&$-0.00246$\\
$0.35$&$0.93166$&$-0.87877$&$-0.00240$\\
$0.36$&$0.92976$&$-0.87880$&$-0.00233$\\
$0.37$&$0.92786$&$-0.87882$&$-0.00226$\\
$0.38$&$0.92597$&$-0.87884$&$-0.00218$\\
$0.39$&$0.92408$&$-0.87885$&$-0.00209$\\
$0.40$&$0.92219$&$-0.87886$&$-0.00200$\\
$0.41$&$0.92030$&$-0.87887$&$-0.00191$\\
$0.42$&$0.91842$&$-0.87887$&$-0.00182$\\
$0.43$&$0.91654$&$-0.87887$&$-0.00172$\\
$0.44$&$0.91467$&$-0.87887$&$-0.00161$\\
$0.45$&$0.91279$&$-0.87886$&$-0.00151$\\
$0.46$&$0.91092$&$-0.87884$&$-0.00140$\\
$0.47$&$0.90906$&$-0.87882$&$-0.00128$\\
$0.48$&$0.90719$&$-0.87880$&$-0.00117$\\
$0.49$&$0.90533$&$-0.87877$&$-0.00105$\\
$0.50$&$0.90347$&$-0.87873$&$-0.00093$\\
$0.51$&$0.90162$&$-0.87869$&$-0.00081$\\
$0.52$&$0.89977$&$-0.87865$&$-0.00069$\\
$0.53$&$0.89792$&$-0.87860$&$-0.00056$\\
$0.54$&$0.89607$&$-0.87855$&$-0.00043$\\
$0.55$&$0.89423$&$-0.87849$&$-0.00031$\\
$0.56$&$0.89239$&$-0.87842$&$-0.00018$\\
$0.57$&$0.89055$&$-0.87836$&$-0.00005$\\
$0.58$&$0.88872$&$-0.87828$&$+0.00007$\\
$0.59$&$0.88689$&$-0.87820$&$+0.00020$\\
$0.60$&$0.88506$&$-0.87812$&$+0.00033$\\
$0.61$&$0.88324$&$-0.87803$&$+0.00045$\\
$0.62$&$0.88141$&$-0.87793$&$+0.00058$\\
$0.63$&$0.87960$&$-0.87783$&$+0.00070$\\
$0.64$&$0.87778$&$-0.87772$&$+0.00083$
\end{tabular}
\end{table}
\newpage
\
\\
\begin{table}[H]
\def\arraystretch{1.2}
\centering
\begin{tabular}{c|ccc}
\hline
\hline
$\xi$&$m^{2}(\xi)/m^{2}_{L}$&$F_{0}(\xi)$&$\theta(\xi)$\\
\hline
$0.65$&$0.87597$&$-0.87761$&$+0.00095$\\
$0.66$&$0.87416$&$-0.87749$&$+0.00107$\\
$0.67$&$0.87235$&$-0.87737$&$+0.00118$\\
$0.68$&$0.87055$&$-0.87724$&$+0.00130$\\
$0.69$&$0.86875$&$-0.87711$&$+0.00141$\\
$0.70$&$0.86696$&$-0.87697$&$+0.00152$\\
$0.71$&$0.86516$&$-0.87682$&$+0.00162$\\
$0.72$&$0.86337$&$-0.87667$&$+0.00173$\\
$0.73$&$0.86158$&$-0.87651$&$+0.00183$\\
$0.74$&$0.85980$&$-0.87635$&$+0.00192$\\
$0.75$&$0.85802$&$-0.87618$&$+0.00201$\\
$0.76$&$0.85624$&$-0.87600$&$+0.00210$\\
$0.77$&$0.85447$&$-0.87582$&$+0.00218$\\
$0.78$&$0.85270$&$-0.87564$&$+0.00226$\\
$0.79$&$0.85093$&$-0.87544$&$+0.00233$\\
$0.80$&$0.84916$&$-0.87524$&$+0.00240$\\
$0.81$&$0.84740$&$-0.87504$&$+0.00246$\\
$0.82$&$0.84564$&$-0.87483$&$+0.00251$\\
$0.83$&$0.84389$&$-0.87461$&$+0.00256$\\
$0.84$&$0.84214$&$-0.87439$&$+0.00261$\\
$0.85$&$0.84039$&$-0.87416$&$+0.00264$\\
$0.86$&$0.83864$&$-0.87393$&$+0.00267$\\
$0.87$&$0.83690$&$-0.87369$&$+0.00270$\\
$0.88$&$0.83516$&$-0.87344$&$+0.00271$\\
$0.89$&$0.83343$&$-0.87319$&$+0.00272$\\
$0.90$&$0.83169$&$-0.87293$&$+0.00272$\\
$0.91$&$0.82996$&$-0.87266$&$+0.00272$\\
$0.92$&$0.82824$&$-0.87239$&$+0.00271$\\
$0.93$&$0.82651$&$-0.87212$&$+0.00268$\\
$0.94$&$0.82479$&$-0.87184$&$+0.00265$\\
$0.95$&$0.82308$&$-0.87155$&$+0.00262$\\
$0.96$&$0.82136$&$-0.87125$&$+0.00257$\\
$0.97$&$0.81965$&$-0.87095$&$+0.00251$\\
$0.98$&$0.81795$&$-0.87064$&$+0.00245$\\
$0.99$&$0.81624$&$-0.87033$&$+0.00238$
\end{tabular}
\end{table}
\newpage
\
\\
\begin{table}[H]
\def\arraystretch{1.2}
\centering
\begin{tabular}{c|ccc}
\hline
\hline
$\xi$&$m^{2}(\xi)/m^{2}_{L}$&$F_{0}(\xi)$&$\theta(\xi)$\\
\hline
$1.00$&$0.81454$&$-0.87001$&$+0.00229$\\
$1.01$&$0.81284$&$-0.86969$&$+0.00220$\\
$1.02$&$0.81115$&$-0.86936$&$+0.00210$\\
$1.03$&$0.80946$&$-0.86902$&$+0.00199$\\
$1.04$&$0.80777$&$-0.86868$&$+0.00186$\\
$1.05$&$0.80608$&$-0.86833$&$+0.00173$\\
$1.06$&$0.80440$&$-0.86798$&$+0.00159$\\
$1.07$&$0.80272$&$-0.86762$&$+0.00144$\\
$1.08$&$0.80105$&$-0.86725$&$+0.00127$\\
$1.09$&$0.79938$&$-0.86688$&$+0.00110$\\
$1.10$&$0.79771$&$-0.86650$&$+0.00091$\\
$1.11$&$0.79604$&$-0.86612$&$+0.00072$\\
$1.12$&$0.79438$&$-0.86573$&$+0.00051$\\
$1.13$&$0.79272$&$-0.86533$&$+0.00029$\\
$1.14$&$0.79106$&$-0.86493$&$+0.00006$\\
$1.15$&$0.78941$&$-0.86452$&$-0.00018$\\
$1.16$&$0.78776$&$-0.86411$&$-0.00044$\\
$1.17$&$0.78611$&$-0.86369$&$-0.00070$\\
$1.18$&$0.78447$&$-0.86327$&$-0.00098$\\
$1.19$&$0.78283$&$-0.86284$&$-0.00127$\\
$1.20$&$0.78119$&$-0.86240$&$-0.00157$
\end{tabular}
\end{table}
\newpage
\thispagestyle{empty}
\
\clearpage


\clearpage
\thispagestyle{empty}
\clearpage

\end{document}